\documentclass[peerreview,11pt]{IEEEtran}
\IEEEoverridecommandlockouts
\newif\ifPeerReviewVersion
\newif\ifTITVersion
\newif\ifDraftVersion
\PeerReviewVersiontrue
\def\hatU{\hat{U}}
\def\hatu{\hat{u}}
\def\hatX{\hat{X}}
\def\hatV{\hat{V}}
\def\hatv{\hat{v}}
\def\hatY{\hat{Y}}
\def\hatK{\hat{K}}
\def\hatk{\hat{k}}
\def\hatA{\hat{A}}
\def\hatb{\hat{b}}
\def\hatB{\hat{B}}
\def\hatsfK{\hat{\sfK}}

\def\boldS{\boldsymbol{S}}
\def\bolds{\boldsymbol{s}}
\def\bolda{\boldsymbol{a}}
\def\boldb{\boldsymbol{b}}
\def\boldX{\boldsymbol{X}}
\def\boldx{\boldsymbol{x}}
\def\boldU{\boldsymbol{U}}
\def\boldu{\boldsymbol{u}}
\def\boldY{\boldsymbol{Y}}
\def\boldy{\boldsymbol{y}}
\def\boldV{\boldsymbol{V}}
\def\boldv{\boldsymbol{v}}

\def\boldA{\boldsymbol{A}}
\def\boldB{\boldsymbol{B}}

\def\boldK{\boldsymbol{K}}
\def\boldk{\boldsymbol{k}}
\def\boldCalV{\boldsymbol{\mathcal{V}}}
\def\boldCalU{\boldsymbol{\mathcal{U}}}
\def\boldCalX{\boldsymbol{\mathcal{X}}}
\def\boldCalY{\boldsymbol{\mathcal{Y}}}
\def\boldCalS{\boldsymbol{\mathcal{S}}}
\def\boldCalK{\boldsymbol{\mathcal{K}}}

\def\boldhatK{\hat{\boldsymbol{K}}}
\def\boldhatk{\hat{\boldsymbol{k}}}
\def\boldhatu{\hat{\boldsymbol{u}}}
\def\boldhatU{\hat{\boldsymbol{U}}}
\def\boldhatS{\hat{\boldS}}
\def\boldhats{\hat{\bolds}}
\def\ulines{\underline{s}}
\def\ulineS{\underline{S}}
\def\ulineB{\underline{B}}
\def\ulineb{\underline{b}}
\def\ulinec{\underline{c}}

\def\ulineA{\underline{A}}
\def\ulineX{\underline{X}}
\def\ulinex{\underline{x}}
\def\ulineU{\underline{U}}
\def\ulineu{\underline{u}}
\def\ulineW{\underline{W}}
\def\ulinew{\underline{w}}
\def\ulineCalS{\underline{\mathcal{S}}}
\def\ulineCalX{\underline{\mathcal{X}}}

\def\ulineCalU{\underline{\mathcal{U}}}
\def\ulineCalV{\underline{\mathcal{V}}}
\def\ulineCalW{\underline{\mathcal{W}}}
\def\ulineCalY{\underline{\mathcal{Y}}}
\def\ulineK{\underline{K}}
\def\ulineZ{\underline{Z}}

\def\ulineY{\underline{Y}}
\def\uliney{\underline{y}}

\def\ulineX{\underline{X}}
\def\ulineY{\underline{Y}}
\def\ulineV{\underline{V}}
\def\ulinev{\underline{v}}
\def\ulinea{\underline{a}}

\def\ulinepi{\underline{\pi}}

\def\ulinemathscrU{\underline{\mathscr{U}}}
\def\ulinemathscrS{\underline{\mathscr{S}}}
\def\ulinemathscrV{\underline{\mathscr{V}}}

\def\ulinemathscrY{\underline{\mathscr{Y}}}

\def\ulinemathscrW{\underline{\mathscr{W}}}

\def\ulinehatA{\underline{\hat{A}}}
\def\ulinehatB{\underline{\hat{B}}}

\def\ulinehatK{\underline{\hat{K}}}

\def\ulinehatU{\underline{\hat{U}}}
\def\ulinehatu{\underline{\hat{u}}}
\def\ulinesfU{\underline{\sfU}}
\def\ulinesfV{\underline{\sfV}}
\def\ulinesfX{\underline{\sfX}}
\def\ulinesfY{\underline{\sfY}}
\def\ulinesfhatU{\underline{\sfhatU}}

\def\sfU{{U}}
\def\sfV{{V}}
\def\sfX{{X}}
\def\sfY{{Y}}
\def\sfK{{K}}
\def\sfhatU{\hat{{U}}}
\def\altu{u}
\def\altv{v}
\def\altx{x}

\def\3To1BC{$3-$to$-1$}

\def\define{:{=}~}

\def\naturals{\mathbb{N}}
\def\reals{\mathbb{R}}

\def\Bernoulli{\mbox{ Ber}}
\def\typical{\overset{\mbox{{\tiny typ}}}{\sim}}
\def\ntypical{\overset{\mbox{{\tiny typ}}}{\nsim}}

\def\mfI{\mathfrak{I}}

\def\cocl{\mbox{cocl}}

\def\Expectation{\mathbb{E}}

\newcommand{\msout}[1]{\text{\sout{\ensuremath{#1}}}}

\newif\ifProofForORDBC
\def\inpsetX{\mathcal{X}_{1}}
\def\inpsetY{\mathcal{X}_{2}}
\def\outset{\mathcal{Y}}
\def\InpX{{X}_{1}}
\def\InpY{{X}_{2}}
\def\Out{{Y}}
\def\mtimesl{m\times l}
\def\ShrdChnl{\mathbb{W}_{Y_{0}|\underline{U}}}
\def\distfn{\mathrm{d}}

\usepackage{amssymb}
\usepackage{amsmath}
\usepackage{mathrsfs}
\usepackage{ulem}
\usepackage{epsf,epsfig}
\usepackage{cite}
\usepackage{color}
\usepackage{dsfont}

\newcommand{\comment}[1]{}
\begin{document}
\sloppy
\newtheorem{remark}{\it Remark}
\newtheorem{thm}{Theorem}
\newtheorem{corollary}{Corollary}
\newtheorem{definition}{Definition}
\newtheorem{lemma}{Lemma}
\newtheorem{example}{Example}
\newtheorem{prop}{Proposition}

\title{Communicating Correlated Sources over MAC and Interference Channels I : 
Separation-based schemes}
\author{\IEEEauthorblockN{Arun Padakandla}\\
\IEEEauthorblockA{Center for Science of Information\\Purdue University}
\thanks{This work was supported by the Center for Science of Information (CSoI), 
an NSF Science and Technology Center, under grant agreement CCF-0939370. This 
work was presented in part at the IEEE International Symposium on Information 
Theory held in Barcelona, Spain (July 2016) and Aachen, Germany (June 2017).}
}
\maketitle
\begin{abstract}
We consider the two scenarios of communicating a pair $S_{1},S_{2}$ of 
distributed correlated sources over $2-$user multiple access (MAC) and 
interference channels (IC) respectively. While in the \textit{MAC problem}, the 
receiver intends to reconstruct both sources losslessly, in the \textit{IC 
problem}, receiver $j$ intends to reconstruct $S_{j}$ losslessly. We undertake 
a Shannon theoretic study and focus on achievability, i.e., characterizing 
sufficient conditions. In the absence of a Ga\'cs-K\"orner-Witsenhausen common 
part, the current known single-letter (S-L) coding schemes are constrained to 
choosing $X_{jt}$ - the symbol input on the channel by encoder $j$ at time $t$ - 
based only $S_{jt}$ - the source symbol observed by it, at time $t$, resulting 
in the pmf $p_{X_{1}X_{2}}$ of the inputs $X_{1},X_{2}$ constrained to the S-L 
long Markov Chain (LMC) $X_{1}-S_{1}-S_{2}-X_{2}$. Taking the lead of Dueck's 
example \cite{198103TIT_Due}, we recognize that the latter constraint is 
debilitating, leading to sub-optimality of S-L coding schemes. The goal of our 
work is to design a coding scheme wherein (i) the choice of $X_{jt}$ is based on 
multiple source symbols  $S_{j}^{l}$, and is yet ii) amenable to 
performance characterization via S-L expressions. In this article, we present 
the first part of our findings. We propose a new separation-based coding scheme 
comprising of (i) a fixed block-length (B-L) code that enables choice of 
$X_{jt}$ based on a generic number $l$ of source symbols, thus permitting 
correlation of the input symbols $X_{1},X_{2}$ through a multi-letter LMC 
$X_{1}-S_{1}^{l}-S_{2}^{l}-X_{2}$, (ii) arbitrarily large B-L codes superimposed 
on multiple \textit{sub-blocks} of the fixed B-L code that communicate the rest 
of the information necessary for source reconstruction at the decoder(s), and 
(iii) a multiplexing unit based on the interleaving technique 
\cite{201406ISIT_ShiPra} that ensures the latter codes of arbitrarily large B-L 
experience a memoryless channel. This careful stitching of S-L coding 
techniques enables us to devise a multi-letter coding scheme that permits 
characterization of sufficient conditions via a S-L expression. We prove that 
the derived inner bound is strictly larger than the current known largest inner 
bounds for both the MAC and IC problems. 

Since the proposed coding scheme is inherently separation based, the derived 
inner bound does not subsume the current known largest. In the second part of 
our work, we propose to enlarge the inner bound derived in this article by 
incorporating the technique of inducing source correlation onto channel inputs 
\cite{198011TIT_CovGamSal}.
\end{abstract}\vspace{-0.1in}
\vspace{-0.2in}
\section{Introduction}
\label{Sec:Introduction}
\begin{figure}\centering
\begin{minipage}{.5\textwidth}
 \centering
\includegraphics[width=2.6in]{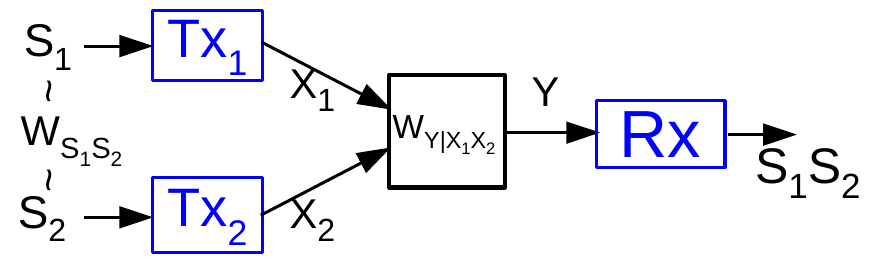}
\caption{Transmission of correlated sources over MAC.}
\label{Fig:GeneralMACProblem}
\end{minipage}\begin{minipage}{.5\textwidth}\centering
\includegraphics[width=2.9in]{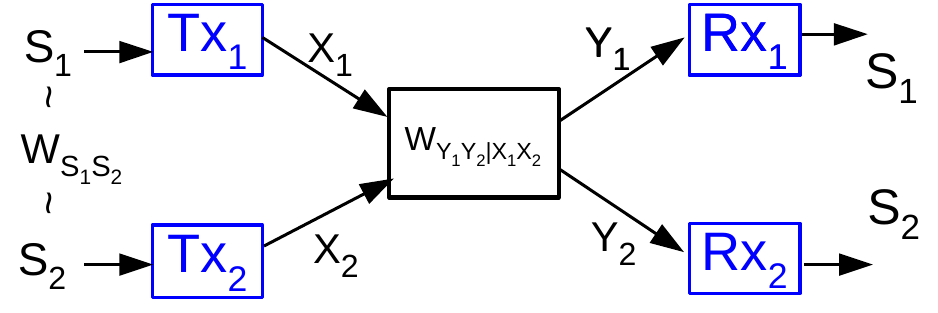}
\caption{Transmission of correlated sources over 2-IC.}
\label{Fig:GeneralICProblem}
\end{minipage}
\end{figure}
Since the pioneering work of Shannon, the problems of deriving single-letter 
(S-L) characterizations for performance limits of communication systems - 
capacity, rate-distortion 
regions as the case maybe - have been regarded to be of fundamental 
importance. In order to derive achievable rate regions, i.e., inner bounds to 
performance limits, a so-called `S-L coding scheme' is analyzed. Informally 
speaking, a random coding scheme is referred to as S-L, if the 
probability mass function (pmf) induced on the $n-$letter Cartesian product of 
the associated 
alphabet sets factors as a product of $n$ identical S-L pmfs. Since 
the performance is characterized in terms of an information functional of the 
induced pmf, the performance of a S-L scheme can be characterized in 
terms of the information functional of this factor pmf which is indeed 
a S-L pmf. Naturally, the goal of providing a S-L characterization for 
the target inner bound has restricted us to analyzing performance of S-L coding 
schemes. In this work, we take a new approach. Recognizing that the 
current known best S-L coding scheme is strictly sub-optimal, we devise a 
multi-letter coding scheme by carefully stitching together S-L coding 
techniques. Indeed, the pmf induced by the devised random coding scheme 
does not factor as a product of S-L pmfs. However, we characterize an 
inner bound to its performance via S-L expression i.e., an expression involving 
information functionals of S-L pmfs. We identify examples for which the 
derived inner bound is strictly larger that the current known largest inner 
bound derived via a S-L coding scheme.

Our primary focus in this article is the Shannon-theoretic study of the two 
scenarios depicted in Figures \ref{Fig:GeneralMACProblem}, 
\ref{Fig:GeneralICProblem}. Figure \ref{Fig:GeneralMACProblem} depicts the 
\textit{MAC problem} wherein a pair $S_{1},S_{2}$ of correlated sources, 
observed at the transmitters (Txs) of a $2-$user multiple access channel (MAC), 
have to be communicated to the receiver (Rx). The Rx intends to reconstruct both 
the sources losslessly. Given a (generic) MAC 
$\mathbb{W}_{Y|X_{1}X_{2}}$, the MAC problem concerns characterizing the set 
$\mathcal{T}(\mathbb{W}_{Y|X_{1}X_{2}}) $ of all \textit{transmissible} source 
pairs 
$\mathbb{W}_{S_{1}S_{2}}$ over the MAC. Figure 
\ref{Fig:GeneralICProblem} depicts the \textit{IC problem} wherein a pair 
$S_{1},S_{2}$ of correlated sources have to be communicated over a $2-$user 
interference channel (IC) $\mathbb{W}_{Y_{1}Y_{2}|X_{1}X_{2}}$. Receiver (Rx) 
$j$ wishes to reconstruct $S_{j}$ losslessly. The IC problem concerns 
characterizing the set $\mathcal{T}(\mathbb{W}_{Y_{1}Y_{2}|X_{1}X_{2}})$ of all 
\textit{transmissible} source pairs $\mathbb{W}_{S_{1}S_{2}}$ over 
the IC $\mathbb{W}_{Y_{1}Y_{2}|X_{1}X_{2}}$. Throughout our work, we restrict 
attention to achievability, i.e., inner bounds. Unless otherwise mentioned, 
we also assume the sources do \textit{not} possess a 
Ga\'cs-K\"orner-Witsenhausen common (GKW) part. In particular, the MAC and IC 
problems in our work refer to characterizing admissible regions  
$\alpha(\mathbb{W}_{Y|X_{1}X_{2}}) \subseteq 
\mathcal{T}(\mathbb{W}_{Y|X_{1}X_{2}})$ and $\alpha( \mathbb{W}_{Y_{1} 
Y_{2}|X_{1} X_{2}}) \subseteq \mathcal{T} ( \mathbb{W}_{Y_{1}Y_{2} | 
X_{1}X_{2}})$ via S-L expressions. This involves characterizing 
sufficient conditions for transmissibility of the sources over MAC and IC.

The central challenge posed by the above problems is to design a coding scheme 
that can optimally transfer/exploit source correlation to enable efficient 
co-ordinated communication. Cover, El Gamal and Salehi 
\cite{198011TIT_CovGamSal} devised an elegant S-L coding scheme, henceforth 
referred to as CES scheme, wherein symbol $X_{jt}$ input on the channel by 
encoder $j$ at time $t$ is chosen based on the source symbol $S_{jt}$ observed 
by it at time $t$. This permitted the channel inputs $X_{1},X_{2}$ to be 
correlated through 
a pmf $p_{X_{1}X_{2}}$ constrained to the S-L LMC $X_{1}-S_{1}-S_{2}-X_{2}$. 
The question of its optimality did not remain open for too long. Within barely 
five months, Dueck \cite{198103TIT_Due} identified a rich example and devised 
an ingenious, though 
very specific, coding scheme for that example to prove sub-optimality of CES 
scheme. A close look at Dueck's finding reveals that the constraint of a S-L LMC 
is debilitating (Remarks \ref{Rem:WhyLCIsSub-optimal}, 
\ref{Rem:Multi-LetterCodingScheme}), and choosing input symbol $X_{jt}$ based on 
multiple source 
symbols $S_{j}^{l}$, not just that at time $t$, permits for richer correlation 
amongst channel inputs that can facilitate more efficient co-ordinated 
communication. In essence, Dueck's finding proves that coding schemes that 
induce a pmf on the channel 
inputs that are constrained to the S-L LMC $X_{1}-S_{1}-S_{2}-X_{2}$ are 
sub-optimal in general. This leads us to the central motivation of our work.

Our goal is to design a coding scheme that, even in the absence of a GKW part, 
is \textit{not} constrained by 
a S-L LMC $X_{1}-S_{1}-S_{2}-X_{2}$, and yet is amenable for performance 
characterization via S-L expressions. Specifically, we intend to design a 
coding scheme, wherein $X_{jt}$ is chosen by encoder $j$ based on a generic 
number $l \in \naturals$ of source symbols $S_{j}^{l}$. We take a clue from the 
CES scheme of coding the GKW part, which is henceforth referred to as GKW 
coding. $X_{jt}$ is chosen based on $S_{jt}$ and $U_{t}-$the $t-$th symbol 
of the common codeword corresponding to the GKW block chosen at both encoders. 
The latter codeword is obtained via \textit{block} mapping of the GKW part, and 
hence the choice of $U_{t}$ is based on the entire block of the GKW part. Since 
$X_{jt}$ is based on $U_{t}$ which inturn is based on the entire block of GKW 
symbols, GKW coding is able to design input symbols based on a block of source 
symbols, while still being amenable to performance characterization via S-L 
expression.

GKW coding using common codes will be the central tool of our work. In the 
absence of a GKW part, it is impossible for the two encoders to agree on 
a common RV $U$ \cite{197501SJAM_Wit}. In fact, as the findings of Witsenhausen 
\cite{197501SJAM_Wit} suggest, in order to extract higher correlation at the 
distributed encoders, it is strictly beneficial to employ codes and maps of 
shorter block-length (B-L). Clearly, $P(S_{1}^{l} \neq S_{2}^{l}) = 1- (1-
P(S_{1}\neq S_{2}))^{l} \rightarrow 1$ as $l \rightarrow \infty$, resulting in 
lesser probability of agreement between the outputs of any non-trivial maps as 
the B-L is increased. We therefore propose \textit{fixed B-L GKW coding}, whose 
B-L is chosen as a function of the problem instance, not the desired 
probability of error. This leads to a fundamental shift. The proposed coding 
scheme will employ GKW coding of 
block-length (B-L) that remain fixed to a generic length $l \in \naturals$, 
irrespective of the desired probability of 
error. The fixed B-L codes induce a mapping from $l-$length \textit{sub-blocks} 
of the source to $l-$length channel inputs, thus permitting extraction and 
transfer of correlation from $l-$length sub-blocks of the source. An outer 
code, whose B-L is chosen arbitrarily large as a function of the desired 
probability of error, is superimposed over multiple sub-blocks of the (inner) 
fixed B-L code. 

The proposed coding scheme leads to challenges in its analysis. Primary among 
them, the outer code being superimposed on multiple $l-$length sub-blocks of 
the fixed B-L code, experiences $l-$length memory. We do not have a 
characterization for the effective channel it experiences, since we do not have 
a characterization of the induced pmf of a good fixed B-L code for a generic 
source-channel pair. Secondly, in the absence of a GKW part, any non-trivial 
GKW coding will result in disagreement between the chosen codewords at the two 
encoders. What then is the effective pmf induced by the fixed B-L GKW coding? 
Thirdly, how do we characterize the performance via a S-L expression, when the 
induced pmf is $l-$letter? As the informed reader will note, these challenges 
have not been addressed in prior work and hence, we do not have the basic 
building blocks of the intended coding scheme.

We present our findings in two parts, the first of 
which is presented in this article. Here, our emphasis is on presenting the 
new tools in a simplified setting and answering the following two central 
questions. How does one multiplex finite and $\infty-$B-L (codes of 
arbitrarily large B-L) information streams 
in a way that permits S-L characterization? and how do we analyze its
performance and derive a S-L expression for the same? In this article, we 
therefore restrict attention to separation based schemes wherein the source 
code encodes the source into two information streams - fixed B-L and 
$\infty-$B-L - and a channel code is designed to communicate these information 
streams. 
We analyze the performance of the proposed coding scheme and derive new 
admissible regions for the MAC and IC problems. By identifying examples, we 
prove that the derived admissible region can be strictly larger than the 
current known largest for the MAC \cite{198011TIT_CovGamSal} and IC problems 
\cite{201112TIT_LiuChe}. Thus having illustrated the power of our tools 
and approach, we build on these findings in the second part of our article, 
where we enlarge the admissible region presented here by incorporating the 
technique of joint source-channel coding proposed by Cover, El Gamal and Salehi 
\cite{198011TIT_CovGamSal}. In particular, as the reader will note, the 
enhancement proposed in the second part is based on leveraging the 
joint source-channel coding technique of inducing the source 
correlation onto channel inputs \cite{198011TIT_CovGamSal} in communicating the 
$\infty-$B-L information stream over the channel. This enables us enlarge the 
admissible region presented in this article to subsume the current known 
largest. Our second part is based on ideas presented in 
\cite{201706ISIT_Pad-MAC}.

Let us briefly comment on the tools we employ. We need channel codes of 
fixed B-L 
whose precise performance is known. The constant composition codes whose 
performance has been elaborately characterized by Csisz\'ar and K\"orner in 
\cite{CK-IT2011}, \cite{199810TIT_Csi} will be employed. As the reader will 
note, its constant composition property will be very useful in our analysis. 
Fano inequality type bounds will enables us upper bound additional information 
that needs to be communicated via outer codes. We leverage the novel technique 
of interleaving devised by Shirani and Pradhan \cite{201406ISIT_ShiPra} in the 
related problem of distributed source coding. Therein, the authors 
\cite{201406ISIT_ShiPra} proposed a pure source coding scheme to communicate 
information streams of different B-Ls to enable the decoder reconstruct 
quantized versions of the distributed sources. The common thread between the 
problem studied herein and \cite{201406ISIT_ShiPra} is the presence of the LMC. 
Our work goes beyond those of \cite{201406ISIT_ShiPra} in the following 
aspects. Firstly, we develop a channel code that involves a 
jointly designed superposition code comprising of a fixed B-L `cloud center 
code' and a satellite code of arbitrary large B-L. Such a joint superposition 
code is not necessary in a purely source coding problem and has therefore not 
been investigated in \cite{201406ISIT_ShiPra}. Secondly, since we cannot pool 
the messages output by the fixed B-L source encoder, our coding scheme is 
indeed crucially different from that proposed in \cite{201406ISIT_ShiPra}. 
Moreover, since these messages have to communicated separately over a 
noisy channel, this results in additional challenges not encountered in 
\cite{201406ISIT_ShiPra}. Thirdly, our use of constant composition code 
provides a much cleaner and elegant approach to characterizing inner bounds.

This article is aimed at presenting the tools necessary for characterizing a 
new admissible regions for the MAC and IC problems. We present these tools in 
three steps. The preliminary step, presented in Section \ref{Sec:DuecksExample} 
demonstrates the core idea of fixed B-L coding and 
its need via examples. In particular, we present generalizations of Dueck's 
example \cite{198103TIT_Due} and design an alternate coding scheme that is 
amenable for generalization. Section \ref{Sec:DuecksExample} provides very 
important intuition and holds the ideas presented in the paper. We then present 
generalization in two steps. In the first step, presented in Section 
\ref{Sec:FBLCodingOverMACAndICStep1}, we decode the fixed and $\infty-$B-L 
information streams separately. This leads new admissible regions (Theorems 
\ref{Thm:MACStep1}, \ref{Thm:ICStep1}) that are 
proven to be strictly larger (Theorems \ref{Thm:FBLBeatsCES}, 
\ref{Thm:FBLBeatsLC}) for specific examples. In the second step, 
presented in Section \ref{Sec:FBLCodingOverMACAndICStep2}, we incorporate joint 
decoding of the fixed and $\infty-$B-L information streams. We conclude with 
remarks in Section \ref{Sec:Conclusion}. We begin with preliminaries in the 
following section.

We conclude this section by summarizing relevant prior work. The technique of 
inducing source correlation onto channel inputs via S-L test channels designed 
by Cover, El Gamal and Salehi \cite{198011TIT_CovGamSal} has found application 
in problems of communicating correlated sources over IC and broadcast channels. 
Han and Costa \cite{198709TIT_HanCos} proposed random source partitioning and 
and employed the above technique, to derive new admissible region for 
communicating correlated sources over broadcast channels, that remains to be the 
current known largest. With regard to the IC problem, the techniques 
of (1) message splitting via superposition coding \cite{198101TIT_HanKob}, (2) 
random source partitioning \cite{198709TIT_HanCos} and (3) inducing source 
correlation onto channel inputs \cite{198011TIT_CovGamSal}, was employed by 
Liu and Chen to derive a set of sufficient (LC) conditions, or equivalently an 
admissible (LC) region. For the general IC problem, the LC region 
remains to be the current known largest. Dueck's example \cite{198103TIT_Due} 
proved that the CES coding scheme is strictly sub-optimal for the MAC problem. 
Dueck's findings \cite{198103TIT_Due} can be used to prove strict 
sub-optimality of the LC technique for the IC problem. Surprisingly, there is no 
mention of this in \cite{201112TIT_LiuChe}.

The presence of the LMC connects the MAC and IC problems to the problem of 
distributed source coding (DSC). Wagner, Kelly and Altug 
\cite{201107TIT_WagKelAlt} prove the sub-optimality of Berger-Tung rate region 
for the DSC problem via a continuity argument. The latter can be traced back to 
\cite{200807ISIT_CheWag}. Their argument can be related to \cite{198103TIT_Due} 
and our findings. Indeed, Dueck's example and our findings are based 
on proving that a sequence of examples, in the limit do not satisfy CES or LC 
conditions, yet are transmissible. Analogous to the inner bounds presented in 
this work, Chaharsooghi, Sahebi and Pradhan \cite{201307ISIT_ChaSahPra-Shrt}, 
followed by Shirani and Pradhan \cite{201406ISIT_ShiPra} propose new coding 
theorems for DSC based on fixed B-L codes. Kang and 
Ulukus \cite{201101TIT_KanUlu} characterize a necessary condition for a pmf 
$p_{X_{1}X_{2}}$ to satisfy an $n-$letter LMC $X_{1}-S_{1}^{n}-S_{2}^{n}-X_{2}$ 
and use that characterization to derive outer bounds for the MAC problem.

The scenario of transmitting correlated sources over multi-user channels is 
quite rich and permits several formulations. 
\cite{201505TIT_SonCheTia, 201007TIT_BroLapTin, 
201006TIT_LapTin,201110TIT_TiaDigSha,201707TIT_TiaCheDigSha} study the 
scenario of reconstructing Gaussian sources subject to distortion constraints. 
In particular, Lapidoth and Tinguely \cite{201006TIT_LapTin} study 
reconstruction of Gaussian sources subject to quadratic distortion constraints 
over Gaussian MAC. Bross Lapidoth and Tinguely \cite{201007TIT_BroLapTin} study 
communication of correlated Gaussians over Gaussian broadcast channel, while 
Tian, Diggavi and Shamai \cite{201110TIT_TiaDigSha} consider communication over 
bandwidth-matched Gaussian broadcast channels. Song, Chen and Tian 
\cite{201505TIT_SonCheTia} study broadcasting vector Gaussian subject to 
distortion constraints. Hybrid coding techniques for communicating Gaussian 
sources over Gaussian channels have been studied by Minero, Lim and Kim in 
\cite{201504TIT_MinLimKim}. Necessary conditions for reconstructing discrete 
memoryless sources subject to distortion constraints at the receiver of a MAC 
are characterized by Lapidoth and Wigger \cite{201607ISIT_LapWig}. 

\section{Preliminaries}
\label{Sec:Preliminaries}
\subsection{Notation}
\label{SubSec:Notation}
We supplement standard information theory notation - upper case for RVs, 
calligraphic letters such as $\mathcal{A},\mathcal{S}$ for finite sets etc. - 
with the 
following. We let an 
\underline{underline} denote an appropriate aggregation of 
related objects. For example, $\underline{S}$ will be used to represent a pair 
$S_{1},S_{2}$ of RVs. $\underline{\mathcal{S}}$ will be used to denote either 
the pair $\mathcal{S}_{1},\mathcal{S}_{2}$ or the Cartesian product 
$\mathcal{S}_{1}\times \mathcal{S}_{2}$, and will be clear from context. If we 
have $3$ components, say $(A_{0},A_{1},A_{2})$, then $\underline{A}$ will denote 
the triple, and we do \textit{not} use an \underline{underline} to denote pairs 
in this case. If $p_{U}$ is a pmf on $\mathcal{U}$, 
$p_{U}^{l}=\prod_{i=1}^{l}p_{U}$ is the product pmf on $\mathcal{U}^{l}$. When 
$j \in \{1,2\}$, then $\msout{j}$ will denote the complement index, i.e., 
$\{j,\msout{j}\}=\{1,2\}$. For $m\in \naturals$, $[m]\define \{1,\cdots,m\}$. 
\ifTITVersion\begin{eqnarray}
 \label{Eqn:TypicalSetDefn}
 T_{\delta}^{n}(U)=\{ u^{n} \in \mathcal{U}^{n} : \left|\frac{N(b|u^{n})}{n} - 
p_{U}(b)\right| \leq \delta p_{U}(b) ~\forall b \in \mathcal{U}\} \nonumber
\end{eqnarray}\fi
\ifPeerReviewVersion\begin{eqnarray}
 \label{Eqn:TypicalSetDefn}
 T_{\delta}^{n}(U)=\{ u^{n} \in \mathcal{U}^{n} : \left|\frac{N(b|u^{n})}{n} - 
p_{U}(b)\right| \leq \delta p_{U}(b) ~\forall b \in \mathcal{U}\} \nonumber
\end{eqnarray}\fi
is our typical set. ``$u^{m}$ is typical with respect to pmf 
$\prod_{t=1}^{m}p_{\mathscr{U}}$'' is abbreviated as 
$u^{m} \typical \prod_{t=1}^{m}p_{\mathscr{U}}$. 
Analogously, $u^{m} \ntypical
\prod_{t=1}^{m}p_{\mathscr{U}}$ abbreviates ``$u^{m}$ is \textit{not} typical 
with respect to pmf $\prod_{t=1}^{m}p_{\mathscr{U}}$'' For a pmf $p_{U}$ on 
$\mathcal{U}$, $b^{*} \in \mathcal{U}$ will denote a symbol with the least 
positive probability wrt $p_{U}$. The underlying pmf $p_{U}$ will be 
clear from context. We let $\tau_{l,\delta}(K) = 
2|\mathcal{K}|\exp\{ -2\delta^{2}p_{K}^{2}(a^{*})l\}$ denote an upper bound on 
$P(K^{l} \notin T_{\delta}^{l}(K))$. For a sequence $x^{n} \in 
\mathcal{X}^{n}$ and an element $a \in \mathcal{X}$, let $N(a|x^{n}) \define 
\sum_{i=1}^{n}\mathds{1}_{\{ x_{i} = a \}}$ denote the number of occurrences of 
$a$ in $x^{n}$. The \textit{type of }$x^{n}$ is the pmf $P_{x^{n}}$ on 
$\mathcal{X}$ defined as $P_{x^{n}}(a) \define \frac{1}{n}N(a|x^{n}) : a \in 
\mathcal{X}$. Given a pmf $p$ on $\mathcal{X}$, the set of all sequences in 
$\mathcal{X}^{n}$ of type $p$ is denoted $T_{p}^{n}$. A pmf $p$ on 
$\mathcal{X}$ is said to be a \textit{type of sequences in }$\mathcal{X}^{n}$ if 
$T_{p}^{n}$ is non-empty. We have used similar notation for typical sequences 
($T^{n}_{\delta}(\cdot)$) and sequences of type $p$ ($T_{p}^{n}$). 
The particular reference will be clear from context.

For a map $f:\mathcal{S} \rightarrow 
\mathcal{K}$, we denote $f^{n}: \mathcal{S}^{n}\rightarrow \mathcal{K}^{n}$ 
denote its $n-$letter extension defined by $f^{n}(s^{n}) \define  
(f(s_{1}),f(s_{2}),\cdots,f(s_{n}))$. While calligraphic letters such as 
$\mathcal{A}$ 
denote finite sets, boldfaced calligraphic letters such as 
$\boldsymbol{\mathcal{A}}$ denote the set of all $m \times l$ 
matrices with entries in $\mathcal{A}$, i.e., 
$\boldsymbol{\mathcal{A}} \define \mathcal{A}^{m \times l}$. Boldfaced letters 
such as $\bold{a},\bold{A}$ denote 
matrices. For a $\mtimesl$ matrix $\bold{a}$, (i) $\bold{a}(t,i)$ denotes the 
entry in row $t$, column $i$, (ii) $\bold{a}(1:m,i)$ denotes the $i^{th}$ 
column, $\bold{a}(t,1:l)$ denotes $t^{th}$ row. ``with high 
probability'', ``single-letter'', ``long Markov 
chain'', ``block-length'' are abbreviated whp, S-L, LMC, B-L respectively. We 
will be employing codes of fixed B-L whose B-L does not depend on the 
desired probability of error. Codes whose B-L will be chosen arbitrarily 
large as a function of the desired probability of error will be informally 
referred to as $\infty-$B-L codes.

For a point-to-point channel (PTP) $(\mathcal{U},\mathcal{Y},\mathbb{W}_{Y|U})$, 
let $E_{r}(R,p_{U},\mathbb{W}_{Y|U})$ denote the random coding exponent for 
constant composition codes of type $p_{U}$ and rate $R$ . Specifically, 
\ifTITVersion\begin{eqnarray}
 \label{Eqn:RandomCodingExponent}
E_{r}(R,p_{U},\mathbb{W}_{Y|U})\define  \min_{V_{Y|U}} 
\left\{D(V_{Y|U}||\mathbb{W}_{Y|U}|p_{U})+|I(p_{U};V_{Y|U})-R|^{+}\right\}
.\nonumber
\end{eqnarray}\fi
\ifPeerReviewVersion\begin{eqnarray}
 \label{Eqn:RandomCodingExponent}
E_{r}(R,p_{U},\mathbb{W}_{Y|U})\define  \min_{V_{Y|U}} 
\left\{D(V_{Y|U}||\mathbb{W}_{Y|U}|p_{U})+|I(p_{U};V_{Y|U})-R|^{+}\right\}
.\nonumber
\end{eqnarray}\fi
For a finite set $\mathcal{B}$ and $\mu \in [0,1]$, we let 
\begin{eqnarray}
\label{Eqn:AdditionalSourceCodingInfo}
\mathcal{L}_{l}(\mu,|\mathcal{A}|) \define
\frac{1}{l}h_{b}(\mu)+\mu \log |\mathcal{B}|\mbox{ and 
}\mathcal{L}(\mu,|\mathcal{B}|) \define 
\mathcal{L}_{1}(\mu,|\mathcal{B}|).\end{eqnarray}
If $A_{1}^{l}\in\mathcal{A}^{l}$ and $A_{2}^{l} \in \mathcal{A}^{l}$ are 
($l-$length) random vectors, we let $\xi^{[l]}(\underline{A}) \define 
P(A_{1}^{l}\neq A_{2}^{l})$, and  $\xi(\underline{A}) \define 
\xi^{[1]}(\underline{A})$. If $(A_{1t},A_{2t}) : t 
\in [l]$ are independent and identically distributed (IID), we 
note\footnote{$(1-x)^{l} \geq 1-xl\mbox{ for }x \in 
[0,1]$.} $\xi^{[l]}(\ulineA)=1-(1-\xi(\ulineA))^{l} \leq l\xi(\ulineA)$.

\subsection{Problem Statement}
\label{SubSec:ProblemStatement}
Consider a $2-$user MAC with input alphabets $\inpsetX,\inpsetY$, output 
alphabet $\outset$ and channel transition probabilities 
$\mathbb{W}_{\Out|\InpX\InpY}$ (Fig. \ref{Fig:GeneralMACProblem}). Let $\ulineS 
\define (S_{1},S_{2})$, taking values over $\ulineCalS \define \mathcal{S}_{1} 
\times \mathcal{S}_{2}$ with pmf $\mathbb{W}_{S_{1}S_{2}}$, denote a pair of 
information sources. For $j\in[2]$, Tx $j$ observes $S_{j}$. The Rx aims to 
reconstruct $\ulineS$ with arbitrarily small probability of error. With regard 
to the MAC problem, our objective is to characterize sufficient conditions for 
transmissibility of sources $(\ulineCalS,\mathbb{W}_{\ulineS})$ over the MAC 
$(\ulineCalX,\outset,\mathbb{W}_{\Out|\ulineX})$. A formal definition follows.

\begin{definition}
 \label{Defn:TransmissibilityOverMAC}
 A pair $(\ulineCalS,\mathbb{W}_{\ulineS})$ is transmissible over MAC 
$(\ulineCalX,\mathcal{Y},\mathbb{W}_{Y|\ulineX})$ if for every $\epsilon > 0$, 
there exists $N_{\epsilon} \in \naturals$ such that, for every $n \geq 
N_{\epsilon}$, there exists encoder maps $e_{j}: \mathcal{S}_{j}^{n} 
\rightarrow \mathcal{X}_{j}^{n} : j \in [2]$ and decoder map $d: 
\mathcal{Y}^{n} \rightarrow \ulineCalS^{n}$ such that
\begin{eqnarray}
 \label{Defn:MACProbOfError}
 \sum_{\ulines^{n}} \mathbb{W}^{n}_{\ulineS}(\ulines^{n}) \sum_{\substack{y^{n} 
\in \mathcal{Y}^{n} : \\d(y^{n}) \neq 
\ulines^{n}}}\mathbb{W}^{n}_{Y|\ulineX}(y^{n}|e_{1}(s_{1}^{n}),e_{2}(s_{2}^{n}
)) \leq \epsilon.\nonumber
\end{eqnarray}
\end{definition}

Consider a $2-$user IC with input alphabets $\inpsetX,\inpsetY$, output 
alphabets $\mathcal{Y}_{1},\mathcal{Y}_{2}$, and transition probabilities 
$\mathbb{W}_{Y_{1}Y_{2}|\InpX\InpY}$ (Fig. \ref{Fig:GeneralICProblem}). Let 
$\ulineS \define (S_{1},S_{2})$, taking values over $\ulineCalS \define 
\mathcal{S}_{1} \times \mathcal{S}_{2}$ with pmf $\mathbb{W}_{S_{1}S_{2}}$, 
denote a pair of information sources. For $j\in[2]$, Tx $j$ observes $S_{j}$, 
and Rx $j$ aims to reconstruct $S_{j}$ with arbitrarily small probability of 
error. If this is possible, we say $\ulineS$ \textit{is transmissible over IC} 
$\mathbb{W}_{\underline{Y}|\ulineX}$. A formal definition follows.
\begin{definition}
\label{Defn:TransmissibilityOverIC}
 A pair $(\ulineCalS,\mathbb{W}_{\ulineS})$ is transmissible over IC 
$(\ulineCalX,\ulineCalY,\mathbb{W}_{\ulineY|\ulineX})$ if for every $\epsilon > 
0$, 
there exists $N_{\epsilon} \in \naturals$ such that, for every $n \geq 
N_{\epsilon}$, there exists encoder maps $e_{j}: \mathcal{S}_{j}^{n} 
\rightarrow \mathcal{X}_{j}^{n} : j \in [2]$ and decoder maps $d_{j}: 
\mathcal{Y}_{j}^{n} \rightarrow \mathcal{S}_{j}^{n}: j \in [2]$ such that
\begin{eqnarray}
 \label{Defn:MACProbOfError}
 \sum_{\ulines^{n}} \mathbb{W}^{n}_{\ulineS}(\ulines^{n}) 
\sum_{\substack{\uliney^{n} 
\in 
\underline{\mathcal{Y}}^{n}}}\mathbb{W}^{n}_{\ulineY^{n}|\ulineX}(y_{1}^{n},y_{2
} ^ { n } |e_ { 1 } (s_{1 } ^{n } ), e_ { 2 } (s_ {2}^{ n }))\mathds{1}_{ 
\left\{ \begin{array}{c} d_{1}(y_{1}^{n})\neq s_{1}^{n}\mbox{ or 
}d_{2}(y_{2}^{n})\neq s_{2}^{n} \end{array} \right\} }
 \leq \epsilon.\nonumber
\end{eqnarray}
\end{definition}
With regard to the IC problem, our 
objective is to characterize sufficient conditions under which 
$(\ulineCalS,\mathbb{W}_{\ulineS})$ is transmissible over IC 
$(\ulineCalX,\ulineCalY,\mathbb{W}_{\underline{Y}|\ulineX})$.

\subsection{Current known coding techniques and sufficient conditions}
\label{SubSec:CurrentKnownTechniques}
The central question posed by the above problems is how does one optimally 
transfer source correlation onto correlated channel inputs that can enable 
efficient communication? The current known techniques are based on the CES 
strategy \cite{198011TIT_CovGamSal} proposed in the context of the MAC problem. 
One key idea of the 
CES strategy is to induce the source correlation onto channel inputs via S-L 
test channels $p_{X_{j}|S_{j}}: j \in [2]$. In other words, the codeword 
assigned for the source block $s_{j}^{n}$ is picked with pmf $\prod_{t=1}^{n} 
p_{X_{j}|S_{j}}(\cdot|s_{jt})$.  While this idea induces correlation across the 
input symbols $X_{1},X_{2}$, their joint pmf is constrained by the S-L 
LMC $X_{1}-S_{1}-S_{2}-X_{2}$.

A second key idea of the CES 
strategy is to exploit the GKW part $K=f_{j}(S_{j}): j \in [2]$ of the sources, 
whenever present, to permit 
a richer class of pmfs for $X_{1},X_{2}$. The GKW part is 
specially coded using a common codebook technique, henceforth 
referred to as \textit{GKW coding}. Specifically, typical sequence $k^{n} \in 
T_{\delta}^{n}(K)$ is mapped to a codeword $U^{n}(k^{n}) \in \mathcal{U}^{n}$ 
that is generated with a generic pmf $\prod_{t=1}^{n}p_{U}$. The codebook 
$\{U^{n}(k^{n}): k^{n} \in T_{\delta}^{n}(K)\}$ and the mapping is shared by 
both encoders. Since 
$K^{n}$ is observed by both encoders, GKW coding ensures each encoder agree on 
the chosen $\mathcal{U}-$codeword, and hence, the symbol at time $t$ 
distributed with pmf $p_{U}$ is 
common information. The codeword $X_{j}^{n}(s_{j}^{n})$ chosen for source block 
$s_{j}^{n}$ is 
picked randomly with pmf $\prod_{t=1}^{n} 
p_{X_{j}|S_{j}U_{j}}(\cdot|s_{jt},U(k^{n})_{t})$, where $U(k^{n})_{t}$ is the 
$t$ -th symbol of codeword $U^{n}(k^{n})$
assigned to the corresponding block of GKW symbols 
$k^{n} = f_{j}^{n}(s_{j}^{n})$. With $X_{j}^{n}(s_{j}^{n}): j \in [2]$ being 
the inputs on the channel corresponding to the pair $(s_{1}^{n},s_{2}^{n})$, it 
can be verified that a generic pair of input symbols $X_{1},X_{2}$ is jointly 
distributed with pmf $\sum_{u \in \mathcal{U}}\sum_{\ulines \in \ulineCalS}
\mathbb{W}_{\ulineS}(\ulines)p_{U}(u)\prod_{j=1}^{2}p_{X_{j}|US_{j}}(x_
{ j } |u , s_ { j})$, and in particular, \textit{not} constrained to a S-L LMC 
$X_{1}-S_{1}-S_{2}-X_{2}$. These two key 
ideas lead to the following sufficient conditions, henceforth referred to as 
CES conditions.
\begin{thm}[Cover, El Gamal and Salehi, \cite{198011TIT_CovGamSal}]
 \label{Thm:CESConditions} A pair of sources $(\ulineCalS,\mathbb{W}_{\ulineS})$ 
is transmissible over a MAC $(\ulineCalX,\mathcal{Y},\mathbb{W}_{Y|\ulineX})$ 
if there exists (i) a finite set $\mathcal{U}$, (ii) a pmf 
$\mathbb{W}_{\ulineS}p_{U}p_{X_{1}|US_{1}}p_{X_{2}|US_{2}}\mathbb{W}_{Y|X_{1} X_ 
{ 2 } } $ on $\ulineCalS\times \mathcal{U}\times \mathcal{X}_{1}\times 
\mathcal{X}_{2} \times \mathcal{Y}$ such that
\begin{eqnarray}
 \label{Eqn:CESConditions}
 H(S_{j}|S_{\msout{j}}) < I(X_{j};Y | X_{\msout{j}}, S_{\msout{j}},U) : j \in 
[2]~, ~H(\ulineS |K ) < I(\ulineX;Y |K, U)~,~ H(\ulineS) < I(\ulineX ; Y).
\end{eqnarray}
where $K = f_{j}(S_{j}) : j \in [2]$ taking values in $\mathcal{K}$ is the GKW 
part of $S_{1},S_{2}$.
\end{thm}
For the sake of completeness, we briefly describe a coding scheme that 
achieves the CES conditions. Let $\mathcal{U}$, pmf 
$\mathbb{W}_{\ulineS}p_{U}p_{X_{1}|US_{1}}p_{X_{2}|US_{2}}\mathbb{W}_{Y|X_{1} 
X_ { 2 } }$ and $K=f_{j}(S_{j})$ be as provided in the theorem statement. 
The codebook generation is as described previously. Encoder $j$ 
observes $S_{j}^{n}$ and inputs $X_{j}^{n}(S_{j}^{n})$ on the channel. Having 
received $Y^{n}$, the decoder looks for all typical pairs 
$(s_{1}^{n},s_{2}^{n}) \in 
T_{\delta}^{n}(S_{1},S_{2})$ such that 
$(s_{1}^{n},s_{2}^{n},k^{n},U^{n}(k^{n}),X_{1}(s_{1}^{n}),X_{2}^{n}(s_{2}^{n}
))$ is jointly typical wrt pmf
$\mathbb{W}_{\ulineS}p_{K|\ulineS}p_{U}p_{X_{1}|US_{1}}p_{X_{2}|US_{2}}\mathbb{W
} _ { Y|X_ { 1 } X_{2}} $, where $p_{K|\ulineS}(k|s_{1},s_{2}) = 
\mathds{1}_{\{ k = f_{j}(s_{j}): j \in [2] \}}$. If it finds a unique such 
pair, the latter is declared as the decoded source pair. Otherwise, an error 
is declared. The reader is referred to \cite[Section 14.1.1]{201201NIT_ElgKim} 
for a proof of Thm \ref{Thm:CESConditions}.
\begin{remark}
 \label{Rem:ConditionalCodingRevisited}
GKW coding crucially relies on identical 
codes and maps at both encoders. In effect, a common source code - 
typical set of $K$ - , a common mapping from its output to the channel code, 
and a common channel code $C_{U}$ ensures both encoders agree on the chosen 
codeword. Since the codebook can be chosen with any pmf $p_{U}$, the encoders 
can agree, distributively, on a common RV with an arbitrary 
pmf.\footnote{The coding scheme does not benefit by choosing $p_{U}$ with 
entropy greater than $H(K)$.} Note that as the B-L $n$ 
increases, the source code effects an efficient compression of $K$ and 
the message index output by this source code can be communicated 
via the best (joint) channel code on the $\mathcal{U}-\mathcal{Y}$ channel.

The current known best coding technique for the IC problem incorporates the 
technique of random source partitioning designed by Han and Costa 
\cite{198709TIT_HanCos}. Random source partitioning facilitates decoding of a 
common message at the two decoders of the IC. The latter technique, being part 
of Han-Kobayashi strategy -the current known best channel coding strategy for 
the IC - provides for a more efficient channel coding strategy for 
communication over the IC. Hence, the CES technique of inducing source 
correlation over channel inputs, coupled with random source partitioning yields 
the LC coding technique which is the current known best technique for the IC 
problem. In the following, we provide a characterization of the LC conditions 
for the specific case when the sources do not possess a GKW part. The reader is 
referred to \cite[Thm. 1]{201112TIT_LiuChe} for the general case.

\begin{thm}[Liu and Chen, \cite{201112TIT_LiuChe}]
 \label{Thm:LCConditions}
 A pair of sources $(\ulineCalS,\mathbb{W}_{\ulineS})$ 
is transmissible over an IC 
$(\ulineCalX,\ulineCalY,\mathbb{W}_{\ulineY|\ulineX})$ 
if there exists (i) finite sets $\mathcal{W}_{1},\mathcal{W}_{2},\mathcal{Q}$, 
(ii) a pmf 
$\mathbb{W}_{\ulineS}p_{Q}p_{W_{1}|Q}p_{W_{2}|Q}p_{ X_{1} | Q W_{1} S_{1} } p_{ 
X_{2} | QW_{2}S_{2}}\mathbb{W}_{\ulineY|\ulineX} $ defined on $\ulineCalS 
\times \mathcal{Q} \times \underline{\mathcal{W}} \times 
\ulineCalX \times \ulineCalY$ such that
\begin{eqnarray}
 \label{Eqn:LCConditions1}
 H(S_{j}) &<& I(S_{j}, X_{j}; Y_{j}|Q ,W_{\msout{j}}) : j \in [2], 
\nonumber\\
\label{Eqn:LCConditions2}
H(S_{1}) + H(S_{2}) &<& \min\left\{ I(S_{j},X_{j}; 
Y_{j}|Q,\underline{W}) + I(W_{j},S_{\msout{j}},X_{\msout{j}};Y_{\msout{j}}|Q) : 
j \in [2]\right\}, \\ 
\label{Eqn:LCConditions3}
\displaystyle H(S_{1}) + H(S_{2}) &<& 
\sum_{j=1}^{2}I(S_{j},W_{\msout{j}},X_{j};Y_{j}|Q, W_{j}) ,\nonumber\\
\label{Eqn:LCConditions4}
2H(S_{j}) + H(S_{\msout{j}}) &<& I(S_{j},X_{j};Y_{j}|Q, \ulineW) + I(S_{j}, 
 W_{\msout{j}},X_{j}; Y_{j}|Q) + 
I(S_{\msout{j}},W_{j},X_{\msout{j}};Y_{j}|Q,W_{j}) : j \in [2] \nonumber
\end{eqnarray}
\end{thm}
The reader is referred to \cite{201112TIT_LiuChe} for a proof.
\end{remark}
\subsection{Tools : Constant composition codes and the Random coding exponent}
\label{SubSec:PrelimsConstantCompositionCodes}
The material presented in this section is made use of only in proofs of the 
theorems in Sections \ref{Sec:FBLCodingOverMACAndICStep1}, 
\ref{Sec:FBLCodingOverMACAndICStep2}. The reader may refer to this material as 
and when needed in those sections.

The ensemble of constant composition codes studied by 
Csisz\'{a}r and K\"orner \cite{CK-IT2011}, \cite{199810TIT_Csi} prove to be a 
very useful tool in our study. The following theorem, due to Csisz\'{a}r and 
K\"orner, guarantee the existence of constant composition codes with guaranteed 
number of codewords and exponentially small error probabilities. In the sequel, 
we let\begin{eqnarray}
\label{Eqn:DefnOfLStar}
l^{*}(\mathcal{A},\mathcal{B},\rho) &\define& \min \left\{ l \in \naturals : 
\exp \left\{ \frac{l\rho}{2} \right\} \geq 
2(l+1)^{2|\mathcal{A}|+2|\mathcal{A}||\mathcal{B}|} \right\} \\&=& 
\min \left\{ l \in \naturals : 
 l\rho \geq \log 4 + (4|\mathcal{A}|+4|\mathcal{A}||\mathcal{B}| )\log (l+1)
 \right\} \nonumber
\end{eqnarray}
where $\mathcal{A},\mathcal{B}$  are finite sets and $\rho >0$.
\begin{thm}
 \label{Thm:ConstCompnCodesForMAC}
 Given any $\alpha > 0$, $\rho > 0$, a memoryless PTP
$(\mathcal{U},\mathcal{Y},p_{Y|U})$, B-L $l \geq 
 l^{*}(\mathcal{U},\mathcal{Y},\rho)$, a type 
$p_{U}$ of sequences in $\mathcal{U}^{l}$, there exists a code 
$(l,M_{u},e_{u},d_{u})$ of B-L $l$, encoder map $e_{u}: [M_{u}] \rightarrow 
\mathcal{U}^{l}$ with codewords $u^{l}(m) \define e_{u}(m)$ for $m \in 
[M_{u}]$, decoder map $d_{u}:\mathcal{Y}^{l} \rightarrow [M_{u}]$ such that (i) 
the codebook contains at least $M_{u} \geq \exp\{ l\alpha\}$ codewords, and 
(ii) 
probability of error of the code, when employed on the memoryless PTP 
$(\mathcal{U},\mathcal{Y}, P_{Y|U})$, is at most
\begin{eqnarray}
 \label{Eqn:ProbOfErrConstCompCodeMAC}
 \sum_{y^{l} \in \mathcal{Y}^{l}} p_{Y|U}^{l}(y^{l}|u^{l}(m)) 
\mathds{1}_{ \left\{ d(y^{l}) \neq m  \right\}} \leq 
(l+1)^{2|\mathcal{U}||\mathcal{Y}|} \exp\left\{ -l 
E_{r}(\alpha+\rho,p_{U},p_{Y|U}) \right\}\mbox{ for every }m \in [M_{u}].
 \nonumber
\end{eqnarray}
\end{thm}
\begin{IEEEproof}
 Follows from \cite[Theorem 10.2]{CK-IT2011}. The lower bound of 
$l^{*}(\mathcal{U},\mathcal{Y},\rho)$ on $l$ can be traced back to the proof of 
\cite[Theorem 10.1]{CK-IT2011} which forms the main ingredient in the proof of 
\cite[Theorem 10.2]{CK-IT2011}. It maybe noted that $\alpha+\rho, \alpha$ in 
our 
statement is equivalent to $R,R-\delta$ in \cite[Theorem 10.2]{CK-IT2011}. 
Lastly, the fact that the maximal probability of error is upper bounded is not 
stated in \cite[Theorem 10.2]{CK-IT2011}, but is evident from the proof.
\end{IEEEproof}

\begin{thm}
 \label{Thm:ConstantCompositionCodes}
 Given any $\alpha > 0$, $\rho > 0$, finite alphabets 
$\mathcal{U},\mathcal{Y}_{1},\mathcal{Y}_{2}$, channel transition 
probabilities $p_{Y_{1}Y_{2}|U}$, B-L $l \geq 
\max \{ l^{*}(\mathcal{U},\mathcal{Y}_{j},\rho) : j \in [2] \}$, a type 
$p_{U}$ of sequences in $\mathcal{U}^{l}$, there exists a code 
$(l,M_{u},e_{U},d_{u,1},d_{u,2})$ with message index set 
$[M_{u}]$ encoder map $e_{u} : [M_{u}] \rightarrow \mathcal{U}^{l}$ with 
codewords $u^{l}(m) \define e_{u}(m):  m \in 
[M_{u}]$ each of type $p_{U}$, and decoder maps $d_{j} : 
\mathcal{Y}_{j} \rightarrow [M_{u}]$ such that, (i) the number of 
codewords $M_{u} \geq \exp\{ l \alpha \}$, and (ii) maximal probability of 
decoding error of decoder $j$ is at most
\begin{eqnarray}
 \label{Eqn:ProbOfErrorConstCompCode}
 \sum_{y_{j}^{l} \in \mathcal{Y}_{j}^{l}} p_{Y_{j}|U}^{l}(y_{j}^{l}|u^{l}(m)) 
\mathds{1}_{ \left\{ d_{j}(y_{j}^{l}) \neq m  \right\}} \leq 
(l+1)^{2|\mathcal{U}||\mathcal{Y}_{j}|}\exp \left\{ -l 
E_{r}(\alpha+\rho,p_{U},p_{Y_{j}|U}) \right\}\mbox{ for }j \in [2],
 \nonumber
\end{eqnarray}
for every $m \in [M_{u}]$ and for any channel transition probabilities 
$p_{Y_{j}|U}$.
\end{thm}
\begin{IEEEproof}
Follows from the fact that the bound in \cite[Theorem 10.2]{CK-IT2011} applies 
to every DMC, and in particular the two DMCs $p_{Y_{1}|U}$ and $p_{Y_{2}|U}$.
\end{IEEEproof}

\section{Fixed B-L coding over isolated channels}
\label{Sec:DuecksExample}
\begin{figure}\centering
\begin{minipage}{.55\textwidth}
 \includegraphics[width=3.9in]{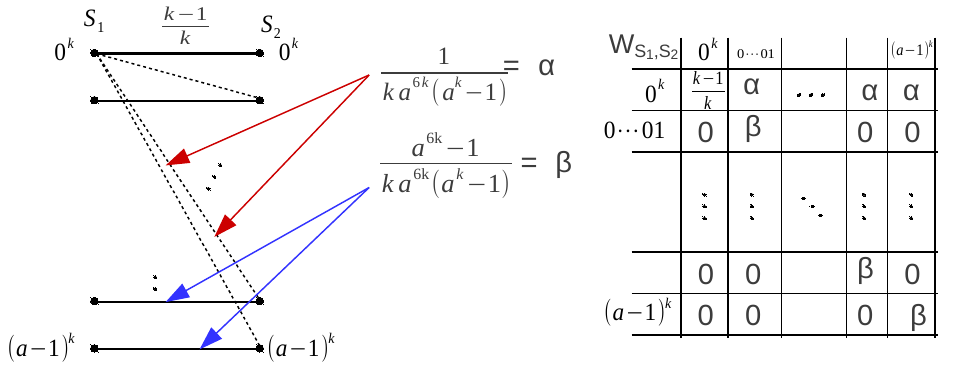}
\caption{On the left, the source pmf is depicted through a bipartite graph. 
Larger probabilities are depicted through edges with thicker lines. On the 
right, we depict the probability matrix.}
\label{Fig:SourceDescription}
\end{minipage}~~~~\begin{minipage}{.55\textwidth}
\includegraphics[width=2.9in]{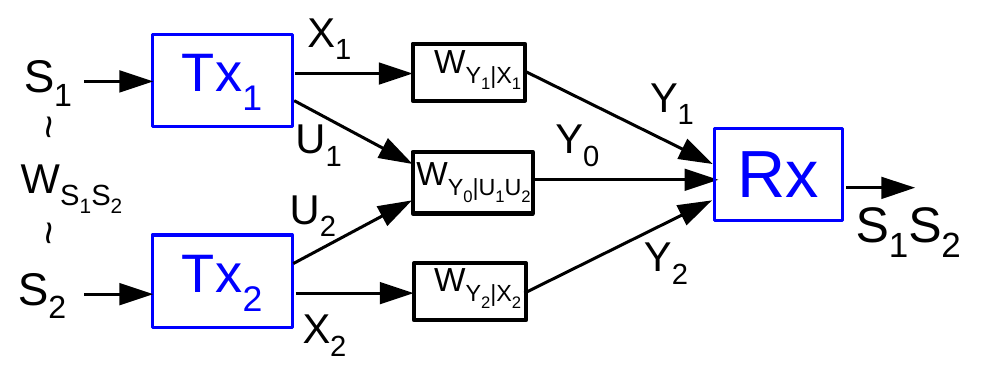}
\caption{MAC of Example \ref{Ex:DuecksExampleForMAC}.}
\label{Fig:MACForDuecksExample}
\end{minipage}
\end{figure}

The coding schemes we develop in this article are applicable for general 
problem instances. To illustrate the power 
of the proposed techniques, we consider specific examples - Example 
\ref{Ex:DuecksExampleForMAC} (MAC problem) and 
Example \ref{Ex:DuecksExampleForIC} (IC problem) - wherein the sources posses 
a \textit{near, but not perfect, GKW part}. These are obtained via a 
simple generalization of Dueck's ingenious example \cite{198103TIT_Due}. 
Following Dueck's argument, we prove that all current known joint 
source-channel coding techniques, in particular CES and LC techniques, are 
incapable of communicating the sources over the corresponding channels. We then 
propose a technique based on fixed B-L codes that enable transmissibility of 
the sources. While the key element of our technique is based on Dueck's fixed 
B-L code, we propose a simpler architecture that is amenable for 
generalization. The proposed technique will be generalized in Sections 
\ref{Sec:FBLCodingOverMACAndICStep1}, 
\ref{Sec:FBLCodingOverMACAndICStep2}. Throughout Section 
\ref{Sec:DuecksExample}, 
$\xi^{[l]}\define \xi^{[l]}(\ulineS), \xi \define \xi(\ulineS)$ and 
$\tau_{l,\delta } \define \tau_{l,\delta}(S_{1})$.
\begin{example}
 \label{Ex:DuecksExampleForMAC}
Source alphabets $\mathcal{S}_{1}=\mathcal{S}_{2} = \{0,1,\cdots,a-1 \}^{k}$. 
Let $\eta\geq 6$ be a positive even integer. The source PMF is
 \ifTITVersion \begin{eqnarray}
  \label{Eqn:SourceDesc}
  \mathbb{W}_{S_{1}S_{2}}(c^{k},d^{k}) = \begin{cases} \frac{k-1}{k} &\mbox{if } 
c^{k}=d^{k}=0^{k} \\
\frac{a^{\eta k}-1}{ka^{\eta k}(a^{k}-1)} & \mbox{if } c^{k}=d^{k}, c^{k}\neq 
0^{k},\\
\frac{1}{ka^{\eta k}(a^{k}-1)}&\mbox{if } c^{k}=0^{k}, d^{k}\neq 0^{k}\mbox{, 
and}\\0 & \mbox{otherwise.}\end{cases}\nonumber
\end{eqnarray}\fi
 \ifPeerReviewVersion \begin{eqnarray}
  \label{Eqn:SourceDesc}
  \mathbb{W}_{S_{1}S_{2}}(c^{k},d^{k}) = \begin{cases} \frac{k-1}{k} &\mbox{if 
} 
c^{k}=d^{k}=0^{k} \\
\frac{a^{\eta k}-1}{ka^{\eta k}(a^{k}-1)} & \mbox{if } c^{k}=d^{k}, c^{k}\neq 
0^{k},\\
\frac{1}{ka^{\eta k}(a^{k}-1)}&\mbox{if } c^{k}=0^{k}, d^{k}\neq 0^{k}\mbox{, 
and}\\0 & \mbox{otherwise.}\end{cases}\nonumber
\end{eqnarray}\fi
Note that in the above eqn. $c^{k},d^{k} \in \mathcal{S}_{1}$ abbreviate the $k$ 
`digits' $c_{1}c_{2}\cdots c_{k}$ and $d_{1}d_{2}\cdots d_{k}$ respectively. 
Fig. \ref{Fig:SourceDescription} depicts the source pmf with $\eta = 6$.

The MAC is depicted in Fig. \ref{Fig:MACForDuecksExample} and described below. 
The input alphabets are $\mathcal{U}\times\mathcal{X}_{1}$ and 
$\mathcal{U}\times\mathcal{X}_{2}$. The output alphabet is 
$\mathcal{Y}_{0}\times \mathcal{Y}_{1}\times \mathcal{Y}_{2}$. $\mathcal{U} = 
\mathcal{Y}_{0} = \{0,1,\cdots,a-1\}$. $(U_{j},X_{j}) \in \mathcal{U} \times 
\mathcal{X}_{j}$ denotes Tx $j$'s input. Moreover, 
$\mathbb{W}_{\ulineY|\underline{U}\underline{X}} = 
\mathbb{W}_{Y_{0}|\ulineU}\mathbb{W}_{Y_{1}|X_{1}}\mathbb{W}_{Y_{2}|X_{2}}$,
where
\ifTITVersion\begin{equation}
\label{Eqn:ChnlDesc}
\mathbb{W}_{Y_{0}|U_{1}U_{2}}(y_{0}|u_{1},u_{2}) = \begin{cases} 1 &\mbox{if } 
y_{0}=u_{1}=u_{2} \\
1 & \mbox{if } u_{1}\neq u_{2}, y_{0}=0,\mbox{ and}\\0 & 
\mbox{otherwise.}\end{cases}\nonumber
 \end{equation}\fi
\ifPeerReviewVersion\begin{equation}
\label{Eqn:ChnlDesc}
\mathbb{W}_{Y_{0}|U_{1}U_{2}}(y_{0}|u_{1},u_{2}) = \begin{cases} 1 &\mbox{if } 
y_{0}=u_{1}=u_{2} \\
1 & \mbox{if } u_{1}\neq u_{2}, y_{0}=0,\mbox{ and}\\0 & 
\mbox{otherwise.}\end{cases}\nonumber
 \end{equation}\fi
 The capacities of the PTPs 
$(\mathcal{X}_{j},\mathcal{Y}_{j},\mathbb{W}_{Y_{j}|X_{j}}) : j =1,2$ are 
$\mathcal{C}_{M}+h_{b}(\frac{2}{k})+\frac{1}{k}\log a$ and 
$\mathcal{C}_{M}+h_{b}(\frac{2}{ka^{\eta k}})$ respectively, where 
$\mathcal{C}_{M}\define\alpha\log a+2 h_{b}(\alpha)+\frac{1}{4k}\log a, 
\alpha = \frac{8k^{4}}{a^{\frac{\eta k}{3}}}$. It can be verified that, 
for sufficiently large $a,k$, the capacity of satellite channel 
$\mathbb{W}_{Y_{j}|X_{j}}$ is at most $\frac{3}{2k}\log a$. For all such $a,k$, 
we choose satellite channels for which $|\mathcal{Y}_{j}| \leq 
a^{\frac{3}{2k}}$.
\end{example}

\begin{figure}
\centering
\includegraphics[width=2.9in]{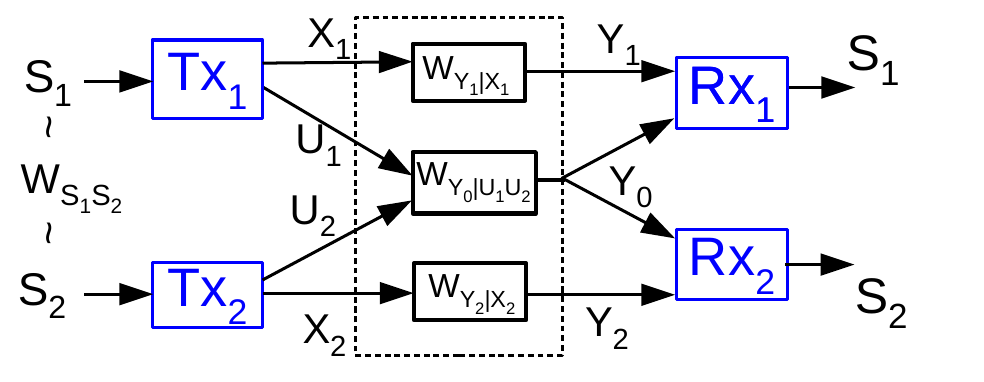}
\caption{IC of Example \ref{Ex:DuecksExampleForIC}.}
\label{Fig:ICForDuecksExample}
\end{figure}

\begin{example}
 \label{Ex:DuecksExampleForIC}
 Let $(\ulineCalS,\mathbb{W}_{\ulineS})$ be the source described in Example 
\ref{Ex:DuecksExampleForMAC}. The IC is depicted in Fig. 
\ref{Fig:ICForDuecksExample} and described below. The input alphabets are 
$\mathcal{U} \times \mathcal{X}_{1}$ and $\mathcal{U}\times \mathcal{X}_{2}$. 
The output alphabets are $\mathcal{Y}_{0} \times \mathcal{Y}_{1}$ and 
$\mathcal{Y}_{0}\times \mathcal{Y}_{2}$. $\mathcal{U} = \mathcal{Y}_{0} = 
\{0,1,\cdots,a-1\}$. $(U_{j},X_{j}) \in \mathcal{U} \times \mathcal{X}_{j}$ 
denotes Tx $j$'s input and $(Y_{0},Y_{j}) \in \mathcal{Y}_{0}\times 
\mathcal{Y}_{j}$ denotes symbols received by Rx $j$. The symbols $Y_{0}$ 
received at both Rxs agree with probability $1$. 
$\mathbb{W}_{Y_{0}Y_{1}Y_{2}|X_{1}U_{1}X_{2}U_{2}} = 
\mathbb{W}_{Y_{1}|X_{1}}\mathbb{W}_{Y_{2}|X_{2}}\mathbb{W}_{Y_{0}|U_{1}U_{2}}$, 
where
\ifTITVersion\begin{equation}
\label{Eqn:ChnlDesc}
\mathbb{W}_{Y_{0}|U_{1}U_{2}}(y_{0}|u_{1},u_{2}) = \begin{cases} 1 &\mbox{if } 
y_{0}=u_{1}=u_{2} \\
1 & \mbox{if } u_{1}\neq u_{2}, y_{0}=0,\mbox{ and}\\0&\mbox{otherwise.}
\end{cases}\nonumber
 \end{equation}\fi
\ifPeerReviewVersion\begin{equation}
\label{Eqn:ChnlDesc}
\mathbb{W}_{Y_{0}|U_{1}U_{2}}(y_{0}|u_{1},u_{2}) = \begin{cases} 1 &\mbox{if } 
y_{0}=u_{1}=u_{2} \\
1 & \mbox{if } u_{1}\neq u_{2}, y_{0}=0,\mbox{ and}\\0&\mbox{otherwise.}
\end{cases}\nonumber
 \end{equation}\fi
 The capacities of PTP channels $\mathbb{W}_{Y_{j}|X_{j}}:j=1,2$ are 
$\mathcal{C}_{I}\define h_{b}(\frac{2}{k})+\frac{2}{k}\log a$ and 
$\mathcal{C}_{I} +h_{b}(\frac{2}{ka^{\eta k}})$ respectively. Just as in 
Example \ref{Ex:DuecksExampleForMAC}, it can be verified that, 
for sufficiently large $a,k$, the capacity of satellite channel 
$\mathbb{W}_{Y_{j}|X_{j}}$ is at most $\frac{5}{2k}\log a$. For all such $a,k$, 
we choose satellite channels for which $|\mathcal{Y}_{j}| \leq 
a^{\frac{5}{2k}}$.
\end{example}
Examples \ref{Ex:DuecksExampleForMAC}, \ref{Ex:DuecksExampleForIC} are very 
similar. To avoid duplication, we provide discussions, describe ideas, coding 
techniques etc. for Example \ref{Ex:DuecksExampleForMAC}, and only indicate the 
differences, where ever present, with regard to Example 
\ref{Ex:DuecksExampleForIC}.

We highlight the key elements of Example \ref{Ex:DuecksExampleForMAC} through 
the following discussion. The parameters we refer to are summarized in Table 
\ref{Table:DueckMACExParameters} for ease of reference. Let $a,k$ be chosen 
sufficiently/quiet large. Firstly, 
the sources do not possess a GKW part, yet agree on most, but not all 
realizations. Verify that $\xi = \frac{1}{ka^{\eta k}}$ is very 
small. Secondly, 
$\mathbb{W}_{S_{j}}, \mathbb{W}_{\ulineS}$ is `very far' from the uniform pmf. 
The symbol pair $(0^{k},0^{k})$ occurs with very high probability 
$1-\frac{1}{k}$ and the rest 
of the symbol pairs occur with exponentially small probabilities, and the sum 
of the latter probabilities is at most $\frac{1}{k}$. Lastly, we have $H(S_{1}), 
H(S_{2}), H(\ulineS) \sim \log a$. In fact, 
\ifTITVersion\begin{eqnarray}
 \label{Eqn:SourceEntropy}
 (1-\frac{1}{a^{\eta k}})\log a - \frac{\log 2}{k} &\leq& H(S_{1}), H(S_{2}),
H(\underline{S}) \nonumber\\&\leq& \log a +h_{b}(\frac{1}{k})+\frac{\log
2}{k}.\nonumber
\end{eqnarray}\fi
\ifPeerReviewVersion\begin{eqnarray}
 \label{Eqn:SourceEntropy}
 (1-\frac{1}{a^{\eta k}})\log a - \frac{\log 2}{k} \leq H(S_{1}), H(S_{2}),
H(\underline{S}) \leq \log a +h_{b}(\frac{1}{k})+\frac{\log
2}{k}.\nonumber
\end{eqnarray}\fi

Moreover,
\ifTITVersion\begin{eqnarray}
 \label{Eqn:BoundOnConditionalEntropy}
 H(S_{1}|S_{2}) \leq \frac{1}{k}h_{b}(\frac{1}{a^{\eta k}}),~ H(S_{2}|S_{1}) 
\leq h_{b}(\frac{2}{ka^{\eta k}}) + \frac{2 \log a}{a^{\eta k}} \nonumber
\end{eqnarray}\fi
\ifPeerReviewVersion\begin{eqnarray}
 \label{Eqn:BoundOnConditionalEntropy}
 H(S_{1}|S_{2}) \leq \frac{1}{k}h_{b}(\frac{1}{a^{\eta k}}),~ H(S_{2}|S_{1}) 
\leq h_{b}(\frac{2}{ka^{\eta k}}) + \frac{2 \log a}{a^{\eta k}} \nonumber
\end{eqnarray}\fi
are very small. We list the consequences of these three 
observations. Firstly, note that the channel input symbols $X_{j},U_{j}:j \in 
[2]$ are constrained to the S-L LMC $X_{1}U_{1}-S_{1}-S_{2}-X_{2}U_{2}$, and in 
particular $U_{1}-S_{1}-S_{2}-U_{2}$. Secondly, any S-L function $g_{j}(S_{j})$ 
will remain considerably non-uniform. Lastly, the Rx benefits a lot by decoding
\textit{either} source. This is in particular true for the IC Example 
\ref{Ex:DuecksExampleForIC} wherein each Rx benefits a lot by decoding 
\textit{either} source.

The MAC has three components - `shared' $\mathbb{W}_{Y_{0}|\ulineU}-$channel,
and two `satellite' PTP channels $\mathbb{W}_{Y_{j}|X_{j}}$. Together, it
supports a sum capacity of at most $\log a +
2\mathcal{C}_{M}+h_{b}(\frac{2}{k})+\frac{1}{k}\log a + h_{b}(\frac{2}{ka^{\eta
k}})$. The sum of the capacities of the `satellite' PTP channels
$\mathbb{W}_{Y_{j}|X_{j}}$ is at most 
\ifTITVersion\begin{equation}
 \label{Eqn:SumOfPTPCapacities}
 \frac{16k^{4}}{a^{\frac{\eta k}{2}}}\left( \log a + 
h_{b}(\frac{8k^{4}}{a^{\frac{\eta k}{2}}})\right)+\frac{3}{2k}\log a 
+h_{b}(\frac{2}{k})+h_{b}(\frac{2}{ka^{\eta k}}).
\end{equation}\fi
\ifPeerReviewVersion\begin{equation}
 \label{Eqn:SumOfPTPCapacities}
 \frac{16k^{4}}{a^{\frac{\eta k}{3}}}\left( \log a + 
h_{b}(\frac{8k^{4}}{a^{\frac{\eta k}{3}}})\right)+\frac{3}{2k}\log a 
+h_{b}(\frac{2}{k})+h_{b}(\frac{2}{ka^{\eta k}}).
\end{equation}\fi
Since (\ref{Eqn:SumOfPTPCapacities}) decays with $a,k$, the bulk of the source 
entropy
($\log a$) has to be communicated via the shared
$\mathbb{W}_{Y_{0}|\ulineU}$-channel. In order to communicate close to a sum
rate of $\log a$ bits via the latter, it is necessary that
\textit{$U_{1}$ must equal $U_{2}$ whp} \underline{and} moreover
\textit{$U_{1}=U_{2}$ must be `close to' uniform}.
\begin{table}[h]
\begin{center}
\begin{tabular}{|c|c|c|}
\hline
Parameter & Value & Comment \\\hline\hline
&&\\
Capacity of satellite channel $\mathbb{W}_{Y_{1}|X_{1}}$& 
\normalsize$\alpha\log a+\alpha h_{b}(\alpha)+\frac{1}{4k}\log 
a+h_{b}(\frac{2}{k})+\frac{1}{k}\log a$ &Shrinks to $0$ for large $a,k$\\
&{\normalsize where $\alpha = \frac{8k^{4}}{a^{\frac{\eta k}{3}}}$}&
\\
&&\\\hline
&&\\
Capacity of satellite channel {\normalsize$\mathbb{W}_{Y_{2}|X_{2}}$}& 
{\normalsize$\alpha\log a+\alpha h_{b}(\alpha)+\frac{1}{4k}\log 
a+h_{b}(\frac{2}{ka^{\eta k}})$ }&Shrinks to $0$ for large $a,k$\\
&{\normalsize where $\alpha = \frac{8k^{4}}{a^{\frac{\eta k}{3}}}$}&
\\
&&\\\hline
&&\\
{\normalsize$H(S_{1},S_{2})$}& 
{\normalsize$\frac{1}{k}\log (a^{k}-1) 
+h_{b}(\frac{1}{k})+\frac{1}{k}h_{b}(\frac{1}{a^{\eta k}})$ }& Scales as $\log 
a$ for large $a,k$
\\
&&\\\hline
&&\\
\normalsize$H(S_{1})$& 
\normalsize$h_{b}(\frac{1}{k}-\frac{1}{ka^{\eta 
k}})+(\frac{1}{k}-\frac{1}{ka^{\eta 
k}})\log (a^{k}-1)$ & Scales as $\log 
a$ for large $a,k$
\\
&&\\\hline&&\\
Uniform upper bound on & 
\normalsize$\log a
+h_{b}(\frac{1}{k})+\frac{1}{k}h_{b}(\frac{1}{a^{\eta k}})$ &Scales as $\log 
a$ for large $a,k$
\\
\normalsize$H(\ulineS),H(S_{j}): j \in [2]$&&\\
&&\\\hline&&\\
Strict upper bound on $H(S_{1})$& 
\normalsize$h_{b}(\frac{1}{k})+\log a$ & Scales as $\log 
a$ for large $a,k$
\\
&&\\\hline
&&\\
Upper bound on $H(S_{2}|S_{1})$& 
\normalsize$h_{b}(\frac{2}{ka^{\eta k}}) + \frac{2 \log a}{a^{\eta k}}$&Shrinks 
to $0$ for large $a,k$
\\
&&\\\hline
&&\\
{\normalsize$\xi = P(S_{1}\neq S_{2})$}& 
{\Large$\frac{1}{ka^{\eta k}}$}&Shrinks 
to $0$ for large $a,k$
\\
&&\\\hline
&&\\
Upper bound on {\normalsize$\xi^{[l]} = P(S_{1}^{l}\neq S_{2}^{l})$}& 
{\normalsize$\xi^{[l]} = 1-(1-\xi^{l}) \leq 1-(1-\l\xi) = \frac{l}{ka^{\eta 
k}}$}&
\\
&&\\\hline\end{tabular}
\end{center}
\caption{Parameters in Example \ref{Ex:DuecksExampleForMAC}}
\label{Table:DueckMACExParameters}
\end{table}

In the following, we prove, following Dueck's argument \cite[Sec. III
C]{198103TIT_Due}, that a S-L CES scheme is incapable of communicating the
sources over the MAC (Example \ref{Ex:DuecksExampleForMAC}). The argument is
based on the fact that, conditioned on the event $\{ \ulineS = (0^{k},0^{k})
\}$, the channel inputs are constrained to be independent, and are therefore
constrained to communicate only a fraction of its co-ordinated capacity of $\log
a$. Since $\{ \ulineS = (0^{k},0^{k}) \}$ occurs with a significant probability
of $1-\frac{1}{k}$, the amount of information that can be communicated over the
MAC via a S-L CES scheme is considerably constrained.

\begin{lemma}
 \label{Lem:Ex1DoesNOTSatisfyCESConditions}
Consider Example \ref{Ex:DuecksExampleForMAC} with any $\eta \in \naturals$. 
There exists an $a_{*} \in \naturals, k_{*} \in \naturals$, such that for any 
$a\geq a_{*}$ and any $k\geq k_{*}$, the sources and the MAC described in 
Example \ref{Ex:DuecksExampleForMAC} do \textit{not} satisfy CES conditions that 
are stated in \cite[Thm. 1]{198011TIT_CovGamSal}.
\end{lemma}
\begin{IEEEproof}
Given any set $\mathcal{Q}$ and any pmf 
$\mathbb{W}_{\ulineS}p_{X_{1}U_{1}|S_{1}Q}p_{X_{2}U_{2}|S_{2}Q}\mathbb{W}_{
\ulineY|\ulineX}$, we will prove
\ifTITVersion\begin{eqnarray}
 \label{Eqn:CESSumRateBoundViolated}
I(\ulineU\ulineX;\ulineY|Q) < H(\ulineS),
\end{eqnarray}\fi
\ifPeerReviewVersion\begin{eqnarray}
 \label{Eqn:CESSumRateBoundViolated}
I(\ulineU\ulineX;\ulineY|Q) < H(\ulineS),
\end{eqnarray}\fi
thereby contradicting (\ref{Eqn:CESConditions}). Towards that 
end, we derive a lower bound on $H(\ulineS)$. By simple substitution, it can be 
verified that
\ifTITVersion\begin{eqnarray}
H(\underline{S}) &\geq& H(S_{2}) =h_{b}(\frac{1}{k})+\frac{1}{k}\log (a^{k}-1) 
\geq \frac{1}{k}\log\frac{ka^{k}}{2}\nonumber\\\
\label{Eqn:LowerBoundOnSourceEntropy}&\geq& \log a +\frac{1}{k}\log 
(\frac{k}{2}) \geq \log a,
\end{eqnarray}\fi
\ifPeerReviewVersion\begin{eqnarray}
H(\underline{S}) &\geq& H(S_{2}) =h_{b}(\frac{1}{k})+\frac{1}{k}\log (a^{k}-1) 
\geq \frac{1}{k}\log\frac{ka^{k}}{2}\nonumber\\\
\label{Eqn:LowerBoundOnSourceEntropy}&\geq& \log a +\frac{1}{k}\log 
(\frac{k}{2}) \geq \log a,
\end{eqnarray}\fi
whenever $a^{k} \geq 2, k \geq 2$. We now consider the LHS of 
(\ref{Eqn:CESSumRateBoundViolated}).

Let $R=\mathds{1}_{\{(S_{1},S_{2})=(0^{k},0^{k}) \}}$.
\ifTITVersion\begin{eqnarray}
 \label{Eqn:DuecksArgumentPart1}
 I({\underline{XU};\underline{Y}|Q}) \leq I({\underline{XU}R;\underline{Y}|Q}) 
\leq \log 2+I({\underline{XU};\underline{Y}|Q,R})\nonumber\\
 \leq \log 2 + \frac{\log 
|\mathcal{Y}_{0}\times\mathcal{Y}_{1}\times\mathcal{Y}_{2}|}{k} 
+(1-\frac{1}{k})I({\underline{XU};\underline{Y}|Q,R=1})\nonumber
\end{eqnarray}\fi
\ifPeerReviewVersion\begin{eqnarray}
 \label{Eqn:DuecksArgumentPart1}
 I({\underline{XU};\underline{Y}|Q}) &\leq& 
I({\underline{XU}R;\underline{Y}|Q}) 
\leq \log 2+I({\underline{XU};\underline{Y}|Q,R})\nonumber\\
 &\leq& \log 2 + \frac{\log 
|\mathcal{Y}_{0}\times\mathcal{Y}_{1}\times\mathcal{Y}_{2}|}{k} 
+(1-\frac{1}{k})I({\underline{XU};\underline{Y}|Q,R=1})\nonumber
\end{eqnarray}\fi
We focus on the third term in the above sum. Conditioned on $R=1$, the sources 
are equal to $(0^{k},0^{k})$. It can be verified that 
$X_{1}U_{1}-S_{1}Q-S_{2}Q-X_{2}U_{2}$. Given $Q=q,R=1$, $(X_{1},U_{1})$ is 
independent of $(X_{2},U_{2})$ and hence 
\ifTITVersion\begin{eqnarray}
I({\underline{XU};\underline{Y}|Q,R=1}) \leq \max_{p_{X_{1}U_{1}}p_{X_{2}U_{2}}} 
I(\underline{XU};\underline{Y})\leq 2\mathcal{C}_{M}+\nonumber\\
\label{Eqn:Step3}
h_{b}(\frac{2}{k})+\frac{1}{k}\log a + h_{b}(\frac{2}{ka^{\eta 
k}})+\max_{p_{U_{1}}p_{U_{2}}\mathbb{W}_{Y_{0}|\ulineU}} H(Y_{0})
\end{eqnarray}\fi
\ifPeerReviewVersion\begin{eqnarray}
\label{Eqn:Step3}
I({\underline{XU};\underline{Y}|Q,R=1}) \leq 
\max_{p_{X_{1}U_{1}}p_{X_{2}U_{2}}} 
I(\underline{XU};\underline{Y})\leq 2\mathcal{C}_{M}+ 
h_{b}(\frac{2}{k})+\frac{1}{k}\log a + h_{b}(\frac{2}{ka^{\eta 
k}})+\max_{p_{U_{1}}p_{U_{2}}\mathbb{W}_{Y_{0}|\ulineU}} H(Y_{0})
\end{eqnarray}\fi
We now evaluate an upper bound on the maximum value of $H(Y_{0})$ subject to 
$U_{1},U_{2}$ being independent. We evaluate the following three possible cases.

Case 1a : For some $u\in\mathcal{U}$, $P(U_{1}=u)\geq \frac{1}{2}$ and 
$P(U_{2}=u)\geq \frac{1}{2}$. Then $P(Y_{0}=u)\geq \frac{1}{4}$ (independence of 
$U_{1},U_{2}$) and hence $H(Y_{0})\leq \log 2 + \frac{3}{4}\log a$.

Case 1b : For some $u\in\mathcal{U}$, $P(U_{1}=u)\geq \frac{1}{2}$ and 
$P(U_{2}=u)\leq \frac{1}{2}$. Then $P(U_{2}\neq u)\geq \frac{1}{2}$ and hence 
$P(Y_{0}=0) \geq \frac{1}{4}$ and hence $H(Y_{0})\leq \log 2 + \frac{3}{4}\log 
a$.

Case 2a : For every $u \in \mathcal{U}$, $P(U_{1}=u)\leq \frac{1}{2}$. Then for 
any $u \in \mathcal{U}$, $P(U_{2} \neq U_{1}) = \sum_{u}\sum_{z\neq 
u}P(U_{2}=u)P(U_{1}=z) \geq \frac{1}{2}\sum_{u}P(U_{2}=u)=\frac{1}{2}$, implying 
$P(Y_{0}=0)\geq \frac{1}{2}$ and hence $H(Y_{0})\leq \log 2 + \frac{3}{4}\log 
a$.

In all cases, we have $H(Y_{0})\leq \log 2 + \frac{3}{4}\log a$. Substituting 
through (\ref{Eqn:Step3}) and above, we conclude
\ifTITVersion\begin{eqnarray}
 \label{Eqn:ConverseLastIneq}
 I({\underline{XU};\underline{Y}|Q}) \leq 
2\log2+2\mathcal{C}_{M}+h_{b}(\frac{2}{k})+\frac{1}{k}\log a \nonumber\\+ 
h_{b}(\frac{2}{ka^{\eta k}})+\frac{3}{4}\log a + \frac{\log |\ulineCalY|}{k}
 < \log a
\end{eqnarray}\fi
\ifPeerReviewVersion\begin{eqnarray}
 \label{Eqn:ConverseLastIneq}
 I({\underline{XU};\underline{Y}|Q}) \leq 
2\log2+2\mathcal{C}_{M}+h_{b}(\frac{2}{k})+\frac{1}{k}\log a + 
h_{b}(\frac{2}{ka^{\eta k}})+\frac{3}{4}\log a + \frac{\log |\ulineCalY|}{k}
 < \log a
\end{eqnarray}\fi
for sufficiently large $k,a$. In (\ref{Eqn:ConverseLastIneq}), we have used the 
fact that for sufficiently large $a,k$ the satellite channels are chosen such 
that $|\mathcal{Y}_{j}| \leq a^{\frac{3}{2k}}$.
\end{IEEEproof}

We leverage the above argument to prove an analogous statement for Example 
\ref{Ex:DuecksExampleForIC}. In particular, we prove that if the LC technique 
enables Rxs $j$ reconstruct $S_{j}$ for $j \in [2]$, then both the Rxs can 
reconstruct $S_{1}$ and $S_{2}$ if each of them is provided $Y_{1}$ \textit{and} 
$Y_{2}$ (and $Y_{0}$). Using the above arguments, we prove that this is not 
permissible.

\begin{lemma}
\label{Lem:Ex2DoesNOTSatisfyLCConditions}
 Consider Example \ref{Ex:DuecksExampleForIC} with any $\eta \in \naturals$. 
There exists an $a_{*} \in \naturals, k_{*} \in \naturals$, such that for any 
$a\geq a_{*}$ and any $k\geq k_{*}$, the sources and the IC described in Example 
\ref{Ex:DuecksExampleForIC} do \textit{not} satisfy LC conditions that are 
stated in \cite[Thm. 1]{201112TIT_LiuChe}.
\end{lemma}
\begin{IEEEproof}
 Since the sources do not have a GKW part, it suffices to prove that Example 
\ref{Ex:DuecksExampleForIC} does not satisfy conditions stated in 
Thm \ref{Thm:LCConditions}. Let $\ulineS 
Q\ulineW\ulineX\ulineU\ulineY$ be \textit{any} collection of RVs whose pmf 
factorizes as 
$\mathbb{W}_{\underline{S}}p_{Q}p_{W_{1}|Q}p_{W_{2}|Q}p_{X_{1}U_{1}|S_{1}W_{1}Q}
p_{X_{2}U_{2}|S_{2}W_{2}Q}\mathbb{W}_{\underline{Y}|\underline{X}\underline{U}}
$. We prove
 \ifTITVersion\begin{eqnarray}
H(\underline{S}) \!> 
\!I(S_{1}X_{1}U_{1};Y_{1}Y_{0}|Q\underline{W})\!+\!I(W_{1}S_{2}X_{2}U_{2};Y_{2}
Y_{0}|Q)\nonumber\\
\label{Eqn:Eqn44OfCorollary1LiuChen}
\!\!\!\!\!-I(S_{1};S_{2})
 \end{eqnarray}\fi
 \ifPeerReviewVersion\begin{eqnarray}
\label{Eqn:Eqn44OfCorollary1LiuChen}
H(\underline{S}) > 
\!I(S_{1}X_{1}U_{1};Y_{1}Y_{0}|Q\underline{W}) + I(W_{1}S_{2}X_{2}U_{2};Y_{2}
Y_{0}|Q) -I(S_{1};S_{2})
 \end{eqnarray}\fi
and thereby contradicting (\ref{Eqn:LCConditions2}). 
The lower bound on $H(\ulineS)$ follows from 
(\ref{Eqn:LowerBoundOnSourceEntropy}). Secondly, the RHS of 
(\ref{Eqn:Eqn44OfCorollary1LiuChen}) can be bounded above by
\ifTITVersion\begin{eqnarray}
 \label{Eqn:SumRateOfICIsSmallerThanMACSumRate} 
\!\!\!\!\!\!\!\!\!\!\!\!\lefteqn{\!\!\!\!\!\!\!\!\!\!\!\!\!I(S_{1}X_{1}U_{1};Y_{
1}Y_{0}|Q\underline{W})\!+\!I(W_{1}S_{2}X_{2}U_{2};Y_{2}Y_{0}|Q)-I(S_{1};S_{2})}
\nonumber\\
 \!\!\!\!\!\!\!\!\!\!\!\!\!\!\!\! \lefteqn{\!\!\!\!\!\!\!\!\!\!\!\!\leq 
I({S_{1}X_{1}U_{1};Y_{1}Y_{0}|Q\underline{W}})+I({\underline{W}S_{2}X_{2}U_{2}
;\underline{Y}|Q})- I({S_{1};S_{2}})} \nonumber\\
 \!\!\!\!&\leq& 
I({\underline{WSXU};\underline{Y}|Q})+I({S_{1}X_{1}U_{1};\underline{Y}
|Q\underline{W}})\nonumber\\
 \!\!\!\!&&-I({S_{1}X_{1}U_{1};\underline{Y}|Q\underline{W}S_{2}X_{2}U_{2}})- 
I({S_{1};S_{2}}) \nonumber\\
 \!\!\!\!&=& 
I({\underline{XU};\underline{Y}|Q})+I({S_{1}X_{1}U_{1};\underline{Y}|Q\underline
{W}})\nonumber\\
 \label{Eqn:ChnlAndSrcMarkovChain}
 \!\!\!\!&&-I({S_{1}X_{1}U_{1};\underline{Y}X_{2}U_{2}|Q\underline{W}S_{2}})- 
I({S_{1};S_{2}}) \\
 \label{Eqn:SrcsIndependentOfW}
 \!\!\!\!&=& 
I({\underline{XU};\underline{Y}|Q})+I({S_{1}X_{1}U_{1};\underline{Y}|Q\underline
{W}})\nonumber\\&&-I({S_{1}X_{1}U_{1};\underline{Y}X_{2}U_{2}S_{2}|Q\underline{W
}})\nonumber\\
 \label{Eqn:EmergenceOFMACSumRateBound}
 &\leq& I({\underline{XU};\underline{Y}|Q})
  \end{eqnarray}\fi
\ifPeerReviewVersion\begin{eqnarray}
 \label{Eqn:SumRateOfICIsSmallerThanMACSumRate} 
\lefteqn{I(S_{1}X_{1}U_{1};Y_{
1}Y_{0}|Q\underline{W})+I(W_{1}S_{2}X_{2}U_{2};Y_{2}Y_{0}|Q)-I(S_{1};S_{2}) }
\nonumber\\
&\leq& I({S_{1}X_{1}U_{1};Y_{1}Y_{0}|Q\underline{W}}) + 
I({\underline{W}S_{2}X_{2}U_{ 2}
;\underline{Y}|Q})- I({S_{1};S_{2}}) \nonumber\\
 &\leq&
I({\underline{WSXU};\underline{Y}|Q})+I({S_{1}X_{1}U_{1};\underline{Y}
|Q\underline{W}})
 -I({S_{1}X_{1}U_{1};\underline{Y}|Q\underline{W}S_{2}X_{2}U_{2}})- 
I({S_{1};S_{2}}) \nonumber\\
 &=&
I({\underline{XU};\underline{Y}|Q})+I({S_{1}X_{1}U_{1};\underline{Y}|Q\underline
{W}})
 \label{Eqn:ChnlAndSrcMarkovChain}
 -I({S_{1}X_{1}U_{1};\underline{Y}X_{2}U_{2}|Q\underline{W}S_{2}})- 
I({S_{1};S_{2}}) \\
 \label{Eqn:EmergenceOFMACSumRateBound}
& = & I({\underline{XU};\underline{Y}|Q})+I({S_{1}X_{1}U_{1};\underline{Y}
|Q\underline{W}}) -I({S_{1}X_{1}U_{1};\underline{Y}X_{2}U_{2}S_{2}|Q\underline{W
}}) \leq I({\underline{XU};\underline{Y}|Q})
  \end{eqnarray}\fi
  Following the steps identical to proof of Lemma 
\ref{Lem:Ex1DoesNOTSatisfyCESConditions}, it can be verified that
\ifTITVersion\begin{eqnarray}
 \label{Eqn:UpperBoundOnMACSumRateBound}
 I({\underline{XU};\underline{Y}|Q}) &\leq& 2\log2+2\mathcal{C}_{I} + 
h_{b}(\frac{2}{ka^{\eta k}})+\frac{3}{4}\log a\nonumber\\ &&+ \frac{\log 
|\ulineCalY|}{k}~~
 < ~\log a
\end{eqnarray}\fi
\ifPeerReviewVersion\begin{eqnarray}
 \label{Eqn:UpperBoundOnMACSumRateBound}
 I({\underline{XU};\underline{Y}|Q}) &\leq& 2\log2+2\mathcal{C}_{I} + 
h_{b}(\frac{2}{ka^{\eta k}})+\frac{3}{4}\log a + \frac{\log 
|\ulineCalY|}{k}~~
 < ~\log a
\end{eqnarray}\fi
for sufficiently large $k,a$. In view of the lower bound on $H(\ulineS)$ 
(\ref{Eqn:LowerBoundOnSourceEntropy}) and the upper bound on the RHS of 
(\ref{Eqn:Eqn44OfCorollary1LiuChen}) via (\ref{Eqn:EmergenceOFMACSumRateBound}) 
and (\ref{Eqn:UpperBoundOnMACSumRateBound}), we are done.
\end{IEEEproof}

\begin{remark}
\label{Rem:WhyLCIsSub-optimal}
Why are S-L LC and CES techniques incapable of communicating $\ulineS$? Any
valid pmf $p_{U_{1}U_{2}}$ induced by a S-L coding scheme is constrained to the
LMC $U_{1}-S_{1}-S_{2}-U_{2}$. For $j\in [2]$, $p_{U_{j}|S_{j}}$ can
equivalently be viewed as $U_{j}=g_{j}(S_{j},W_{j})$, for some function $g_{j}$
and RV $W_{1},W_{2}$ that are \textit{independent}. Owing to independence,
$W_{1}$ and/or $W_{2}$ being \textit{non-trivial} RVs, reduces $P(U_{1}=U_{2})$.
If we let, $W_{1},W_{2}$ be deterministic, the only way to make $U_{j}$ uniform
is to pool less likely symbols. However, the source is `highly' non-uniform, and
even by pooling \textit{all} the less likely symbols, we can gather a
probability, of at most, $\frac{1}{k}$. Consequently, any $p_{U_{1}U_{2}}$
induced via a S-L coding scheme is sufficiently far from any pmf that satisfies
$U_{1}=U_{2}$ whp \textit{and} $U_{1}=U_{2}$ close to uniform. When constrained
to a S-L coding technique, the shared channel - the main communication 
resource in communicating $\ulineS$ to the decoder - cannot be utilized 
efficiently leading to the incapability of communicating $\ulineS$ over the MAC 
and IC.
\end{remark}
\begin{remark}
 \label{Rem:Multi-LetterCodingScheme} An $l-$letter (multi-letter with $l>1$) 
coding scheme is constrained by an $l-$letter LMC 
$U_{1}^{l}-S_{1}^{l}-S_{2}^{l}-U_{2}^{l}$. Suppose we choose $l$ 
\textit{reasonably} large such that 1) $\xi^{[l]}$ is not high, and 2) 
$S_{j}^{l}$ is \textit{reasonably} uniform on its typical set 
$T_{\delta}^{l}(S_{j})$, and define $U_{j}:j \in [2]$ through 
\textit{identical} 
functions $U_{j}^{l}=g(S_{j}^{l}):j \in [2]$, then one can easily visualize the 
existence of $g$ such that $p_{U_{1}^{l}U_{2}^{l}}$ satisfies the twin 
objectives of $U_{1}^{l}=U_{2}^{l}$ whp and $U_{1}^{l}=U_{2}^{l}$ is close to 
uniform. Our coding scheme, will in fact, identify such $g$ maps. This portrays 
the sub-optimality of S-L schemes for joint source-channel coding.
\end{remark}
\subsection{Fixed B-L coding over isolated noiseless channels}
\label{SubSec:FixedB-LCodingOverIsolatedChnls}
We propose a coding techniques based on fixed B-L codes that enables $\ulineS$
to be communicated over the corresponding channels in Examples
\ref{Ex:DuecksExampleForMAC}, \ref{Ex:DuecksExampleForIC} (for all $a,k$
sufficiently large). The coding technique we propose for both examples are
identical. In the following, we describe the same in the context of Example
\ref{Ex:DuecksExampleForMAC}. Recall $\xi^{[l]}\define \xi^{[l]}(\ulineS), \xi 
\define \xi(\ulineS)$ and 
$\tau_{l,\delta } \define \tau_{l,\delta}(S_{1})$ throughout this section.

Our goal is to exploit the presence of near GKW parts to co-ordinate 
($U_{1}=U_{2}$) and communicate efficiently ($U_{1}=U_{2}$ is close to uniform) 
over the $\mathbb{W}_{Y_{0}|\ulineU}-$channel. We take a cue from Remark 
\ref{Rem:ConditionalCodingRevisited} and employ the same codes and mappings 
at both encoders to communicate on the latter channel. Specifically, we 
encode both $S_{1},S_{2}$ with the same source code, and map their output 
message indices identically to a common channel code that produces codewords for 
the $\mathbb{W}_{Y_{0}|\ulineU}-$channel. However, we note that as the B-L of 
these codes increase, the probability that the source code produces 
different message indices at the two encoders, increases. This is because 
the source blocks disagree with a probability $\xi^{[n]}= 
P(S_{1}^{n}\neq S_{2}^{n}) = 1-(1-\xi)^{n} \displaystyle\rightarrow 1$ as $n 
\rightarrow \infty$. We therefore \textit{fix} the B-L of these codes to $l$, 
irrespective of the desired probability of error. $l$ is chosen large enough 
such that the source can be \textit{reasonably} efficiently compressed, and yet 
small enough, to ensure $\xi^{[l]}$ is \textit{reasonably} small. We 
refer to these $l-$length blocks as \textit{sub-blocks}. Since $l$ is fixed, 
there is a non-vanishing probability that these source sub-blocks will be 
decoded erroneously. An outer code, operating on an arbitrarily large number $m$ 
of these sub-blocks, will carry information to correct for these `errors' and 
communicate rest of the necessary information. The outer code will operate over 
satellite channel $\mathbb{W}_{Y_{j}|X_{j}}$. We begin with a description of 
the fixed B-L (inner) codes.

We employ a simple fixed B-L (inner) code. Let $T_{\delta}^{l}(S_{1})$ be the 
source code, and let $C_{U}=\mathcal{U}^{l}$ be the channel code. Let $l\alpha 
=\lfloor\log a^{l}\rfloor$ bits, of the $\lceil\log |T_{\delta}^{l}(S_{1})| 
\rceil$ bits output by the source code, be mapped to $C_{U}$. Both encoders use 
the same source code, channel code and mapping. We reiterate that encoder $2$ 
also employs source code 
$T_{\delta}^{l}(S_{1})$, (and not $T_{\delta}^{l}(S_{2})$).

Suppose we communicate an \textit{arbitrarily large} number $m$ of these 
sub-blocks on $\ShrdChnl$ as above. Moreover, suppose encoder 
$1$ communicates the rest of the $l\beta = \lceil\log |T_{\delta}^{l}(S_{1})| 
\rceil-l\alpha$ bits output by its source code to Rx on its satellite channel 
$\mathbb{W}_{Y_{1}|X_{1}}$.\footnote{Through our description, we assume 
communication over $\mathbb{W}_{Y_{j}|X_{j}}$ is noiseless. In the end, we prove 
that the rate we demand of $\mathbb{W}_{Y_{j}|X_{j}}$ is lesser than its 
capacity, justifying this assumption.} How much more information needs to be 
communicated to Rx, to enable it reconstruct $\ulineS^{lm}$? We do a simple 
analysis that suggests a natural coding technique.

We employ a matrix notation in the sequel. View the $m$ sub-blocks of the source 
$S_{j}$ as the rows of the matrix $\bold{S}_{j}(1:m,1:l) \in \mathcal{S}_{j}^{m 
\times l}$. Let $\bold{\hatK}(1:m,1:l) \in \mathcal{S}_{1}^{ m \times l}$ denote 
Rx's reconstruction.\footnote{1) Encoder $j$ could input any arbitrary 
$C_{U}-$codeword when its sub-block $S_{j}^{l} \notin T_{\delta}^{l}(S_{1})$, 
and decoder $j$ could declare an arbitrary reconstruction when it observes 
$Y_{0}^{l}=0^{l}$. Our probability of error analysis handles these events.} The 
$m$ sub-blocks
\ifTITVersion\begin{equation}
 \label{Eqn:Sub-BlocksAreIID}
 \left\{ \left(\bold{S}_{j}(t,1:l),\bold{\hatK}(t,1:l):j=1,2\right):t \in [m] \right\}
\end{equation}\fi
\ifPeerReviewVersion\begin{equation}
 \label{Eqn:Sub-BlocksAreIID}
 \left\{ \left(\bold{S}_{j}(t,1:l),\bold{\hatK}(t,1:l):j=1,2\right):t \in [m] 
\right\}
\end{equation}\fi
are IID with an $l-$length distribution $\mathbb{W}_{S_{1}^{l}S_{2}^{l}} 
p_{\hatK^{l}|S_{1}^{l}S_{2}^{l}} = \mathbb{W}_{S_{1}S_{2}}^{l} 
p_{\hatK^{l}|S_{1}^{l}S_{2}^{l}}$. This suggests 
that we can treat the $l-$length sub-blocks as \textit{super-symbols} and employ 
a standard binning technique. It suffices for encoder $j:j \in [2]$ to send 
$H(S^{l}_{j}|\hat{K}^{l},S^{l}_{\msout{j}})$ bits \textit{per} source 
\textit{sub-block}, so long as their sum rate is at least 
$H(S_{1}^{l},S_{2}^{l}|\hatK^{l})$. We do 
not have a characterization of $p_{\hatK^{l}|S_{1}^{l}S_{2}^{l}}$ and we 
therefore derive an upper bound. We have
\ifPeerReviewVersion\begin{eqnarray}
  \label{Eqn:DueckExExUpperBoundOnBinningRates}
%==============newline
 \lefteqn{H(S_{j}^{l}|\hat{K}^{l},S_{\msout{j}}^{l}) \leq 
H(S_{j}^{l},\mathds{1}_{\{\hat{K}^{l} \neq 
S_{1}^{l}\}}|\hat{K}^{l},S_{\msout{j}}^{l}) \leq  h_{b}(P(\hat{K}^{l} \neq 
S_{1}^{l}))+H(S_{j}^{l}|\hat{K}^{l},S_{\msout{j}}^{l},\mathds{1}_{\{ 
\hat{K}^{l} \neq S_{1}^{l}\}}) } \\
%==============newline
&\leq&h_{b}(P(\hat{K}^{l} \neq 
S_{1}^{l}))+P(\hat{K}^{l} \neq S_{1}^{l})\log 
|\mathcal{S}_{j}^{l}|+P(\hat{K}^{l} = 
S_{1}^{l})H(S_{j}^{l}|S_{1}^{l},S_{\msout{j}}^{l})\nonumber\\
%==============newline
&\leq&  l\mathcal{L}_{l}(P(\hat{K}^{l} \neq 
S_{1}^{l}),|\mathcal{S}_{j}|)+lH(S_{j}|S_{1},S_{\msout{j}})
\label{Eqn:DueckExExUpperBoundOnIndividualBinningRate}
,\mbox{ for }j \in [2], \mbox{ and}\\
 \label{Eqn:DueckExExUpperBoundOnSumBinningRate}
%==============newline
\lefteqn{H(\ulineS^{l}|\hat{K}^{l}) \leq 
H(\ulineS^{l},\mathds{1}_{\{\hat{K}^{l} \neq 
S_{1}^{l}\}}|\hat{K}^{l}) \leq h_{b}(P(\hat{K}^{l} \neq 
S_{1}^{l}))+H(\ulineS^{l}|\hat{K}^{l},\mathds{1}_{\{ 
\hat{K}^{l} \neq S_{1}^{l}\}}) } \nonumber\\
%==============newline
&\leq&h_{b}(P(\hat{K}^{l} \neq 
S_{1}^{l}))+P(\hat{K}^{l} \neq S_{1}^{l})\log 
|\ulineCalS^{l}|+P(\hat{K}^{l} = 
S_{1}^{l})H(\ulineS^{l}|S_{1}^{l})\nonumber\\
%==============newline
\label{Eqn:DueckExExActUpperBoundOnSumBinningRate}
&\leq&  l\mathcal{L}_{l}(P(\hat{K}^{l} \neq 
S_{1}^{l}),|\ulineCalS|)+lH(S_{2}|S_{1}),\mbox{ where 
}\mathcal{L}_{l}(\cdot,\cdot) \mbox{, as defined in 
(\ref{Eqn:AdditionalSourceCodingInfo})}
\end{eqnarray}\fi
\ifTITVersion\begin{eqnarray}
 H(S_{j}^{l}|\hat{K}^{l}) \leq H(S_{j}^{l},\mathds{1}_{\{\hat{K}^{l} \neq 
S_{1}^{l}\}}|\hat{K}^{l}) \leq h_{b}(P(\hat{K}^{l} \neq S_{1}^{l}))+\nonumber\\
 \label{Eqn:BoundOnAddInformation}
\!\!\!+P(\hat{K}^{l} \neq S_{1}^{l})\log |\mathcal{S}_{j}^{l}|+P(\hat{K}^{l} = 
S_{1}^{l})H(S_{j}^{l}|S_{1}^{l}).\!\!\!\\
\leq l\mathcal{L}_{l}(P(\hat{K}^{l} \neq 
S_{1}^{l}),|\mathcal{S}_{j}|)+lH(S_{j}|S_{1}),\mbox{ where}\nonumber\\
\label{Eqn:AdditionalInformationToGoOnSatelliteChannels}
 \mathcal{L}_{l}(\phi,|\mathcal{K}|) \define \frac{1}{l}h_{b}(\phi) + 
\phi\log|\mathcal{K}|~~~~~~~
\end{eqnarray}\fi
represents the additional source coding rate needed to 
compensate for the errors in the fixed B-L decoding. 

It suffices to prove that the above rates are supported by the satellite 
channels. Specifically, it suffices to prove
\ifPeerReviewVersion\begin{eqnarray}
 \label{Eqn:SatelliteChannelsSupportTheRatesCondition}
\mathcal{L}_{l}(P(\hat{K}^{l} \neq 
S_{1}^{l}),|\mathcal{S}_{j}|)+\beta\mathds{1}_{\{ 
j=1\}}+H(S_{j}|S_{1})\mathds{1}_{\{ j=2\}}
 \leq
 \mathcal{C}_{M}+h_{b}(\frac{2}{ka^{\eta k}})\mathds{1}_{\{ j=2\}}
 + [h_{b}(\frac{2}{k})+\frac{1}{k}\log a]\mathds{1}_{\{j=1\}},\\
 \label{Eqn:DueckExampleSumRateCondn}
 \mathcal{L}_{l}(P(\hat{K}^{l} \neq 
S_{1}^{l}),|\ulineCalS|)+\beta +H(S_{2}|S_{1}) \leq 
2\mathcal{C}_{M}+h_{b}(\frac{2}{ka^{\eta k}})
 + [h_{b}(\frac{2}{k})+\frac{1}{k}\log a]
\end{eqnarray}\fi
\ifTITVersion\begin{equation}\scalebox{0.91}{$
 \label{Eqn:SatelliteChannelsSupportTheRatesCondition}
 \begin{array}{l}
 \!\!\!\!\!\mathcal{L}_{l}(P(\hat{K}^{l} \neq 
S_{1}^{l}),|\mathcal{S}_{j}|)\\\!\!\!\!\!+\beta\mathds{1}_{\{ 
j=1\}}+H(S_{j}|S_{1})\mathds{1}_{\{ j=2\}}
 \end{array} \!\!\!\leq
 \begin{array}{c}
 \mathcal{C}_{M}+h_{b}(\frac{2}{ka^{\eta k}})\mathds{1}_{\{ j=2\}}\\
 + [h_{b}(\frac{2}{k})+\frac{1}{k}\log a]\mathds{1}_{\{j=1\}},
 \end{array}\!\!\!\!\!\!$}
\end{equation}\fi
where the RHSs in (\ref{Eqn:SatelliteChannelsSupportTheRatesCondition}), 
(\ref{Eqn:DueckExampleSumRateCondn}) are the capacities of 
$\mathbb{W}_{Y_{j}|X_{j}}$ and the `MAC' $\mathbb{W}_{Y_{1}Y_{2}|X_{1}X_{2}}$ 
comprised of the two satellite channels $\mathbb{W}_{Y_{1}|X_{1}}, 
\mathbb{W}_{Y_{2}|X_{2}}$, respectively. Since
$\mathcal{L}_{l}(\mu,|\mathcal{K}|)$ is non-decreasing in $\mu$ if $\mu \leq
\frac{1}{2}$, we bound $P(\hat{K}^{l} \neq S_{1}^{l})$ by a quantity that is
less than $\frac{1}{2}$, and substitute the same to derive an upper bound on
$\mathcal{L}_{l}(P(\hat{K}^{l} \neq S_{1}^{l}),|\mathcal{S}_{j}|)$. Towards
that end, note that $\{ S_{1}^{l} \neq \hat{K}^{l}\} \subseteq \{ S_{1}^{l}\neq
S_{2}^{l}\} \cup \{ S_{1}^{l} \notin T_{\delta}^{l}(S_{1})\}$. Indeed,
$S_{1}^{l}=S_{2}^{l} \in T_{\delta}^{l}(S_{1})$ implies both encoders input same
$C_{U}-$codeword and agree on the $l\beta$ bits communicated by encoder 1. 
Therefore
$P( S_{1}^{l} \neq \hat{K}^{l})\leq \phi$, where 
$\phi=\xi^{[l]}+\tau_{l,\delta}$, 
\ifTITVersion\begin{eqnarray}\!\!\!\!\!\!\tau_{l,\delta} = 1
2|\mathcal{S}_{}|\exp \left\{ -2\delta^{2}p_{S_{1}}^{2}(a^{*})l \right\} 
\label{Eqn:BoundOnTau}
\leq 2a^{k}\exp\{ -\frac{\delta^{2}l}{2k^{2}a^{2k}}\}\mbox{ and }\xi^{[l] } \leq \frac{l}{ka^{\eta k}}.
\end{eqnarray}\fi
\ifPeerReviewVersion\begin{eqnarray}\!\!\!\!\!\!\tau_{l,\delta} = 
2|\mathcal{S}|\exp 
\left\{ -2\delta^{2}p_{S_{1}}^{2}(a^{*})l \right\} 
\label{Eqn:BoundOnTau}
\leq 2a^{k}\exp\{ -\frac{\delta^{2}l}{2k^{2}a^{2k}}\}\mbox{ and }\xi^{[l] } \leq 
\frac{l}{ka^{\eta k}}.
\end{eqnarray}\fi
Choose $l= k^{4}a^{\frac{\eta k}{2}}, \delta = \frac{1}{k}$, 
substitute in
(\ref{Eqn:BoundOnTau}) and verify
\begin{eqnarray}
 \label{Eqn:BoundOnTauAndXi}
 \tau_{l,\delta} \leq 2a^{k} \exp \left\{ -\frac{1}{2}a^{\left( 
\frac{\eta}{2}-2  \right)k}  \right\}, \xi^{[l]} \leq 
\frac{k^{3}}{a^{\frac{\eta k}{2}}}. \mbox{ Since }\eta\geq 6,\mbox{ we have 
}\phi = \xi^{[l]}+\tau_{l,\delta} \leq 2k^{3}a^{-\frac{\eta k}{2}}<\frac{1}{2} 
\mbox{ for sufficiently}
\end{eqnarray}
large $a,k$. Verify
\ifTITVersion\begin{equation}\label{Eqn:FinalBoundOnExtraSourceCodingRate}
\mathcal { L } ^ { S } _ { l }
(2k^{3}a^{-5k},|\mathcal{S}_{j}|) \leq \alpha
(h_{b}(\alpha)+\log a),\mbox{ with }\alpha =
\frac{8k^{4}}{a^{\frac{\eta k}{3}}}\end{equation}\fi
\ifPeerReviewVersion\begin{equation}
\label{Eqn:FinalBoundOnExtraSourceCodingRate}
\mathcal { L } ^ { S } _ { l }
(2k^{3}a^{-\frac{\eta}{2}k},|\mathcal{S}_{j}|) \leq \alpha
(h_{b}(\alpha)+\log a),\mbox{ with }\alpha =
\frac{8k^{4}}{a^{\frac{\eta k}{3}}}\end{equation}\fi
for sufficiently large 
$a,k$.
Substituting $\delta=\frac{1}{k}$, verify\footnote{Use $H(S_{1})\leq \log a
+h_{b}(\frac{1}{k})$ and $|T_{\delta}(S_{1})|\leq \exp\{l(1+\delta)H(S_{1})\}$.}
 \ifTITVersion\begin{equation}
 \label{Eqn:UpperBoundOnB}
 \beta \leq (2/l)+(1/k)\log a + (1+(1/k))h_{b}(1/k).
 \end{equation}\fi
\ifPeerReviewVersion\begin{equation}
 \label{Eqn:UpperBoundOnB}
 \beta \leq (2/l)+(1/k)\log a + (1+(1/k))h_{b}(1/k).
 \end{equation}\fi
 Since $h_{b}(\frac{2}{k})-(1+\frac{1}{k})h_{b}(\frac{1}{k}) \geq 
\frac{1}{2k}\log\frac{k}{256}$ for large enough $k$, we have $\beta\leq 
h_{b}(\frac{2}{k})+\frac{5}{4k}\log a$ for sufficiently large $a,k$. Lastly, 
note that $H(S_{2}|S_{1})\leq h_{b}(\frac{1}{(k-1)a^{\eta k}})+\frac{2}{a^{\eta 
k}}\log a \leq h_{b}(\frac{2}{ka^{\eta k}})+\frac{1}{4k}\log a$ for sufficiently 
large $a,k$. The validity of 
(\ref{Eqn:SatelliteChannelsSupportTheRatesCondition}), 
(\ref{Eqn:DueckExampleSumRateCondn}) for sufficiently large 
$a,k$, can now be verified by substituting 
(\ref{Eqn:FinalBoundOnExtraSourceCodingRate}) and the above derived bounds.

A few details with regard to the above coding technique is worth
mentioning. $p_{\hatK_{1}^{l}\hatK_{2}^{l}|S_{1}^{l}S_{2}^{l}}$ can in principle
be computed, once the fixed B-L codes, encoding and decoding maps are
chosen. $S_{j}^{lm}$ will be binned at rate $H(S_{j}^{l}|\hatK_{j}^{l})$ and the
 decoder can employ a joint-typicality based decoder using the computed
$p_{S_{j}^{l}|\hatK_{j}^{l}}$. With respect to the MAC problem, we conclude the
following.

\begin{thm}
 \label{Thm:StrictSub-OptimalityOFCESCondns}
The CES conditions stated in \cite[Thm. 1]{198011TIT_CovGamSal} are not 
necessary. 
\end{thm}

An analogous statement holds with regard to the IC problem.

\begin{thm}
 \label{Thm:StrictSub-OptimalityOfLC}
 The LC conditions stated in \cite[Thm. 1]{201112TIT_LiuChe} are not 
necessary. Refer to Example \ref{Ex:DuecksExampleForIC}. There exists $a^{*} \in 
\naturals$ and $k^{*}\in \naturals$ such that for any $a\geq a^{*}$ and any $k 
\geq k^{*}$, $S_{1},S_{2}$ and the IC 
$\mathbb{W}_{\underline{Y}|\underline{X}\underline{U}}$ do not satisfy LC 
conditions, and yet, $\ulineS$ is transmissible over IC 
$\mathbb{W}_{\underline{Y}|\underline{X}\underline{U}}$.
\end{thm}
\begin{IEEEproof}
In view of Lemma \ref{Lem:Ex2DoesNOTSatisfyLCConditions}, we only need to prove
the latter statement. A coding technique identical to that proposed for Example
\ref{Ex:DuecksExampleForMAC} works. Following are the only differences. Each
encoder has to communicate the $l\beta$ bits, corresponding to each source
sub-block, to its decoder on the satellite channel. The decoder reconstructions
$\hat{\boldsymbol{K}}_{j}(t,1:l) \in \mathcal{S}_{j}^{l}: j \in [2]$
corresponding to the $t-$th source sub-block are not-necessarily identical.
Following the above sequence of steps, it can be verified that decoder $j$ needs
to be communicated
 \ifTITVersion\begin{equation}
 \label{Eqn:ICDecoderjNeedsToBeCommunicated}
 \mathcal{L}_{l}(P(\hat{K}_{j}^{l} \neq 
S_{1}^{l}),|\mathcal{S}_{j}|)+\beta+H(S_{j}|S_{1})\nonumber
 \end{equation}\fi
\ifPeerReviewVersion\begin{equation}
 \label{Eqn:ICDecoderjNeedsToBeCommunicated}
 \mathcal{L}_{l}(P(\hat{K}_{j}^{l} \neq 
S_{1}^{l}),|\mathcal{S}_{j}|)+\beta+H(S_{j}|S_{1})\nonumber
 \end{equation}\fi
units per channel use over the satellite channel 
$\mathbb{W}_{Y_{j}|X_{j}}$.
Substituting
the above choice of $l= k^{4}a^{\frac{\eta k}{2}}, \delta = \frac{1}{k}$, it
can be verified that bounds (\ref{Eqn:BoundOnTau}),
(\ref{Eqn:FinalBoundOnExtraSourceCodingRate}),
(\ref{Eqn:UpperBoundOnB}) hold and the above quantity is
dominated by the capacity of
$\mathbb{W}_{Y_{j}|X_{j}}$ for large enough $a,k$. The reader is referred to
\cite[Section III. A]{201706ISIT_Pad-IC} where the details are provided.
\end{IEEEproof}

\begin{remark}
\label{Rem:PerformanceWithl} The coding techniques proposed in this section
crucially relies on the choice of $l$ being neither too big, nor too small. This
is elegantly captured as follows. As $l$ increases,
$\xi^{[l]}(\ulineS)\rightarrow 1$, $\tau_{l,\delta}\rightarrow 0$.
As $l$ decreases, $\xi^{[l]}(\ulineS) \rightarrow \xi(\ulineS)$, and
$\tau_{l,\delta}\rightarrow 1$. If $\phi \rightarrow 0.5$,
$\mathcal{L}_{l}(\phi,|\mathcal{S}_{j}|)\rightarrow
0.5\log|\mathcal{S}_{j}|=\frac{k}{2}\log a$.
\end{remark}

The core idea of the fixed B-L coding is to let each symbol $U_{j}:j \in [2]$ 
input on the $\mathbb{W}_{Y_{0}|\ulineU}-$channel be determined by a fixed 
number $l$ of source symbols. $l$ is neither too big, nor too small, and most 
importantly remains fixed irrespective of the desired probability of error. In 
Sections \ref{Sec:FBLCodingOverMACAndICStep1}, 
\ref{Sec:FBLCodingOverMACAndICStep2}, we generalize the coding scheme proposed 
herein for a general problem instance.

\section{Fixed B-L coding over arbitrary MAC and IC Step 1 : Separate Decoding}
\label{Sec:FBLCodingOverMACAndICStep1}
The examples provided in Section \ref{Sec:DuecksExample} and the fixed 
B-L coding scheme proposed in Section 
\ref{SubSec:FixedB-LCodingOverIsolatedChnls} demonstrate the central idea of 
fixed B-L coding. Generalizing this scheme for an arbitrary problem instance 
involving arbitrarily correlated sources and general MAC or IC involves 
fundamental challenges. These 
challenges arise from the fact that two information streams - fixed B-L and 
$\infty-$B-L (i.e, B-L's arbitrarily large, chosen as a function of the 
desired probability of error) - have to be multiplexed through a single 
channel. In other words, a generic MAC or IC does not provide isolated channels 
to communicate fixed B-L and $\infty-$B-L codes separately, as was provided in 
Examples \ref{Ex:DuecksExampleForMAC}, \ref{Ex:DuecksExampleForIC}. Evidently, 
most of the new elements in the following sections concern the channel coding 
module.

Following are three primary challenges. Firstly, how does one multiplex 
codes of different B-Ls, particularly with one of them being fixed to B-L $l$, 
in a way that permits performance characterization via S-L expressions? The 
second and third challenges concerns performance characterization. Performance 
characterization requires an explicit description of the pmf induced on the 
associated alphabets by the coding technique. In conventional S-L coding 
schemes via IID codebooks, this can be obtained as a particular combination of 
the marginal S-L pmfs chosen for each of the codebooks. Refer to Section 
\ref{SubSec:FixedB-LCodingOverIsolatedChnls} and note that we are unable to 
characterize the joint $l-$letter pmf of the pair of $C_{U}-$codewords chosen 
by the two encoders and hence we do not have an explicit 
characterization of the joint $l-$letter pmf of the inputs on 
$\mathcal{U}^{l}\times \mathcal{U}^{l}$.\footnote{This ignorance did not 
inhibit us since $\mathcal{U}-\mathcal{Y}_{0}$-channel is isolated and we could 
compute the performance of $C_{U}$ with bare hands.} In the coding scheme we 
propose, we will be unable to characterize the joint pmf of the message indices 
output by the fixed B-L source code, and moreover, since we pick an 
off-the-shelf constant composition code for a generic channel, we do not have a 
characterization of its codewords. The natural question is : How do we 
characterize the information-theoretic performance of the proposed coding 
scheme? Finally, how do we characterize performance of a multi-letter coding 
scheme via a S-L expression?

These challenges, not encountered in previous work, will require new tools. To 
facilitate a step-by-step description of the new elements, we present our 
coding theorems in two steps. In the first step, presented in this section, we 
analyze separate decoding of the fixed B-L and $\infty-$B-L information 
streams. In the second step, presented in Section 
\ref{Sec:FBLCodingOverMACAndICStep2}, we analyze conditional decoding of the 
$\infty-$B-L information stream. In addition, for our first main coding theorem 
(Theorem \ref{Thm:MACStep1}), we provide an outline of the coding scheme and 
the analysis which will aid the reader recognize the outline of the new tools.
\subsection{MAC Problem}
\label{SubSec:MACStep1GeneralizationSeparateDecoding}
We state and prove our first set of sufficient conditions for the MAC problem. 
In Theorem \ref{Thm:FBLBeatsCES}, we prove these conditions are strictly weaker 
than CES 
conditions.
\begin{thm}
 \label{Thm:MACStep1}
A pair of sources $(\ulineCalS,\mathbb{W}_{\ulineS})$ is transmissible over a 
MAC $(\ulineCalX,\outset,\mathbb{W}_{\Out|\ulineX})$ if there exists 
\begin{enumerate}
 \item[(i)] finite sets 
$\mathcal{K},\mathcal{U},\mathcal{V}_{1},\mathcal{V}_{2}$,
 \item[(ii)] maps $f_{j}:\mathcal{S}_{j}\rightarrow \mathcal{K}$, 
with $K_{j}=f_{j}(S_{j})$ for $j \in [2]$,
\item[(iii)] $\alpha,\beta \geq 0$, $\rho > 0 $, $\delta > 0$, 
\item[(iv)] $l \in 
\naturals, l \geq l^{*}(\rho,\mathcal{U},\mathcal{Y})$, where 
$l^{*}(\cdot,\cdot,\cdot)$ is defined in (\ref{Eqn:DefnOfLStar}),
\item[(v)] pmf $p_{U}p_{V_{1}}p_{V_{2}}p_{X_{1}|UV_{1}}p_{X_{2}|UV_{2}} 
\mathbb{W}_{Y|\ulineX}$ defined on $\mathcal{U}\times 
\ulineCalV\times\ulineCalX\times \mathcal{Y}$, where $p_{U}$ is a type of 
sequences in $\mathcal{U}^{l}$, such that
\end{enumerate}\vspace{-0.15in}
\ifPeerReviewVersion
 \begin{eqnarray}
  (1+\delta)H(K_{1}) &<& \alpha+\beta ,\nonumber\\ 
 \label{Eqn:StepIMACIndvdualRateBound}
H(S_{j}|S_{\msout{j}},K_{1}) + 
\mathcal{L}_{l}(\phi,|\mathcal{S}_{j}|) &<& 
I(V_{j};\Out|V_{\msout{j}})-\mathcal{L}(\phi,|\mathcal{V}_{j}|)
\mbox{ for }j \in [2]\mbox{ and} \\
\label{Eqn:StepIMACSumRateBound}
\beta + H(\ulineS|K_{1})+ 
\mathcal{L}_{l}(\phi,|\ulineCalS|) &<& I(\ulineV;Y) - 
\mathcal{L}(\phi,|\ulineCalV|),\\
\lefteqn{
\!\!\!\!\!\!\!\!\!\!\!\!\!\!\!\!\!\!\!\!\!\!\!\!\!\!\!\!\!\!\!\!\!\!\!\!\!\!\!\!
\!\!\!\!\!\!\!\!\!\!\!\!\!\!\!\!\!\!\!\!\!\!\!\!\!\!\!\!\!\!\!\!\!\!\!\!\!\!\!\!
\!\!\!\!\!\!\!\!\!\!\!\!\!\!\!\!\!\!\!\!\!\!\!\!\!\!\!\!\!\!\!\!\!\!\!
\phi \in [0,0.5)\mbox{ where }\phi \define 
g(\alpha+\rho,l)+\xi^{[l]}(\ulineK)+\tau_{l,\delta}(K_{1}),~ g(R,l) 
\define 
(l+1)^{2|\mathcal{U}||\mathcal{Y}|}\exp\{-lE_{r}(R,p_{U},p_{Y|U})\}}
\label{Eqn:Step1MACLastBnd}
\end{eqnarray}
$\mathcal{L}_{l}(\cdot,\cdot), \mathcal{L}(\cdot,\cdot)$ is as defined 
in (\ref{Eqn:AdditionalSourceCodingInfo}).\fi
\ifTITVersion
 \begin{eqnarray}
  (1+\delta)H(K_{1}) \!\!\!\!&\leq&\!\!\!\! \alpha+\beta ,\nonumber\\ 
\label{Eqn:StepIMACIndvdualRateBound}
H(S_{j}|S_{\msout{j}},K_{1}) + 
\mathcal{L}_{l}(\phi,|\mathcal{S}_{j}|) \!\!\!\!&<&\!\!\!\!
I(V_{j};\Out|V_{\msout{j}})-\mathcal{L}(\phi,
|\mathcal{V}_{j} |)\nonumber\\&&\!\!\!\!~~~
\mbox{ for }j \in [2]\mbox{ and} \\
\label{Eqn:StepIMACSumRateBound}
\beta + H(\ulineS|K_{1})+ 
\mathcal{L}_{l}(\phi,|\ulineCalS|) \!\!\!\!&<&\!\!\!\! I(\ulineV;Y) - 
\mathcal{L}(\phi,|\ulineCalV|),
\end{eqnarray}
$\phi \in [0,0.5)$ where
\begin{eqnarray}\phi &\define &
g_{\rho,l}+\xi^{[l]}(\ulineK)+\tau_{l,\delta}(K_{1}),\nonumber\\
g(\alpha+\rho,l) &\define &
\exp\{-l(E_{r}(\alpha+\rho,p_{U},p_{Y|U})-\rho)\}\nonumber .\end{eqnarray}\fi
\end{thm}
\begin{remark}
 \label{Rem:Single-LetterCharacterization}
 The characterization provided here and those 
in Thms. \ref{Thm:ICStep1} 
\ref{Thm:ICStep2}, \ref{Thm:MACStep2} is via S-L PMFs and S-L expressions.
\end{remark}
\begin{remark}
\label{Rem:ClarifyingNearGKWPart}
In this Thm. \ref{Thm:MACStep1} as in the rest of Thms. 
\ref{Thm:ICStep1}, \ref{Thm:ICStep2}, \ref{Thm:MACStep2}, $K_{j}=f_{j}(S_{j}): 
j \in [2]$ can be arbitrary functions of $S_{j}: j \in [2]$. For 
comprehension, it helps to visualize $K_{1},K_{2} \in 
\mathcal{K}$ as the \textit{near GKW parts} of the sources $\ulineS$. We 
informally refer to $K_{1},K_{2}$ here and in proof of Theorem 
\ref{Thm:ICStep2} as near GKW parts of $\ulineS$. This is only to aid 
intuition.
\end{remark}

We provide an informal description of the coding scheme and outline the main 
steps in the analysis. The latter serves as a high level 
justification/explanation for the bounds (\ref{Eqn:StepIMACIndvdualRateBound}) - 
(\ref{Eqn:Step1MACLastBnd}). A formal 
proof follows the outline.

\textit{Outline of the coding scheme :} Let $\mathcal{K},K_{1},\cdots,\rho$ be 
provided as in theorem statement. $K_{1},K_{2} \in \mathcal{K}$ represent the 
near GKW parts of the sources 
$\ulineS$. The rest of the parameters will be described as and when they appear. 
The coding scheme we propose 
is designed to exploit the presence of near GKW parts and is, at an 
architectural level identical/similar to that proposed Section 
\ref{SubSec:FixedB-LCodingOverIsolatedChnls}. The CES technique of 
coding the GKW part via a common code, informally referred to as GKW-coding, is 
employed to code the near 
GKW parts $K_{1},K_{2} \in \mathcal{K}$. Specifically, a common source code 
encodes $K_{1},K_{2}$. The index output by this source code is mapped to a 
common channel code 
$C_{U}$ built over $\mathcal{U}$ - the auxiliary alphabet set provided in the 
theorem statement. Both encoders employ identical maps. The common source code 
is chosen to be a good (lossless) source code for $K_{1}$. $C_{U}$ is chosen 
to be a good PTP constant composition channel code of type $p_{U}$ (Theorem 
\ref{Thm:ConstCompnCodesForMAC}) for the induced PTP $\mathcal{U} - 
\mathcal{Y}$. As in GKW-coding, the receiver, in an attempt to recover a common 
message, employs the 
(PTP) decoder of $C_{U}$. This decoded message is input to the decoder of the 
(common) source code whose output serves as a `reconstruction' of the near GKW 
parts.

Increasing the B-L of this common source and channel code has a detrimental 
effect. Note that the effectiveness of GKW-coding 
crucially relies on both encoders choosing the same $C_{U}-$codeword. As the 
B-L $n \rightarrow \infty$, the 
$n-$length $K_{1},K_{2}$ blocks disagree with probability $\xi^{[n]}(\ulineK) 
=1-(1-\xi(\ulineK))^{n}$ that increases to $1$, resulting in different 
indices output at the two encoders by the common source code. We therefore 
\textit{fix} the B-L of the common source and 
channel codes operating over the near GKW parts, to $l$, irrespective 
of the desired probability of error. $l$ is as 
provided in the theorem statement. These $l-$length 
blocks are referred to as \textit{sub-blocks}. We emphasize that the messages 
output by the source code corresponding to different sub-blocks are 
\textit{not} pooled together, but instead communicated separately by mapping 
them to codewords from $C_{U}$ (that are also of B-L $l$). The mapping from 
these message indices to $C_{U}$ are identical at both encoders and across the 
sub-blocks.

The above \textit{fixed B-L} coding, owing to its B-L being fixed, results in 
sub-block errors. Information, necessary to correct for these sub-block errors, 
and moreover to recover the sources, needs to 
be communicated. We propose an \textit{outer code} operating over an 
arbitrarily large number $m$ of these sub-blocks to communicate rest of the 
necessary information. This outer code, also referred to as the 
\textit{$\infty-$B-L} code, is a simple separation based code involving a 
Slepian-Wolf distributed (lossless) source encoder followed by a MAC channel 
code built over input alphabets $\mathcal{V}_{1},\mathcal{V}_{2}$ of the 
induced MAC $\mathcal{V}_{1},\mathcal{V}_{2}-\mathcal{Y}$. Here, 
$\mathcal{V}_{1},\mathcal{V}_{2}$ are auxiliary alphabet sets as provided in 
the theorem statement. The Slepian Wolf decoder, in an attempt to recover the 
pair of $m$ sub-blocks of the source, utilizes the 
reconstructions of the near GKW parts, output by the fixed B-L decoding, as side 
information. Since the outer code operates over multiple sub-blocks of the 
inner code, we encounter two challenges. We employ a matrix notation to 
describe these and the design of the outer code to overcome the same.

View the $m$ sub-blocks of the source $S_{j}$, near GKW parts $K_{j}$ as 
the rows of the matrix $\bold{S}_{j} \in \mathcal{S}_{j}^{m \times 
l}$, $\boldK_{j} \in \mathcal{K}^{m \times l}$. Let 
$\bold{\hatK} \in \mathcal{K}^{ m \times l}$ denote decoder's 
reconstruction. The reconstruction $\boldhatK(t,1:l)$, being the 
output of an $l-$length coding 
scheme, is not IID. By coding the sub-blocks separately and 
identically, we ensure the $m$ sub-block reconstructions $\boldhatK(t,1:l): t 
\in [m]$ to be IID with an $l-$letter pmf. This suggests treating each sub-block 
as a \textit{super-symbol} and employ a Slepian-Wolf code operating over $m$ 
super-symbols. Our Slepian-Wolf source encoder partitions 
$\mathcal{S}_{j}^{lm}$ into $2^{mR_{j}}$ bins, treating each $l-$length 
sub-block as \textit{super-symbols}, and communicates the bin index of 
$\boldS_{j}$ to the decoder via the MAC channel code. A joint-typicality based 
decoder finds within the indexed pair of bins, a (unique) pair of $lm-$length 
source sequences that are jointly typical with the $m$ reconstructed sub-blocks 
$\boldhatK(t,1:l) : t \in [m]$.

The second challenge concerns multiplexing a codeword of the outer MAC code 
with $m$ codewords of $C_{U}$. If a single codeword from the former code is 
multiplexed with $m$ codewords of $C_{U}$, it experiences a channel with 
$l-$length memory. Since we seek an efficient technique based on S-L codes and 
a S-L characterization, we seek sub-vectors of this block of $lm$ symbols that 
are IID. The idea is to multiplex codewords of the outer MAC code along these 
sub-vectors, so that these codewords experience a memoryless channel. We are 
led to the elegant technique of interleaving devised by Shirani and Pradhan 
\cite{201406ISIT_ShiPra} in the related work of distributed source coding. 
Let rows of $\boldU_{j}$ denote codewords of $C_{U}$ obtained by encoding 
corresponding rows of $\boldK_{j}$ via the fixed B-L source encoder and 
$C_{U}$. Since the $m$ sub-blocks $\boldK_{j}(t,1:l): t \in [m]$ are separately 
and identically coded, the $m$ pairs 
$\left(\boldU_{1}(t,1:l),\boldU_{2}(t,1:l) \right): t \in [m]$, that 
constitute rows of $\boldU_{1},\boldU_{2}$, are IID with an $l-$letter pmf 
$p_{U_{1}^{l}U_{2}^{l}}$. If one were to randomly, independently and uniformly 
choose column numbers $\Pi_{1},\cdots \Pi_{m} \in [l]$ from each of the rows, 
then the $m$ pairs
$\left( \boldU_{1}(t,\Pi_{t}),\boldU_{2}(t,\Pi_{t})\right) : t \in [m] $ are 
IID $p_{\mathscr{U}_{1}\mathscr{U}_{2}} \define \sum_{i=1}^{l} p_{U_{1i}U_{2i}} 
$ (Lemma \ref{Lem:SimpleInterleavingLemma}). This 
leads us to the following idea. Suppose $\Pi_{t} : 
[l]\rightarrow [l]: t \in [m]$ is a collection of $m$ random 
independent and uniformly chosen surjective maps, then for every $i \in [l]$, 
the sub-vector 
$\left(\boldU_{1}(t,\Pi_{t}(i)),\boldU_{2}(t,\Pi_{t}(i)) : t \in [m]\right)$ has 
pmf 
$\prod_{t=1}^{m}p_{\mathscr{U}_{1}\mathscr{U}_{2}}$. One 
can therefore multiplex codewords 
chosen from $l$ \textit{different} outer MAC codes with these $l$ 
sub-vectors and guarantee 
that each codeword experiences a memoryless channel.

\textit{Outline of the analysis :} Conceptually, our analysis has three parts. 
The first part 
involves 
characterizing/quantifying a lower bound on the amount of information that is 
communicated via fixed B-L codes. This involves characterizing an upper 
bound $\phi$ on
\ifPeerReviewVersion\begin{eqnarray}
\label{Eqn:FixedBLMappingErrorEvent}
P\left(\boldhatK(t,1:l) \neq 
\boldK_{1}(t,1:l) \right).
\end{eqnarray}\fi
\ifTITVersion\begin{eqnarray}
\label{Eqn:FixedBLMappingErrorEvent}
 P\left(\boldhatK(t,1:l) \neq \boldK_{1}(t,1:l)
\right).\end{eqnarray}\fi
Since
\ifPeerReviewVersion
\begin{eqnarray}
\phi = P\left(\!\!\!\begin{array}{c}
               \mbox{Source code output corresponding 
to}\\K_{1}^{l},K_{2}^{l}\mbox{ disagree}
              \end{array}\!\!
  \right)+
\label{Eqn:TwoTermsMakeUpPhi}
  P\left( 
\!\!\begin{array}{c}
               \mbox{Common message index}\\\mbox{incorrectly 
decoded}\\\mbox{by }C_{U}-\mbox{code decoder}
              \end{array}  \!\!\!\left|\begin{array}{c}
               \mbox{Source code output}\\\mbox{corresponding 
to }\\K_{1}^{l},K_{2}^{l}\mbox{ agree}
              \end{array}\!\!\right.\right)~~
\end{eqnarray}
\fi
\ifTITVersion
\begin{eqnarray}
\lefteqn{\phi = P\left(\!\!\!\begin{array}{c}
               \mbox{Source code output corresponding 
to}\\K_{1}^{l},K_{2}^{l}\mbox{ disagree}
              \end{array}\!\!
  \right)+}\nonumber\\ 
\label{Eqn:TwoTermsMakeUpPhi}
  &&\!\!\!\!\!\!\!\!\!\!\!\!P\left( 
\!\!\begin{array}{c}
               \mbox{Common message index}\\\mbox{incorrectly 
decoded}\\\mbox{by }C_{U}-\mbox{code decoder}
              \end{array}  \!\!\!\left|\begin{array}{c}
               \mbox{Source code output}\\\mbox{corresponding 
to }\\K_{1}^{l},K_{2}^{l}\mbox{ agree}
              \end{array}\!\!\right.\right)~~
\end{eqnarray}
\fi
is an upper bound on (\ref{Eqn:FixedBLMappingErrorEvent}), we investigate the 
latter two terms. Suppose we employ the $l-$length typical set 
$T_{\delta}^{l}(K_{1})$ of $K_{1}$ as the source code at both 
encoders\footnote{We reiterate that the encoder $2$ also employs the typical 
set $T_{\delta}^{l}(K_{1})$ of $K_{1}$ (and not $K_{2}$) to compress $l-$length 
sub-blocks of $K_{2}$.}, then the first term in (\ref{Eqn:TwoTermsMakeUpPhi}) is 
at most
\ifPeerReviewVersion\begin{eqnarray}
 \label{Eqn:SourceCodeIndicesNotMatching}
 P(\boldK_{1}(t,1:l)\neq\boldK_{2}(t,1:l))+P(\boldK_{1}(t,1:l) \notin 
T_{\delta}^{l}(K_{1})) \leq \xi^{[l]}(\ulineK)+\tau_{\delta,l}(K_{1}).\nonumber
\end{eqnarray}\fi
\ifTITVersion\begin{eqnarray}
 \label{Eqn:SourceCodeIndicesNotMatching}
  \label{Eqn:SourceCodeIndicesNotMatching}
 &P(\boldK_{1}(t,1:l)\neq\boldK_{1}(t,1:l))+P(\boldK_{1}(t,1:l) \notin 
T_{\delta}^{l}(K_{1}))&\nonumber\\& \leq ~
\xi^{[l]}(\ulineK)+\tau_{\delta,l}(K_{1})&\nonumber
\end{eqnarray}\fi
To compute the second term, we specify how message indices are communicated 
over the channel, followed by the choice and 
performance of $C_{U}$.

We propose that each encoder splits the message index output by 
$T_{\delta}^{l}(K_{1})$ into two \textit{sub-message} indices taking values in 
index sets $[M_{u}]\define [\exp\{ l\alpha\}]$, $[\exp\{ l\beta\}]$ and 
communicate the first sub-message index through the constant composition 
channel code $C_{U}$ of type $p_{u}$. 
$\alpha,\beta$ are as provided in the theorem 
statement. We appeal to Theorem \ref{Thm:ConstCompnCodesForMAC} (\cite[Thm 
10.2]{CK-IT2011}) for the choice of $C_{U}$. 
In 
particular, we choose $C_{U}$ to be constant composition code of type $p_{U}$ 
with $M_{u}=\exp\{ l\alpha\}$ codewords each of B-L $l$ and maximal probability 
of decoding error $g(\alpha+\rho,l)$ when employed over a channel with 
transition 
probabilities $p_{Y|U}$. Note that $p_{Y|U}$ is the induced channel from the 
input alphabet $\mathcal{U}$ to the output alphabet $\mathcal{Y}$, 
corresponding to the pmf provided in the theorem statement.\footnote{In the 
proof, we formally establish that the chosen $C_{U}-$codewords, whenever are 
agreed upon by both the encoders, experience a memoryless $p_{Y|U}-$channel. 
In this outline, we assume this and proceed with computing the rates.} 
$g(\alpha+\rho,l)$ 
is an 
upper bound on the second term in (\ref{Eqn:TwoTermsMakeUpPhi}) and we have 
$\phi = 
g(\alpha+\rho,l)+\xi^{[l]}(\ulineK)+\tau_{l,\delta}(K_{1})$ as an upper bound 
on 
(\ref{Eqn:FixedBLMappingErrorEvent}). In stating 
$g(\alpha+\rho,l)+\xi^{[l]}(\ulineK)+\tau_{l,\delta}(K_{1})$ as an upper bound 
on 
(\ref{Eqn:FixedBLMappingErrorEvent}), we have implicitly assumed that 
the second sub-message index taking values in $[\exp\{l\beta\}]$ are 
communicated to the decoder error free. This will be proven to be true since 
this sub-message index is communicated by one of the encoders via the outer 
code - a code of arbitrarily large B-L operating on the 
$\mathcal{V}_{1},\mathcal{V}_{2}-\mathcal{Y}$ 
channel.\footnote{This explains the occurrence of $\beta$ in the sum rate bound 
(\ref{Eqn:StepIMACSumRateBound}) of the 
$\mathcal{V}_{1},\mathcal{V}_{2}-\mathcal{Y}$ MAC.}

The second part involves quantifying how much 
information \textit{needs} to be communicated via the outer codes. In 
particular, this 
involves deriving lower bounds on the rates of the Slepian-Wolf codes. Since we 
view 
the decoded 
sub-blocks $\boldhatK(t,1:l): t \in [m]$ as side information, and moreover the 
sub-blocks 
are proven to be IID, it suffices to characterize 
$H(S_{j}^{l}|\hat{K}^{l},S_{\msout{j}}^{l}): j \in [2]$ and 
$H(\ulineS^{l}|\hat{K}^{l})$, where $S_{j}^{l},K_{j}^{l}: j \in [2], 
\hat{K}^{l}$ are distributed with pmf 
$\left\{\prod_{i=1}^{l}\mathbb{W}_{\ulineS}p_{K_{1}|S_{1}}p_{K_{2}|S_{2}} 
\right\} p_{\hatK^{l}|K_{1}^{l}K_{2}^{l}}$. Owing the fixed B-L code, we are 
unable to characterize 
$p_{\hatK^{l}|K_{1}^{l}K_{2}^{l}}$, and hence we derive an upper bound on 
$H(S_{j}^{l}|\hat{K}^{l},S_{\msout{j}}^{l}): j \in [2]$ and 
$H(\ulineS^{l}|\hat{K}^{l})$. By following steps similar to 
(\ref{Eqn:DueckExExUpperBoundOnBinningRates}) - 
(\ref{Eqn:DueckExExActUpperBoundOnSumBinningRate}), it maybe verified that
\ifTITVersion\begin{eqnarray}
 \label{Eqn:UpperBoundOnBinningRates}
%==============newline
 \lefteqn{H(S_{j}^{l}|\hat{K}^{l},S_{\msout{j}}^{l}) l\left(  
\mathcal{L}_{l}(\phi,|\mathcal{S}_{j}|)+H(S_{j}|K_{1},S_{\msout{j}}) 
\right),\mbox{ for }j \in [2], \mbox{ and}} \\
%==============newline
\lefteqn{H(\ulineS^{l}|\hat{K}^{l}) \leq l\left(  
\mathcal{L}_{l}(\phi,|\ulineCalS|)+H(\ulineS|K_{1}) 
\right)} \nonumber\\
%==============newline
&\leq & h_{b}(P(\hat{K}^{l} \neq 
K_{1}^{l}))+H(\ulineS^{l}|\hat{K}^{l},\mathds{1}_{\{ 
\hat{K}^{l} \neq K_{1}^{l}\}}) \nonumber\\
%==============newline
&\leq&h_{b}(P(\hat{K}^{l} \neq 
K_{1}^{l}))+P(\hat{K}^{l} \neq K_{1}^{l})\log 
|\ulineCalS^{l}|\nonumber\\&&~~+P(\hat{K}^{l} = 
K_{1}^{l})H(\ulineS^{l}|K_{1}^{l})\nonumber\\
%==============newline
&\leq&  l\mathcal{L}_{l}(P(\hat{K}^{l} \neq 
K_{1}^{l}),|\ulineCalS|)+lH(\ulineS|K_{1})\nonumber\\
\label{Eqn:ActUpperBoundOnSumBinningRate}
&\leq &l\left(  
\mathcal{L}_{l}(\phi,|\ulineCalS|)+H(\ulineS|K_{1}) 
\right),
\end{eqnarray}\fi
\ifPeerReviewVersion\begin{eqnarray}
 \label{Eqn:UpperBoundOnIndividualBinningRate}
H(S_{j}^{l}|\hat{K}^{l},S_{\msout{j}}^{l}) \leq 
l\mathcal{L}_{l}(P(\hat{K}^{l} \neq 
K_{1}^{l}),|\mathcal{S}_{j}|)+lH(S_{j}|K_{1},S_{\msout{j}}) &\leq &
l\left(  
\mathcal{L}_{l}(\phi,|\mathcal{S}_{j}|)+H(S_{j}|K_{1},S_{\msout{j}}) 
\right),\mbox{ for }j \in [2],~~ \\
%==============newline
\label{Eqn:ActUpperBoundOnSumBinningRate}
 \mbox{and }H(\ulineS^{l}|\hat{K}^{l}) \leq l\mathcal{L}_{l}(P(\hat{K}^{l} 
\neq 
K_{1}^{l}),|\ulineCalS|)+lH(\ulineS|K_{1}) &\leq & l \left(  
\mathcal{L}_{l}(\phi,|\ulineCalS|)+H(\ulineS|K_{1}) 
\right)
\end{eqnarray}\fi
where $\mathcal{L}_{l}(\cdot,\cdot)$, as defined 
in (\ref{Eqn:AdditionalSourceCodingInfo}) represents the additional source 
coding rate. Since $\mathcal{L}_{l}(\mu,|\mathcal{S}_{j}|)$ is increasing in 
$\mu \in [0,\frac{1}{2})$ and the first part yields an upper bound of $\phi 
\geq 
P(\hat{K}^{l} \neq K_{1}^{l})$, 
the upper bounds on (\ref{Eqn:UpperBoundOnIndividualBinningRate}), 
(\ref{Eqn:ActUpperBoundOnSumBinningRate}) are true, so long as $\phi \leq 
\frac{1}{2}$. This analysis indicates 
that if the bin index output by the Slepian Wolf code at Tx $j$ has rate 
at least the 
RHS of (\ref{Eqn:UpperBoundOnIndividualBinningRate}), and the pair of bin 
indices has rate at least the RHS of 
(\ref{Eqn:ActUpperBoundOnSumBinningRate}), then the 
decoder will be able to reconstruct the source matrices $\underline{\boldS}$ if 
it is provided with the bin indices. This justifies/explains the LHSs 
of (\ref{Eqn:StepIMACIndvdualRateBound}), 
(\ref{Eqn:StepIMACSumRateBound}) which are 
indeed RHSs of (\ref{Eqn:UpperBoundOnIndividualBinningRate}), 
(\ref{Eqn:ActUpperBoundOnSumBinningRate}).

The third part involves quantifying how much information can be communicated 
via the outer MAC channel code. Recall that we split this information into $l$ 
different streams and communicate the same through $l$ different codebooks. The 
central question here is : What is the effective MAC channel experienced by 
these codewords? To answer this, we investigate the joint distribution induced 
by the coding scheme on the Cartesian product $\mathcal{U} \times 
\mathcal{U}\times\mathcal{V}_{1} \times \mathcal{V}_{2}\times\mathcal{X}_{1} 
\times \mathcal{X}_{2}\times\mathcal{Y}$. Suppose $C_{U}$ is made of message 
index set $[M_{u}]$ and codewords $(u^{l}(m) : m \in [M_{u}])$. At Tx $j$, the 
fixed B-L typical set source code encodes the $t-$th sub-block 
$\boldK_{j}(t,1:l)$ into a message, part of which indexes $C_{U}$. Let $A_{jt}$ 
denote this latter part. We therefore have the chosen $C_{U}$ codeword in the 
$t-$th sub-block to be $\boldU_{j}(t,1:l) = u^{l}(A_{jt})$. We note 
that pmf of $(A_{1t},A_{2t})$ is invariant with $t$, and hence let 
$(A_{1},A_{2}) \in [M_{u}] \times [M_{u}]$ have the same pmf of 
$(A_{1t},A_{2t})$. The rows of $\boldU_{1},\boldU_{2}$ are IID with pmf
\begin{eqnarray}
 \label{Eqn:Step1MACPMFOfRowsU1U2}
 p_{U_{1}^{l}U_{2}^{l}}(u_{1}^{l},u_{2}^{l}) =
\sum_{\substack{(a_{1},a_{2}) \in \\ 
[M_{u}]\times [M_{u}]}} 
\!\!\!\!\!\!P\left(\!\!\!
 \begin{array}{c}
 A_{1}=a_{1},A_{2}=a_{2}
 \end{array}\!\!\!\right)\mathds{1}_{\left\{ \begin{array}{c}
u^{l}(a_{j})= u_{j}^{l}:j \in [2]\end{array}\right\}} \nonumber
\end{eqnarray}
By choosing the codewords of the $l$ outer codes IID with pmf 
$\prod_{t=1}^{m}p_{V_{j}}$ and the mapping from $\mathcal{U} \times 
\mathcal{V}_{j} \rightarrow \mathcal{X}_{j}$ IID with pmf 
$\prod_{t=1}^{m}p_{X_{j}|U,V_{j}}$, we ensure that the distribution of the 
$l-$length sub-blocks on $\mathcal{U}^{l}\times \mathcal{U}^{l}\times 
\mathcal{V}_{1}^{l}\times \mathcal{V}_{2}^{l}\times \mathcal{X}_{1}^{l}\times 
\mathcal{X}_{2}^{l}\times \mathcal{Y}^{l}$ is
\begin{eqnarray}
 p_{\ulineU^{l}\ulineV^{l}\ulineX^{l}Y^{l}}(\ulineu^{l},\ulinev^{l},\ulinex^{l},
y^{l}) = \left[ \sum_{\substack{(a_{1},a_{2}) \in \\ [M_{u}]\times [M_{u}]}} 
\!\!\!\!\!\!\!P(
 \begin{array}{c}
 A_{1}=a_{1}\\A_{2}=a_{2}
 \end{array})\mathds{1}_{\left\{\substack{ 
u^{l}(a_{j})=\\u_{j}^{l}:j \in [2]}\right\}}\right]
 \times \left[ \prod_{j=1}^{2} \left\{ 
\prod_{i=1}^{l}p_{V_{j}}(v_{ji})p_{X_{j}|UV_{j}}(x_{ji}|u_{ji},v_{ji}) \right\} 
\right]\nonumber\\
 \label{Eqn:Step1MACOutlinePMFForDecodingRule}
 \times \left[ \prod_{i=1}^{l} 
\mathbb{W}_{Y|X_{1}X_{2}}(y_{i}|x_{1i},x_{2i})\right].
\end{eqnarray}
In other words, our coding scheme of B-L $lm$ which maybe viewed as $m$ 
sub-blocks of length $l$ induces a pmf
$\prod_{t=1}^{m}p_{\ulineU^{l}\ulineV^{l}\ulineX^{l}Y^{l}}(\ulineu^{l},\ulinev^{
l},\ulinex^{l},y^{l})$ on $\mathcal{U}^{lm} \times 
\mathcal{U}^{lm}\times\mathcal{V}_{1}^{lm} \times 
\mathcal{V}_{2}^{lm}\times\mathcal{X}_{1}^{lm} 
\times \mathcal{X}_{2}^{lm}\times\mathcal{Y}^{lm}$. Each of the $l$ outer 
codes, operating on interleaved columns of these $m$ sub-blocks will experience 
a MAC with channel transition probabilities 
$p_{\mathscr{Y}|\mathscr{V}_{1}\mathscr{V}_2}$, where
\begin{eqnarray}
\label{Eqn:Step1MACOutlinePMFOfIntrlvdVec}
p_{\mathscr{U}_{1}\mathscr{U}_{2}\mathscr{V}_{1}\mathscr{V}_{2}\mathscr{X}_{
1}\mathscr{X}_{2}\mathscr{Y}}({\ulinea},{\ulineb},{
\ulinec},{d}) \define 
\displaystyle\frac{1}{l}\sum_{i=1}^{l}p_{U_{1i}U_{2i}V_{1i}V_{2i}X_{
1i}X_{2i}Y_{i}}(a_{1},a_{2},b_{1},b_{2},c_{1},c_{2},d).
\end{eqnarray}
and $p_{U_{1i}U_{2i}V_{1i}V_{2i}X_{
1i}X_{2i}Y_{i}}$ is the pmf of the $i-$th co-ordinate of the Cartesian product 
of vectors $U_{1}^{l},U_{2}^{l},V_{1}^{l},V_{2}^{l},X_{1}^{l},X_{2}^{l},Y^{l}$ 
which is distributed with pmf (\ref{Eqn:Step1MACOutlinePMFForDecodingRule}). 
The rates of the $i$th MAC outer code is therefore constrained to lie within 
the achievable region of the MAC 
$(\mathcal{V}_{1},\mathcal{V}_2,\mathcal,{Y},p_{\mathscr{Y}|\mathscr{V}_{1}
\mathscr{V}_2})$ corresponding to the pmf 
$p_{\mathscr{V}_{1}}p_{\mathscr{V}_2}$.\footnote{It can be verified that 
marginal $p_{V_{1}^{l}V_{2}^{l}}$ corresponding to pmf 
(\ref{Eqn:Step1MACOutlinePMFForDecodingRule}) factors as 
$\prod_{i=1}^{l}p_{V_{1}}p_{V_{2}}$ and hence 
$p_{\mathscr{V}_{1}\mathscr{V}_2} =  p_{\mathscr{V}_{1}}p_{\mathscr{V}_2} = 
p_{V_{1}}p_{V_{2}}$. These and other properties of 
(\ref{Eqn:Step1MACOutlinePMFForDecodingRule}) can be found in Lemma 
\ref{Lem:SimplePropDecodingLem}.} We are left to quantify 
$I(\mathscr{V}_{j};\mathscr{Y}|\mathscr{V}_{\msout{j}})$ and 
$I(\mathscr{V}_{1},\mathscr{V}_{2};\mathscr{Y})$ in terms of the pmf 
$p_{U}p_{V_{1}}p_{V_{2}}p_{X_{1}|UV_{1}}p_{X_{2}|UV_{2}}\mathbb{W}_{Y|\ulineX}$ 
provided in the theorem statement. We derive lower bounds on the above 
quantities. The reader is referred to the material following 
(\ref{Eqn:MACStep1SummaryIndBound}) through till 
(\ref{Eqn:MACSTep1SumChnlBnd-5}) where we prove that if $\frac{1}{2} \geq 
\epsilon \geq P(A_{1}\neq A_{2})$ and $C_{U}$ is a constant composition code of 
type $P_{U}$, then (\ref{Eqn:MACSTep1IndChnlBnd-5}), 
(\ref{Eqn:MACSTep1SumChnlBnd-5}) are lower bounds on
$I(\mathscr{V}_{j};\mathscr{Y}|\mathscr{V}_{\msout{j}})$ and 
$I(\mathscr{V}_{1},\mathscr{V}_{2};\mathscr{Y})$ respectively.\footnote{The 
necessary notation is provided therein and the arguments can be easily 
followed.} Recognize that (\ref{Eqn:MACSTep1IndChnlBnd-5}) is 
$I(V_{j};Y|V_{\msout{j}}) -\mathcal{L}(\epsilon,|\mathcal{V}_{j}|)$ and 
(\ref{Eqn:MACSTep1SumChnlBnd-5}) is $I(V_{1},V_{2};Y) 
-\mathcal{L}(\epsilon,|\ulineCalV|)$. Since $\epsilon = 
\tau_{l,\delta}(K_{1})+\xi^{[l]}(\ulineK) < \phi$ and 
$\mathcal{L}(\mu,|\mathcal{A}|)$ is increasing in $\mu$ for $\mu \in 
[0,\frac{1}{2}]$, we are led to the sufficient condition that (i) the RHS 
of (\ref{Eqn:UpperBoundOnIndividualBinningRate}) must be less than $l( 
I(V_{j};Y|V_{\msout{j}}) -\mathcal{L}(\phi,|\mathcal{V}_{j}|)) $ for 
$j \in [2]$, and the (ii) sum of $\l\beta$ and RHS of 
(\ref{Eqn:ActUpperBoundOnSumBinningRate}) must be less than $I(V_{1},V_{2};Y) 
-\mathcal{L}(\phi,|\ulineCalV|)$. These are indeed the sufficient 
conditions characterized in Theorem \ref{Thm:MACStep1}.
\begin{IEEEproof}
Let $\mathcal{K},K_{1},\cdots,\rho$ be provided as in theorem statement. For 
simplicity we assume $\beta = 0$ and $\alpha > (1+\delta)H(K_{1})$. 
$K_{1},K_{2} \in \mathcal{K}$ represent the near GKW parts of the sources 
$\ulineS$. The rest of the parameters will be described as and when they appear 
in the proof.

\textit{Coding Scheme}: We propose a separation based scheme that 
communicates information via two 
streams - fixed B-L and $\infty-$B-L (arbitrarily large B-L).
The B-L of the coding scheme is $lm$. We will view this block as an $m 
\times l$ matrix and our coding technique is best viewed as matrix encoding 
and decoding. $l$ is as provided in the theorem statement and will remain 
fixed, while $m$ will be chosen sufficiently 
large, as a function of the desired probability of error. Let $\boldS_{j} \in 
\mathcal{S}_{j}^{m \times l}$ denote the block of 
source symbols observed by encoder $j$. For $(t,i) \in 
[m]\times [l]$, $\boldS_{j}(t,i)$ is the symbol 
observed during $(t-1)l+i$ -th symbol interval. Let $\boldK_{j} \in 
\mathcal{K}^{m \times l}$ be defined as $\boldK_{j}(t,i) = 
f_{j}(\boldS_{j}(t,i))$ for $(t,i) \in [m] \times [l]$, where 
$f_{j}:\mathcal{S}_{j} \rightarrow \mathcal{K}$ is as specified in the theorem 
statement. Rows of $\boldK_{j}$ will be encoded by a B-L $l$ source encoder. 
We employ the $l-$length typical set $T_{\delta}^{l}(K_{1})$ as the 
source encoder at both Txs. Here $\delta$ is as provided in the theorem 
statement. We emphasize that Tx $2$ also employs the 
the $l-$length typical set $T_{\delta}^{l}(K_{1})$ of $K_{1}$ to encode rows of 
$\boldK_{2}$. We let this source code be defined through 
message index set $[|T_{\delta}^{l}(K_{1})|]$, encoder map 
$e_{K}: \mathcal{K}^{l} \rightarrow [|T_{\delta}^{l}(K_{1})|]$ and 
decoder map $d_{K} : 
[|T_{\delta}^{l}(K_{1})|] \rightarrow \mathcal{K}^{l}$ such that 
$d_{K}(e_{K}(k^{l}))=k^{l}$ for $k^{l} \in T_{\delta}^{l}(K_{1})$. Let $A_{jt} 
\define e_{K}(\boldK_{j}(t,1:l))$ denote the message output by this $l-$length 
source encoder corresponding to the $t$-th row of $\boldK_{j}$. In particular, 
if the $t^{\small\mbox{th}}$ row $\boldK_{j}(t,1:l) \in T_{\delta}^{l}(K_{1})$, 
$A_{jt}$ is set to the corresponding index in the typical set. If 
$\boldK_{j}(t,1:l) \notin T_{\delta}^{l}(K_{1})$, then $A_{jt}$ is set to $1$.
These common fixed B-L maps ensure
\ifPeerReviewVersion
\begin{eqnarray}
 \label{Eqn:FixedBLSourceMessEqual}
&\left\{A_{1t} \neq A_{2t} \right\} \subseteq \left\{ 
\boldK_{1}(t,1:l) \notin T_{\delta}^{l}(K_{1})\right\}\bigcup \left\{ 
\boldK_{1}(t,1:l) \neq \boldK_{2}(t,1:l)\right\}\mbox{ , and hence}&\nonumber\\
\label{Eqn:ProbOfSub-BlocksDiffering}
&P(A_{1t} \neq A_{2t}) \leq \epsilon \define 
\xi^{[l]}(\ulineK)+\tau_{\delta,l}(K_{1}).&
\end{eqnarray}\fi
\ifTITVersion
\begin{eqnarray}
 \left\{A_{1t} \neq A_{2t} \right\} \subseteq \left\{ 
\boldK_{1}(t,1:l) \notin T_{\delta}^{l}(K_{1})\right\}~~~~~~~~~~~~~~~~~~~~~
\nonumber\\\bigcup \left\{ 
\boldK_{1}(t,1:l) \neq \boldK_{2}(t,1:l)\right\}\mbox{ 
for } t \in [m],\nonumber\\
\label{Eqn:FixedBLSourceMessEqual}
\mbox{ and hence }P(A_{1t} \neq A_{2t}) \leq \epsilon \define 
\xi^{[l]}(\ulineK)+\tau_{\delta,l}(K_{1}).
\end{eqnarray}\fi
Without pooling these messages, Tx $j$ communicates $A_{jt}: t \in [m]$ 
via B-L $l$ channel code $C_{U}$ built over $\mathcal{U}$, where $\mathcal{U}$ 
is as provided in the theorem statement. In particular, a constant composition 
code $C_{U} = (l,M_{u},e_{u},d_{u})$ of B-L $l$ is built over $\mathcal{U}$, 
consisting of $M_{u} \geq |T_{\delta}^{l}(K_{1})|$ codewords each of type 
$p_{U}$, that is characterized via encoder map $e_{u}:[M_{u}]\rightarrow 
\mathcal{U}^{l}$ and decoder map $d_{u}:\mathcal{Y}^{l}\rightarrow [M_{u}]$.
Both Txs $1,2$ employ $C_{U}$ to communicate their messages $A_{jt}: t 
\in [m]$ to the decoder. We let $u^{l}(a)\define e_{u}(a): a \in [M_{u}]$, and 
for $\ulinea \in [M_{u}]^{m}$, we let $\bold{\altu}\{\ulinea\} \in \mathcal{U}^{ 
m \times l}$ be defined through $\bold{\altu}\{\ulinea\}(t,1:l)=u^{l}(a_{t}):t 
\in [m]$. We have thus described $C_{U}$ - the first component of our channel 
code. When we analyze probability of error, we do not randomize over the choice 
of fixed B-L codes - $T_{\delta}^{l}(K_{1})$ and $C_{U}$. The corresponding 
encoder mappings $\mathcal{K}^{l} \rightarrow [M_{u}] \rightarrow 
\mathcal{U}^{l}$ and decoder mappings $\mathcal{Y}^{l} \rightarrow [M_{u}] 
\rightarrow \mathcal{K}^{l}$ will remain fixed throughout our study.

The second component of our channel code are the $l$ channel codes $C_{V_{j},i} 
: i \in [l]$ employed by Tx $j$ to communicate the rest of the information. 
$C_{V_{j},1},\cdots, C_{V_{j},l}$ is built over $\mathcal{V}_{j}$ - the 
auxiliary alphabet set provided in the theorem statement. We 
allude the reader to the outline of the coding scheme, wherein the need for 
splitting the rest of information into $l$ streams and communicating them via 
$l$ different codebooks was discussed. The $l-$pairs of codebooks 
$C_{V_{1},i},C_{V_{2},i}: i \in [l]$ constitute the $l$ MAC codes. 
Specifically, Tx $j$ employs codebooks $C_{V_{ji}} = (m,M_{V_{j}}, 
e_{V_{ji}}, d_{V_{ji}}): i \in [l]$, each built over $\mathcal{V}_{j}$. 
$C_{V_{ji}}$ is of B-L $m$, has message index set $[M_{V_{j}}]$ and is 
characterized via encoder map $e_{V_{ji}}: 
[M_{V_{j}}] \rightarrow \mathcal{V}_{j}^{m}$, decoder map 
$d_{V_{ji}}:\mathcal{Y}^{m}\rightarrow [M_{V_{1}}]\times [M_{V_{2}}]$. We let 
$(v_{ji}^{m}(b): b \in [M_{V_{j}}])$ denote the codewords of $C_{V_{j},i}$.

$C_{V_{j},i}: i \in [l]: j \in [2]$ are used to communicate the bin index 
output by a Slepian Wolf lossless distributed source encoder compressing 
$\mathcal{S}_{j}: j \in [2]$. In particular, $\mathcal{S}_{j}^{lm}$ is 
partitioned into $M_{V_{j}}^{l}$ bins. Let $\beta_{j}: 
\mathcal{S}_{j}^{lm}\rightarrow [M_{V_{j}}]^{l}$ denote the partition map 
effected by the Slepian Wolf code, and let $\ulineB_{j} = (B_{j1}, \cdots, 
B_{jl}) \in [M_{V_{j}}]^{l}$ denote the bin index of $\boldS_{j}$. Tx $j$ has 
to multiplex the collection $v_{ji}^{m}(B_{ji}) : i \in [l]$ of codewords with  
$\boldU_{j} \define \boldu \{ \ulineA_{j}\}$, the rows of which are the $m$ 
codewords $u^{l}(A_{jt}) : t \in [m]$ chosen from the constant composition code 
$C_{U}$. The third and fourth components of our channel code constitute the 
multiplexing unit.

The third component of our channel code are the $m$ surjective maps 
$\pi_{t}:[l] \rightarrow [l] : t \in [m]$ which enable us identify $l$ 
sub-vectors of $\boldU_{j}$ along which the $l$ codewords $v_{ji}^{m}(B_{ji}) : 
i \in [l]$ will be multiplexed. Specifically, $v_{ji}^{m}(B_{ji})$ will be 
multiplexed with the $i$-th \textit{interleaved column} 
$(\boldU_{j}(t,\pi_{t}(i)):t \in [m])$. We employ the following notation in the 
sequel which greatly simplify exposition in relation to \textit{interleaving}.

For $\bold{A} \in \mathcal{A}^{m \times l}$, and a collection $\lambda_{t}:[l] 
\rightarrow [l]: t \in [m]$ of surjective maps, we let 
$\bold{A}^{\underline{\lambda}} \in \mathcal{A}^{ m\times l}$ be such that 
$\bold{A}^{\underline{\lambda}}(t,i) \define \bold{A}(t,\lambda_{t}(i))$ for 
each $(t,i)\in [m]\times [l]$. To reduce clutter, we let 
$\bold{A}^{{\lambda}}=\bold{A}^{\underline{\lambda}}$. If $\boldA \in 
\mathcal{A}^{m \times l},\boldB \in \mathcal{B}^{m \times l}$, then 
$[\boldA\boldB]^{\lambda}(1:m,i) \define 
(\boldA^{\lambda}(1:m,i),\boldB^{\lambda}(1:m,i))$.

The above notation helps us specify multiplexing of the chosen codewords 
$v_{ji}^{m}(B_{ji}): i \in [l]$ with $\boldU_{j}$. For $j \in 
[2]$, $\ulineb_{j} \in [M_{V_{j}}]^{l}$, we let 
$\bold{\altv_{j}}\{\ulineb_{j}\}\in \mathcal{V}_{j}^{m \times l}$ be defined 
through 
$\bold{\altv_{j}}\{\ulineb_{j}\}^{\ulinepi}(1:m,i)=v_{ji}^{m}(b_{ji}): i 
\in 
[l]$ where $\pi_{t}:[l]\rightarrow [l] : t \in [m]$ are the surjective 
maps that make up our channel code. Our last step in the encoding rule is to 
map $\boldU_{j} \define \boldu\{ 
\ulineA_{j}\}, \boldV_{j} \define \boldv\{ \ulineB_{j} \}$ into channel inputs 
on $\mathcal{X}_{j}$. This leads us to the fourth and last component of our 
channel code. For $j \in [2]$, $\boldu 
\in \mathcal{U}^{m \times l}$, $\boldv_{j} \in \mathcal{V}_{j}^{ m \times l}$, 
we 
let $\boldx_{j}(\boldu,\boldv_{j}) \in \mathcal{X}_{j}^{m \times l}$ be 
predefined 
$m \times l$ matrices in $\mathcal{X}_{j}$. Encoder $j$ 
maps $(\ulineA_{j},\ulineB_{j}) \in [M_{U}]^{m}\times 
[M_{V_{j}}]^{l}$ into $\boldx_{j}(\boldu\{ \ulineA_{j}\},\boldv\{ 
\ulineB_{j}\}) \in \mathcal{X}_{j}^{m \times l}$. For $(t,i) \in [m] \times 
[l]$, the encoder inputs symbol $\boldx_{j}(\ulineA_{j}, \ulineB_{j})(t,i)$ 
on the channel during symbol interval $(t-1)l+i$.

\begin{table}[h]
\begin{center}
\begin{tabular}{|c|c|c|}
\hline
Symbol & Description & Comment \\\hline
$lm$&B-L of coding scheme&Block viewed as $m \times l$ matrix.\\&&$l$ 
remains fixed. $m$ chosen arbitrarily large.\\
\hline\hline
$T_{\delta}^{l}(K_{1})$&Source code employed at both encoders & 
$T_{\delta}^{l}(K_{1})$ is used to encode rows of $\boldK_{1},\boldK_{2}$.\\
\hline
$e_{K}:\mathcal{K}^{l}\rightarrow [|T_{\delta}^{l}(K_{1})|]$&Encoder map of 
(common) typical set source code&fixed B-L $l$
\\\hline
$A_{jt}$&Index output by $T_{\delta}^{l}(K_{1})$ corresponding to 
$\boldK_{j}(t,1:l)$& $\ulineA_{j} \define (A_{j1},\cdots, A_{jm})$\\
\hline
$d_{K}:[|T_{\delta}^{l}(K_{1})|]\rightarrow T_{\delta}^{l}(K_{1})$&Decoder 
map of 
(common) typical set source code&$d_{K}(e_{K}(k^{l}))=k^{l}$ for $k^{l} \in 
 T_{\delta}^{l}(K_{1})$
\\
\hline\hline
$C_{U}$&The common channel code over $\mathcal{U}$ of fixed B-L $l$.&$C_{U}$ is 
constant composition of type $p_{U}$\\&Employed at both encoders.&\\
\hline
$[M_{u}]$&Message Index set of $C_{U}$&Assuming $\beta = 0$, we have $M_{u} 
\geq |T_{\delta}^{l}(K_{1})|$\\
\hline
$e_{u}: [M_{U}] \rightarrow \mathcal{U}^{l}$&Encoder map of $C_{U}$&\\
\hline
$d_{u}: \mathcal{Y}^{l} \rightarrow [M_{u}]$&Decoder map of $C_{U}$&\\
\hline
$u^{l}(a) = e_{u}(a)$&$C_{U}$ codeword corresponding to $a \in [M_{u}]$&\\
\hline
$\boldu\{ \ulinea\} \in \boldCalU$&$\boldu\{ \ulinea\}(t,1:m) \define 
u^{l}(a_{t}) : t \in [m]$&\\
\hline\hline
$C_{V_{j},i}$&Outer channel code $i$ employed by encoder $j$.&B-L\\
\hline
$[M_{V_{j}}]$&Message index set of $C_{V_{j},i}$.&\\
\hline
$e_{V_{ji}}:[M_{V_{j}}]\rightarrow \mathcal{V}_{j}^{m}$&Encoder map of 
$C_{V_{j},i}$&Codewords $v_{ji}^{m}(b) \define e_{V_{j,i}}(b): b \in 
[M_{V_{j}}]$\\
\hline
$d_{V_{ji}}\!\!:\mathcal{Y}^{m}\rightarrow [M_{V_{1}}]\times 
[M_{V_{2}}]$&Decoder map of MAC channel code $C_{V_{1},i},C_{V_{2},i}$& Joint 
typicality decoding wrt pmf $p_{\mathscr{V}_{1}\mathscr{V}_{2}\mathscr{Y}}$\\
\hline\hline
$\beta_{j}: \mathcal{S}_{j}^{lm} \rightarrow [M_{V_{j}}]^{l}$&Encoder map of 
Slepian Wolf code&Message output by this code lies in $[M_{V_{j}}]^{l}$
\\\hline
$\ulineB_{j} \define \beta_{j}(\boldS_{j})$&Message Index $\ulineB_{j} = 
(B_{j1},\cdots, B_{jl})$ output by&\\&Slepian Wolf code at Encoder $j$&\\
\hline\hline
$\pi_{t}:[l]\rightarrow [l]$&Surjective maps employed for multiplexing&\\
\hline
$\bolda^{\pi}(t,i) = \bolda(t,\pi_{t}(i))$&Interleaving notation&\\
\hline
$\boldv_{j}\{\ulineb_{j}\} 
$&$\boldv_{j}\{ \ulineb_{j}\}^{\pi}(t,1:m) = v_{ji}^{m}(b_{ji})$&$v_{ji}^{m}(b) 
\define e_{V_{j,i}}(b): b \in 
[M_{V_{j}}]$\\
\hline
$\boldx_{j}(\cdot,\cdot):\boldCalU\times \boldCalV_{j} \rightarrow  
\boldCalX_{j}
$&Predefined matrices employed for mapping&\\& $\boldu \in 
\boldCalU$, $\boldv_{j}\in \boldCalV_{j}$ into 
$\boldx_{j}(\boldu,\boldv_{j})$&\\
\hline
$\boldx_{j}(\ulinea_{j},\ulineb_{j}) \define 

$&$\boldu\{ \cdot\}$ and $\boldv_{j}\{ \cdot \}$ and 
$\boldx_{j}(\cdot,\cdot):\boldCalU\times \boldCalV_{j} \rightarrow  
\boldCalX_{j}$&\\$\boldx_{j}(\boldu \{ 
\ulinea_{j}\},\boldv_{j}\{\ulineb_{j}\})$& is as defined in above rows&\\
\hline
\end{tabular}
\end{center}
\caption{Description of elements that constitute the coding scheme}
\label{Table:Step1MACSymbolTable}
\end{table}

\begin{figure}
\centering
\includegraphics[width=9.4in,height=7.3in,angle=90]{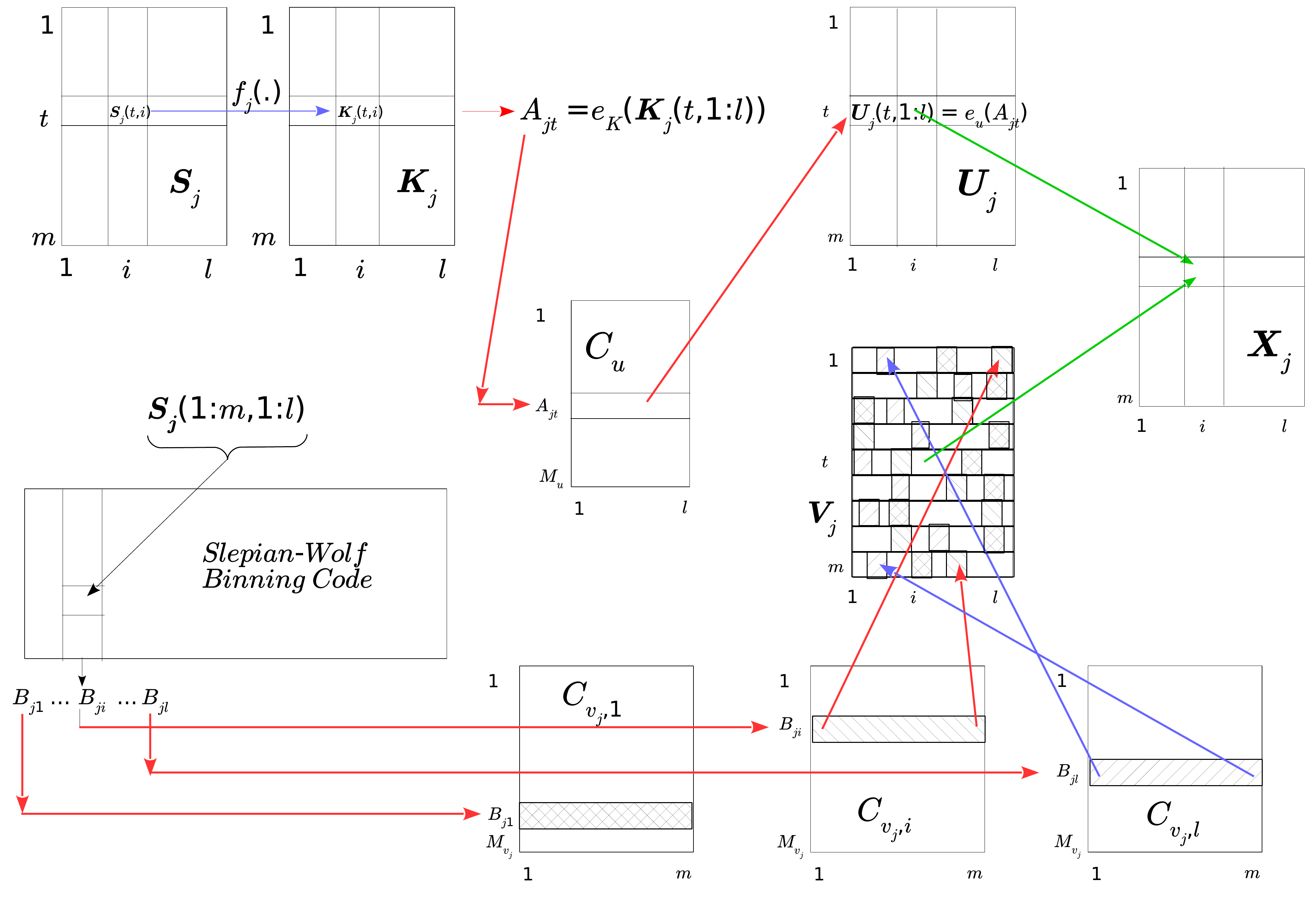}
\caption{Coding Scheme.}
\label{Fig:CodingScheme}
\end{figure}

We summarize the operations of the channel coding module. Refer to Table 
\ref{Table:Step1MACSymbolTable} wherein the components of the code are mentioned 
and 
Figure \ref{Fig:CodingScheme} wherein the encoding technique is depicted 
pictorially. Encoding rule : Tx $j$ observes $\ulineA_{j} = (A_{jt}: t \in [m]) 
\in [M_{u}]^{m}$ and $\ulineB_{j} \in [M_{V_{j}}]^{l}$. For $t \in [m]$, let 
$\boldU_{j}(t,1:l)\define e_{u}(A_{jt})=u^{l}(A_{jt})$. For $i \in [l]$, let 
$\bold{V}_{j}^{\pi}(1:m,i) = v_{ji}^{m}(B_{ji})$.\footnote{Note 
$\bold{v}_{j}^{\pi}(1:m,i) = \bold{v}_{j}(1,\pi_{1}(i))\cdots 
\bold{v}_{j}(m,\pi_{m}(i))$.} Let 
$\boldX_{j}=\boldx_{j}(\boldU_{j},\boldV_{j})$. $\boldX_{j}(t,i)$ is input on 
the channel during symbol-interval $(t-1)l+i$. The encoding rule can therefore 
be specified as follows. Tx $j$ observes observes $\ulineA_{j} = (A_{jt}: t \in 
[m]) \in [M_{u}]^{m}$ and $\ulineB_{j} \in [M_{V_{j}}]^{l}$. Let $\boldU_{j} = 
\bold{\altu}\{\ulineA_{j}\}$, $\boldV_{j} = \bold{\altv}_{j}\{\ulineB_{j}\}$ 
and $\boldX_{j}=\bold{\altx}_{j}\{\ulineA_{j},\ulineB_{j}\}$. Input symbol 
$\boldX_{j}(t,i)$ during symbol-interval $(t-1)l+i$.

\textit{Decoding rule} : Let $\boldY \in \mathbb{Y}^{m \times l}$ denote the 
matrix of 
received symbols 
with $\boldY(t,i)$ being the symbol received during symbol interval $(t-1)l+i$. 
The channel code decoder attempts to recover 
$(\ulineA_{1},\ulineB_{1},\ulineB_{2}) \in [M_{u}]^{m}\times [M_{V_{1}}]^{l} 
\times [M_{V_{2}}]^{l}$. In the first step, $d_{u}$ - the 
decoder of $C_{U}$ - operates on rows of $\boldY$ separately to output 
$\hat{A}_{t} \define d_{u}(\boldY(t,1:l)) : t \in [m]$ - the decoded messages 
corresponding to $\ulineA_{1}$. In the second step, the decoder attempts to 
recover $(\ulineB_{1},\ulineB_{2})$. Since, in this step, we propose separate 
decoding of the fixed B-L and arbitrarily large B-L codes, $\ulinehatA \define 
(\hatA_{t}: t \in [m])$ is not used in decoding $\ulineB_{1},\ulineB_{2}$. The 
latter collection of messages are decoded via a 
standard joint-typicality decoder that operates on the interleaved columns of 
$\boldY$. The point to note is that joint typicality is with respect to 
$p_{\mathscr{V}_{1}\mathscr{V}_{2}\mathscr{Y}}$ and \textit{not} the pmf 
$p_{V_{1}V_{2}Y}$ that is induced by the pmf provided in the theorem statement. 
We now define pmf $p_{\mathscr{V}_{1}\mathscr{V}_{2}\mathscr{Y}}$. To begin 
with, let $(A_{1},A_{2})$ have the same pmf as $(A_{1t},A_{2t})$ for any $t \in 
[m]$.\footnote{Since each sub-block $\boldK_{j}(t,1:l)$ is encoded 
separately and identically by the fixed B-L 
typical set source code $T_{\delta}^{l}(K_{1})$, pmf of $(A_{1t},A_{2t})$ is 
invariant with $t$.} Let
\ifTITVersion\begin{eqnarray}
\lefteqn{p_{\ulineU^{l}\ulineV^{l}\ulineX^{l}Y^{l}}\!\!\left(\!\!\!\begin{array}
{c }
\ulineu^ { l
}, \ulinev^ { l } ,\\ \ulinex^ {l},
y^{l}\end{array}\!\!\!\right) \!=\! \left[ \sum_{\substack{(a_{1},a_{2}) \in \\ 
[M_{u}]\times [M_{u}]}} 
\!\!\!\!\!\!\!\!\!P\left(\!\!\!
 \begin{array}{c}
 A_{1}=a_{1}\\A_{2}=a_{2}
 \end{array}\!\!\!\right)\mathds{1}_{\left\{\substack{ 
u^{l}(a_{j})=\\u_{j}^{l}:j \in [2]}\right\}}\right]\!\!\!\!\!\!\!\!} \nonumber\\
&& \times \left[ \prod_{j=1}^{2} \left\{ 
\prod_{i=1}^{l}p_{V_{j}}(v_{ji})p_{X_{j}|UV_{j}}(x_{ji}|u_{ji},v_{ji}) \right\} 
\right]~~~~~~~~~~~~~~~\nonumber\\
 \label{Eqn:Step1MACPMFForDecodingRule}
&& \times \left[ \prod_{i=1}^{l} 
\mathbb{W}_{Y|X_{1}X_{2}}(y_{i}|x_{1i},x_{2i})\right]
\end{eqnarray}\fi
\ifPeerReviewVersion\begin{eqnarray}
p_{\ulineU^{l}\ulineV^{l}\ulineX^{l}Y^{l}}(\ulineu^{l},\ulinev^{l},\ulinex^{l},
y^{l}) = \left[ \sum_{\substack{(a_{1},a_{2}) \in \\ [M_{u}]\times [M_{u}]}} 
\!\!\!\!\!\!\!P(
 \begin{array}{c}
 A_{1}=a_{1}\\A_{2}=a_{2}
 \end{array})\mathds{1}_{\left\{\substack{ 
u^{l}(a_{j})=\\u_{j}^{l}:j \in [2]}\right\}}\right]
 \times \left[ \prod_{j=1}^{2} \left\{ 
\prod_{i=1}^{l}p_{V_{j}}(v_{ji})p_{X_{j}|UV_{j}}(x_{ji}|u_{ji},v_{ji}) \right\} 
\right]\nonumber\\
 \label{Eqn:Step1MACPMFForDecodingRule}
 \times \left[ \prod_{i=1}^{l} 
\mathbb{W}_{Y|X_{1}X_{2}}(y_{i}|x_{1i},x_{2i})\right]
\end{eqnarray}\fi
be a pmf\footnote{In (\ref{Eqn:Step1MACPMFForDecodingRule}), 
$\ulineU^{l}\ulineV^{l}\ulineX^{l}Y^{l}$ abbreviates 
$U_{1}^{l}U_{2}^{l}V_{1}^{l}V_{2}^{l}X_{1}^{l} X_{2}^{l}Y^{l} $ and similarly 
$\ulineu^{l},\ulinev^{l},\ulinex^{l},
y^{l}$ abbreviates $u_{1}^{l},u_{2}^{l},v_{1}^{l},v_{2}^{l},x_{1}^{l}, 
x_{2}^{l},y^{l}$.} on $\ulineCalU^{l}\times 
\ulineCalV^{l}\times\ulineCalX^{l}\times \mathcal{Y}^{l}$, where 
$p_{V_{j}},p_{X_{j}|U V_{j}} : j \in [2]$ are as provided in theorem statement. 
Moreover, let
\ifTITVersion\begin{eqnarray}
&p_{\mathscr{U}_{1}\mathscr{U}_{2}\mathscr{V}_{1}\mathscr{V}_{2}\mathscr{X}_{
1}\mathscr{X}_{2}\mathscr{Y}}({\ulinea},{\ulineb},{
\ulinec},{d}) \define& 
\nonumber\\&\label{Eqn:Step1MACPMFOfInterleavedVector}\displaystyle\frac{1}{l}
\sum_ { i=1 }
^{l}p_{U_{1i}U_{2i}V_{1i}V_{2i}X_{
1i}X_{2i}Y_{i}}(a_{1},a_{2},b_{1},b_{2},c_{1},c_{2},d).& 
\end{eqnarray}\fi
\ifPeerReviewVersion\begin{eqnarray}
\label{Eqn:Step1MACPMFOfInterleavedVector}
p_{\mathscr{U}_{1}\mathscr{U}_{2}\mathscr{V}_{1}\mathscr{V}_{2}\mathscr{X}_{
1}\mathscr{X}_{2}\mathscr{Y}}({\ulinea},{\ulineb},{
\ulinec},{d}) \define 
\displaystyle\frac{1}{l}\sum_{i=1}^{l}p_{U_{1i}U_{2i}V_{1i}V_{2i}X_{
1i}X_{2i}Y_{i}}(a_{1},a_{2},b_{1},b_{2},c_{1},c_{2},d)
\end{eqnarray}\fi
be a pmf on $\ulineCalU\times 
\ulineCalV\times\ulineCalX\times \mathcal{Y}$. In Appendix 
\ref{AppSec:PropOfDecodingPMF}, we list and prove certain 
simple properties of pmfs (\ref{Eqn:Step1MACPMFForDecodingRule}), 
(\ref{Eqn:Step1MACPMFOfInterleavedVector}) that we will have opportunity to 
leverage in the sequel. We now specify the decoding rule for recovering 
$\ulineB_{1}, \ulineB_{2}$. For $i \in [l]$, populate
\ifTITVersion\begin{eqnarray}
 \label{Eqn:Step1MACDecodingSelectionRule}
 \mathcal{D}_{i}(\boldY) \define \left\{ (\hatb_{1i},\hatb_{2i}): 
\begin{array}{c}(v_{1i}^{m}(\hatb_{1i}),v_{2i}^{m}(\hatb_{2i}),\boldY^{\pi}(1:m,
i))\\\mbox{ is jointly typical wrt } 
\displaystyle\prod_{t=1}^{m}p_{\mathscr{V}_{1}\mathscr{V}_{2}\mathscr{Y}}
\end{array} 
\right\}. \nonumber
\end{eqnarray}\fi
\ifPeerReviewVersion\begin{eqnarray}
 \label{Eqn:Step1MACDecodingSelectionRule}
 \mathcal{D}_{i}(\boldY) \define \left\{ (\hatb_{1i},\hatb_{2i}): 
\begin{array}{c}(v_{1i}^{m}(\hatb_{1i}),v_{2i}^{m}(\hatb_{2i}),\boldY^{\pi}(1:m,
i))\mbox{ is jointly typical wrt } 
\displaystyle\prod_{t=1}^{m}p_{\mathscr{V}_{1}\mathscr{V}_{2}\mathscr{Y}}
\end{array} 
\right\}. \nonumber
\end{eqnarray}\fi
For $i \in [l]$, such that $\mathcal{D}_{i}(\boldY)$ is empty, set 
$\hatB_{1i}=\hatB_{2i} = (1,1)$. For $i \in [l]$ such that 
$\mathcal{D}_{i}(\boldY)$ is \textit{not} empty, choose one among the pairs  in
$\mathcal{D}_{i}(\boldY)$ uniformly at random, 
and set $(\hatB_{1i},\hatB_{2i})$ to be that pair. Note that if 
$\mathcal{D}_{i}(\boldY)$ is a singleton for each $i \in [l]$, there is a 
unique choice for $\ulinehatB_{1},\ulinehatB_{2}$. The channel code decoder 
forwards $\ulinehatA,\ulinehatB_{1},\ulinehatB_{2}$ to the source code decoder.

In the first step, the decoded messages $\hat{A}_{t}: t \in [m]$ is mapped to 
the corresponding typical 
sequences in $T_{\delta}^{l}(K_{1})$. Let $\boldhatK(t,1:l): t \in [m]$ denote 
the 
corresponding sequences. The map from $\ulinehatK$ to $\boldhatK$ is via the 
decoder of the fixed B-L typical set source code $T_{\delta}^{l} (K_{1} )$. In 
the second step, the Slepian Wolf decoder performs a
standard joint-typicality decoding within the indexed pair 
$\ulinehatB_{1},\ulinehatB_{2}$ of bins, treating the rows of $\boldhatK$ as 
$m$ 
super-symbols of side-information. Specifically,
\begin{eqnarray}
\mathcal{D}(\boldhatK,\ulinehatB_{1},\ulinehatB_{2}) \define \left\{ 
(\boldhats_{1},\boldhats_{2}) : 
\beta(\boldhats_{j}) = 
\ulinehatB_{j} : j \in [2], \mbox{ and 
}(\boldhatK,\boldhats_{1},\boldhats_{2})\mbox{ is 
jointly typical wrt }\prod_{t=1}^{m} p_{\hatK^{l}S_{1}^{l}S_{2}^{l}} \right\}, 
\nonumber
\end{eqnarray}
where, for $s_{1}^{l} = s_{11}s_{12}\cdots s_{1l}$, $s_{2}^{l} = 
s_{21}s_{22}\cdots s_{2l}$, 
\begin{eqnarray}
\label{Eqn:Step1MACSourceCodeDecodingPMF}
p_{S_{1}^{l}S_{2}^{l}
\hatK^{l}}(s_{1}^{l},s_{
2 }^{l},\hatk^{l}) = 
p_{\hatsfK^{l}|\sfK_{1}^{l}\sfK_{2}^{l}}  \left(  \hatk^{
l} \left| \!\!\!\begin{array}{c} f_{1}(s_{11})f_{1}(s_{12}),\cdots 
f_{1}(s_{1l})\\  f_{2}(s_{21})f_{2}(s_{22})\cdots f_{2}(s_{2l}) \end{array} 
\!\!  \right.\right) 
\prod_{i=1}^{l}\mathbb{W}_{S_{1}S_{2}}(s_{1i},s_{2i}), \\
\label{Eqn:Step1MACSourceCodeDecodingAuxPMF}
p_{\hatsfK^{l} 
 | \sfK_{1}^{l} \sfK_{2}^{l}} (\hatk^{l} |k_{1}^{l},k_{2}^{l}) = 
\!\!\!\sum_{\substack{y^{l} \in\mathcal{Y}^{l} 
}}\!p_{\sfY^{l}|\sfU_{1}^{l},\sfU_{2}^{l}}(y^{ l} | 
e_{u}(e_{K}(k_{1}^{l})),e_{u}(e_{K}(k_{2}^{l})))\mathds{1}_{  
\left\{ \!\!\!\begin{array}{c} \hatk^{l} = d_{K}(d_{u}(y^{l}))
\end{array}\!\!\!\right\} },
\end{eqnarray}
and $p_{Y^{l}|U_{1}^{l}U_{2}^{l}}$ is given by the corresponding 
conditional marginal in (\ref{Eqn:Step1MACPMFForDecodingRule}). If 
$\mathcal{D}(\boldhatK,\ulinehatB_{1},\ulinehatB_{2})$ is empty, set 
$(\boldhatS_{1},\boldhatS_{2})$ to a predefined pair in $\boldCalS_{1}\times 
\boldCalS_{2}$ that is arbitrarily fixed upfront. Otherwise, choose one among 
the pairs in $\mathcal{D}(\boldhatK,\ulinehatB_{1},\ulinehatB_{2})$ uniformly at 
random and set $(\boldhatS_{1},\boldhatS_{2})$ to be that pair. Declare 
$(\boldhatS_{1},\boldhatS_{2})$ as the decoded matrix of source symbols.

\textit{Error event}: Let us characterize the error event $\mathscr{E}$. 
Suppose
\begin{eqnarray}
\mathscr{E}_{1} \define \bigcup_{i=1}^{l}\left\{ (B_{1i},B_{2i}) 
\neq (\hatB_{1i},\hatB_{2i}) \right\}, 
\mathscr{E}_{2} = \left\{  (\boldhatK,\boldS_{1},\boldS_{2}) \mbox{ is not 
typical wrt} \prod_{t=1}^{m} p_{ \hatK^{l} S_{1}^{l} S_{2}^{l}} \right\} 
\nonumber\\
\mathscr{E}_{3} = \bigcup_{\substack{(\boldhats_{1},\boldhats_{2}) \\\in 
\boldCalS_{1} \times 
\boldCalS_{2}}}\left\{\!\!\!\begin{array}{c}  (\boldS_{1},\boldS_{2}) \neq 
(\boldhats_{1},\boldhats_{2}), 
\beta_{j}(\boldhats_{j}) = \ulineB_{j}: j \in [2]  
\\(\boldhatK,\boldhats_{1},\boldhats_{2}) \mbox{ is typical wrt } 
\prod_{t=1}^{m} p_{ \hatK^{l} S_{1}^{l} S_{2}^{l}}  \end{array}\!\!\! 
\right\}\mbox{, then note that } \mathscr{E} \subseteq \mathscr{E}_{1} \cup 
\mathscr{E}_{2} \cup \mathscr{E}_{3}.
\nonumber
\end{eqnarray}
Indeed, if the Slepian Wolf decoder is provided the pair of bin indices that 
contain the observed source matrices, $(\boldhatK,\boldS_{1},\boldS_{2})$ 
is typical wrt $\prod_{t=1}^{m} p_{ \hatK^{l} S_{1}^{l} S_{2}^{l}}$, and there 
exists \textit{no} other pair $(\boldhats_{1},\boldhats_{2})$ in the indexed 
bin pair that is jointly typical with reconstructions $\boldhatK$, then the 
Slepian Wolf decoder will declare $(\boldhatS_{1},\boldhatS_{2}) = 
(\boldS_{1},\boldS_{2})$, thus confirming $\overline{\mathscr{E}_{1} 
\cap 
\mathscr{E}_{2} \cap \mathscr{E}_{3}} \subseteq \overline{\mathscr{E} }$. 
$\mathscr{E}_{1}$ corresponds to erroneous decoding into one of the pairs 
$C_{V_{1},i},C_{V_{2},i}$ of codebooks. $\mathscr{E}_{2},\mathscr{E}_{3}$ are 
error events concerning the Slepian Wolf code. In the following, we derive 
upper bounds on $P(\mathscr{E}_{1}), P(\mathscr{E}_{2}),P(\mathscr{E}_{3})$.

\textit{Probability of Error Analysis} : We analyze error probability of a 
random code. Towards that end, let us describe its distribution. As we 
mentioned, we do not randomize over the choice of fixed B-L typical set source 
code and the constant composition code $C_{U}$. In other words, the marginal of 
the pmf of the random code corresponding to these components is singular. This 
leaves us with having to specify the distribution of 
random (i) binning indices $\beta_{j}(\bolds_{j}^{lm}): s_{j}^{lm} \in 
\mathcal{S}_{j}^{lm}:j \in [2]$ that constitute the $\infty-$B-L Slepian Wolf 
source 
code, (ii) codewords $V_{ji}^{m}(b_{j}): b_{j} \in [M_{V_{j}}]: i \in [l]$, 
(iii) surjective maps $\pi_{t}: [l] \rightarrow [l] : t \in [m]$, and (iv) 
$\boldx_{j}(\boldu,\boldv) \in \mathcal{X}^{ m \times l} : (\boldu,\boldv) \in 
\mathcal{U}^{m \times l} \times \mathcal{V}_{j}^{m \times l}$. The four
elements (i) $(\beta_{j}(\bolds_{j}^{lm}):\bolds_{j} \in \mathcal{S}_{j}^{lm}): 
j \in [2]$, (ii) $(V_{ji}^{m}(b_{j})\in 
\mathcal{V}_{j}^{m}:b_{j}\in 
[M_{V_{j}}],i \in [l], j\in [2])$, (iii) $(\Pi_{t}: t \in [m])$ and (iv) 
$(\boldX_{j}(u,v_{j})\in 
\mathcal{X}_{j}^{m \times l}:u \in \mathcal{U}^{m \times l}, v_{j} \in 
\mathcal{V}_{j}^{m \times l})$ are mutually independent. With regard to the 
bin indices, the collections 
$(\beta_{1}(\bolds_{1}^{lm}): \bolds_{1}^{lm} \in \mathcal{S}_{1}^{m \times 
l})$ and $(\beta_{2}(\bolds_{2}^{lm}): \bolds_{2}^{lm} \in 
\mathcal{S}_{2}^{m \times 
l})$ are mutually independent. Moreover, for each $j 
\in [2]$, the bin indices $\beta_{j}(\bolds_{j}^{lm}): \bolds_{j}^{lm} \in 
\mathcal{S}_{j}^{lm}$ are uniformly and independently chosen from 
$[M_{V_{j}}]^{l}$. The $m$ surjective 
maps 
$\Pi_{t}:t \in [m]$ are mutually independent and uniformly distributed over the 
entire collection of surjective maps over $[l]$. Each codeword in the 
collection 
$(V_{ji}^{m}(b_{j})\in \mathcal{V}_{j}^{m}:b_{j}\in [M_{V_{j}}],i \in [l], j\in 
[2])$ is mutually independent of the others and $V_{ji}^{m}(b_{ji}) \sim 
\prod_{t=1}^{m}p_{V_{j}}(\cdot)$, where $p_{V_{j}}$ corresponds to the chosen 
test channel. The collection $(\boldX_{j}(u,v_{j})\in \mathcal{X}_{j}^{m \times 
l}:u \in \mathcal{U}^{m \times l}, v_{j} \in \mathcal{V}_{j}^{m \times l})$ is 
mutually independent and $\boldX_{j}(u,v_{j}) \sim 
\prod_{t=1}^{m}\prod_{i=1}^{l}p_{X_{j}|UV_{j}}(\cdot|u(t,i),v_{j}(t,i))$. This 
defines the distribution of our random code. We employ an analogous notation 
for our random code. For example, given $\ulineb_{j} = (b_{ji}: i \in [l])$, we 
let $\boldV_{j}\{ \ulineb_{j} \} \in \boldCalV_{j}$ be defined through 
$\boldV_{j}\{ \ulineb_{j}\}^{\Pi}(1:m,i) = V_{ji}^{m}(b_{ji}) : i \in [l]$, and 
similarly $\boldX_{j}\{ \ulinea_{j},\ulineb_{j} \} \define 
\boldX_{j}(\boldu\{ \ulinea_{j} \} ,\boldV\{\ulineb_{j}\})$.

Before we analyze $P(\mathscr{E}_{i}):i \in [3]$, we prove that (i) the rows of 
the collection \[\boldU_{j} \define \boldu \{\ulineA_{j} \}, 
\boldV_{j} \define \boldV_{j}\{ \ulineB_{j} \}, \boldX_{j} \define \boldX_{j}\{ 
\ulineA_{j},\ulineB_{j}\} : j \in [2],\boldY\] are IID with pmf 
$p_{\ulineU^{l}\ulineV^{l}\ulineX^{l}Y^{l}}$ defined in 
(\ref{Eqn:Step1MACPMFForDecodingRule}), and (ii) the rows in 
$\boldS_{1},\boldS_{2},\boldhatK$ are IID with pmf 
$p_{S_{1}^{l}S_{2}^{l}\hatK^{l}}$ - the corresponding marginal of the pmf in 
(\ref{Eqn:Step1MACSourceCodeDecodingPMF}). As the informed reader will note, 
this forms a key step in deriving upper bounds on $P(\mathscr{E}_{i}):i \in 
[3]$. We note that

\begin{eqnarray}
 \label{Eqn:Step1MACPrelim2-1}
 P\left(\!\!\!  
 \begin{array}{c}
\boldu\{ \ulineA_{j} \} = \boldu_{j}, \boldV_{j}\{ \ulineB_{j} \} = \boldv_{j}\\
\boldX_{j}\{ \ulineA_{j},\ulineB_{j} 
\}=\boldx_{j}: j \in [2]\\ \boldY=\boldy
 \end{array}
 \!\!\!\right) =
\sum_{\substack{\ulinea_{1},\ulinea_{2}\\\ulineb_{1},\ulineb_{2}}}\!
P \left(  \!\!\!
\begin{array}{c}
\ulineA_{j}=\ulinea_{j}\\\ulineB_{j}=\ulineb_{j} \\: j \in [2]
\end{array}
 \!\!\! \right)
 P\left(\!\!\!  \left.
 \begin{array}{c}
\boldu\{ \ulinea_{j} \} = \boldu_{j}, \boldV_{j}\{ \ulineb_{j} \} = \boldv_{j} 
\\
\boldX_{j}\{ \ulinea_{j},\ulineb_{j} 
\}=\boldx_{j} : j \in [2]\\
\boldY=\boldy
 \end{array}
 \!\!\!\right| \!\!\!
\begin{array}{c}
\ulineA_{j}=\ulinea_{j}\\\ulineB_{j}=\ulineb_{j} \\:j \in [2]
\end{array}
 \!\!\!
 \right) \\
 \label{Eqn:Step1MACPrelim2-2}
 = \sum_{\substack{\ulinea_{1},\ulinea_{2}\\\ulineb_{1},\ulineb_{2}}}\!
P \left(  \!\!\!
\begin{array}{c}
\ulineA_{j}=\ulinea_{j}\\\ulineB_{j}=\ulineb_{j} \\: j \in [2]
\end{array}
 \!\!\! \right) \prod_{t=1}^{m}\left(  \mathds{1}_{\left\{\!\!\! 
\begin{array}{c} \boldu_{j}(t,1:l) =\\ u^{l}(a_{jt}): j \in [2] 
\end{array}\!\!\!\right\}}\left[ \prod_{i=1}^{l}\left\{ \prod_{j=1}^{2}
p_{V_{j}}(\boldv_{j}(t,\Pi_{t}(i)))p_{X_{j}|UV_{j}}
\left(
\boldx_{j}(t, 
i)\left|\!\!\!\begin{array}{c}
\boldu_{j}(t,i)\\\boldv_{j}(t,i)
\end{array}\right.
\!\!\!\right) 
\right\} \right.  \right.\nonumber\\
\label{Eqn:Step1MACPrelim2-3}
\left. 
\left. 
\mathbb{W}_{Y | \ulineX} ( \boldy(t,i) 
| \boldx_{1}(t, i), \boldx_{2}(t,i))\right] 
\right)\\
 \label{Eqn:Step1MACPrelim2-4}
 = \sum_{\substack{\ulinea_{1},\ulinea_{2}\\\ulineb_{1},\ulineb_{2}}}\!
P \left(  \!\!\!
\begin{array}{c}
\ulineA_{j}=\ulinea_{j}\\\ulineB_{j}=\ulineb_{j} \\: j \in [2]
\end{array}
 \!\!\! \right) \prod_{t=1}^{m}\left(  \mathds{1}_{\left\{\!\!\! 
\begin{array}{c} \boldu_{j}(t,1:l) =\\ u^{l}(a_{jt}): j \in [2] 
\end{array}\!\!\!\right\}}\left[ \prod_{i=1}^{l}\left\{ \prod_{j=1}^{2}
p_{V_{j}}(\boldv_{j}(t,i))p_{X_{j}|UV_{j}}(\boldx_{j}(t, 
i)|\boldu_{j}(t,i)\boldv_{j}(t,i)) \right\} \right.  \right.\nonumber\\
\label{Eqn:Step1MACPrelim2-5}
\left. 
\left. 
\mathbb{W}_{Y | \ulineX} ( \boldy(t,i) 
| \boldx_{1}(t, i), \boldx_{2}(t,i))\right] 
\right)
\end{eqnarray}
where, in stating (\ref{Eqn:Step1MACPrelim2-3}) we have used
\begin{eqnarray}
 \label{Eqn:Step1MACPrelim1-Prelim1}
 P\left( \!\!\!
 \begin{array}{c}
\boldu \{\ulinea_{j} \} = \boldu_{j},
\boldV_{j} \{ \ulineb_{j} \} = \boldv_{j}\\: j \in [2]
 \end{array}\!
 \left| \!
 \begin{array}{c}
  \ulineA_{j}=\ulinea_{j},\ulineB_{j}=\ulineb_{j} \\: j \in [2]
 \end{array}\!\!\!
 \right. \right) &=& P \left(  \!\!
 \begin{array}{c}
 \boldV_{ji}^{m}(b_{ji}) = \boldv_{j}^{\Pi}(1:m,i)
:i \in [2],\\j \in [2],\boldu \{\ulinea_{j} \} = \boldu_{j}: j \in [2]
 \end{array}\!  \right) 
 \\
  \label{Eqn:Step1MACPrelim1-Prelim1Mid}
 &=& \prod_{t=1}^{m}  \mathds{1}_{\left\{\!\!\! 
\begin{array}{c} \boldu_{j}(t,1:l) =\\ u^{l}(a_{jt}): j \in [2] 
\end{array}\!\!\!\right\}} \prod_{i=1}^{l} \prod_{j=1}^{2}
p_{V_{j}}(\boldv_{j}(t,\Pi_{t}(i))),~~~~\\
 \label{Eqn:Step1MACPrelim1-Prelim2}
 P\left( \!\!\!
 \begin{array}{c}
\boldX_{j}\{ \ulineA_{j},\ulineB_{j} 
\}=\boldx_{j}\\: j \in [2]
 \end{array}\!
 \left| \!
 \begin{array}{c}
  \boldu \{\ulinea_{j} \} = 
\boldu_{j},\ulineA_{j}=\ulinea_{j}\\
\boldV_{j} \{ \ulineb_{j} \} = \boldv_{j},\ulineB_{j}=\ulineb_{j} \\: j \in [2]
 \end{array}\!\!\!
 \right. \right) &=& \prod_{t=1}^{m}   \prod_{i=1}^{l} \prod_{j=1}^{2}
p_{X_{j}|UV_{j}}(\boldx_{j}(t, 
i)|\boldu_{j}(t,i)\boldv_{j}(t,i)), 
\end{eqnarray}
which follow from the distribution of the code, and 
in arriving at (\ref{Eqn:Step1MACPrelim2-5}) we used
$\prod_{t=1}^{m}\prod_{i=1}^{l}p_{V_{j}}(\boldv_{j}(t,\Pi_{t}(i))) = 
\prod_{t=1}^{m}\prod_{i=1}^{l}p_{V_{j}}(\boldv_{j}(t,i))$. Expression 
(\ref{Eqn:Step1MACPrelim2-5}) is given by
\begin{eqnarray}
\label{Eqn:Step1MACPrelim2-6}
\sum_{\substack{\ulinea_{1},\ulinea_{2}}}\!
P \left(  \!\!\!
\begin{array}{c}
\ulineA_{j}=\ulinea_{j}\\: j \in [2]
\end{array}
 \!\!\! \right) \prod_{t=1}^{m}\left(  \mathds{1}_{\left\{\!\!\! 
\begin{array}{c} \boldu_{j}(t,1:l) =\\ u^{l}(a_{jt}): j \in [2] 
\end{array}\!\!\!\right\}}\left[ \prod_{i=1}^{l}\left\{ \prod_{j=1}^{2}
p_{V_{j}}(\boldv_{j}(t,i))p_{X_{j}|UV_{j}}(\boldx_{j}(t, 
i)|\boldu_{j}(t,i)\boldv_{j}(t,i)) \right\} \right.  \right.\nonumber\\
\label{Eqn:Step1MACPrelim2-7}
\left. 
\left. 
\mathbb{W}_{Y | \ulineX} ( \boldy(t,i) 
| \boldx_{1}(t, i), \boldx_{2}(t,i))\right] \right)\nonumber\\
\label{Eqn:Step1MACPrelim2-8}
=
\sum_{\substack{\ulinea_{1},\ulinea_{2}}}\!
\prod_{t=1}^{m}\left(P\left(\!\!\!
\begin{array}{c}A_{j}=a_{jt}\\: j \in [2]
\end{array}
 \!\!\! \right)  
\mathds{1}_{\left\{\!\!\! 
\begin{array}{c} \boldu_{j}(t,1:l) =\\ u^{l}(a_{jt}): j \in [2] 
\end{array}\!\!\!\right\}}\left[ \prod_{i=1}^{l}\left\{ \prod_{j=1}^{2}
p_{V_{j}}(\boldv_{j}(t,i))p_{X_{j}|UV_{j}}(\boldx_{j}(t, 
i)|\boldu_{j}(t,i)\boldv_{j}(t,i)) \right\} \right.  \right.\nonumber\\
\label{Eqn:Step1MACPrelim2-9}
\left. 
\left. 
\mathbb{W}_{Y | \ulineX} ( \boldy(t,i) 
| \boldx_{1}(t, i), \boldx_{2}(t,i))\right] 
\right)\end{eqnarray}\begin{eqnarray}
\label{Eqn:Step1MACPrelim2-10}
=
\prod_{t=1}^{m}\left(
\left[ \sum_{\substack{a_{1} , a_{2}}}
P\left(\!\!\!
\begin{array}{c}A_{j}=a_{j}\\: j \in [2]
\end{array}
 \!\!\! \right)  
\mathds{1}_{\left\{\!\!\! 
\begin{array}{c} \boldu_{j}(t,1:l) =\\ u^{l}(a_{j}): j \in [2] 
\end{array}\!\!\!\right\}}\right]\left[ \prod_{i=1}^{l}\left\{ \prod_{j=1}^{2}
p_{V_{j}}(\boldv_{j}(t,i))p_{X_{j}|UV_{j}}\left(\boldx_{j}(t, 
i)\left|\!\!\!\begin{array}{c}\boldu_{j}(t,i)\\\boldv_{j}(t,i) 
\end{array}\!\!\!\right)\right\}\right. \right.  \right.\nonumber\\
\label{Eqn:Step1MACPrelim2-11}
\left. 
\left. 
\mathbb{W}_{Y | \ulineX} ( \boldy(t,i) 
| \boldx_{1}(t, i), \boldx_{2}(t,i))\right] \right)\\
\label{Eqn:Step1MACPrelim2-12}
= \prod_{t=1}^{m} p_{\ulinesfU^{l}\ulinesfV^{l}\ulinesfX^{l}\ulinesfY^{l}} 
 \!\!\left(\!\!\!\begin{array}{c}\boldu_{1}(t,1:l),\boldu_{2}(t,1:l),
 \boldv_{1}(t,1:l),\boldv_{2}(t,1:l),\\
 \boldx_{1}(t,1:l),\boldx_{2}(t,1:l),
 \boldy(t,1:l)\end{array} \!\!\!\right)
\end{eqnarray}
where (\ref{Eqn:Step1MACPrelim2-9}) follows from the invariance of the 
distribution of $A_{jt} = d_{K}(\boldk_{j}(t,1:l))$ with $t \in [m]$. Recall 
that $(A_{1},A_{2})$ is identically distributed as $(A_{1t},A_{2t})$ for any 
$t \in [m]$. This was stated prior to (\ref{Eqn:Step1MACPMFForDecodingRule}). 
In arriving at (\ref{Eqn:Step1MACPrelim2-11}), we (i) leveraged the sum over 
$\ulinea_{1},\ulinea_{2}$ being over all of $[M_{u}]^{m} \times [M^{m}]^{m}$, 
(ii) the rest of the terms not depending on $\ulinea_{1},\ulinea_{2}$, and 
(iii) $u^{l}(\cdot )$ being invariant with $t$. Finally, 
(\ref{Eqn:Step1MACPrelim2-12}) follows from definition of 
(\ref{Eqn:Step1MACPMFForDecodingRule}). Following from
(\ref{Eqn:Step1MACPrelim2-1}) to (\ref{Eqn:Step1MACPrelim2-12}), we conclude
\begin{eqnarray}
\label{Eqn:Step1MACPrelim2-13Conclusion}
 P\left(\!\!\!  
 \begin{array}{c}
\boldu\{ \ulineA_{j} \} = \boldu_{j}, \boldV_{j}\{ \ulineB_{j} \} = \boldv_{j}\\
\boldX_{j}\{ \ulineA_{j},\ulineB_{j} 
\}=\boldx_{j}: j \in [2]\\ \boldY=\boldy
 \end{array}
 \!\!\!\right) = \prod_{t=1}^{m} 
p_{\ulinesfU^{l}\ulinesfV^{l}\ulinesfX^{l}Y^{l}} 
 \!\!\left(\!\!\!\begin{array}{c}\boldu_{1}(t,1:l),\boldu_{2}(t,1:l),
 \boldv_{1}(t,1:l),\boldv_{2}(t,1:l),\\
 \boldx_{1}(t,1:l),\boldx_{2}(t,1:l),
 \boldy(t,1:l)\end{array} \!\!\!\right).
\end{eqnarray}

We now characterize pmf of $\boldS_{1},\boldS_{2},\boldhatK$. Towards that end, 
our first step is the following. Suppose for $j \in [2], t \in [m], i \in [l]$, 
we have $\boldk_{j}(t,i) = 
f_{j}(\bolds_{j}(t,i))$ and $a_{jt} = e_{k}(\boldk_{j}(t,1:l))$, then
\begin{eqnarray}
 \label{Eqn:Step1MACPrelim1-1}
 P\left(\!\!\! 
 \begin{array}{c}
\boldS_{j}=\bolds_{j},\boldX_{j}\{ \ulinea_{j},\ulineb_{j} \} = 
\boldx_{j},\ulineB_{j}=\ulineb_{j}\\
\boldV_{j}\{ 
\ulineb_{j} \} = \boldv_{j}: j \in [2],\boldY = \boldy, 
\boldhatK = \boldhatk
\end{array}
 \!\!\!\right) = P\left(\!\!\!\begin{array}{c} 
\boldS_{1}=\bolds_{1}\\\boldS_{2}=\bolds_{2}
\end{array}\!\!\!\right) 
\frac{1}{M_{V_{1}}^{l}M_{V_{2}}^{l}}\prod_{t=1}^{m}\left\{  \left\{ 
\prod_{i=1}^{l}  \mathbb{W}_{Y|\ulineX}\left(\!\!\! 
\begin{array}{c} \boldy(t,i) \end{array} 
\!\!\!\left|\!\!\! \begin{array}{c}
\boldx_{1}(t,i)\\ \boldx_{2}(t,i) \end{array}\!\!\!\right.\right)\right.  
\right.\\ \left. \left.
\label{Eqn:Step1MACPrelim1-2}
\left\{\prod_{j=1}^{2}p_{V_{j}}(\boldv_{j}(t,\Pi_{t}(i))) 
  p_{X_{j}|UV_{j}}( 
\boldx_{j} (t,i) | e_{u}(a_{jt})_{i},\boldv_{j}(t,i) )\right\} 
\right\}\mathds{1}_{\left\{  \!\!\! 
\begin{array}{c}
\boldhatk(t,1:l) =\\d_{k}( d_{u}( \boldy(t,1:l) ) )
\end{array}
\!\!\!\right\} }\right\}
\end{eqnarray}
wherein $e_{u}(a_{jt})_{i} 
$ denotes the 
$i$-th symbol in $e_{u}(a_{jt}) = u^{l} (a_{jt}) \in \mathcal{U}^{l}$. The 
truth of (\ref{Eqn:Step1MACPrelim1-2}) follows from (i)
\begin{eqnarray}
 \label{Eqn:Step1MACPrelim1Prelim1}
 P\left(\!\!\!
 \begin{array}{c}
 \ulineB_{j} = \ulineb_{j}:j \in [2]
 \end{array}\!\!\!\left|
 \!\!
 \begin{array}{c}
 \boldS_{1}=\bolds_{1},\boldS_{2}=\bolds_{2}
 \end{array}\!\!\!\right.
 \right) = P\left(\!\!\!
 \begin{array}{c}
 \beta_{j}(\bolds_{j}) = \ulineb_{j}:j \in [2]
 \end{array}\!\!\!\left|
 \!\!
 \begin{array}{c}
 \boldS_{1}=\bolds_{1},\boldS_{2}=\bolds_{2}
 \end{array}\!\!\! \right.
 \right) = \frac{1}{M_{V_{1}}^{l}M_{V_{2}}^{l}}
  \nonumber
\end{eqnarray}
owing to the uniform distribution of $\beta_{j}(\ulines_{j}) : j \in [2]$ and 
its independence from the source realization, (ii)
\begin{eqnarray}
 P\left( \!\!\!
 \begin{array}{c}
 \boldV_{j}\{\ulineb_{j}\} = \boldv_{j} :j \in [2]
 \end{array}\!\!\!\left|
 \!\!
 \begin{array}{c}
 \boldS_{j}=\bolds_{j},\ulineB_{j} = \ulineb_{j}: j \in [2]
 \end{array}\!\!\!\right.
 \right) = \prod_{t=1}^{m}\prod_{i=1}^{l}p_{V_{j}}(\boldv_{j}(t,\Pi_{t}(i))) 
\nonumber
\end{eqnarray}
since $\boldV_{j}\{ \ulineb_{j} \}$ is independent of the 
$\boldS_{j},\beta_{j}(\boldS_{j}): j \in [2]$, (iii)
\begin{eqnarray}
 P\left(  \!\!\!
 \begin{array}{c}
 \boldX_{j}\{\ulinea_{j},\ulineb_{j}\} = \boldx_{j} \\:j \in [2]
 \end{array}\!\!\!\left|
 \!\!
 \begin{array}{c}
 \boldV_{j}\{ \ulineb_{j}\} = \boldv_{j},\boldS_{j}=\bolds_{j},\ulineB_{j} = 
\ulineb_{j}\\: j \in [2]
 \end{array}\!\!\!\right.
 \right) = \prod_{t=1}^{m}\prod_{i=1}^{l}p_{X_{j}|UV_{j}}( 
\boldx_{j} (t,i) | e_{u}(a_{jt})_{i},\boldv_{j}(t,i) ) \nonumber
\end{eqnarray}
since  $a_{jt} = 
e_{k}(\boldk_{j}(t,1:l)) : t \in [m]$ and $e_{u}(a_{jt})_{i} 
$ denotes the $i$-th symbol in $e_{u}(a_{jt}) = u^{l} (a_{jt}) \in 
\mathcal{U}^{l}$, and most importantly, (iv) 
\begin{eqnarray}
 P\left(  \!\!\!
 \begin{array}{c}
 \boldY = \boldy
 \end{array}\!\!\!\left|
 \!\!
 \begin{array}{c}
 \boldV_{j}\{ \ulineb_{j}\} = \boldv_{j},\boldS_{j}=\bolds_{j},\ulineB_{j} = 
\ulineb_{j}\\ \boldX_{j}\{\ulinea_{j},\ulineb_{j}\} = \boldx_{j}: j \in [2]
 \end{array}\!\!\!\right.
 \right) = \prod_{t=1}^{m}\prod_{i=1}^{l}\mathbb{W}_{Y|\ulineX}\left(\!\!\! 
\begin{array}{c} \boldy(t,i) \end{array} 
\!\!\!\left|\!\!\! \begin{array}{c}
\boldx_{1}(t,i)\\ \boldx_{2}(t,i) \end{array}\!\!\!\right.\right)\nonumber
\end{eqnarray}
since $a_{jt} = e_{k}(\boldk_{j}(t,1:l)) : t \in [m]$. Expression 
(\ref{Eqn:Step1MACPrelim1-2}) is equal to
\begin{eqnarray}
\label{Eqn:Step1MACPrelim1-3}
P\left(\!\!\!\begin{array}{c} 
\boldS_{1}=\bolds_{1}\\\boldS_{2}=\bolds_{2}\end{array}\!\!\!\right)
\frac{1}{M_{V_{1}}^{l}M_{V_{2}}^{l}}\prod_{t=1}^{m}\left\{  \left\{ 
\prod_{i=1}^{l}  
\mathbb{W}_{Y|\ulineX}\left(\!\!\! 
\begin{array}{c} \boldy(t,i) \end{array} 
\!\!\!\left|\!\!\! \begin{array}{c}
\boldx_{1}(t,i)\\ \boldx_{2}(t,i) \end{array}\!\!\!\right.\right)\right.  
\!\!\!\left\{\prod_{j=1}^{2}p_{V_{j}}(\boldv_{j}(t,i)) 
  p_{X_{j}|UV_{j}}\left( 
\boldx_{j} (t,i) \left| \!\!\! 
\begin{array}{c} e_{u}(a_{jt})_{i}\\\boldv_{j}(t,i) \end{array}\!\!\!\right. 
\right)\right\}\right\} \nonumber\\ 
\label{Eqn:Step1MACPrelim1-4}
\left.
\mathds{1}_{\left\{  \!\!\! 
\begin{array}{c}
\boldhatk(t,1:l) =d_{k}( d_{u}( \boldy(t,1:l) ) ) 
\end{array}
\!\!\!\right\} }\right\} \\
\label{Eqn:Step1MACPrelim1-5}
= P\left(\!\!\!\begin{array}{c} 
\boldS_{1}=\bolds_{1}\\\boldS_{2}=\bolds_{2}\end{array}\!\!\!\right)\!\frac{1}{
M_{ V_{1}}^{l}M_{V_{2}}^{l}}\!\prod_{t=1}^{m}\! p_{\ulineV^{l}\ulineX^{l} 
Y^{l}|\ulineU^{l}}\!\left( \!\!\!\! \left.
\begin{array}{c}
\boldv_{1}(t,1:l),\boldv_{2}(t,1:l),\boldx_{1}(t,1:l)\\
\boldx_{2}(t,1:l),\boldy(t,1:l)
\end{array}\!\!\!\right| \!\!\!
\begin{array}{c}
 e_{u}(a_{1t})\\e_{u}(a_{2t})
\end{array}
\!\!\!
\right)
\mathds{1}_{\left\{  \!\!\! 
\begin{array}{c}
\boldhatk(t,1:l) =\\d_{k}( d_{u}( \boldy(t,1:l) ) )
\end{array}
\!\!\!\right\} }
\end{eqnarray}
where (i) (\ref{Eqn:Step1MACPrelim1-4}) is obtained by re-ordering the product 
$\prod_{i=1}^{l}\prod_{t=1}^{m}p_{V_{j}}(\boldv_{j}(t,\Pi_{t}(i))) = 
\prod_{i=1}^{l}\prod_{t=1}^{m}p_{V_{j}}(\boldv_{j}(t,i))$, and (iv) 
(\ref{Eqn:Step1MACPrelim1-5}) follows from noting that the marginal 
$p_{\ulineU^{l}}$ wrt pmf in (\ref{Eqn:Step1MACPMFForDecodingRule}) is given by
\begin{eqnarray}
\label{Eqn:Step1MACPU1U2Marginal}
p_{\ulineU^{l}}(u_{1}^{l},u_{2}^{l}) = 
\sum_{\substack{(a_{1},a_{2}) \in \\ 
[M_{u}]\times [M_{u}]}} 
\!\!P(
 \begin{array}{c}
 A_{1}=a_{1},A_{2}=a_{2}
  \end{array})\mathds{1}_{\left\{\substack{ 
 u^{l}(a_{j})=u_{j}^{l}:j \in [2]}\right\}}\mbox{ and hence }
 \nonumber\\
 p_{\ulineV^{l}\ulineX^{l} 
Y^{l}|\ulineU^{l}}(\ulinev^{l},\ulinex^{l},y^{l}|\ulineu^{l}) = 
\left[ \prod_{j=1}^{2} \left\{ 
\prod_{i=1}^{l}p_{V_{j}}(v_{ji})p_{X_{j}|UV_{j}}(x_{ji}|u_{ji},v_{ji})\right\} 
\right]\left[ \prod_{i=1}^{l} 
\mathbb{W}_{Y|X_{1}X_{2}}(y_{i}| x_{1i},x_{2i}) \right].
\nonumber
 \end{eqnarray}
Following from (\ref{Eqn:Step1MACPrelim1-2}) to (\ref{Eqn:Step1MACPrelim1-5}), 
we conclude that if
\begin{eqnarray}
\label{Eqn:Step1MACPrelim1-6Conclusion}
\lefteqn{ \begin{array}{c}
\boldk_{j}(t,i) = 
f_{j}(\bolds_{j}(t,i))\mbox{ and }a_{jt} = 
e_{k}(\boldk_{j}(t,1:l))\\ 
\mbox{for }j \in [2], t \in [m], i \in [l],\mbox{ 
we have}\end{array}\mbox{ 
then }
P\left(\!\!\! 
 \begin{array}{c}
\boldS_{j}=\bolds_{j},\boldX_{j}\{ \ulinea_{j},\ulineb_{j} \} = 
\boldx_{j},\ulineB_{j}=\ulineb_{j}\\
\boldV_{j}\{ 
\ulineb_{j} \} = \boldv_{j}: j \in [2],\boldY = \boldy, 
\boldhatK = \boldhatk
\end{array}
 \!\!\!\right) =}\nonumber\\&&\!\!\!\!\!\! \!\!P\left(\!\!\!\begin{array}{c} 
\boldS_{1}=\bolds_{1}\\\boldS_{2}=\bolds_{2}\end{array}\!\!\!\right)\!\frac{1}{
M_{ V_{1}}^{l}M_{V_{2}}^{l}}\!\prod_{t=1}^{m}\! p_{\ulineV^{l}\ulineX^{l} 
Y^{l}|\ulineU^{l}}\!\left( \!\!\!\! \left.
\begin{array}{c}
\boldv_{1}(t,1:l),\boldv_{2}(t,1:l),\boldx_{1}(t,1:l)\\
\boldx_{2}(t,1:l),\boldy(t,1:l)
\end{array}\!\!\!\right| \!\!\!
\begin{array}{c}
 e_{u}(a_{1t})\\e_{u}(a_{2t})
\end{array}
\!\!\!
\right)
\mathds{1}_{\left\{  \!\!\! 
\begin{array}{c}
\boldhatk(t,1:l) =\\d_{k}( d_{u}( \boldy(t,1:l) ) )
\end{array}
\!\!\!\right\} }\!.~~
\end{eqnarray}
Equipped with (\ref{Eqn:Step1MACPrelim1-6Conclusion}), we now characterize pmf 
of $\boldS_{1},\boldS_{2},\boldhatK$. Note that if $\boldk_{j}(t,i) = 
f_{j}(\bolds_{j}(t,i))$ and $a_{jt} = e_{k}(\boldk_{j}(t,1:l))$ for $j \in [2], 
t \in [m], i \in [l]$, we have 
\begin{eqnarray}
 \label{Eqn:Step1MACEpsilon1}
 P\left(\!\!\!  
 \begin{array}{c}
\boldS_{j}=\bolds_{j}: j \in [2]\\
\boldhatK = \boldhatk
 \end{array}
 \!\!\!\right) = 
\sum_{\ulineb_{1},\ulineb_{2}}
\sum_{\substack{\boldv_{1} \in 
\boldCalV_{1} \\ \boldv_{2} \in \boldCalV_{2}}}
\sum_{\substack{\boldx_{1} \in 
\boldCalX_{1} \\ \boldx_{2} \in \boldCalX_{2}}}
\sum_{\substack{\boldy \in 
\boldCalY }}
P \left(
\!\!\!  
\begin{array}{c}
\boldS_{j}=\bolds_{j},\boldX_{j}\{ \ulinea_{j},\ulineb_{j} \} = \boldx_{j}, 
\boldhatK = \boldhatk\\
\ulineB_{j}=\ulineb_{j}, \boldV_{j}\{ 
\ulineb_{j} \} = \boldv_{j}: j \in [2],\boldY = \boldy
\end{array}
 \!\!\!
\right)\nonumber\\
= \sum_{\ulineb_{1},\ulineb_{2}}
\sum_{\substack{\boldv_{1} \in 
\boldCalV_{1} \\ \boldv_{2} \in \boldCalV_{2}}}
\sum_{\substack{\boldx_{1} \in 
\boldCalX_{1} \\ \boldx_{2} \in \boldCalX_{2}}}
\sum_{\substack{\boldy \in 
\boldCalY}}
\!\!\!
\frac{P\left(\!\!\!\begin{array}{c} 
\boldS_{1}=\bolds_{1}\\\boldS_{2}=\bolds_{2}\end{array}\!\!\!\right)}{
M_{ V_{1}}^{l}M_{V_{2}}^{l}}\!\prod_{t=1}^{m}\! p_{\ulineV^{l}\ulineX^{l} 
Y^{l}|\ulineU^{l}}\!\left( \!\!\!\! \left.
\begin{array}{c}
\boldv_{1}(t,1:l),\boldv_{2}(t,1:l)\\\boldx_{1}(t,1:l)
\boldx_{2}(t,1:l)\\\boldy(t,1:l)
\end{array}\!\!\!\right| \!\!\!
\begin{array}{c}
 e_{u}(a_{1t})\\e_{u}(a_{2t})
\end{array}
\!\!\!
\right)
\mathds{1}_{\left\{  \!\!\! 
\begin{array}{c}
\boldhatk(t,1:l) =\\d_{k}( d_{u}( \boldy(t,1:l) ) )
\end{array}
\!\!\!\right\} }\nonumber\\
\label{Eqn:Step1MACPMFOfS1S2Khat}
= \sum_{\substack{\boldy \in 
\boldCalY}}P\left(\!\!\!\begin{array}{c} 
\boldS_{1}=\bolds_{1}\\\boldS_{2}=\bolds_{2}\end{array}\!\!\!\right)\prod_{t=1}^
{ m } p_{Y^{l}|\ulineU^{l}}\!\left( \!\!\!\!
\begin{array}{c}
\boldy(t,1:l)
\end{array}\!\!\left| \!\!
\begin{array}{c}
 e_{u}(e_{k}(\boldk_{1}(t,1:l)))\\e_{u}(e_{k}(\boldk_{2}(t,1:l)))
\end{array}\right.
\!\!\!
\right)
\mathds{1}_{\left\{  \!\!\! 
\begin{array}{c}
\boldhatk(t,1:l) =d_{k}( d_{u}( \boldy(t,1:l) ) ) 
\end{array}
\!\!\!\right\} }.\end{eqnarray}
Since the above sum is over all of 
$\boldCalY$, we rename dummy variables 
$\boldy(t,1:l)$ and we use 
(\ref{Eqn:Step1MACSourceCodeDecodingPMF}), 
(\ref{Eqn:Step1MACSourceCodeDecodingAuxPMF}) to conclude that 
(\ref{Eqn:Step1MACPMFOfS1S2Khat}) is equal to
\begin{eqnarray}
\lefteqn{\sum_{\substack{\boldy \in 
\boldCalY}}\prod_{t=1}^
{ m } \left\{ \prod_{i=1}^{l} \mathbb{W}_{S_{1}S_{2}}\left(\!\!\!  
 \begin{array}{c}\bolds_{1}(t,i)\\\bolds_{2} (t ,i)\end{array}
 \!\!\!
 \right)\right\} 
p_{Y^{l}
|\ulineU^{l}}\!\left( \!\!\!\! 
\begin{array}{c}
\boldy(t,1:l)
\end{array}\!\!\left| \!\!
\begin{array}{c}
 e_{u}(e_{k}(\boldk_{1}(t,1:l)))\\e_{u}(e_{k}(\boldk_{2}(t,1:l)))
\end{array}\right.
\!\!\!
\right)
\mathds{1}_{\left\{  \!\!\! 
\begin{array}{c}
\boldhatk(t,1:l) =\\d_{k}( d_{u}( \boldy(t,1:l) ) ) 
\end{array}
\!\!\!\right\} }}
\nonumber\\
&=& \prod_{t=1}^
{ m } \left[\left\{ \prod_{i=1}^{l} \mathbb{W}_{S_{1}S_{2}}\left(\!\!\!  
 \begin{array}{c}\bolds_{1}(t,i)\\\bolds_{2} (t ,i)\end{array}
 \!\!\!
 \right)\right\}  \left\{ \sum_{y^{l} \in \mathcal{Y}^{l}}
p_{Y^{l}
|\ulineU^{l}}\!\left( \!\!\!\! 
\begin{array}{c}
y^{l}
\end{array}\!\!\left| \!\!
\begin{array}{c}
 e_{u}(e_{k}(\boldk_{1}(t,1:l)))\\e_{u}(e_{k}(\boldk_{2}(t,1:l)))
\end{array}\right.
\!\!\!
\right)
\mathds{1}_{\left\{  \!\!\! 
\begin{array}{c}
\boldhatk(t,1:l) =d_{k}( d_{u}( y^{l} ) ) 
\end{array}
\!\!\!\right\} } \right\}\right]
\nonumber\nonumber\\
&=& \prod_{t=1}^
{ m } \left[\left\{ \prod_{i=1}^{l} \mathbb{W}_{S_{1}S_{2}}\!\!\left(\!\!\!  
 \begin{array}{c}\bolds_{1}(t,i)\\\bolds_{2} (t ,i)\end{array}
 \!\!\!
 \right)\right\}  \left\{  p_{\hatK^{l}|K_{1}^{l}K_{2}^{l}} 
\left(\!\boldhatk(t,1:l)\!\left| \!\!\begin{array}{c} 
f_{1}(\bolds_{1}(t,1))\cdots f_{1}(\bolds_{1}(t,l))\\ 
f_{2}(\bolds_{2}(t,1))\cdots f_{2}(\bolds_{2}(t,l))  \end{array}\!\! \!\right. 
\right) \right\}\right]
\nonumber\nonumber\end{eqnarray}\begin{eqnarray}
&=& \prod_{t=1}^
{ m } p_{S_{1}^{l}S_{2}^{l}\hatK^{l}}\left(\!\!\!  
 \begin{array}{c}\bolds_{1}(t,1:l),\bolds_{2} (t , 
1:l),\boldhatk(t,1:l)\end{array}
 \!\!\!
 \right). \nonumber
\end{eqnarray}
We therefore have
\begin{eqnarray}
 \label{Eqn:Step1S1S2KHatPMF}
  P\left(\!\!\!  
 \begin{array}{c}
\boldS_{1}=\bolds_{1}, \boldS_{2}=\bolds_{2},
\boldhatK = \boldhatk
 \end{array}
 \!\!\!\right)
 &=& \prod_{t=1}^
{ m } p_{S_{1}^{l}S_{2}^{l}\hatK^{l}}\left(\!\!\!  
 \begin{array}{c}\bolds_{1}(t,1:l),\bolds_{2} (t , 
1:l),\boldhatk(t,1:l)\end{array}
 \!\!\!
 \right)
\end{eqnarray}
We have established that the $m$ sub-blocks of the 
source and the reconstructions are IID with pmf 
$p_{S_{1}^{l}S_{2}^{l}\hatK^{l}}$. We can now appeal to standard arguments 
pertaining to Slepian Wolf decoding. In particular, using techniques presented 
in \cite[Chap 10]{201201NIT_ElgKim}, it can be verified that 
there exists $\xi > 0$, such that\footnote{Refer to 
\cite[Problem 10.9]{201201NIT_ElgKim}.}
\begin{eqnarray}
 \max{\{ P(\mathscr{E}_{2}),P(\mathscr{E}_{3})\}} \leq \exp \{ -m\xi \}\mbox{ if 
}\frac{\log M_{V_{j}}^{l}}{m} > H(S_{j}^{l}|\hatK^{l},S_{\msout{j}}^{l}) : j \in 
[2]~,~~ \frac{\log M_{V_{1}}^{l}M_{V_{2}}^{l}}{m} > 
H(S_{1}^{l},S_{2}^{l}|\hatK^{l}).
\end{eqnarray}

We now analyze $P(\mathscr{E}_{1})$, and in particular derive an upper bound on 
$ \sum_{i=1}^{l}P((B_{1i},B_{2i}) \neq (\hatB_{1i},\hatB_{2i}))$. Towards that 
end, let us focus on one of the terms in the previous sum. Furthermore, since
\begin{eqnarray}
 \lefteqn{P ( \!
 \begin{array}{c}
  (B_{1i},B_{2i}) \neq  (\hatB_{1i},\hatB_{2i})
 \end{array}\!
 )  \leq 
 P (
  (\!
 \begin{array}{c}
  V_{1i}^{m}(B_{1i}),V_{2i}^{m}(B_{2i}) ,
  \boldY^{\Pi}(1:m,i)
 \end{array}\!) \ntypical \prod_{t=1}^{m} 
p_{\mathscr{V}_{1}\mathscr{V}_{2}\mathscr{Y}}
 ) }\nonumber\\ 
  \label{Eqn:Step1MAC-E1}
  &&+ \displaystyle P \left( \bigcup_{\substack{\hatb_{1i},\hatb_{2i} \in 
\\M_{V_{1}} \times M_{V_{2}} }}  \! \! \! \! \!\left\{
 \!
 \begin{array}{c}
  (B_{1i},B_{2i})  \neq (\hatb_{1i},\hatb_{2i})
 \end{array}\!, ~( \! \!
 \begin{array}{c}
  V_{1i}^{m}(\hatb_{1i}),V_{2i}^{m}(\hatb_{2i}) ,
  \boldY^{\Pi}(1:m,i)
 \end{array}  \! \!) \typical \prod_{t=1}^{m} 
p_{\mathscr{V}_{1}\mathscr{V}_{2}\mathscr{Y}} \right\}
 \right)
\end{eqnarray}
With regard to the first term in (\ref{Eqn:Step1MAC-E1}), it suffices to prove 
\begin{eqnarray}\label{Eqn:Step1MAC-E1FirstEventSuff1}
\left( V_{1i}^{m}(B_{1i}),V_{2i}^{
m } (B_ { 2i } ) ,  \boldY^{\Pi}(1:m,i) \right) \mbox{ is 
distributed with pmf }\prod_{t=1}^{m} 
p_{\mathscr{V}_{1}\mathscr{V}_{2}\mathscr{Y}}.
\end{eqnarray}
Since 
$V_{1i}^{m}(B_{1i}),V_{2i}^{m}(B_{2i}) ,  \boldY^{\Pi}(1:m,i) = 
[\boldV_{1}\{ \ulineB_{1} \} ~\boldV_{2}\{ \ulineB_{2}\}~ 
\boldY]^{\Pi}(1:m,i)$, (\ref{Eqn:Step1MAC-E1FirstEventSuff1}) holds if 
\begin{eqnarray}\label{Eqn:Step1MAC-E1FirstEventSuff2}
\left(\boldV_{1}\{ \ulineB_{1} \},\boldV_{2}\{ \ulineB_{2}\}, 
\boldY \right) \!\!\begin{array}{c}\mbox{is 
distributed}\\\mbox{ with pmf }\end{array}\!\!\prod_{t=1}^{m} 
p_{V_{1}^{l}V_{2}^{l}Y^{l}}\mbox{ and }\Pi_{1},\cdots,\Pi_{m}\mbox{ is 
independent of } \boldV_{1}\{ \ulineB_{1} \},\boldV_{2}\{ \ulineB_{2}\}, 
\boldY.
\end{eqnarray}
Indeed, sufficiency of (\ref{Eqn:Step1MAC-E1FirstEventSuff2}) follows from 
Lemma \ref{Lem:FullInterleavingLemma} (Appendix 
\ref{AppSec:PMFOfInterleavedVector}). Our proof of 
(\ref{Eqn:Step1MAC-E1FirstEventSuff2}) will follow steps similar to those that 
got us from (\ref{Eqn:Step1MACPrelim2-1}) to 
(\ref{Eqn:Step1MACPrelim2-13Conclusion}). Note that
\begin{eqnarray}
\label{Eqn:Step1MAC-E1First--1}
 P\left(\!\!\!  
 \begin{array}{c}
\boldu\{ \ulineA_{j} \} = \boldu_{j}, \boldV_{j}\{ \ulineB_{j} \} = \boldv_{j}\\
\boldX_{j}\{ \ulineA_{j},\ulineB_{j} 
\}=\boldx_{j}: j \in [2]\\ \boldY=\boldy,\Pi_{t} =\pi_{t}: t \in [m]
 \end{array}
 \!\!\!\right) =
\sum_{\substack{\ulinea_{1},\ulinea_{2}\\\ulineb_{1},\ulineb_{2}}}\!
P \left(  \!\!\!
\begin{array}{c}
\ulineA_{j}=\ulinea_{j}\\\ulineB_{j}=\ulineb_{j} \\: j \in [2]
\end{array}
 \!\!\! \right)
 P\left(\!\!\!  \left.
 \begin{array}{c}
\boldu\{ \ulinea_{j} \} = \boldu_{j}, \boldV_{j}\{ \ulineb_{j} \} = \boldv_{j} 
\\
\boldX_{j}\{ \ulinea_{j},\ulineb_{j} 
\}=\boldx_{j} : j \in [2]\\
\boldY=\boldy,\Pi_{t} =\pi_{t}: t \in [m]
 \end{array}
 \!\!\!\right| \!\!\!
\begin{array}{c}
\ulineA_{j}=\ulinea_{j}\\\ulineB_{j}=\ulineb_{j} \\:j \in [2]
\end{array}
 \!\!\!
 \right),
 \end{eqnarray}
and we break up the second factor of a generic term in the sum above, just as 
we did in (\ref{Eqn:Step1MACPrelim1-Prelim1}) - 
(\ref{Eqn:Step1MACPrelim1-Prelim2}). In particular,
\begin{eqnarray}
 \label{Eqn:Step1MAC-E1First-0}
 P(\Pi_{t} = \pi_{t}: t \in [m] | 
\ulineA_{j}=\ulinea_{j},\ulineB_{j}=\ulineb_{j} :j \in [2] ) = \frac{1}{l!}^{m} 
\end{eqnarray}
\begin{eqnarray}
 \label{Eqn:Step1MAC-E1First-1}
 P\left( \!\!\!
 \begin{array}{c}
\boldV_{j} \{ \ulineb_{j} \} = \boldv_{j}\\\boldu \{\ulinea_{j} \} = 
\boldu_{j},: j \in [2]
 \end{array}\!
 \left| \!
 \begin{array}{c}
  \ulineA_{j}=\ulinea_{j},\ulineB_{j}=\ulineb_{j} \\: j \in 
[2],\Pi_{t}=\pi_{t}: t \in [m]
 \end{array}\!\!\!
 \right. \right) &=& P \left(  \!\!
 \begin{array}{c}
 \boldV_{ji}^{m}(b_{ji}) = \boldv_{j}^{\pi}(1:m,i)
:i \in [2],\\j \in [2],\boldu \{\ulinea_{j} \} = \boldu_{j}: j \in [2]
 \end{array}\!  \right) 
 \\
 \label{Eqn:Step1MAC-E1FirstMid}
 &=& \prod_{t=1}^{m}  \mathds{1}_{\left\{\!\!\! 
\begin{array}{c} \boldu_{j}(t,1:l) =\\ u^{l}(a_{jt}): j \in [2] 
\end{array}\!\!\!\right\}} \prod_{i=1}^{l} \prod_{j=1}^{2}
p_{V_{j}}(\boldv_{j}(t,\pi_{t}(i))),~~~~~~\\
 \label{Eqn:Step1MAC-E1First-2}
 P\left( \!\!\!
 \begin{array}{c}
\boldX_{j}\{ \ulineA_{j},\ulineB_{j} 
\}=\boldx_{j}\\: j \in [2]
 \end{array}\!
 \left| \!
 \begin{array}{c}
 \label{Eqn:Step1MAC-E1First-3}
  \boldu \{\ulinea_{j} \} = 
\boldu_{j},\ulineA_{j}=\ulinea_{j}\\
\boldV_{j} \{ \ulineb_{j} \} = \boldv_{j},\ulineB_{j}=\ulineb_{j} \\: j \in 
[2],\Pi_{t}=\pi_{t}: t \in [m]
 \end{array}\!\!\!
 \right. \right) &=& \prod_{t=1}^{m}   \prod_{i=1}^{l} \prod_{j=1}^{2}
p_{X_{j}|UV_{j}}(\boldx_{j}(t, 
i)|\boldu_{j}(t,i)\boldv_{j}(t,i)),
\end{eqnarray}
where (\ref{Eqn:Step1MAC-E1First-1}) - (\ref{Eqn:Step1MAC-E1First-3}) are 
analogous to (\ref{Eqn:Step1MACPrelim1-Prelim1}) - 
(\ref{Eqn:Step1MACPrelim1-Prelim2}). Substituting (\ref{Eqn:Step1MAC-E1First-1}) 
- (\ref{Eqn:Step1MAC-E1First-3}), rewriting 
$\prod_{t=1}^{m}\prod_{i=1}^{l}p_{V_{j}}(\boldv_{j}(t,\pi_{t}(i)))$ as 
$\prod_{t=1}^{m}\prod_{i=1}^{l}p_{V_{j}}(\boldv_{j}(t,i))$, 
(\ref{Eqn:Step1MAC-E1First--1}) is given by
\begin{eqnarray}
 P\left(\!\!\!  
 \begin{array}{c}
\boldu\{ \ulineA_{j} \} = \boldu_{j}, \boldV_{j}\{ \ulineB_{j} \} = \boldv_{j}\\
\boldX_{j}\{ \ulineA_{j},\ulineB_{j} 
\}=\boldx_{j}: j \in [2]\\ \boldY=\boldy,\Pi_{t} =\pi_{t}: t \in [m]
 \end{array}
 \!\!\!\right) = 
\sum_{\substack{\ulinea_{1},\ulinea_{2}\\\ulineb_{1},\ulineb_{2}}}\!
P \left(  \!\!\!
\begin{array}{c}
\ulineA_{j}=\ulinea_{j},\ulineB_{j}=\ulineb_{j} : j \in [2]
\end{array}
 \!\!\! \right)
  \prod_{t=1}^{m}\left(  \mathds{1}_{\left\{\!\!\! 
\begin{array}{c} \boldu_{j}(t,1:l) = u^{l}(a_{jt}): j \in [2] 
\end{array}\!\!\!\right\}}\right.\nonumber\\ \left. \times\left[ 
\prod_{i=1}^{l}\left\{ \prod_{j=1}^{2}
p_{V_{j}}(\boldv_{j}(t,i))p_{X_{j}|UV_{j}}(\boldx_{j}(t, 
i)|\boldu_{j}(t,i)\boldv_{j}(t,i)) \right\} 
\mathbb{W}_{Y | \ulineX} ( \boldy(t,i) 
| \boldx_{1}(t, i), \boldx_{2}(t,i))\right] 
\right)\left(\frac{1}{l!}\right)^{m}.
\label{Eqn:Step1MAC-E1-FirstEvent}
\end{eqnarray}
Verify that (\ref{Eqn:Step1MAC-E1-FirstEvent}), when substituted in 
(\ref{Eqn:Step1MAC-E1First--1}) yields an expression identical to 
(\ref{Eqn:Step1MACPrelim2-5}) scaled by a factor 
$\left(\frac{1}{l!}\right)^{m}$. Following steps identical to those from 
(\ref{Eqn:Step1MACPrelim2-5}) to (\ref{Eqn:Step1MACPrelim2-12}), we conclude
\begin{eqnarray}
\label{Eqn:Step1MACE1-FirstAlmConc}
 P\left(\!\!\!  
 \begin{array}{c}
\boldu\{ \ulineA_{j} \} = \boldu_{j}, \boldV_{j}\{ \ulineB_{j} \} = \boldv_{j}\\
\boldX_{j}\{ \ulineA_{j},\ulineB_{j} 
\}=\boldx_{j}: j \in [2]\\ \boldY=\boldy,\Pi_{t}=\pi_{t}: t \in [m]
 \end{array}
 \!\!\!\right)\!\!\! &=&\!\!\!\!\! \left( 
\frac{1}{l!}\right)^{m}\prod_{t=1}^{m} 
p_{\ulinesfU^{l}\ulinesfV^{l}\ulinesfX^{l}Y^{l}} 
 \!\!\left(\!\!\!\begin{array}{c}\boldu_{1}(t,1:l),\boldu_{2}(t,1:l),
 \boldv_{1}(t,1:l),\boldv_{2}(t,1:l)\\
 \boldx_{1}(t,1:l),\boldx_{2}(t,1:l),
 \boldy(t,1:l)\end{array} \!\!\!\right)\\
 \!\!\!&=&\!\!\!\!\!P\left(\!\!\!  
 \begin{array}{c}
\boldu\{ \ulineA_{j} \} = \boldu_{j}, \boldV_{j}\{ \ulineB_{j} \} = \boldv_{j}\\
\boldX_{j}\{ \ulineA_{j},\ulineB_{j} 
\}=\boldx_{j}: j \in [2], \boldY=\boldy
 \end{array}
 \!\!\!\right) P\left(\!\!\!  
 \begin{array}{c}
\Pi_{t}=\pi_{t}: t \in [m]\end{array} \!\!\!  \right),
\end{eqnarray}
and in particular

\begin{eqnarray}
\label{Eqn:Step1MACE1-FirstAlmConcAg}
 P\left(\!\!\!  
 \begin{array}{c}
\boldV_{j}\{ \ulineB_{j} \} = \boldv_{j}
: j \in [2]\\ \boldY=\boldy,\Pi_{t}=\pi_{t}: t \in [m]
 \end{array}
 \!\!\!\right)= 
\prod_{t=1}^{m} \frac{1}{l!}
p_{\ulinesfV^{l}Y^{l}} 
 \!\!\left(\!\!\!\begin{array}{c}
 \boldv_{j}(t,1:l): j \in [2]\\
  \boldy(t,1:l)\end{array} \!\!\!\right) = P\left(\!\!\!  
 \begin{array}{c}
\boldV_{j}\{ \ulineB_{j} \} = \boldv_{j}\\: j \in [2], \boldY=\boldy
 \end{array}
 \!\!\!\right) P\left(\!\!\!  
 \begin{array}{c}
\Pi_{t}=\pi_{t}\\: t \in [m]\end{array} \!\!\!  \right),
\end{eqnarray}
(\ref{Eqn:Step1MACE1-FirstAlmConcAg}) proves 
(\ref{Eqn:Step1MAC-E1FirstEventSuff2}). We therefore conclude existence of a 
$\xi > 0$ such that the first term in the RHS of (\ref{Eqn:Step1MAC-E1}) 
\begin{eqnarray}
\label{Eqn:Step1MAC-E1Bnd1stEvent}
 P (
  (\!
 \begin{array}{c}
  V_{1i}^{m}(B_{1i}),V_{2i}^{m}(B_{2i}) ,
  \boldY^{\Pi}(1:m,i)
 \end{array}\!) \ntypical \prod_{t=1}^{m} 
p_{\mathscr{V}_{1}\mathscr{V}_{2}\mathscr{Y}}
 ) \leq \exp\{ -m\xi \}
\end{eqnarray}
and hence can be made arbitrarily small by choosing $m$ sufficiently large. In 
addition to the analysis provided herein, we point the reader to Appendix 
\ref{AppSec:PE2Event1FromFirstPrin} wherein we analyze the first term in 
(\ref{Eqn:Step1MAC-E1}) from first principles. We now analyze the second term 
in (\ref{Eqn:Step1MAC-E1}). To begin with, we 
derive an upper bound on
\ifTITVersion\begin{eqnarray}
 \label{Eqn:BothIllegitMsgsDecoded}\!
 P\!\left(\! \bigcup_{\substack{b_{1i}\\b_{2i}}} ~
\bigcup_{\substack{\hatb_{1i}\neq b_{1i} \\\hatb_{2i} \neq b_{2i} 
}}\!\!\!\!\! \left\{\!\!\! \begin{array}{l} B_{1i}=b_{1i}\\B_{2i} = b_{2i} 
\end{array}\!\!\!,\!\left(\!\!\! \begin{array}{c} 
V_{1i}^{m}(\hatb_{1i}),V_{2i}^{m}(\hatb_{2i})\\,\boldY^{\Pi}(1:m,i) \end{array} 
\!\!\! \right)\!\! \in\! T_{\beta}^{m}(p_{\underline{\mathscr{V}}\mathscr{Y}})  
\!\right\} 
\! \right)
\nonumber
\end{eqnarray}\fi
\ifPeerReviewVersion\begin{eqnarray}
 \label{Eqn:BothIllegitMsgsDecoded}\!
 P\!\left(\! \bigcup_{\substack{b_{1i}, b_{2i}}} ~
\bigcup_{\substack{\hatb_{1i}\neq b_{1i} 
}}\bigcup_{\hatb_{2i} \neq b_{2i}}\!\!\! \left\{\!\!\! \begin{array}{l} 
B_{1i}=b_{1i},B_{2i} = b_{2i} 
\end{array}\!\!\!,\!\left(\!\!\! \begin{array}{c} 
V_{1i}^{m}(\hatb_{1i}),V_{2i}^{m}(\hatb_{2i}),\boldY^{\Pi}(1:m,i) \end{array} 
\!\!\! \right)\!\! \in\! T_{\beta}^{m}(p_{\underline{\mathscr{V}}\mathscr{Y}})  
\!\right\} 
\! \right)
\nonumber
\end{eqnarray}\fi
By the union bound and the law of total probability, the above quantity is 
bounded on the above by
\ifTITVersion\begin{eqnarray}
\sum_{\substack{\ulinea_{1} \in 
[M_{u}]^{m},\\\ulinea_{2} \in [M_{u}]^{m}}}~
\sum_{\substack{\ulineb_{1} \in [M_{V_{1}}]^{l}\\\ulineb_{2}\in 
[M_{V_{2}}]^{l}}}~
%  \sum_{\substack{\ulinea_{1},\ulineb_{1}\\\ulinea_{2},\ulineb_{2}}}
 \sum_{\substack{\hatb_{1i} : \\\hatb_{1i}\neq b_{1i} }}~
\sum_{\substack{\hatb_{2i}:\\\hatb_{2i} \neq b_{2i} }}~
 \sum_{\substack{v_{1}^{m},v_{2}^{m}\\x_{1}^{m},x_{2}^{m}}}~
 \sum_{\substack{ (\hatv_{1}^{m},\hatv_{2}^{m},y^{m})\\ \in  
T_{\beta}^{m}(p_{\underline{\mathscr{V}}\mathscr{Y}})}}\nonumber\\
P\left( \!\!\!\begin{array}{c}
\ulineA_{j}=\ulinea_{j},V_{ji}^{m}(\hatb_{ji})= 
\hatv_{ j}^{m},\boldV_{j}\{ \ulineb_{j} \}^{\Pi}(1:m,i)=v_{j}^{m}\\
\ulineB_{j}=\ulineb_{j},\boldX_{j}\{ \ulinea_{j},\ulineb_{j} \}^{\Pi}(1:m,i) = 
x_{j}^{m} 
: j \in [2]\\\boldY^{\Pi}(1:m,i) = y^{m}\\
\end{array}\!\!\!  \right)\!\!\!.
\label{Eqn:PE2BothIlleAfterLOTP}
\end{eqnarray}\fi
\ifPeerReviewVersion\begin{eqnarray}
 \sum_{\substack{\ulinea_{1},\ulineb_{1}\\\ulinea_{2},\ulineb_{2}}}~
  \sum_{\substack{\hatb_{1i} : \\\hatb_{1i}\neq b_{1i} }}~
\sum_{\substack{\hatb_{2i}:\\\hatb_{2i} \neq b_{2i} }}~
\sum_{\substack{v_{1}^{m}\\v_{2}^{m}}}~
 \sum_{\substack{x_{1}^{m}\\x_{2}^{m}}}~
 \sum_{\substack{ (\hatv_{1}^{m},\hatv_{2}^{m},y^{m})\\ \in  
T_{\beta}^{m}(p_{\underline{\mathscr{V}}\mathscr{Y}})}}\!\!\!
P\left( \!\!\!\begin{array}{c}
\ulineA_{j}=\ulinea_{j},\ulineB_{j}=\ulineb_{j},V_{ji}^{m}(\hatb_{ji})= 
\hatv_{ j}^{m},\boldV_{j}\{ \ulineb_{j} \}^{\Pi}(1:m,i)=v_{j}^{m}\\
\boldX_{j}\{ \ulinea_{j},\ulineb_{j} \}^{\Pi}(1:m,i) = x_{j}^{m} 
: j \in [2],\boldY^{\Pi}(1:m,i) = y^{m}\\
\end{array}\!\!\!  \right).
\label{Eqn:PE2BothIlleAfterLOTP}
\end{eqnarray}\fi
Consider a generic term in the above sum. Firstly, the triple 
$ \ulineA_{j}, \ulineB_{j},\boldV_{j}\{ \ulineb_{j}\}^{\Pi}(1:m,i): j \in [2]$ 
is independent of $V_{ji}^{m}(\hatb_{ji}):j \in [2]$. This is because (i) the 
codebook generation process is independent of the messages, and (ii) 
$\boldV_{j}\{ \ulineb_{j}\}^{\Pi}(1:m,i): j \in [2]$ is a function of 
$V_{ji}(b_{ji}): j \in [2], i \in [l]$ and $\Pi_{t}: t \in [m]$, and these 
random objects are mutually independent of $V_{ji}^{m}(\hatb_{ji}): j \in [2]$. 
Secondly, $\boldX_{j}\{ \ulinea_{j},\ulineb_{j} \}^{\Pi}(1:m,i) : j \in [2]$ is 
conditionally independent of $V_{ji}^{m}(\hatb_{ji}): j \in [2]$ given 
$\ulineA_{j},\ulineB_{j},\boldV_{j}\{ \ulineb_{j} \}^{\Pi}(1:m,i) : j \in [2]$. 
This is true because (i) $\boldX_{j}\{ \ulinea_{j},\ulineb_{j} \}^{\Pi}(1:m,i) 
: j \in [2]$ is conditionally independent of the rest of the random objects, 
given $\boldV_{j}\{ \ulineb_{j} \}^{\Pi}(1:m,i), \boldu\{ 
\ulineA_{j}\}^{\Pi}(1:m,i): j \in [2]$, where $\boldu \{ \ulineA_{j}\}$ is a 
deterministic function\footnote{We do not randomize over the fixed B-L code 
$C_{U}$.} of $\ulineA_{j}$, and (ii) $\Pi_{t}: t \in [m]$ is independent of 
$V_{ji}^{m}(\hatb_{ji}): j \in [2]$. Finally, $\boldY^{\Pi}(1:m,i)$ is 
conditionally independent of $V_{ji}^{m}(\hatb_{ji}): j \in [2]$ given 
$\boldX_{j}\{ \ulinea_{j},\ulineb_{j} \}^{\Pi}(1:m,i) 
: j \in [2]$. These observations lead us to 
\ifTITVersion\begin{eqnarray}
 \label{Eqn:PE2SecEventViaIndependence}
 P\left( \!\!\!\begin{array}{c}
\left[ 
\boldV_{j}\{ \ulineb_{j} \} 
\boldX_{j}\{ \ulinea_{j},\ulineb_{j} \}
\boldY \right]^{\Pi}(1:m,i) = 
(v_{j}^{m},x_{j}^{m} 
,y^{m})\\\ulineA_{j}=\ulinea_{j},\ulineB_{j}=\ulineb_{j},V_{ji}^{m}(\hatb_{ji}
)= 
\hatv_{j}^{m}: j \in [2]
\end{array}\!\!\!  \right)\nonumber\\
% newline
\!\!= P\left( \!\!\!\!\begin{array}{c}
\left[ 
\boldV_{j}\{ \ulineb_{j} \} 
\boldX_{j}\{ \ulinea_{j},\ulineb_{j} \}
\boldY \right]^{\Pi}\!\!(1:m,i)\\ = 
(v_{j}^{m},x_{j}^{m} 
,y^{m}),\ulineA_{j}=\ulinea_{j},\ulineB_{j}=\ulineb_{j}\\: j \in [2]
\end{array}\!\!\!\!  \right)\!\!\!\prod_{j=1}^{2}\prod_{t=1}^{m} 
p_{V_{j}}(\hatv_{jt})
\label{Eqn:OneGenericTerm1}
\\
\!\!= P\left( \!\!\!\!\begin{array}{c}
\left[ 
\boldV_{j}\{ \ulineB_{j} \} 
\boldX_{j}\{ \ulineA_{j},\ulineB_{j} \}
\boldY \right]^{\Pi}\!\!(1:m,i)\\ = 
(v_{j}^{m},x_{j}^{m} 
,y^{m}),\ulineA_{j}=\ulinea_{j},\ulineB_{j}=\ulineb_{j}\\: j \in [2]
\end{array}\!\!\!\!  
\right)\!\!\!\prod_{j=1}^{2}\prod_{t=1}^{m}p_{V_{j}}(\hatv_{jt})
\label{Eqn:OneGenericTerm2}
\end{eqnarray}\fi
\ifPeerReviewVersion\begin{eqnarray}
 \label{Eqn:PE2SecEventViaIndependence}
 \lefteqn{P\left( \!\!\!\begin{array}{c}
\left[ 
\boldV_{j}\{ \ulineb_{j} \} 
\boldX_{j}\{ \ulinea_{j},\ulineb_{j} \}
\boldY \right]^{\Pi}(1:m,i) = 
(v_{j}^{m},x_{j}^{m} 
,y^{m}),\ulineA_{j}=\ulinea_{j},\ulineB_{j}=\ulineb_{j},V_{ji}^{m}(\hatb_{ji}
)= 
\hatv_{j}^{m}: j \in [2]
\end{array}\!\!\!  \right)}\nonumber\\
% newline
&=& P\left( \!\!\begin{array}{c}
\left[ 
\boldV_{j}\{ \ulineb_{j} \} 
\boldX_{j}\{ \ulinea_{j},\ulineb_{j} \}
\boldY \right]^{\Pi}\!\!(1:m,i) = 
(v_{j}^{m},x_{j}^{m} 
,y^{m}),\ulineA_{j}=\ulinea_{j},\ulineB_{j}=\ulineb_{j}: j \in [2]
\end{array}\!\!  \right)\!\prod_{j=1}^{2}\prod_{t=1}^{m} 
p_{V_{j}}(\hatv_{jt})
\label{Eqn:OneGenericTerm1}
\\
&=& P\left( \!\!\begin{array}{c}
\left[ 
\boldV_{j}\{ \ulineB_{j} \} 
\boldX_{j}\{ \ulineA_{j},\ulineB_{j} \}
\boldY \right]^{\Pi}\!\!(1:m,i) = 
(v_{j}^{m},x_{j}^{m} 
,y^{m}),\ulineA_{j}=\ulinea_{j},\ulineB_{j}=\ulineb_{j}: j \in [2]
\end{array}\!\!  
\right)\!\prod_{j=1}^{2}\prod_{t=1}^{m}p_{V_{j}}(\hatv_{jt})
\label{Eqn:OneGenericTerm2}
\end{eqnarray}\fi
Substituting (\ref{Eqn:OneGenericTerm2}) in (\ref{Eqn:PE2BothIlleAfterLOTP}) 
and summing over 
$\ulinea_{j},\ulineb_{j},v_{j}^{m},x_{j}^{m}: j \in [2]$, we obtain
\ifTITVersion\begin{eqnarray}
   \sum_{\substack{\hatb_{1i} : \\\hatb_{1i}\neq b_{1i} }}~
\sum_{\substack{\hatb_{2i}:\\\hatb_{2i} \neq b_{2i} }}
 \sum_{\substack{ (\hatv_{1}^{m},\hatv_{2}^{m},y^{m})\\ \in  
T_{\beta}^{m}(p_{\underline{\mathscr{V}}\mathscr{Y}})}} \!\!\!\!\!\!
P\left(\!\!\!\begin{array}{c}\boldY^{\Pi}(1:m,i) \\= 
y^{m}\end{array}\!\!\!\right)\prod_{j=1}^{2}\prod_{t=1}^{m}p_{V_{j}}(\hatv_{jt}
)\nonumber\\
= M_{V_{1}}M_{V_{2}}\!\!\!\!\!\!\!\sum_{\substack{ 
(\hatv_{1}^{m},\hatv_{2}^{m},y^{m})\\ \in  
T_{\beta}^{m}(p_{\underline{\mathscr{V}}\mathscr{Y}})}}\prod_{
t=1}^{m}p_{\hatY}(y_{t})p_{\hatV_{1}}(\hatv_{1t})p_{\hatV_{2}}(\hatv_{2t}
\label{Eqn:Step1MACBothIncorrectAfterSum}
\end{eqnarray}\fi
\ifPeerReviewVersion\begin{eqnarray}
 \sum_{\substack{\hatb_{1i} : \\\hatb_{1i}\neq b_{1i} }}~
\sum_{\substack{\hatb_{2i}:\\\hatb_{2i} \neq b_{2i} }}
 \sum_{\substack{ (\hatv_{1}^{m},\hatv_{2}^{m},y^{m})\\ \in  
T_{\beta}^{m}(p_{\underline{\mathscr{V}}\mathscr{Y}})}} 
\!\!\!\!\!\!\!\!P(\boldY^{\Pi}(1:m,i) = 
y^{m})\prod_{t=1}^{m}p_{V_{1}}(\hatv_{1t})p_{V_{2}}(\hatv_{2t}) 
= M_{V_{1}}M_{V_{2}}\!\!\!\!\!\!\!\sum_{\substack{ 
(\hatv_{1}^{m},\hatv_{2}^{m},y^{m})\\ \in  
T_{\beta}^{m}(p_{\underline{\mathscr{V}}\mathscr{Y}})}}\!\prod_{
t=1}^{m}p_{\mathscr{Y}}(y_{t})p_{\mathscr{V}_{1}}(\hatv_{1t})p_{\mathscr{V}_{2}}
(\hatv_ {2t}
)
\label{Eqn:Step1MACBothIncorrectAfterSum}
\end{eqnarray}\fi
as an upper bound on (\ref{Eqn:PE2BothIlleAfterLOTP}), if (i) pmf of 
$\boldY^{\Pi}(1:m,i)$ is $\prod_{t=1}^{m}p_{\mathscr{Y}}$, (ii) $p_{V_{j}} = 
p_{\mathscr{V}_{j}}$ for $ j \in [2]$. (i) follows from 
(\ref{Eqn:Step1MAC-E1FirstEventSuff1}), or 
(\ref{Eqn:Step1MAC-E1FirstEventSuff2}) in conjunction with Lemma 
\ref{Lem:FullInterleavingLemma}. We have proved 
(\ref{Eqn:Step1MAC-E1FirstEventSuff1}) and equivalently 
(\ref{Eqn:Step1MAC-E1FirstEventSuff2}) through the sequence of steps from 
(\ref{Eqn:Step1MAC-E1First--1}) through (\ref{Eqn:Step1MACE1-FirstAlmConcAg}). 
With regard to (ii), note that marginal $p_{V_{j}^{l}}$ wrt 
(\ref{Eqn:Step1MACPMFForDecodingRule}) is indeed equal to 
$\prod_{i=1}^{l}p_{V_{j}}$ (where $p_{V_{j}}$ is as provided in Thm statement). 
It is therefore straightforward to verify $p_{V_{j}} = 
p_{\mathscr{V}_{j}}$ for $ j \in [2]$.

Following standard typicality argument, for example lemma 
\cite[Lemma 3.1]{201201NIT_ElgKim}, it can be proved that given any $\eta > 0$, 
there exists a choice for $\beta > 0$ and $m_{\beta,\eta} \in \naturals$ such 
that for all $m \geq m_{\beta,\eta}$, (\ref{Eqn:Step1MACBothIncorrectAfterSum}) 
is at most 
$\eta$ if $\frac{\log M_{V_{1}}M_{V_{2}}}{m} < 
I(\mathscr{V}_{1}\mathscr{V}_{2};\mathscr{Y})$

We now consider the event when a codeword corresponding to an illegitimate 
message for one of the users is jointly typical. In particular, we derive an 
upper bound on
\ifTITVersion\begin{eqnarray}
 \label{Eqn:OneIllegMsgsDecoded}\!
 P\!\left(\! \bigcup_{\substack{b_{1i}\\b_{2i}}} ~
\bigcup_{\substack{\hatb_{1i}\neq b_{1i} }}\!\!\!\!\! \left\{\!\!\! 
\begin{array}{l} B_{1i}=b_{1i}\\B_{2i} = b_{2i} 
\end{array}\!\!\!,\!\left(\!\!\! \begin{array}{c} 
V_{1i}^{m}(\hatb_{1i}),V_{2i}^{m}(b_{2i})\\,\boldY^{\Pi}(1:m,i) \end{array} 
\!\!\! \right)\!\! \in\! T_{\beta}^{m}(p_{\underline{\mathscr{V}}\mathscr{Y}})  
\!\right\} 
\! \right)
\nonumber
\end{eqnarray}\fi
\ifPeerReviewVersion\begin{eqnarray}
 \label{Eqn:OneIllegMsgsDecoded}\!
 P\!\left(\! \bigcup_{\substack{b_{1i}\\b_{2i}}} ~
\bigcup_{\substack{\hatb_{1i}\neq b_{1i} }}\!\!\!\!\! \left\{\!\!\! 
\begin{array}{l} B_{1i}=b_{1i}\\B_{2i} = b_{2i} 
\end{array}\!\!\!,\!\left(\!\!\! \begin{array}{c} 
V_{1i}^{m}(\hatb_{1i}),V_{2i}^{m}(b_{2i})\\,\boldY^{\Pi}(1:m,i) \end{array} 
\!\!\! \right)\!\! \in\! T_{\beta}^{m}(p_{\underline{\mathscr{V}}\mathscr{Y}})  
\!\right\} 
\! \right)
\nonumber
\end{eqnarray}\fi
By the union bound and the law of total probability, the above quantity is 
bounded on the above by
\ifTITVersion\begin{eqnarray}
 \sum_{\substack{\ulinea_{1} \in 
[M_{u}]^{m},\\\ulinea_{2} \in [M_{u}]^{m}}}~
\sum_{\substack{\ulineb_{1} \in [M_{V_{1}}]^{l}\\\ulineb_{2}\in 
[M_{V_{2}}]^{l}}}~
%  \sum_{\substack{\ulinea_{1},\ulineb_{1}\\\ulinea_{2},\ulineb_{2}}}
 \sum_{\substack{\hatb_{1i} : \\\hatb_{1i}\neq b_{1i} }}~
 \sum_{\substack{v_{1}^{m}}}
 \sum_{\substack{x_{1}^{m}\\x_{2}^{m}}}
 \sum_{\substack{ (\hatv_{1}^{m},v_{2}^{m},y^{m})\\ \in  
T_{\beta}^{m}(p_{\underline{\mathscr{V}}\mathscr{Y}})}}\nonumber\\
P\left( \!\!\!\begin{array}{c}
\ulineA_{j}=\ulinea_{j},\boldV_{j}\{ \ulineb_{j} \}^{\Pi}(1:m,i)=v_{j}^{m}\\
\ulineB_{j}=\ulineb_{j},\boldX_{j}\{ \ulinea_{j},\ulineb_{j} \}^{\Pi}(1:m,i) = 
x_{j}^{m} 
: j \in [2]\\V_{1i}^{m}(\hatb_{1i})=
\hatv_{1}^{m},\boldY^{\Pi}(1:m,i) = y^{m}
\end{array}\!\!\!  \right).
\label{Eqn:PE2OneIlleAfterLOTP}
\end{eqnarray}\fi
\ifPeerReviewVersion\begin{eqnarray}
 \sum_{\substack{\ulinea_{1},\ulineb_{1}\\\ulinea_{2},\ulineb_{2}}}~
  \sum_{\substack{\hatb_{1i} : \\\hatb_{1i}\neq b_{1i} }}~
 \sum_{\substack{v_{1}^{m}}}
 \sum_{\substack{x_{1}^{m}\\x_{2}^{m}}}
 \sum_{\substack{ (\hatv_{1}^{m},v_{2}^{m},y^{m})\\ \in  
T_{\beta}^{m}(p_{\underline{\mathscr{V}}\mathscr{Y}})}}\!\!\!\!\!\!\!
P\left( \!\!\!\begin{array}{c}
\ulineA_{j}=\ulinea_{j},\ulineB_{j}=\ulineb_{j},V_{1i}^{m}(\hatb_{1i})= 
\hatv_{1}^{m},\boldV_{j}\{ \ulineb_{j} \}^{\Pi}(1:m,i)=v_{j}^{m}\\
\boldX_{j}\{ \ulinea_{j},\ulineb_{j} \}^{\Pi}(1:m,i) = x_{j}^{m} 
: j \in [2],\boldY^{\Pi}(1:m,i) = y^{m}\\
\end{array}\!\!\!  \right).
\label{Eqn:PE2OneIlleAfterLOTP}
\end{eqnarray}\fi
Consider a generic term in the above sum. We make three observations similar 
to the ones we made following (\ref{Eqn:PE2BothIlleAfterLOTP}). Firstly, the 
triple 
$ \ulineA_{j}, \ulineB_{j},\boldV_{j}\{ \ulineb_{j}\}^{\Pi}(1:m,i): j \in [2]$ 
is independent of $V_{1i}^{m}(\hatb_{1i})$. This is because (i) the 
codebook generation process is independent of the messages, and (ii) 
$\boldV_{j}\{ \ulineb_{j}\}^{\Pi}(1:m,i): j \in [2]$ is a function of 
$V_{ji}(b_{ji}): j \in [2], i \in [l]$ and $\Pi_{t}: t \in [m]$, and these 
random objects are mutually independent of $V_{1i}^{m}(\hatb_{1i})$. 
Secondly, $\boldX_{j}\{ \ulinea_{j},\ulineb_{j} \}^{\Pi}(1:m,i) : j \in [2]$ is 
conditionally independent of $V_{1i}^{m}(\hatb_{1i})$ given 
$\ulineA_{j},\ulineB_{j},\boldV_{j}\{ \ulineb_{j} \}^{\Pi}(1:m,i) : j \in [2]$. 
This is true because (i) $\boldX_{j}\{ \ulinea_{j},\ulineb_{j} \}^{\Pi}(1:m,i) 
: j \in [2]$ is conditionally independent of the rest of the random objects, 
given $\boldV_{j}\{ \ulineb_{j} \}^{\Pi}(1:m,i), \boldu\{ 
\ulineA_{j}\}^{\Pi}(1:m,i): j \in [2]$, where $\boldu \{ \ulineA_{j}\}$ is a 
deterministic function of $\ulineA_{j}$, and (ii) $\Pi_{t}: t \in [m]$ is 
independent of 
$V_{1i}^{m}(\hatb_{1i})$. Finally, $\boldY^{\Pi}(1:m,i)$ is 
conditionally independent of $V_{1i}^{m}(\hatb_{1i})$ given 
$\boldX_{j}\{ \ulinea_{j},\ulineb_{j} \}^{\Pi}(1:m,i) 
: j \in [2]$. These observations lead us to
\ifTITVersion\begin{eqnarray}
 \label{Eqn:PE2SecEventViaIndependence}
 P\left( \!\!\!\begin{array}{c}
\left[ 
\boldV_{j}\{ \ulineb_{j} \} 
\boldX_{j}\{ \ulinea_{j},\ulineb_{j} \}
\boldY \right]^{\Pi}(1:m,i) = 
(v_{j}^{m},x_{j}^{m} 
,y^{m})\\\ulineA_{j}=\ulinea_{j},\ulineB_{j}=\ulineb_{j}: j \in 
[2],V_{1i}^{m}(\hatb_{1i}
)= 
\hatv_{1}^{m}
\end{array}\!\!\!  \right)\nonumber\\
% newline
\!\!= P\left( \!\!\!\!\begin{array}{c}
\left[ 
\boldV_{j}\{ \ulineb_{j} \} 
\boldX_{j}\{ \ulinea_{j},\ulineb_{j} \}
\boldY \right]^{\Pi}\!\!(1:m,i)\\ = 
(v_{j}^{m},x_{j}^{m} 
,y^{m}),\ulineA_{j}=\ulinea_{j},\ulineB_{j}=\ulineb_{j}\\: j \in [2]
\end{array}\!\!\!\!  \right)\!\prod_{t=1}^{m} 
p_{V_{1}}(\hatv_{1t})
\label{Eqn:OneGenericTerm1Again}
\\
\!\!= P\left( \!\!\!\!\begin{array}{c}
\left[ 
\boldV_{j}\{ \ulineB_{j} \} 
\boldX_{j}\{ \ulineA_{j},\ulineB_{j} \}
\boldY \right]^{\Pi}\!\!(1:m,i)\\ = 
(v_{j}^{m},x_{j}^{m} 
,y^{m}),\ulineA_{j}=\ulinea_{j},\ulineB_{j}=\ulineb_{j}\\: j \in [2]
\end{array}\!\!\!\!  
\right)\!\!\prod_{t=1}^{m}p_{V_{j}}(\hatv_{1t})
\label{Eqn:OneGenericTerm2Again}
\end{eqnarray}\fi
\ifPeerReviewVersion\begin{eqnarray}
 \label{Eqn:PE2SecEventViaIndependence}
 \lefteqn{P\left( \!\!\!\begin{array}{c}
\left[ 
\boldV_{j}\{ \ulineb_{j} \} 
\boldX_{j}\{ \ulinea_{j},\ulineb_{j} \}
\boldY \right]^{\Pi}(1:m,i) = 
(v_{j}^{m},x_{j}^{m} 
,y^{m}),\ulineA_{j}=\ulinea_{j},\ulineB_{j}=\ulineb_{j}: j \in 
[2],V_{1i}^{m}(\hatb_{1i}
)= 
\hatv_{1}^{m}
\end{array}\!\!\!  \right)}\nonumber\\
% newline
&=& P\left( \!\!\!\!\begin{array}{c}
\left[ 
\boldV_{j}\{ \ulineb_{j} \} 
\boldX_{j}\{ \ulinea_{j},\ulineb_{j} \}
\boldY \right]^{\Pi}\!\!(1:m,i) = 
(v_{j}^{m},x_{j}^{m} 
,y^{m}),\ulineA_{j}=\ulinea_{j},\ulineB_{j}=\ulineb_{j}: j \in [2]
\end{array}\!\!\!\!  \right)\!\prod_{t=1}^{m} 
p_{V_{1}}(\hatv_{1t})
\label{Eqn:OneGenericTerm1Again}
\\
&=& P\left( \!\!\!\!\begin{array}{c}
\left[ 
\boldV_{j}\{ \ulineB_{j} \} 
\boldX_{j}\{ \ulineA_{j},\ulineB_{j} \}
\boldY \right]^{\Pi}\!\!(1:m,i) = 
(v_{j}^{m},x_{j}^{m} 
,y^{m}),\ulineA_{j}=\ulinea_{j},\ulineB_{j}=\ulineb_{j}: j \in [2]
\end{array}\!\! 
\right)\prod_{t=1}^{m}p_{V_{j}}(\hatv_{1t})
\label{Eqn:OneGenericTerm2Again}
\end{eqnarray}\fi
Substituting (\ref{Eqn:OneGenericTerm2Again}) in 
(\ref{Eqn:PE2OneIlleAfterLOTP}) 
and summing over 
$\ulinea_{j},\ulineb_{j},x_{j}^{m}: j \in [2],v_{1}^{m}$, we obtain
\ifTITVersion\begin{eqnarray}
 \sum_{\substack{\hatb_{1i} : \\\hatb_{1i}\neq b_{1i} }}\sum_{\substack{ 
(\hatv_{1}^{m},v_{2}^{m},y^{m})\\ \in  
T_{\beta}^{m}(p_{\underline{\mathscr{V}}\mathscr{Y}})}} 
\!\!\!\!\!\!\!\!\!\!P\left(\!\!\begin{array}{c}\left[ \boldV_{2}\{ 
\ulineB_{2} \} \boldY\right]^{\Pi}\!\!(1:m,i) \\= 
(v_{2}^{m},y^{m})\end{array} \!\!
\right)\prod_{t=1}^{m}p_{V_{1}}(\hatv_{1t})\nonumber\\
= M_{V_{1}}\sum_{\substack{ 
(\hatv_{1}^{m},v_{2}^{m},y^{m})\\ \in  
T_{\beta}^{m}(p_{\underline{\mathscr{V}}\mathscr{Y}})}} 
\prod_{t=1}^{m}p_{\hatV_{1}}(\hatv_{1t})p_{\hatV_{2}\hatY}(v_{2t},y_{2t})
 \label{Eqn:PE2AfterSumming}
\end{eqnarray}\fi
\ifPeerReviewVersion\begin{eqnarray}
 \label{Eqn:PE2AfterSumming}
 \sum_{\substack{\hatb_{1i} : \\\hatb_{1i}\neq b_{1i} }}\sum_{\substack{ 
(\hatv_{1}^{m},v_{2}^{m},y^{m})\\ \in  
T_{\beta}^{m}(p_{\underline{\mathscr{V}}\mathscr{Y}})}} \!\!\!\!\!\!P(\left[ 
\boldV_{2}\{ 
\ulineB_{2} \} \boldY\right]^{\Pi}\!\!(1:m,i) = 
(v_{2}^{m},y^{m}))\prod_{t=1}^{m}p_{V_{1}}(\hatv_{1t})= 
M_{V_{1}}\!\!\!\sum_{\substack{ 
(\hatv_{1}^{m},v_{2}^{m},y^{m})\\ \in  
T_{\beta}^{m}(p_{\underline{\mathscr{V}}\mathscr{Y}})}} 
\prod_{t=1}^{m}p_{\mathscr{V}_{1}}(\hatv_{1t})p_{\mathscr{V}_{2}\mathscr{Y}}(v_{
2t } , y_ { 2t } )
\end{eqnarray}\fi
as an upper bound on (\ref{Eqn:PE2OneIlleAfterLOTP}), where the last equality 
follows from arguments identical to those that established truth of 
(\ref{Eqn:Step1MACBothIncorrectAfterSum}). Once again, based on standard 
typicality argument, for 
example lemma 
\cite[Lemma 3.1]{201201NIT_ElgKim}, it can be proved that given any $\eta > 0$, 
there exists a choice for $\beta > 0$ and $m_{\beta,\eta} \in \naturals$ such 
that for all $m \geq m_{\beta,\eta}$, (\ref{Eqn:PE2AfterSumming}) is at most 
$\eta$ if $\frac{\log M_{V_{1}}}{m} < I(\mathscr{V}_{1};\mathscr{Y} 
\mathscr{V}_{2})$.

We summarize our proof thus far. We have proved that if
\begin{eqnarray}
 \label{Eqn:MACStep1SummaryIndBound}
 H(S_{j}^{l} | \hatK^{l}, S_{\msout{j}}^{l}) < \frac{\log M_{V_{j}}^{l}}{m} <  
l I(\mathscr{V}_{j};\mathscr{Y} \mathscr{V}_{\msout{j}}) : j \in [2],\mbox{ and 
} 
H(S_{1}^{l},S_{2}^{l} | \hatK^{l}) < \frac{\log 
M_{V_{1}}^{l}M_{V_{2}}^{l}}{m} <  
l I(\mathscr{V}_{1}\mathscr{V}_{2};\mathscr{Y})
\end{eqnarray}
where $S_{1}^{l},S_{2}^{l},\hatK^{l}$ and 
$\mathscr{V}_{1},\mathscr{V}_{2},\mathscr{Y}$ are distributed as in 
(\ref{Eqn:Step1MACSourceCodeDecodingPMF}) and 
(\ref{Eqn:Step1MACPMFForDecodingRule}) respectively, then the proposed coding 
scheme can enable the decoder recover $\boldS_{1},\boldS_{2}$ with arbitrarily 
high reliability by choosing $m$ sufficiently large. Our last step involves 
characterizing the upper and lower bounds above in terms of the pmf 
$p_{\ulineU\ulineV\ulineX Y}$ provided in the Theorem statement. We begin with 
the channel coding bounds.

\textit{Lower bounds on $I( 
\mathscr{V}_{j} ; \mathscr{Y} | \mathscr{V}_{\msout{j}}) $ and 
$I(\mathscr{V}_{1},\mathscr{V}_{2};\mathscr{Y})$}: Suppose 
$(\ulineU^{l},\ulineV^{l},\ulineX^{l},Y^{l}) 
= (U_{1}^{l},U_{2}^{l},V_{1}^{l},V_{2}^{l } ,X_ { 1 } ^{l } , X_ { 2 } ^ { l}, 
Y^ { l }) $ 
is distributed with pmf (\ref{Eqn:Step1MACPMFForDecodingRule}), and $I \in 
\{1,\cdots, l \}$ is a random index independent of the collection 
$\ulineU^{l},\ulineV^{l},\ulineX^{l},Y^{l}$, then 
$U_{1I},U_{2I},V_{1I},V_{2I},X_{1I},X_{2I},Y_{I}$ is
distributed with PMF (\ref{Eqn:Step1MACPMFOfInterleavedVector}). Hence we study 
$I(V_{jI};Y_{I},V_{\msout{j}I}) = I(\mathscr{V}_{j};\mathscr{Y} 
\mathscr{V}_{\msout{j}})$ 
and $I(V_{1I},V_{2I};Y_{I}) = I(\mathscr{V}_{1}\mathscr{V}_{2};\mathscr{Y})$. 
From 
(\ref{Eqn:FixedBLSourceMessEqual}), $\frac{1}{2}\geq \epsilon \geq 
P(A_{1}\neq A_{2})\geq P(U_{1}^{l}\neq U_{2}^{l})$, and hence
\ifTITVersion
\begin{eqnarray}
\lefteqn{I(V_{jI};Y_{I},V_{\msout{j}I}) = 
H(V_{jI})+H(V_{\msout{j}I},Y_{I})-H(V_{1I}
V_{2I},Y_{I})}
\nonumber\\
&=& 
H(V_{j})+H(V_{\msout{j}I},Y_{I})-H(V_{1I},{V}_{2
I} ,Y_{I})
\nonumber\\
&\geq& 
H(V_{j}) + H( V_{\msout{j}I} ,Y_{I}| 
\mathds{1}_{\{U_{1}^{l}=U_{2}^{l}\}}) 
- H(V_{1I},V_{2I}, Y_{I}, 
\mathds{1}_{\{U_{1}^{l}=U_{2}^{l}\}} )
\nonumber\\
&\geq& 
H(V_{j}) + 
H(V_{\msout{j}I},Y_{I}|\mathds{1}_{\{U_{1}^{l}=U_{2}^{l 
} \}})-h_{b}(\epsilon)
\nonumber\\
&&- H(V_{1I},V_{2I} , Y_{I}| \mathds{1}_{
\{U_{1}^{l}=U_{2}^{l}\} } ) 
\nonumber\\
&\geq&H(V_{j})-h_{b}(\epsilon)+ (1-\epsilon )
H(V_{\msout{j}I},Y_{I}|\mathds{1}_{\{U_{1}^{l}=U_{2}^{l}\}}
=1)
\nonumber\\
&&\!\!\!\!\!\!\!\!\!\!\! -(1-\epsilon)H(V_{1I},V_{2I} , 
Y_{I}|\mathds{1}_{\{U_{1}^{l}=U_{2}^{l}\}}=1 ) -\epsilon 
\log|\mathcal{V}_{1}||\mathcal{V}_{2}||\mathcal{Y}|
\nonumber\\
&=& H(V_{j}) +(1-\epsilon)\left[ H(V_{\msout{j}},Y)-H(\ulineV,Y) \right] 
\nonumber\\
&& -h_{b}(\epsilon)-\epsilon 
\log|\mathcal{V}_{1}||\mathcal{V}_{2}||\mathcal{Y}| 
\nonumber\\
&\geq& \!\!\!\!I(V_{j};Y,V_{\msout{j}})+\epsilon 
H(V_{\msout{j}}|V_{\msout{j}},Y)-h_{b}(\epsilon)-\epsilon 
\log|\ulineCalV||\mathcal{Y}|
\end{eqnarray}
\fi
\ifPeerReviewVersion
\begin{eqnarray}
\lefteqn{I(V_{jI};Y_{I},V_{\msout{j}I}) = 
H(V_{jI})-H(V_{jI}|V_{\msout{j}I},Y_{I})\geq 
H(V_{jI})-H(V_{jI},\mathds{1}_{\{U_{1}^{l}=U_{2}^{l 
} \}}|V_{\msout{j}I},Y_{I})}
\label{Eqn:MACSTep1IndChnlBnd-1}
\nonumber\\
&\geq& 
H(V_{j}) -H(V_{jI}|V_{\msout{j}I},Y_{I},\mathds{1}_{\{U_{1}^{l}=U_{2}^{l 
} \}}) -h_{b}(\epsilon)
\label{Eqn:MACSTep1IndChnlBnd-3}
\\
&=&H(V_{j})- P(U_{1}^{l} = 
U_{2}^{l})  
\left [ H(V_{jI} , V_{\msout{j}I} , Y_{I}| \mathds{1}_{\{U_{1}^{l}=U_{2}^{l}\}}
=1) - H(V_{\msout{j}I} , Y_{I}| \mathds{1}_{\{U_{1}^{l}=U_{2}^{l}\}}
=1 ) \right] \nonumber\\&&-P(U_{1}^{l} \neq U_{2}^{l})  
H(V_{jI}|V_{\msout{j}I},Y_{I},\mathds{1}_{\{U_{1}^{l}=U_{2}^{l}\}}
=0) -h_{b}(\epsilon) \label{Eqn:MACSTep1IndChnlBnd-4}
\\
&\geq& H(V_{j}) -\left[ H(\ulineV,Y) - H(V_{\msout{j}},Y) \right] 
-\epsilon 
\log|\mathcal{V}_{j}| -h_{b}(\epsilon)
\label{Eqn:MACSTep1IndChnlBnd-5}
=I(V_{j};Y,V_{\msout{j}})-\epsilon 
\log|\mathcal{V}_{j}| -h_{b}(\epsilon)
\end{eqnarray}
\fi
where (\ref{Eqn:MACSTep1IndChnlBnd-3}) follows from 
$p_{V_{j}}=p_{V_{jI}}=p_{\mathscr{V}_{j}}$ (Lemma 
\ref{Lem:SimplePropDecodingLem}) and $\frac{1}{2} \geq \epsilon \geq 
P(U_{1}^{l}\neq U_{2}^{l})$, (\ref{Eqn:MACSTep1IndChnlBnd-5}) follows from 
Lemma 
\ref{Lem:PMFOfUnifAndRandCo-Ordinate} in Appendix 
\ref{AppSec:PropOfDecodingPMF} and $\frac{1}{2} \geq \epsilon \geq 
P(U_{1}^{l}\neq U_{2}^{l})$. 
Indeed, note that Lemma \ref{Lem:PMFOfUnifAndRandCo-Ordinate} states
\begin{eqnarray}
 \label{Eqn:ChnlCodingBoundLemmaImport}
 P(U_{1I} =u , U_{2I}=u,V_{1I}=v_{1} 
, V_{2I}=v_{2},X_{1I} = x_{1},X_{2I}=x_{2},Y_{I} =y|\mathds{1}_{\left\{ 
U_{1}^{l} = U_{2}^{l} \right\}}=1) = 
p_{U\ulineV\ulineX Y}(u,\ulinev,\ulinex,y) 
\end{eqnarray}
for every $u,\ulinev,\ulinex,y \in \mathcal{U} \times \ulineCalV\times 
\ulineCalX \times \mathcal{Y}$ and hence, any functional of the pmf on the LHS 
of (\ref{Eqn:ChnlCodingBoundLemmaImport}) is equal to any functional of the pmf 
on the RHS of (\ref{Eqn:ChnlCodingBoundLemmaImport}), and in particular the 
entropy functional. Following an analogous sequence of steps, we have
\ifTITVersion
\begin{eqnarray}
\lefteqn{I(V_{jI};Y_{I},V_{\msout{j}I}) = 
H(V_{jI})+H(V_{\msout{j}I},Y_{I})-H(V_{1I}
V_{2I},Y_{I})}
\nonumber\\
&=& 
H(V_{j})+H(V_{\msout{j}I},Y_{I})-H(V_{1I},{V}_{2
I} ,Y_{I})
\nonumber\\
&\geq& 
H(V_{j}) + H( V_{\msout{j}I} ,Y_{I}| 
\mathds{1}_{\{U_{1}^{l}=U_{2}^{l}\}}) 
- H(V_{1I},V_{2I}, Y_{I}, 
\mathds{1}_{\{U_{1}^{l}=U_{2}^{l}\}} )
\nonumber\\
&\geq& 
H(V_{j}) + 
H(V_{\msout{j}I},Y_{I}|\mathds{1}_{\{U_{1}^{l}=U_{2}^{l 
} \}})-h_{b}(\epsilon)
\nonumber\\
&&- H(V_{1I},V_{2I} , Y_{I}| \mathds{1}_{
\{U_{1}^{l}=U_{2}^{l}\} } ) 
\nonumber\\
&\geq&H(V_{j})-h_{b}(\epsilon)+ (1-\epsilon )
H(V_{\msout{j}I},Y_{I}|\mathds{1}_{\{U_{1}^{l}=U_{2}^{l}\}}
=1)
\nonumber\\
&&\!\!\!\!\!\!\!\!\!\!\! -(1-\epsilon)H(V_{1I},V_{2I} , 
Y_{I}|\mathds{1}_{\{U_{1}^{l}=U_{2}^{l}\}}=1 ) -\epsilon 
\log|\mathcal{V}_{1}||\mathcal{V}_{2}||\mathcal{Y}|
\nonumber\\
&=& H(V_{j}) +(1-\epsilon)\left[ H(V_{\msout{j}},Y)-H(\ulineV,Y) \right] 
\nonumber\\
&& -h_{b}(\epsilon)-\epsilon 
\log|\mathcal{V}_{1}||\mathcal{V}_{2}||\mathcal{Y}| 
\nonumber\\
&\geq& \!\!\!\!I(V_{j};Y,V_{\msout{j}})+\epsilon 
H(V_{\msout{j}}|V_{\msout{j}},Y)-h_{b}(\epsilon)-\epsilon 
\log|\ulineCalV||\mathcal{Y}|
\end{eqnarray}
\fi
\ifPeerReviewVersion
\begin{eqnarray}
\lefteqn{I(V_{1I},V_{2I};Y_{I}) = 
H(V_{1I},V_{2I})-H( V_{1I},V_{2I} |Y_{I})\geq 
H(V_{1I},V_{2I})-H(V_{1I},V_{2I},\mathds{1}_{\{U_{1}^{l}=U_{2}^{l 
} \}}|Y_{I})}
\label{Eqn:MACSTep1SumChnlBnd-1}
\nonumber\\
&\geq& 
H(V_{1},V_{2}) 
-H(V_{1I},V_{2I}|Y_{I},\mathds{1}_{\{U_{1}^{l}=U_{2}^{l 
} \}}) -h_{b}(\epsilon)
\label{Eqn:MACSTep1SumChnlBnd-3}
\\
&=&H(V_{1},V_{2})- P(U_{1}^{l} = 
U_{2}^{l})  
\left [ H(V_{1I},V_{2I} , Y_{I}| \mathds{1}_{\{U_{1}^{l}=U_{2}^{l}\}}
=1) - H(Y_{I}| \mathds{1}_{\{U_{1}^{l}=U_{2}^{l}\}}
=1 ) \right] \nonumber\\&&-P(U_{1}^{l} \neq U_{2}^{l})  
H(V_{1I},V_{2I}|Y_{I},\mathds{1}_{\{U_{1}^{l}=U_{2}^{l}\}}
=0) -h_{b}(\epsilon) \label{Eqn:MACSTep1SumChnlBnd-4}
\\
&\geq& H(V_{1},V_{2}) -\left[ H(V_{1},V_{2},Y) - H(Y) \right] 
-\epsilon 
\log|\ulineCalV| -h_{b}(\epsilon)
\label{Eqn:MACSTep1SumChnlBnd-5}
=I(V_{1},V_{2};Y)-\epsilon 
\log|\ulineCalV| -h_{b}(\epsilon)
\end{eqnarray}
\fi
We now seek upper bounds on $H(S_{j}^{l}|\hatK^{l},S_{\msout{j}}^{l}), 
H(S_{1}^{l}S_{2}^{l}|\hat{K}^{l})$. Recall from 
(\ref{Eqn:Step1S1S2KHatPMF}) that $p_{S_{1}^{l},S_{2}^{l},\hatK^{l}}$ is the pmf 
of any row of the triplet $\boldS_{1},\boldS_{2},\boldhatK$ of matrices. 
Appealing to the sequence of steps from 
(\ref{Eqn:DueckExExUpperBoundOnBinningRates}) 
through (\ref{Eqn:DueckExExUpperBoundOnSumBinningRate}) we 
recognize that it suffices to characterize an upper bound $\phi$ on 
$P(\boldhatK(t,1:l) \neq \boldK_{1}(t,1:l))$, that is at most $\frac{1}{2}$. 
Towards that end, recall that our typical set 
source code ensures $d_{k}(e_{k}(k_{1}^{l})) = k_{1}^{l}$ for every $k_{1}^{l} 
\in T_{\delta}^{l}(K_{1})$. This guarantees $\{ \boldhatK (t,1:l) \neq 
\boldK_{1} (t,1:l) \} \subseteq \{ A_{1t} \neq \hatA_{t} \}$. In 
order to derive an upper bound on the latter event, we are required to 
characterize the channel $p_{Y^{l}|U^{l}}$ experienced by codewords of $C_{U}$. 
In particular, since \begin{eqnarray}\label{Eqn:Step1MACFixedBLCodeError}P( 
A_{1t} \neq \hatA_{t}) \leq P(A_{1t} \neq A_{2t}) + P(\hatA_{t} \neq A_{1t}, 
A_{1t} = A_{2t}) \leq \epsilon + P(\hatA_{t} \neq A_{1t}, A_{1t} = 
A_{2t}),\end{eqnarray}
we are required to characterize the channel $p_{Y^{l}|U^{l}}$ experienced by 
those commonly selected codewords. In the sequel, we will prove that if the two 
transmitters choose a common $C_{U}-$codeword, then the latter experiences a 
memoryless $\prod_{i=1}^{l}p_{Y|U}$ channel. By our choice of the constant 
composition code \cite[Thm. 10.2]{CK-IT2011}, we conclude that the latter 
event has probability at most $g_{\rho,l}$. Towards that end, we note that
\begin{eqnarray}
\label{Eqn:Step1MACFixedB-LChnlCodeError2}
P\left(\!\!\!  
 \begin{array}{c}
\boldu\{ \ulinea_{j} \} = \boldu_{j}, \boldV_{j}\{ \ulineB_{j} \} = \boldv_{j}\\
\boldX_{j}\{ \ulineA_{j},\ulineB_{j} 
\}=\boldx_{j},\ulineA_{j} = \ulinea_{j}\\: j \in [2], \boldY=\boldy
 \end{array}
 \!\!\!\right) =
\sum_{\substack{\ulineb_{1},\ulineb_{2}}}\!
P \left(  \!\!\!
\begin{array}{c}
\ulineA_{j}=\ulinea_{j}\\\ulineB_{j}=\ulineb_{j} \\: j \in [2]
\end{array}
 \!\!\! \right)
 P\left(\!\!\!  \left.
 \begin{array}{c}
\boldu\{ \ulinea_{j} \} = \boldu_{j}, \boldV_{j}\{ \ulineb_{j} \} = \boldv_{j} 
\\
\boldX_{j}\{ \ulinea_{j},\ulineb_{j} 
\}=\boldx_{j} : j \in [2]\\
\boldY=\boldy
 \end{array}
 \!\!\!\right| \!\!\!
\begin{array}{c}
\ulineA_{j}=\ulinea_{j}\\\ulineB_{j}=\ulineb_{j} \\:j \in [2]
\end{array}
 \!\!\!
 \right)
\end{eqnarray}
where (\ref{Eqn:Step1MACFixedB-LChnlCodeError2}) is identical to 
(\ref{Eqn:Step1MACPrelim2-1}) except for the range of the summation. Using 
(\ref{Eqn:Step1MACPrelim1-Prelim1}), 
(\ref{Eqn:Step1MACPrelim1-Prelim1Mid}), (\ref{Eqn:Step1MACPrelim1-Prelim2}) and 
following a sequence of steps analogous to those that took us from 
(\ref{Eqn:Step1MACPrelim2-1}) to (\ref{Eqn:Step1MACPrelim2-5}), we have 
(\ref{Eqn:Step1MACFixedB-LChnlCodeError2}) equal to
\begin{eqnarray}
\label{Eqn:Step1MACFixedB-LChnlCodeError3}
\lefteqn{P\left(\!\!\!
\begin{array}{c}\ulineA_{j}=\ulinea_{j}\\: j \in [2]
\end{array}
 \!\!\! \right) \!\prod_{t=1}^{m} 
\mathds{1}_{\left\{\!\!\!\! 
\begin{array}{c} \boldu_{j}(t,1:l) =\\ u^{l}(a_{jt}): j \in [2] 
\end{array}\!\!\!\!\right\}}\!\!\!\left[ \prod_{i=1}^{l}\!\!\left\{ \!
\prod_{j=1}^{2}
p_{V_{j}}(\boldv_{j}(t,i))p_{X_{j}|UV_{j}}\left(\boldx_{j}(t, 
i)\left|\!\!\! \begin{array}{c}\boldu_{j}(t,i)\\\boldv_{j}(t,i) 
\end{array}\!\!\!\right.\! \right) \!\!\right\} \!
\mathbb{W}_{Y | \ulineX} \left(\! \boldy(t,i) 
\left|\!\!\! \begin{array}{c} \boldx_{1}(t, 
i)\\\boldx_{2}(t,i)\end{array}\!\!\!\right.\right)\!\right]} \nonumber\\
\label{Eqn:Step1MACFixedB-LChnlCodeError4}
&\!\!\!\!\!\!=&\!\!\!\!P\!\left(\!\!\!
\begin{array}{c}\ulineA_{j}=\ulinea_{j}\\: j \in [2]
\end{array}
 \!\!\! \right) \!\prod_{t=1}^{m}\! \mathds{1}_{\!\left\{\!\!\!\! 
\begin{array}{c} \boldu_{j}(t,1:l) =\\ u^{l}(a_{jt}): j \in [2] 
\end{array}\!\!\!\!\right\}}\!\!\!\!
\left[ \prod_{i=1}^{l}\left\{ \prod_{j=1}^{2}
p_{V_{j}}(\boldv_{j}(t,i))p_{X_{j}|UV_{j}}\!\left(\boldx_{j}(t, 
i)\left|\!\!\! \begin{array}{c}u^{l}(a_{jt})_{i}\\\boldv_{j}(t,i) 
\end{array}\!\!\!\right.\! \right) \!\!\right\} \!
\mathbb{W}_{Y | \ulineX} \left(\!\! \boldy(t,i) \!
\left|\!\!\! \begin{array}{c} \boldx_{1}(t, 
i)\\\boldx_{2}(t,i)\end{array}\!\!\!\right.\right)\!\right]
\nonumber\\
\label{Eqn:Step1MACFixedB-LChnlCodeError5}
&\!\!\!\!\!\!=&\!\!\!\! \left(\!\!\!
\begin{array}{c}\ulineA_{j}=\ulinea_{j}\\: j \in [2]
\end{array}
 \!\!\! \right) \!\prod_{t=1}^{m} \mathds{1}_{\left\{\!\!\!\! 
\begin{array}{c} \boldu_{j}(t,1:l) =\\ u^{l}(a_{jt}): j \in [2] 
\end{array}\!\!\!\!\right\}}p_{\ulineV^{l}\ulineX^{l}Y^{l}|\ulineU^{l}} 
 \left(\!\!\!\begin{array}{c} 
\boldv_{1}(t,1:l),\boldv_{2}(t,1:l),\boldx_{1}(t,1:l)\\\boldx_{2}(t,1:l), 
\boldy(t,1:l)\end{array}\!\!\!\left|u^{l}(a_{1t}),u^{l}(a_{2t}) \right. 
\right)\end{eqnarray}
where $u^{l}(a_{jt})_{i}$ is the $i$-th symbol of $u^{l}(a_{jt})$ and 
(\ref{Eqn:Step1MACFixedB-LChnlCodeError5}) follows from 
(\ref{Eqn:AppSecPropDecPMFCondU1lU2}). Since LHS of 
(\ref{Eqn:Step1MACFixedB-LChnlCodeError2}) is equal to the RHS of 
(\ref{Eqn:Step1MACFixedB-LChnlCodeError5}), summing the latter
(\ref{Eqn:Step1MACFixedB-LChnlCodeError5}), we have
\begin{eqnarray}
\label{Eqn:Step1MACFixedB-LChnlCodeError6}
 \lefteqn{\!\!\!\!\!\!\!\!\!\!\!\!\!\!\!\!\!\!\!\!\!\!\!\!\sum_{\boldu_{1}} 
\sum_{\boldu_{2}} \sum_{\boldv_{1}} \sum_{\boldv_{2}}
\sum_{\boldx_{1}} \sum_{\boldx_{2}}
P\left(\!\!\!  
 \begin{array}{c}
\boldu\{ \ulinea_{j} \} = \boldu_{j}, \boldV_{j}\{ \ulineB_{j} \} = \boldv_{j}\\
\boldX_{j}\{ \ulineA_{j},\ulineB_{j} 
\}=\boldx_{j},\ulineA_{j} = \ulinea_{j}\\: j \in [2], \boldY=\boldy
 \end{array}
 \!\!\!\right) = P\left(\!\!\!
\begin{array}{c}\ulineA_{j}=\ulinea_{j}\\: j \in [2]
\end{array}
 \!\!\! \right)\prod_{t=1}^{m} p_{Y^{l}|\ulineU^{l}} 
 \left(\!\!\!\begin{array}{c} 
\boldy(t,1:l)\end{array}\!\!\!\left| 
\!\!\!\begin{array}{c}u^{l}(a_{1t}) \\ u^{l}(a_{2t}) \end{array}\!\!\!\right. 
\right)}\nonumber\\
\label{Eqn:Step1MACFixedB-LChnlCodeError7}
&&\mbox{and hence }P(\boldY = \boldy | 
\ulineA_{1}=\ulinea_{1},\ulineA_{2}=\ulinea_{2}) = \prod_{t=1}^{m} 
p_{Y^{l}|\ulineU^{l}} 
 \left(\!\!\!\begin{array}{c} 
\boldy(t,1:l)\end{array}\!\!\!\left| 
\!\!\!\begin{array}{c}u^{l}(a_{1t}), u^{l}(a_{2t}) \end{array}\!\!\!\right. 
\right)~~~~~~~\nonumber\\
\label{Eqn:Step1MACFixedB-LChnlCodeError8}
&&~~~~~~~~~~~~~~=  \prod_{\substack{t \in [m] : \\
a_{1t} =a_{2t}}} \left\{
\prod_{i=1}^{l}p_{Y|U}(\boldy(t,i)|u^{l}(a_{t})_{i}) \right\} \times  
\prod_{\substack{t \in [m] : \\
a_{1t} \neq a_{2t}}} p_{Y^{l}|\ulineU^{l}} 
 \left(\!\!\!\begin{array}{c} 
\boldy(t,1:l)\end{array}\!\!\!\left| 
\!\!\!\begin{array}{c}u^{l}(a_{1t}), u^{l}(a_{2t}) \end{array}\!\!\!\right. 
\right)
\end{eqnarray}
where (\ref{Eqn:Step1MACFixedB-LChnlCodeError8}) follows from 
(\ref{Eqn:AppSecPropDecPMFCommonCuCode}).
\end{IEEEproof}

\begin{remark}
 \label{Rem:CanDecodeK2}
 In the coding scheme presented above, the fixed B-L codes attempted to 
communicate $K_{1}$ to the decoder. A simple alternate is to attempt 
communication of $K_{2}$ to the decoder via the fixed B-L codes. For the sake 
of completeness, we provide the corresponding sufficient conditions.
\end{remark}

\begin{corollary}
 \label{Cor:MACStep1}
A pair of sources $(\ulineCalS,\mathbb{W}_{\ulineS})$ is transmissible over a 
MAC $(\ulineCalX,\outset,\mathbb{W}_{\Out|\ulineX})$ if there exists 
\begin{enumerate}
 \item[(i)] finite sets 
$\mathcal{K},\mathcal{U},\mathcal{V}_{1},\mathcal{V}_{2}$,
 \item[(ii)] maps $f_{j}:\mathcal{S}_{j}\rightarrow \mathcal{K}$, 
with $K_{j}=f_{j}(S_{j})$ for $j \in [2]$,
\item[(iii)] $\alpha,\beta \geq 0$, $\rho > 0 $, $\delta > 0$, 
\item[(iv)] $l \in 
\naturals, l \geq l^{*}(\rho,\mathcal{U},\mathcal{Y})$, where 
$l^{*}(\cdot,\cdot,\cdot)$ is defined in (\ref{Eqn:DefnOfLStar}),
\item[(v)] pmf $p_{U}p_{V_{1}}p_{V_{2}}p_{X_{1}|UV_{1}}p_{X_{2}|UV_{2}} 
\mathbb{W}_{Y|\ulineX}$ defined on $\mathcal{U}\times 
\ulineCalV\times\ulineCalX\times \mathcal{Y}$, where $p_{U}$ is a type of 
sequences in $\mathcal{U}^{l}$, such that
\end{enumerate}
\ifPeerReviewVersion
 \begin{eqnarray}
  (1+\delta)H(K_{a}) &<& \alpha+\beta ,\nonumber\\ 
 \label{Eqn:CorStepIMACIndvdualRateBound}
H(S_{j}|S_{\msout{j}},K_{a}) + 
\mathcal{L}_{l}(\phi,|\mathcal{S}_{j}|) &<& 
I(V_{j};\Out|V_{\msout{j}})-\mathcal{L}(\phi,|\mathcal{V}_{j}|)
\mbox{ for }j \in [2]\mbox{ and} \\
\label{Eqn:CorStepIMACSumRateBound}
\beta + H(\ulineS|K_{1})+ 
\mathcal{L}_{l}(\phi,|\ulineCalS|) &<& I(\ulineV;Y) - 
\mathcal{L}(\phi,|\ulineCalV|),\\
\lefteqn{
\!\!\!\!\!\!\!\!\!\!\!\!\!\!\!\!\!\!\!\!\!\!\!\!\!\!\!\!\!\!\!\!\!\!\!\!\!\!\!\!
\!\!\!\!\!\!\!\!\!\!\!\!\!\!\!\!\!\!\!\!\!\!\!\!\!\!\!\!\!\!\!\!\!\!\!\!\!\!\!\!
\!\!\!\!\!\!\!\!\!\!\!\!\!\!\!\!\!\!\!\!\!\!\!\!\!\!\!\!\!\!\!\!\!\!\!
\phi \in [0,0.5)\mbox{ where }\phi \define 
g(\alpha+\rho,l)+\xi^{[l]}(\ulineK)+\tau_{l,\delta}(K_{a}),~ g(R,l) 
\define 
(l+1)^{2|\mathcal{U}||\mathcal{Y}|}\exp\{-lE_{r}(R,p_{U},p_{Y|U})\}}
\end{eqnarray}
for some $a \in [2]$, where $\mathcal{L}_{l}(\cdot,\cdot), 
\mathcal{L}(\cdot,\cdot)$ is as defined 
in (\ref{Eqn:AdditionalSourceCodingInfo}).\fi
\ifTITVersion
 \begin{eqnarray}
  (1+\delta)H(K_{1}) \!\!\!\!&\leq&\!\!\!\! \alpha+\beta ,\nonumber\\ 
\label{Eqn:StepIMACIndvdualRateBound}
H(S_{j}|S_{\msout{j}},K_{1}) + 
\mathcal{L}_{l}(\phi,|\mathcal{S}_{j}|) \!\!\!\!&<&\!\!\!\!
I(V_{j};\Out|V_{\msout{j}})-\mathcal{L}(\phi,
|\mathcal{V}_{j} |)\nonumber\\&&\!\!\!\!~~~
\mbox{ for }j \in [2]\mbox{ and} \\
\label{Eqn:StepIMACSumRateBound}
\beta + H(\ulineS|K_{1})+ 
\mathcal{L}_{l}(\phi,|\ulineCalS|) \!\!\!\!&<&\!\!\!\! I(\ulineV;Y) - 
\mathcal{L}(\phi,|\ulineCalV|),
\end{eqnarray}
$\phi \in [0,0.5)$ where
\begin{eqnarray}\phi &\define &
g_{\rho,l}+\xi^{[l]}(\ulineK)+\tau_{l,\delta}(K_{1}),\nonumber\\
g(\alpha+\rho,l) &\define &
\exp\{-l(E_{r}(\alpha+\rho,p_{U},p_{Y|U})-\rho)\}\nonumber .\end{eqnarray}\fi
\end{corollary}
We now prove that the admissible region characterized in Theorem 
\ref{Thm:MACStep1} can be strictly larger than the CES region.

\begin{thm}
 \label{Thm:FBLBeatsCES}
There exists a source pair $(\ulineCalS,\mathbb{W}_{\ulineS})$ and a MAC 
$(\ulineX,\mathcal{Y},\mathbb{W}_{Y|\ulineX})$ that do not satisfy CES 
conditions \cite[Thm. 1]{198011TIT_CovGamSal} (Theorem 
\ref{Thm:CESConditions} here) and yet satisfy conditions stated in 
Theorem \ref{Thm:MACStep1}. In particular, consider Example 
\ref{Ex:DuecksExampleForMAC}. There exists $a^{*} 
\in \naturals$ and $k^{*}\in \naturals$ such that for any $a\geq a^{*}$ and any 
$k \geq k^{*}$, $\ulineS$ and MAC 
$\mathbb{W}_{\underline{Y}|\underline{U}\underline{X}}$ 1) do not satisfy CES 
conditions \cite[Thm. 1]{198011TIT_CovGamSal} (Theorem 
\ref{Thm:CESConditions} here), and 2) satisfy conditions 
stated in Theorem \ref{Thm:MACStep1}.
\end{thm}
\begin{IEEEproof}
In view of Lemma \ref{Lem:Ex1DoesNOTSatisfyCESConditions} we only need to prove 
the second statement. Towards that end, consider the 
following assignment for the auxiliary parameters in Theorem 
\ref{Thm:MACStep1}.

Let $\mathcal{K}= \mathcal{S}_{1}=\mathcal{S}_{2}$, $\mathcal{U}$ be the input 
alphabet of the shared channel $\mathbb{W}_{Y_{0}|\ulineU}$, 
$\mathcal{V}_{j}=\mathcal{X}_{j}: j \in [2]$ be the input alphabet of the 
satellite channels $\mathbb{W}_{Y_{j}|X_{j}} : j \in [2]$ respectively.
Let $f_{j}(s) = s$ for $s \in \mathcal{S}_{j}$ be the identity map, and hence 
$K_{j}=S_{j}$ for $j \in [2]$.
Let $\alpha = \left(1-\frac{1}{4k} \right)\log a$, $\beta = \frac{5}{4k}\log a 
+ \left( 1+\frac{1}{k} \right)h_{b}(\frac{1}{k})$, $\rho = 
\frac{1}{4k}\log\frac{a}{4}$, $\delta = \frac{1}{k}$.
Let $l = k^{4}a^{\frac{\eta k}{2}}$.
Let $p_{U}$ be the uniform pmf on $\mathcal{U} = \{ 0,\cdots,a-1\}$. Let 
$p_{V_{j}} : j \in [2]$ be the capacity achieving distribution on satellite 
channels $\mathbb{W}_{Y_{j}|X_{j}} : j \in [2]$ respectively.
Note that, for any $u \in \mathcal{U}$, $l p_{U}(u) = k^{4}a^{\frac{\eta 
k}{2}-1}$ is a natural number since $\eta \geq 6$ is an even integer. For the 
above assignment, note that (\ref{Eqn:DefnOfLStar}) is
\begin{eqnarray}
  l^{*}(\frac{1}{4k}\log\frac{a}{4},\mathcal{U},\ulineCalY) &=& \min\left\{ l 
: \frac{l}{4k}\log \frac{a}{4} \geq \log 4 + 
4a(1+a|\mathcal{Y}_{1}||\mathcal{Y}_{2}|) \log (l+1)\right\} \nonumber\\
\label{Eqn:LStarForAssignment}
&\leq& \min\left\{ l : 
\frac{l}{4k}\log \frac{a}{4} \geq \log 4 + 
4a(1+a|\mathcal{Y}_{1}||\mathcal{Y}_{2}|) \log 2l\right\}.
\end{eqnarray}
Recall that for sufficiently large $a,k$, the satellite channels defined in 
Example \ref{Ex:DuecksExampleForMAC} have $|\mathcal{Y}_{j}| \leq 
a^{\frac{3}{2k}}$. It can be verified that the RHS of 
(\ref{Eqn:LStarForAssignment}) is lesser than or equal to $k^{4}a^{\frac{\eta 
k}{2}}$ for sufficiently large $a,k$. Therefore, the assignment $l = 
k^{4}a^{\frac{\eta k }{2}} \geq 
l^{*}(\frac{1}{4k}\log\frac{a}{4},\mathcal{U},\ulineCalY)$ for sufficiently 
large $a,k$.

From Table \ref{Table:DueckMACExParameters}, verify that $(1+\delta)H(K_{1}) = 
(1+\delta)H(S_{1}) < (1+\frac{1}{k})\log a + (1+\frac{1}{k})h_{b}(\frac{1}{k}) 
= \alpha + \beta$. Since $p_{U}$ is uniform and $p_{Y_{0}|U}$ induced by the 
chosen pmf is deterministic, it can be verified that 
$E_{r}(\alpha+\rho,p_{U},p_{Y|U}) = \log a - (\alpha + \rho) = \frac{1}{4k}\log 
4$. Hence \begin{eqnarray}
\label{Eqn:BndOnTauAndXi}
g(\alpha+\rho,l) = 
(l+1)^{2|\mathcal{U}||\ulineCalY|}\exp \left\{- \frac{l}{4k}\log 4\right\} 
\leq 4^{-\frac{l}{4k}}(l+1)^{2a^{\frac{6}{2k}+2}} \leq 
4^{ - \frac{l}{4k}}(2l)^{2a^{3}} \leq 
\frac{k^{3}}{a^{\frac{\eta k}{2}}}\end{eqnarray}
for sufficiently large $a,k$. Since our choice of $\delta = \frac{1}{k}, l = 
k^{4}a^{\frac{\eta k}{2}}$ are identical to that in Section 
\ref{SubSec:FixedB-LCodingOverIsolatedChnls}, we appeal to 
(\ref{Eqn:BoundOnTau}), (\ref{Eqn:BoundOnTauAndXi}) and conclude
\begin{eqnarray}
 \label{Eqn:XiAndTau}
 \tau_{l,\delta}(\ulineK) + \xi^{[l]}(\ulineK)\leq \frac{2k^{3}}{a^{\frac{\eta 
k}{2}}},\mbox{ and in conjunction with (\ref{Eqn:BndOnTauAndXi}) we have, }\phi 
\leq \frac{3k^{3}}{a^{\frac{\eta k}{2}}} \leq \frac{1}{2}.
\end{eqnarray}
for sufficiently large $a,k$. Substituting this 
upper bound in $\mathcal{L}_{l}(\cdot,\cdot)$ and 
$\mathcal{L}(\cdot,\cdot)$, it can be verified that
\begin{eqnarray}
 \label{Eqn:BoundOnErrTerms}
 \mathcal{L}_{l}(\phi,|\mathcal{S}_{j}|) \leq 
\mathcal{L}_{l}(\phi,|\ulineCalS|) \leq \frac{1}{l}h_{b}(\alpha) + 
\frac{\alpha}{2}\log a\mbox{ where } \alpha = \frac{8k^{4}}{a^{\frac{\eta 
k}{3}}}, \mathcal{L}(\phi,|\mathcal{V}_{j}|) \leq  
\mathcal{L}(\phi,|\ulineCalV|) \leq h_{b}(\alpha) + \frac{\alpha}{2} \log a,
\end{eqnarray}
where we have used that fact that for large $a,k$, we have $|\mathcal{V}_{j}|
\leq a^{\frac{3}{2k}}$. We are now set to prove the remaining inequalities 
(\ref{Eqn:StepIMACIndvdualRateBound}), (\ref{Eqn:StepIMACSumRateBound}). This 
follows by simple substitution of $\beta = \frac{5}{4k}\log a 
+ \left( 1+\frac{1}{k} \right)h_{b}(\frac{1}{k})$, upper bound of 
$h_{b}(\frac{2}{ka^{\eta k}}) + \frac{2 \log a}{a^{\eta k}}$ on 
$H(S_{2}|S_{1})$ (Table \ref{Table:DueckMACExParameters}), capacities of 
$\mathbb{W}_{Y_{j}|X_{j}}$ for $I(V_{j};Y|V_{\msout{j}})$, the sum of these 
capacities for $I(\ulineV;Y)$, (\ref{Eqn:BoundOnErrTerms}) and is left to 
the reader.
\end{IEEEproof}
\subsection{IC Problem}
\label{SubSec:ICStep1GeneralizationSeparateDecoding}
Our results in this section are analogous to those presented in Section 
\ref{SubSec:MACStep1GeneralizationSeparateDecoding} for the MAC problem. We 
provide a new set of sufficient conditions for the IC problem in Theorem 
\ref{Thm:ICStep1}, and prove in Theorem \ref{Thm:FBLBeatsLC}, that these are 
strictly weaker 
than the LC conditions.

\begin{thm}
\label{Thm:ICStep1}
A pair of sources $(\ulineCalS,\mathbb{W}_{\ulineS})$ is transmissible over an 
IC $(\ulineCalX,\ulineCalY,\mathbb{W}_{\ulineY|\ulineX})$ if there exists 
\begin{enumerate}
 \item[(i)]  finite sets 
$\mathcal{K},\mathcal{U},\mathcal{V}_{1},\mathcal{V}_{2}$,
 \item[(ii)] maps $f_{j}:\mathcal{S}_{j}\rightarrow 
\mathcal{K}$, with $K_{j}=f_{j}(S_{j})$ for $j \in [2]$,
 \item[(iii)] $\alpha,\beta \geq 0$, $\rho > 0 $, $\delta > 0$, 
 \item[(iv)] $l \in 
\naturals, l \geq \max\{ l^{*}(\rho,\mathcal{U},\mathcal{Y}_{j}): j \in [2] 
\}$, where $l^{*}(\cdot,\cdot,\cdot)$ is defined in (\ref{Eqn:DefnOfLStar}),
\item[(v)] pmf $p_{U}p_{V_{1}}p_{V_{2}}p_{X_{1}|UV_{1}}p_{X_{2}|UV_{2}} 
\mathbb{W}_{\ulineY|\ulineX}$ defined on $\mathcal{U}\times 
\ulineCalV\times\ulineCalX\times \ulineCalY$, where $p_{U}$ is a type of 
sequences in $\mathcal{U}^{l}$, such that
\end{enumerate}
\ifPeerReviewVersion
 \begin{eqnarray}
&&\!\!\!\!\!\!\!\!\!\!\!\!\!\!\!\!\!\!\!\!\!\!\!\!\!\!\!\!\!\!\!\!
 \!\!\!(1+\delta)H(K_{a}) \leq \alpha+\beta ,~~
 \label{Eqn:Step1ICIndvdualRateBound}
H(S_{j}|K_{a}) + \beta +
\mathcal{L}_{l}(\phi_{j},|\mathcal{S}_{j}|)<
I(V_{j};Y_{j})-\mathcal{L}(\phi_{j},|\mathcal{V}_{j}|)
\mbox{ for }j \in [2],~~\phi_{j} \leq \frac{1}{2}\\
\label{Eqn:ICBoundStep1}
\lefteqn{\!\!\!\!\!\!\!\!\!\!\!\!\!\!\!\!\!\!\!\!\!\!\!\!\!
\mbox{where }\phi_{j} \!\define \!
g_{j}(\alpha+\rho,l)+\xi^{[l]}(\ulineK)+\tau_{l,\delta}(K_{a}),
g_{j}(R,l) \define 
(l+1)^{2|\mathcal{U}||\mathcal{Y}_{j}|}\exp\{-lE_{r}(R,p_{U},p_{Y_{j}|U})\}
\mbox{ for }j \in [2]}
\end{eqnarray}
and some $a \in [2]$, where $\mathcal{L}_{l}(\cdot,\cdot), 
\mathcal{L}(\cdot,\cdot)$ is 
as defined 
in (\ref{Eqn:AdditionalSourceCodingInfo}).\fi
\ifTITVersion
 \begin{eqnarray}
 \label{Eqn:TypicalSetSize}
 (1+\delta)H(K_{1}) \!\!\!\!&\leq&\!\!\!\! \alpha+\beta ,\mbox{ and 
for } j \in [2]\nonumber\\ 
\label{Eqn:Step1ICIndvdualRateBound}
H(S_{j}|K_{1}) + \beta +
\mathcal{L}_{l}(\phi_{j},|\mathcal{S}_{j}|) \!\!\!\!&\leq&\!\!\!\!
I(V_{j};Y_{j})-\mathcal{L}(\phi_{j},
|\mathcal {V}_{j} |),
\end{eqnarray}
$\phi_{j} \in [0,0.5)$ where
\begin{eqnarray}\phi_{j} &\define &
\label{Eqn:ICStep1Phij}
g_{\rho,l}^{j}+\xi^{[l]}(\ulineK)+\tau_{l,\delta}(K_{1}),\\
\label{Eqn:ICStep1gRhoLj}
g_{\rho,l}^{j} &\define &
\exp\{-l(E_{r}(\alpha+\rho,p_{U},p_{Y_{j}|U})-\rho)\} 
.\end{eqnarray}\fi
\end{thm}
The proof contains no new elements beyond those presented in Section 
\ref{SubSec:MACStep1GeneralizationSeparateDecoding}. Moreover, in Section 
\ref{SubSec:ICStep2GeneralizationJointDecoding}, we provide a proof of a more 
general theorem for the IC problem. In view of these, we omit a proof of the 
above theorem. We only provide an informal outline of the coding scheme 
and the analysis.

\textit{Outline of the coding scheme:} Let us fix $a=1$. The reader is 
encouraged to revisit the 
coding scheme presented in Section 
\ref{SubSec:MACStep1GeneralizationSeparateDecoding} for the MAC. Encoding is 
identical except for the following (minor) differences. Recall that if 
$M_{u}$ - the number of codewords in $C_{U}$ - is less than 
$|T_{\delta}^{l}(K_{1})|$ - the 
range of the index output by the fixed B-L common source code 
$T_{\delta}^{l}(K_{1})$, then the latter index is 
split into two sub-message indices taking values in $[M_{u}] = \exp 
\{l\alpha\}, \exp\{ l \beta \}$. In contrast to the MAC, where only one of the 
Txs communicated the second message index via the outer code, we require that 
\textit{both} Txs communicate their second sub-message indices to their 
respective receivers via the outer code. Secondly, the outer code we employ to 
communicate over the IC $\mathcal{V}_{1},\mathcal{V}_{2} \rightarrow 
\mathcal{Y}_{1},\mathcal{Y}_{2}$ is simply a pair of PTP codes for the 
channels $\mathcal{V}_{j}-\mathcal{Y}_{j}$. Indeed, the coding scheme does not 
build any resilience to interference. In fact, as in Section 
\ref{SubSec:MACStep1GeneralizationSeparateDecoding}, self interference between 
parallel $\mathcal{U},\mathcal{V}_{j}$ streams is also ignored.

In relation to decoding, each receiver employs the PTP decoders of $C_{U}$ and 
the source code $T_{\delta}^{l}(K_{1})$ to reconstruct the $l-$length sub-block 
of $K_{1}$. Let $\boldhatK_{j}(t,1:l)$ denote the $t$-th sub-block of decoder 
$j$'s reconstruction of $\boldK_{1}(t,1:l)$. We emphasize that decoder 2's 
reconstruction $\boldhatK_{2}(t,1:l)$ is also viewed as a reconstruction of 
$\boldK_{1}(t,1:l)$. In recovering $\boldS_{j}$
 via the joint typicality based decoder, it employs $\boldhatK_{j}$ as the side 
information.

\textit{Outline of the analysis:} Since our analysis proceeds through steps 
identical to that provided for the MAC in Section 
\ref{SubSec:MACStep1GeneralizationSeparateDecoding}, we highlight only the 
differences in the three steps we mentioned in the outline therein. In the 
first step, with regard to quantifying the amount of information communicated 
through fixed B-L coding, observe that decoder $j$ can recover a common message 
encoded through $C_{U}$ with maximal error probability $g_{j}(\alpha+\rho,l)$ 
as 
defined in (\ref{Eqn:ICBoundStep1}). Hence an upper bound on 
$P(\boldhatK_{j}(t,1:l)\neq \boldK_{1}(t,1:l))$ is $\phi_{j}$ as defined in 
(\ref{Eqn:ICBoundStep1}). $\phi_{j}$, as the reader will recall/note, 
quantifies 
the amount of information communicated via the fixed B-L code. In the second 
step, we have to only take into account that decoder is attempting to recover 
$\boldS_{j}$ and has reconstructed $\boldhatK_{j}$. Following a sequence of 
steps analogous to those that took us from 
(\ref{Eqn:DueckExExUpperBoundOnBinningRates}) 
to (\ref{Eqn:DueckExExUpperBoundOnIndividualBinningRate}), one can prove
\ifTITVersion\begin{eqnarray}
 \label{Eqn:ICStep1AnalBoundOnBinningRate}
 H(S_{j}^{l}|\hatK_{j}^{l}) \leq 
l(\mathcal{L}_{l}(\phi_{j},|\mathcal{S}_{j}|)+H(S_{j}|K_{1})).
\end{eqnarray}\fi
\ifPeerReviewVersion\begin{eqnarray}
 \label{Eqn:ICStep1AnalBoundOnBinningRate}
 H(S_{j}^{l}|\hatK_{j}^{l}) \leq 
l(\mathcal{L}_{l}(\phi_{j},|\mathcal{S}_{j}|)+H(S_{j}|K_{1})).
\end{eqnarray}\fi
Recall that each encoder must communicate the second sub-message index taking 
values in $\exp\{ l \beta\}$ through the outer channel code. The sum of $\beta$ 
and the RHS of (\ref{Eqn:ICStep1AnalBoundOnBinningRate}) is indeed the LHS 
of (\ref{Eqn:Step1ICIndvdualRateBound}).

In the third part, we have to characterize the effective IC experienced by the 
outer PTP codes. Following the description provided in Section 
\ref{SubSec:MACStep1GeneralizationSeparateDecoding}, it is straight forward to 
note that the IC channel experienced by the $i$-th pair of outer codes is 
$p_{\mathscr{Y}_{1}\mathscr{Y}_{2}|\mathscr{V}_{1}\mathscr{V}_{2}}$, where 
\begin{eqnarray}
p_{\mathscr{U}_{1}\mathscr{U}_{2}\mathscr{V}_{1}\mathscr{V}_{2}\mathscr{X}_{1}  
\mathscr{X}_{2}\mathscr{Y}_{1}\mathscr{Y}_{2}}\left( \!\!\!
\begin{array}{c} 
u_{1},u_{2},v_{1},v_{2},\\x_{1},x_{2},y_{1},y_{2}
\end{array}\!\!\!\right)
= \frac{1}{l} 
\sum_{i=1}^{l}p_{U_{1i} U_{2i} V_{1i} V_{2i} X_{1i} X_{2i}Y_{1i}
Y_{2i} } 
\left( \!\!\!
\begin{array}{c} 
u_{1},u_{2},v_{1},v_{2},\\x_{1},x_{2},y_{1},y_{2}
\end{array}\!\!\!\right)\mbox{ and}
\label{Eqn:Step1ICPMFInterleavedVec}\nonumber\\
p_{\ulinesfU^{l}\ulinesfV^{l}\ulinesfX^{l}
\ulinesfY^{l}} 
 \!\!\left(\!\!\!\begin{array}{c}\ulineu^{l},
 \ulinev^{l } ,\\ \ulinex^ {l} ,
\uliney^{l}\end{array} \!\!\!\right) = \left[ 
\sum_{\substack{(a_{1},a_{2}) \in \\ 
[M_{u}]\times [M_{u}]}} 
\!\!\!\!\!\!\!P(
 \begin{array}{c}
 A_{1}=a_{1}\\A_{2}=a_{2}
  \end{array})\mathds{1}_{\left\{\substack{ 
 u^{l}(a_{j})=\\u_{j}^{l}:j \in [2]}\right\}}\right]
  \times \left[ \prod_{j=1}^{2} \left\{ 
\prod_{i=1}^{l}p_{V_{j}}(v_{ji})p_{X_{j}|UV_{j}}(x_{ji}|u_{ji},v_{ji})\right\} 
\right]\nonumber\\
 \label{Eqn:Step1ICPMFForDecRule}
 \times \left[ \prod_{i=1}^{l} 
\mathbb{W}_{Y_{1}Y_{2}|X_{1}X_{2}}(y_{1i},y_{2i}| x_{1i},x_{2i}) 
\right]. \nonumber\end{eqnarray}
$i-$th outer codebook of Tx $j$ can have rate at most 
$I(\mathscr{V}_{j};\mathscr{Y}_{j})$. Following steps identical to those in 
(\ref{Eqn:MACSTep1IndChnlBnd-1}) - (\ref{Eqn:MACSTep1SumChnlBnd-5}), it can be 
proved that $I(\mathscr{V}_{j};\mathscr{Y}_{j}) \geq 
I(V_{j};Y_{j})-\mathcal{L}(\phi,|\mathcal{V}_{j}|)$ which is indeed the 
RHS of (\ref{Eqn:Step1ICIndvdualRateBound}). This concludes our outline. The 
interested reader is invited to peruse through proof of Theorem 
\ref{Thm:ICStep2} is more general than Theorem \ref{Thm:ICStep1}.
\begin{thm}
 \label{Thm:FBLBeatsLC}
There exists a source pair $(\ulineCalS,\mathbb{W}_{\ulineS})$ and an IC 
$(\ulineCalX,\ulineCalY,\mathbb{W}_{\ulineY|\ulineX})$ that do not satisfy LC 
conditions \cite[Thm. 1]{201112TIT_LiuChe} (Theorem 
\ref{Thm:LCConditions} here) and yet satisfy conditions stated in 
Theorem \ref{Thm:ICStep1}. In particular, consider Example 
\ref{Ex:DuecksExampleForIC}. There exists $a^{*} 
\in \naturals$ and $k^{*}\in \naturals$ such that for any $a\geq a^{*}$ and any 
$k \geq k^{*}$, source pair $\ulineS$ and IC 
$\mathbb{W}_{\underline{Y}|\underline{U}\underline{X}}$ 1) do not satisfy LC 
conditions \cite[Thm. 1]{201112TIT_LiuChe} (Theorem 
\ref{Thm:LCConditions} here), and 2) satisfy conditions 
stated in Theorem \ref{Thm:ICStep1}.
\end{thm}
\begin{IEEEproof}
We only need to prove the second statement. Naturally, our assignment for the 
auxiliary parameters is identical to that is proof of Lemma 
\ref{Thm:FBLBeatsCES}. We provide the same for the sake of completeness.

Let $\mathcal{K}= \mathcal{S}_{1}=\mathcal{S}_{2}$, $\mathcal{U}$ be the input 
alphabet of the shared channel $\mathbb{W}_{Y_{0}|\ulineU}$, 
$\mathcal{V}_{j}=\mathcal{X}_{j}: j \in [2]$ be the input alphabet of the 
satellite channels $\mathbb{W}_{Y_{j}|X_{j}} : j \in [2]$ respectively.
Let $f_{j}(s) = s$ for $s \in \mathcal{S}_{j}$ be the identity map, and hence 
$K_{j}=S_{j}$ for $j \in [2]$.
Let $\alpha = \left(1-\frac{1}{4k} \right)\log a$, $\beta = \frac{5}{4k}\log a 
+ \left( 1+\frac{1}{k} \right)h_{b}(\frac{1}{k})$, $\rho = 
\frac{1}{4k}\log\frac{a}{4}$, $\delta = \frac{1}{k}$.
Let $l = k^{4}a^{\frac{\eta k}{2}}$.
Let $p_{U}$ be the uniform pmf on $\mathcal{U} = \{ 0,\cdots,a-1\}$. Let 
$p_{V_{j}} : j \in [2]$ be the capacity achieving distribution on satellite 
channels $\mathbb{W}_{Y_{j}|X_{j}} : j \in [2]$ respectively.
Note that, for any $u \in \mathcal{U}$, $l p_{U}(u) = k^{4}a^{\frac{\eta 
k}{2}-1}$ is a natural number since $\eta \geq 6$ is an even integer. We refer 
to the arguments in proof of Theorem \ref{Thm:FBLBeatsCES} that proves the 
choice $l = k^{4}a^{\frac{\eta k}{2}} \geq l^{*}\left( 
\frac{1}{4k}\log\frac{a}{4},\mathcal{U},\ulineCalY \right)$ for sufficiently 
large $a,k$.

Note that
\begin{eqnarray}
 \label{Eqn:LemFBLBeatsCES1}
 \beta + H( S_{j}|S_{1} ) +\mathcal{L}_{l} ( \phi, | \mathcal{S}_{j}|)+ 
\mathcal{L} (\phi,|\mathcal{V}_{j}|) &\leq &\frac{5}{4k}\log a + 
(1+\frac{1}{k})h_{b}(\frac{1}{k})+ \mathds{1}_{\{ j=2\}}h_{b}(\frac{2}{ka^{\eta 
k}})+ \phi(1+k) \log a \nonumber\\
 \label{Eqn:LemFBLBeatsCES2}
 &&+ (1+\frac{1}{l})h_{b}(\phi) \leq 
\frac{2}{k}\log a + h_{b}(\frac{2}{k})+\mathds{1}_{\{ 
j=2\}}h_{b}(\frac{2}{ka^{\eta 
k}})\nonumber
\end{eqnarray}
for sufficiently large $a,k$ because with the above choice for $\delta, l, 
\alpha, \rho$, we have from (\ref{Eqn:XiAndTau}) $\phi \leq 
\frac{8k^{4}}{a^{\frac{\eta k}{3}}}$ for sufficiently large $a,k$. The RHS of 
(\ref{Eqn:LemFBLBeatsCES1}) is $I(V_{j};Y_{j})$ and we have therefore proved 
(\ref{Eqn:Step2ICIndvdualRateBound}) for the choice of $p_{V_{j}}$ being the 
capacity achieving pmf.
\end{IEEEproof}

\section{Fixed B-L coding over arbitrary MAC and IC Step 2 : Conditional 
Decoding}
\label{Sec:FBLCodingOverMACAndICStep2}
We enhance the coding scheme presented in Step 1 (Section 
\ref{Sec:FBLCodingOverMACAndICStep1}) via the well known technique of 
conditional (joint) decoding. In Step 1, the fixed B-L and $\infty-$B-L 
information streams caused interference to each other, when multiplexed through 
the separate channel codes. The interference from the former can be nullified 
by conditional decoding of the latter. Step 2 builds on this approach.

The central challenge in conditional decoding arises from the fact that a 
non-vanishing fraction $\phi > 0$ of the fixed B-L codewords have been decoded 
erroneously. We overcome this challenge by the technique of interleaving and 
treating the decoded $C_{U}$ codewords as providing soft 
information\footnote{akin to noisy channel state information at the decoder}. 
Recall that the fixed B-L decoder of $C_{U}$ operates separately and 
identically on each of the $m$ received sub-blocks $\boldY_{j}(t,1:l)$ and 
declares $\hatA_{jt}$ as the corresponding decoded message. This indicates that 
the $m$ sub-blocks $u^{l}(A_{t}),u^{l}(\hatA_{t}): t \in [m]$, where $u^{l}(a)$ 
is the $C_{U}-$codeword corresponding to message $a \in [M_{u}]$, are 
distributed with an $l-$letter pmf. As we noted in the proof of Thm 
\ref{Thm:MACStep1} (Appendix \ref{AppSec:PMFOfInterleavedVector}), interleaving 
enables us extract IID sub-vectors, and moreover since the outer code is 
multiplexed along interleaved columns, the corresponding interleaved columns 
$\hat{\boldu}^{\pi}(1:m,i)$, where $\hat{\boldu}(t,1:l) = u^{l}(\hatA_{t}): t 
\in [m]$ is treated as soft information for conditional decoding of the outer 
code. Based on these ideas, we derive sufficient conditions for the IC 
(Sections \ref{SubSec:ICStep2GeneralizationJointDecoding}, 
\ref{SubSec:ICStep2HKDecoding}) and MAC (Section 
\ref{SubSec:MACStep2GeneralizationJointDecoding}).
\subsection{IC Problem Step II: Joint decoding of Fixed and $\infty-$B-L 
information 
streams}
\label{SubSec:ICStep2GeneralizationJointDecoding}
It is natural to expect the sufficient conditions to take the form of 
(\ref{Eqn:Step1ICIndvdualRateBound}) with $I(V_{j};Y_{j})$ on the RHS replaced 
by $I(V_{j};Y_{j}|U)$ ignoring the change in correction terms 
$\mathcal{L}_{l}(\cdot,\cdot)$. Indeed, as we will see, all of the 
sufficient conditions presented for the IC will involve corresponding 
substitutions. We present our first set of sufficient conditions for the IC 
based on conditional decoding of the outer code.

\begin{thm}
\label{Thm:ICStep2}
A pair of sources $(\ulineCalS,\mathbb{W}_{\ulineS})$ is transmissible over an 
IC $(\ulineCalX,\ulineCalY,\mathbb{W}_{\ulineY|\ulineX})$ if there exists 
\begin{enumerate}
 \item[(i)]  finite sets 
$\mathcal{K},\mathcal{U},\mathcal{V}_{1},\mathcal{V}_{2}$,
 \item[(ii)] maps $f_{j}:\mathcal{S}_{j}\rightarrow 
\mathcal{K}$, with $K_{j}=f_{j}(S_{j})$ for $j \in [2]$,
 \item[(iii)] $\alpha,\beta \geq 0$, $\rho > 0$, $\delta > 0$, 
 \item[(iv)] $l \in 
\naturals, l \geq \max\{ l^{*}(\rho,\mathcal{U},\mathcal{Y}_{j}): j \in [2] 
\}$, where $l^{*}(\cdot,\cdot,\cdot)$ is defined in (\ref{Eqn:DefnOfLStar}),
\item[(v)] pmf $p_{U}p_{V_{1}}p_{V_{2}}p_{X_{1}|UV_{1}}p_{X_{2}|UV_{2}} 
\mathbb{W}_{\ulineY|\ulineX}$ defined on $\mathcal{U}\times 
\ulineCalV\times\ulineCalX\times \ulineCalY$, where $p_{U}$ is a type of 
sequences in $\mathcal{U}^{l}$, such that for some $a \in [2]$, we have
\end{enumerate}
\ifPeerReviewVersion
 \begin{eqnarray}
\!\!(1+\delta)H(K_{a}) \leq \alpha+\beta ,~~
 \label{Eqn:Step2ICIndvdualRateBound}
H(S_{j}|K_{a}) + \beta +
\mathcal{L}_{l}(\phi_{j},|\mathcal{S}_{j}|) < 
I(V_{j};Y_{j}|U)-\mathcal{L}(\phi_{j},|\mathcal{V}|)
\mbox{ for }j \in [2],~~ \phi_{j} \leq \frac{1}{2}\\
\mbox{where }\phi_{j} \define 
g_{j}(\alpha+\rho,l)+\xi^{[l]}(\ulineK)+\tau_{l,\delta}(K_{a}),
g_{j}(R,l) \define 
(l+1)^{2|\mathcal{U}||\mathcal{Y}_{j}|}\exp\{-lE_{r}(R,p_{U},p_{Y_{j}|U})\}, 
\mbox{for }j \in [2],
\end{eqnarray}
where $\mathcal{L}_{l}(\cdot,\cdot), 
\mathcal{L}(\cdot,\cdot)$ is as 
defined 
in (\ref{Eqn:AdditionalSourceCodingInfo}).\fi
\ifTITVersion
 \begin{eqnarray}
 \label{Eqn:TypicalSetSize}
 (1+\delta)H(K_{1}) \!\!\!\!&\leq&\!\!\!\! \alpha+\beta ,\mbox{ and 
for } j \in [2]\nonumber\\ 
\label{Eqn:Step2ICIndvdualRateBound}
H(S_{j}|K_{1}) + \beta +
\mathcal{L}_{l}(\phi_{j},|\mathcal{S}_{j}|) \!\!\!\!&\leq&\!\!\!\!
I(V_{j};Y_{j}|U)-\mathcal{L}(\phi_{j},
|\mathcal {V}_{j} |),
\end{eqnarray}
$\phi_{j} \in [0,0.5)$ where
\begin{eqnarray}\phi_{j} &\define &
\label{Eqn:ICStep1Phij}
g_{\rho,l}^{j}+\xi^{[l]}(\ulineK)+\tau_{l,\delta}(K_{1}),\\
\label{Eqn:ICStep1gRhoLj}
g_{\rho,l}^{j} &\define &
\exp\{-l(E_{r}(\alpha+\rho,p_{U},p_{Y_{j}|U})-\rho)\} 
.\end{eqnarray}\fi
\end{thm}
\begin{remark}
\label{Rem:Step1ICSubsumesSimpleSeparation}
If the sources have a GKW part $K=K_{1}=K_{2}$, then $\xi(\ulineK)=0$. One can 
choose $l$ arbitrarily large such that $\phi_{1},\phi_{2}$ can be made 
arbitrarily small. The resulting inner bound corresponds to a very simple 
separation based scheme involving a common message communicated over the IC. 
\end{remark}
\begin{IEEEproof}
We assume above conditions are satisfied for $a=1$, and
$\mathcal{K},K_{1},\cdots,\rho$ be provided as in theorem statement. We assume 
$\beta = 0$ and hence $\alpha > (1+\delta)H(K_{1})$. $K_{1},K_{2} \in 
\mathcal{K}$ represent the near GKW parts of the sources 
$\ulineS$ (see Remark \ref{Rem:ClarifyingNearGKWPart}). The rest of the 
parameters will be described as and when they 
appear. We begin with a description of the coding scheme.

\textit{Coding Scheme:} The (only) difference in the coding scheme presented 
here, in comparison to those presented in Section 
\ref{Sec:FBLCodingOverMACAndICStep1} is that the channel code decoder of the 
outer code utilizes the decoded codewords of $C_{U}$ - the fixed B-L channel 
code - and conditionally decodes into the outer code. Since the decoded 
codewords of $C_{U}$ (i) are incorrect with a non-zero and non-vanishing 
probability, and (ii) have an $l-$letter pmf, randomly and uniformly chosen 
symbols from $m$ such decoded codewords are treated as soft (noisy) information 
in decoding the outer code. A formal description of the coding scheme follows.

The B-L of the coding scheme is $lm$, where $l$ is as provided in the theorem 
statement. A block is viewed as an $m \times l$ matrix with $l-$length rows 
referred to as \textit{sub-blocks}. The encoding and decoding rules at both 
encoders $j 
\in [2]$ are identical, and we describe the same in terms of a generic index 
$j$. The source coding module comprises of two source codes - a fixed B-L 
typical set code and an $\infty-$B-L Slepian-Wolf binning code. Let $\boldS_{j} 
\in \mathcal{S}_{j}^{m \times l}, \boldK_{j} \in \mathcal{K}^{ m \times l}$ 
denote the matrix of source and near GKW part observed by encoder $j$. For 
$(t,i) \in [m] \times [l]$, $\boldS_{j}(t,i)$ and $\boldK_{j}(t,i) \define 
f_{j}(\boldS_{j}(t,i))$ are the symbols of the source and near GKW part observed 
during $(t-1)l+i$ -th symbol interval, where $f_{j}:\mathcal{S}_{j} \rightarrow 
\mathcal{K}$ is as specified in the theorem statement. The 
fixed B-L typical set code operates separately and identically on the rows of 
$\boldK_{j}(t,1:l): t \in [m]$. In particular, the index of $\boldK_{j} (t,1:l)$ 
in the typical set $T_{\delta}^{l}(K_{1})$ is output by the fixed B-L source 
code.\footnote{$\delta$ is as provided in the theorem statement.} Formally, the 
fixed B-L typical set code is defined by an index set 
$[M_{K}]$ with $M_{K} = |T_{\delta}^{l}(K_{1})|$, encoder map $e_{K}: 
\mathcal{K}^{l} \rightarrow [M_{K}]$ and decoder map $d_{K}: [M_{K}] \rightarrow 
\mathcal{K}^{l}$ such that $d_{K}(e_{K}(k^{l}))=k^{l}$ for every $k^{l} \in 
T_{\delta}^{l}(K_{1})$. Let $A_{jt} \define e_{K}(\boldK_{j}(t,1:l)) : t \in 
[m]$ denote the $m$ messages output by the fixed B-L typical set 
code corresponding $\boldK_{j}$ and we let $\ulineA_{j} = (A_{jt}: t \in [m]) 
\in [M_{k}]^{m}$. We emphasize that both transmitters employ the same fixed B-L 
typical set code corresponding to $K_{1}$.

The $\infty-$B-L Slepian Wolf binning code operates over the entire block of 
$lm$ source symbols and outputs a bin index corresponding to the bin in which 
$\boldS_{j}$ lies\footnote{Here $\boldS_{j}$ is referencing the $lm-$length 
vector $(\boldS_{j}(t,1:l): t \in [m])$.}. In particular, let $\beta_{j}: 
\mathcal{S}_{j}^{lm} 
\rightarrow [  M_{V_{j}}]^{l}$ define a partition of the $lm-$length 
source sequences into $ M_{V_{j}}^{l}  $ bins. The $\infty-$B-L Slepian 
Wolf binning code outputs the index $\ulineB_{j} 
\define (B_{j1},\cdots, B_{jl}) \define \beta(\boldS_{j}) \in [ 
M_{V_{j}}]^{l}$ of the bin in 
which $\boldS_{j}$ lies. The pair $(\ulineA_{j},\ulineB_{j}) \in 
[M_{K}]^{m} \times [ M_{V_{j}}]^{l}$ of messages 
constitute of the output of the source coding 
module.

The channel coding module of encoder $j$ comprises of a 
fixed B-L constant composition code built over $\mathcal{U}$, $l$ codes of 
B-L $m$, referred as $\infty-$B-L codes, built over $\mathcal{V}_{j}$, and a 
multiplexing unit. $\mathcal{U}, \mathcal{V}_{1},\mathcal{V}_{2}$ are as 
provided in theorem statement. Let $C_{U}$ denote a constant composition code 
$(l,M_{u},e_{u},d_{u,1},d_{u,2})$ of B-L $l$, with message index set $[M_{u}]$, 
$M_{u} \geq M_{K}$, encoder 
map $e_{u}: [M_{u}] \rightarrow \mathcal{U}^{l}$ and decoder maps $d_{u,j}: 
\mathcal{Y}_{j}^{l} \rightarrow [M_{u} ]$ such that the maximal probability of 
error, when employed over the memoryless PTP 
$(\mathcal{U},\mathcal{Y}_{j},p_{Y_{j}|U})$, 
is at most $g_{j}(\rho,l) \leq 
\exp\{-l(E_{r}(\frac{\log M_{u}}{l},p_{U},p_{Y_{j}|U})-\rho)\} $. The existence 
of such 
a code is guaranteed 
by \cite[Thm. 10.2]{CK-IT2011}. We let $u^{l}(a) = e_{u}(a): a \in [M_{u}]$ 
denote the codewords of $C_{U}$. $C_{U}$ will be used to communicate 
$\ulineA_{j} \in [M_{u}]^{m}$ output by the source coding module. For $\ulinea 
\in [M_{u}]^{m}$, we let $\boldu \{ \ulinea \} \in \mathcal{U}^{m \times l}$ 
denote the matrix whose rows $\boldu\{\ulinea\}(t,1:l) \define 
u^{l}(a_{t}): t \in [m]$ are codewords of $C_{U}$ corresponding to messages 
$\ulinea$. We let $\boldU_{j} \define \boldu \{ \ulineA_{j}\}$ denote the 
matrix of $C_{U}-$codewords corresponding to messages $\ulineA_{j}$ output by 
the fixed B-L typical set code. We let $d_{u,j}^{l}: 
\mathcal{Y}_{j}^{l}\rightarrow \mathcal{U}^{l}$ 
be defined as $d_{u,j}^{l}(y_{j}^{l}) \define u^{l}(d_{u,j}(y_{j}^{l})) = 
e_{u}(d_{u,j}(y_{j}^{l}))$ denote 
the codeword corresponding to the decoded message.

The $\infty-$B-L channel code comprises of $l$ channel codes, each of B-L $m$ 
built 
over alphabet set $\mathcal{V}_{j}$. For $i \in [l]$, let $C_{V_{j},i}$ denote 
code with message index set $[M_{V_{j}}]$ and codewords 
$v_{ji}^{m}(b_{j}) : b_{j} \in 
[M_{V_{j}}]$. For $i \in [l]$, codebook $C_{V_{j},i}$ will be used to 
communicate $B_{ji}$ 
output by the $\infty-$B-L source code.

The multiplexing unit maps $m$ codewords chosen from $C_{U}$, $l$ codewords 
chosen from $C_{V_{j},i}: i \in [l]$ into a matrix of input symbols. It 
comprises of $m$ surjective maps $\pi_{t}: [l] \rightarrow 
[l]: t \in [m]$ and a map $\boldx_{j}: \mathcal{U}^{ m \times l} \times 
\mathcal{V}_{ j}^{m \times l} \rightarrow \mathcal{X}_{j}^{ m \times l}$. 
Suppose (i) $u^{l}(a_{jt}): t \in [m]$ are the $m$ codewords chosen 
from $C_{U}$ to form the matrix $\boldu\{ \ulinea\}$, and (ii) 
$v_{ji}^{m}(b_{ ji }): i \in [l]$ are the $l$ codewords chosen from $l$ 
codebooks $C_{V_{j},i}$. We let $\boldv_{j}\{ \ulineb \} \in 
\mathcal{V}_{j}^{m \times l}$ be defined through $\boldv_{j}\{ \ulineb 
\}^{\ulinepi}(1:m,i) = (\boldv_{j}(t,\pi_{t}(i)): t \in [m]) = 
v_{ji}^{m}(b_{ji})$ for $i \in [l]$ and $\boldx_{j}\{ \ulinea_{j},\ulineb_{j} 
\} = \boldx_{j}(\boldu \{ \ulinea_{j} \},\boldv_{j}\{ \ulineb_{j}\})$.

We now state the encoding rule. $\boldx_{j}\{ \ulineA_{j},\ulineB_{j} \}(t,i) $ 
is input on the channel during symbol interval $(t-1)l+i$, where 
$(\ulineA_{j},\ulineB_{j}) \in 
[M_{K}]^{m} \times [ M_{V_{j}}]^{l}$ are the messages 
output by the source coding module.

Before we state the decoding rule, we characterize the following pmfs that will 
be necessary to state the joint-typicality decoding rules. Let $A_{j} \define 
e_{K}(K_{j}^{l}) : j \in [2]$, where $K_{j1},\cdots K_{jl}$ are $l$ IID 
symbols of the near GKW part $K_{j}$, and let

\begin{eqnarray}
p_{\ulinesfU^{l}\ulinesfV^{l}\ulinesfX^{l}\ulinesfY^{l} \ulinesfhatU^{l}} 
 \!\!\left(\!\!\!\begin{array}{c}\ulineu^{l},
 \ulinev^{l } ,\\ \ulinex^ {l} ,
\uliney^{l},\\\ulinehatu^{l}\end{array} \!\!\!\right) = \left[ 
\sum_{\substack{(a_{1},a_{2}) \in \\ 
[M_{u}]\times [M_{u}]}} 
\!\!\!\!\!\!\!P(
 \begin{array}{c}
 A_{1}=a_{1}\\A_{2}=a_{2}
  \end{array})\mathds{1}_{\left\{\substack{ 
 u^{l}(a_{j})=\\u_{j}^{l}:j \in [2]}\right\}}\right]
  \times \left[ \prod_{j=1}^{2} \left\{ 
\prod_{i=1}^{l}p_{V_{j}}(v_{ji})p_{X_{j}|UV_{j}}(x_{ji}|u_{ji},v_{ji})\right\} 
\right]\nonumber\\
 \label{Eqn:Step2ICPMFForDecRule}
 \times \left[ \prod_{i=1}^{l} 
\mathbb{W}_{Y_{1}Y_{2}|X_{1}X_{2}}(y_{1i},y_{2i}| x_{1i},x_{2i}) \right] 
\mathds{1}_{\left\{ \hatu_{j}^{l}=d^{l}_{u,j}(y_{j}^{l}): j \in [2] \right\}},\\
p_{\mathscr{U}_{1}\mathscr{U}_{2}\mathscr{V}_{1}\mathscr{V}_{2}\mathscr{X}_{1}  
\mathscr{X}_{2}\mathscr{Y}_{1}\mathscr{Y}_{2}\hat{\mathscr{U}}_{1}\hat{\mathscr{
U}}_{2}}\left( \!\!\!
\begin{array}{c} 
u_{1},u_{2},v_{1},v_{2},\\x_{1},x_{2},y_{1},y_{2},\\\hatu_{1},\hatu_ { 2 } 
\end{array}\!\!\!\right)
= \frac{1}{l} 
\sum_{i=1}^{l}p_{U_{1i} U_{2i} V_{1i} V_{2i} X_{1i} X_{2i}Y_{1i}
Y_{2i} \hatU_{1i} 
\hatU_{2i}} 
\left( \!\!\!
\begin{array}{c} 
u_{1},u_{2},v_{1},v_{2},\\x_{1},x_{2},y_{1},y_{2},\\\hatu_{1},\hatu_ { 2 } 
\end{array}\!\!\!\right)
\label{Eqn:Step2ICPMFInterleavedVec}\end{eqnarray}\begin{eqnarray}
p_{\hatsfK_{1}^{l}\hatsfK_{2}^{l}|\sfK_{1}^{l}\sfK_{2}^{l}}(\hatk_{1}^{l},\hatk_
{ 2 } ^ { l } |k_{1}^{l},k_{2}^{l}) = 
\!\!\!\sum_{\substack{(y_{1}^{l},y_{2}^{l}) \in\\\mathcal{Y}_{1}^{l}\times 
\mathcal{Y}_{2}^{l} 
}}\!\!\!\!\!p_{\sfY_{1}^{l}\sfY_{2}^{l}|\sfU_{1}^{l},\sfU_{2}^{l}}(y_{1}^{ l}, 
y_{ 2 } ^ { l } | e_{u}(e_{K}(k_{1}^{l})),e_{u}(e_{K}(k_{2}^{l})))\mathds{1}_{  
\left\{ \!\!\!\begin{array}{c} \hatk_{j}^{l} = d_{K}(d_{u,j}(y_{j}^{l})): j \in 
[2]  \end{array}\!\!\!\right\} },\label{Eqn:Step2ICSourceCodeDecodingPMFAux}\\
p_{S_{1}^{l}S_{2}^{l}K_{1}^{l}K_{2}^{l}\hatK_{1}^{l}\hatK_{2}^{l}}(s_{1}^{l},s_{
2 }^{l},k_{1}^{l},k_{2}^{l},\hatk_{1}^{l},\hatk_{2}^{l}) = \left\{ 
\prod_{i=1}^{l}\mathbb{W}_{S_{1}S_{2}}(s_{1,i},s_{2,i})\mathds{1}_{ \left\{ 
k_{j} = f_{j}(s_{j,i}) \right\}  } 
\right\}p_{\hatsfK_{1}^{l}\hatsfK_{2}^{l}|\sfK_{1}^{l}\sfK_{2}^{l}}(\hatk_{1}^{l 
},\hatk_
{ 2 } ^ { l } |k_{1}^{l},k_{2}^{l}),\\
\mbox{and hence } \label{Eqn:Step2ICSourceCodeDecodingPMF}
p_{S_{1}^{l}S_{2}^{l}
\hatK_{1}^{l}\hatK_{2}^{l}}(s_{1}^{l},s_{
2 }^{l},\hatk_{1}^{l},\hatk_{2}^{l} ) = 
p_{\hatsfK_{1}^{l},\hatsfK_{2}^{l}|\sfK_{1}^{l}\sfK_{2}^{l}}  \left(  
\hatk_{1}^{l}  , \hatk_{2}^{l}  \left| \!\!\!\begin{array}{c} 
f_{1}(s_{11})f_{1}(s_{12}),\cdots 
f_{1}(s_{1l})\\  f_{2}(s_{21})f_{2}(s_{22})\cdots f_{2}(s_{2l}) \end{array} 
\!\!  \right.\right) 
\prod_{i=1}^{l}\mathbb{W}_{S_{1}S_{2}}(s_{1i},s_{2i}),
\end{eqnarray}
where $s_{1}^{l} = s_{11} s_{12}\cdots s_{1l}$ and $s_{2}^{l} = s_{21} 
s_{22}\cdots s_{2l}$.\footnote{The reader will recognize that 
(\ref{Eqn:Step2ICPMFForDecRule}), (\ref{Eqn:Step2ICPMFInterleavedVec}), 
(\ref{Eqn:Step2ICSourceCodeDecodingPMFAux}) 
(\ref{Eqn:Step2ICSourceCodeDecodingPMF}) are analogous to 
(\ref{Eqn:Step1MACPMFForDecodingRule}), 
(\ref{Eqn:Step1MACPMFOfInterleavedVector}), 
(\ref{Eqn:Step1MACSourceCodeDecodingAuxPMF}) and 
(\ref{Eqn:Step1MACSourceCodeDecodingPMF}) respectively.}

\textit{Decoding rule}: We now describe the decoding rule. Let 
$\boldY_{j} \in \mathcal{Y}_{j}^{ m \times l}$ denote the matrix of received 
symbols with $\boldY_{j}(t,i)$ being the symbol received during symbol interval 
$(t-1)l+i$. The channel-code decoding module comprises of the $C_{U}-$decoder 
and the $C_{V_{j},i}-$decoders. The $C_{U}-$decoder decodes rows of 
$\boldY_{j}$ separately and identically into $\hatA_{jt} \define 
d_{u,j}(\boldY_{j}(t,1:l)): t \in [m]$ and reconstructs $\boldhatU_{j} 
\define \boldu 
\{ \ulinehatA_{j} \}$. For each $i \in [l]$, the $C_{V_{j},i}-$decoder looks 
for all messages $\hatb_{ji} \in [M_{V_{j}}]$ such that the corresponding 
codeword is jointly typical with $\boldY_{j}^{\pi}(1:m,i),\boldu\{ 
\ulinehatA_{j}\}^{\pi}(1:m,i)$. Specifically for $i \in [l]$, populate
\begin{eqnarray}
 \label{Eqn:VjCodebookDecodingSet}
 \mathcal{D}_{j}(\boldY_{j},\ulinehatA_{j}) \define \left\{ \hatb_{ji} \in 
[M_{V_{j}}] : \left( 
v_{ji}^{m}(\hatb_{ji}),\boldY_{j}^{\pi}(1:m,i),\boldu\{ 
\ulinehatA_{j}\}^{\pi}(1:m,i) \right) \mbox{ is jointly typical wrt 
}\prod_{t=1}^{m}p_{\mathscr{V}_{j}\mathscr{Y}_{j}\hat{\mathscr{U}_{j}}} 
\right\}. 
\end{eqnarray}
For $i \in [l]$, such that $\mathcal{D}_{j}(\boldY_{j},\ulinehatA_{j})$ is 
empty, set $\hatB_{ji} = 1$. For $i \in [l]$ such that 
$\mathcal{D}_{j}(\boldY_{j},\ulinehatA_{j})$ is \textit{not} empty, choose one 
among 
the elements in
$\mathcal{D}_{j}(\boldY_{j},\ulinehatA_{j})$ uniformly at random, 
and set $\hatB_{ji}$ to be that element. Note that if 
$\mathcal{D}_{j}(\boldY_{j},\ulinehatA_{j})$ is a singleton for each $i \in 
[l]$, there is a unique choice for $\ulinehatB_{j}$. The channel code decoder 
furnishes $(\ulinehatA_{j},\ulinehatB_{j}) \in [M_{u}]^{m} \times 
[M_{V_{j}}]^{l}$ to the decoder of the source-coding module.

Let $\boldhatK_{j}(t,1:l) = d_{K}(\hatA_{jt})$ be the reconstructions output 
by the fixed B-L typical set decoder. The decoder of the Slepian-Wolf 
code looks for
\begin{eqnarray}
 \label{Eqn:SWDecodingSet}
 \mathcal{D}(\boldhatK_{j},\ulinehatB_{j}) \define \left\{  \bolds_{j} \in 
\mathcal{S}_{j}^{m \times l} : \beta_{j}(\bolds_{j}) = 
\hat{\ulineB}_{j}\mbox{ and }(\bolds_{j},\boldhatK_{j})\mbox{ is jointly 
typical wrt }\prod_{t=1}^{m}p_{S_{j}^{l}\hatK_{j}^{l}} \right\}.
\end{eqnarray}
If $\mathcal{D}(\boldhatK_{j},\ulinehatB_{j})$ is empty, set 
$\boldhatS_{j}$ to a predefined matrix in $\boldCalS_{j}$ that is arbitrarily 
fixed upfront. Otherwise, choose one among 
the matrices in $\mathcal{D}(\boldhatK,\ulinehatB_{1},\ulinehatB_{2})$ 
uniformly at random and set $\boldhatS_{j}$ to be that element. Declare 
$\boldhatS_{j}$ as the decoded matrix of source symbols.

\textit{Error event}: Let us characterize the error event $\mathscr{E}$. 
Suppose
\begin{eqnarray}
\mathscr{E}_{1j} \define \bigcup_{i=1}^{l}\left\{ 
B_{ji} 
\neq \hatB_{ji} \right\}, 
\mathscr{E}_{2} = \left\{  (\boldS_{1},\boldS_{2},\boldhatK_{1},\boldhatK_{2}) 
\mbox{ is \underline{not} 
typical wrt} \prod_{t=1}^{m} p_{  S_{1}^{l} S_{2}^{l} \hatK_{1}^{l} 
\hatK_{2}^{l} } \right\} 
\nonumber\\
\mathscr{E}_{3j} = \bigcup_{\substack{\boldhats_{j} \in 
\boldCalS_{j}}}\left\{\!\!\!\begin{array}{c}  \boldS_{j} \neq 
\boldhats_{j}, 
\beta_{j}(\boldhats_{j}) = \ulineB_{j}  
\\(\boldhats_{j},\boldhatK_{j}) \mbox{ is typical wrt } 
\prod_{t=1}^{m} p_{ S_{j}^{l} \hatK_{j}^{l}}  \end{array}\!\!\! 
\right\}\mbox{, then note that } \mathscr{E} \subseteq \mathscr{E}_{2} \cup 
\bigcup_{j=1}^{2} \mathscr{E}_{1j} 
 \cup \mathscr{E}_{3j}.
\nonumber
\end{eqnarray}
$\mathscr{E}_{1j}$ corresponds to erroneous decoding into one of $l$ codebooks
$C_{V_{j},i}: i \in [l]$. $\mathscr{E}_{2},\mathscr{E}_{3j}$ are 
error events concerning the Slepian Wolf code. In the following, we derive 
upper bounds on $P(\mathscr{E}_{1j}), P(\mathscr{E}_{2}),P(\mathscr{E}_{3j})$.

\textit{Probability of Error Analysis} : We analyze error probability of a 
random code. With respect to the distribution of the random code, we employ 
the same distribution as of that in proof of Theorem \ref{Thm:MACStep1}. We 
restate the same for completeness and ease of reference. The fixed B-L 
typical set code comprising of $M_{K}, e_{K}, d_{K}$ and the fixed B-L constant 
composition code characterized by $M_{u},e_{u},d_{u}$ remain fixed throughout 
our analysis. This leaves us with having to specify the distribution of 
random (i) binning indices $\beta_{j}(\bolds_{j}^{lm}): s_{j}^{lm} \in 
\mathcal{S}_{j}^{lm}:j \in [2]$ that constitute the $\infty-$B-L Slepian Wolf 
source 
code, (ii) codewords $V_{ji}^{m}(b_{j}): b_{j} \in [M_{V_{j}}]: i \in [l]$, 
(iii) surjective maps $\pi_{t}: [l] \rightarrow [l] : t \in [m]$, and (iv) 
$\boldx_{j}(\boldu,\boldv) \in \mathcal{X}^{ m \times l} : (\boldu,\boldv) \in 
\mathcal{U}^{m \times l} \times \mathcal{V}_{j}^{m \times l}$. The four
elements (i) $(\beta_{j}(\bolds_{j}^{lm}):\bolds_{j} \in \mathcal{S}_{j}^{lm}): 
j \in [2]$, (ii) $(V_{ji}^{m}(b_{j})\in 
\mathcal{V}_{j}^{m}:b_{j}\in 
[M_{V_{j}}],i \in [l], j\in [2])$, (iii) $(\Pi_{t}: t \in [m])$ and (iv) 
$(\boldX_{j}(u,v_{j})\in 
\mathcal{X}_{j}^{m \times l}:u \in \mathcal{U}^{m \times l}, v_{j} \in 
\mathcal{V}_{j}^{m \times l})$ are mutually independent. With regard to the 
bin indices, the collections 
$(\beta_{1}(\bolds_{1}^{lm}): \bolds_{1}^{lm} \in \mathcal{S}_{1}^{m \times 
l})$ and $(\beta_{2}(\bolds_{2}^{lm}): \bolds_{2}^{lm} \in 
\mathcal{S}_{2}^{m \times 
l})$ are mutually independent. Moreover, for each $j 
\in [2]$, the bin indices $\beta_{j}(\bolds_{j}^{lm}): \bolds_{j}^{lm} \in 
\mathcal{S}_{j}^{lm}$ are uniformly and independently chosen from 
$[M_{V_{j}}]^{l}$. The $m$ surjective 
maps 
$\Pi_{t}:t \in [m]$ are mutually independent and uniformly distributed over the 
entire collection of surjective maps over $[l]$. Each codeword in the 
collection 
$(V_{ji}^{m}(b_{j})\in \mathcal{V}_{j}^{m}:b_{j}\in [M_{V_{j}}],i \in [l], j\in 
[2])$ is mutually independent of the others and $V_{ji}^{m}(b_{ji}) \sim 
\prod_{t=1}^{m}p_{V_{j}}(\cdot)$, where $p_{V_{j}}$ corresponds to the chosen 
test channel. The collection $(\boldX_{j}(u,v_{j})\in \mathcal{X}_{j}^{m \times 
l}:u \in \mathcal{U}^{m \times l}, v_{j} \in \mathcal{V}_{j}^{m \times l})$ is 
mutually independent and $\boldX_{j}(u,v_{j}) \sim 
\prod_{t=1}^{m}\prod_{i=1}^{l}p_{X_{j}|UV_{j}}(\cdot|u(t,i),v_{j}(t,i))$. This 
defines the distribution of our random code. We employ an analogous notation 
for our random code. For example, given $\ulineb_{j} = (b_{ji}: i \in [l])$, we 
let $\boldV_{j}\{ \ulineb_{j} \} \in \boldCalV_{j}$ be defined through 
$\boldV_{j}\{ \ulineb_{j}\}^{\Pi}(1:m,i) = V_{ji}^{m}(b_{ji}) : i \in [l]$, and 
similarly $\boldX_{j}\{ \ulinea_{j},\ulineb_{j} \} \define 
\boldX_{j}(\boldu\{ \ulinea_{j} \} ,\boldV\{\ulineb_{j}\})$.

Our analysis will closely follow the steps provided in proof of Theorem 
\ref{Thm:MACStep1}. Our first step is to prove rows of \[\boldU_{j} \define 
\boldu 
\{\ulineA_{j} \}, \boldV_{j} \define 
\boldV_{j}\{ \ulineB_{j} \}, \boldX_{j} \define \boldX_{j}\{ 
\ulineA_{j},\ulineB_{j}\},\boldY_{j},\boldhatU_{j} \define \boldu \{ 
\ulinehatA_{j}\} : j \in [2]\] are IID with 
pmf $p_{\ulineU^{l}\ulineV^{l}\ulineX^{l}\ulineY^{l}\ulinehatU^{l}}$ defined in 
(\ref{Eqn:Step2ICPMFForDecRule}). This can be done by following a sequence of 
steps analogous to those that took us from (\ref{Eqn:Step1MACPrelim2-1}) to 
(\ref{Eqn:Step1MACPrelim2-12}). For the sake of completeness, we provide these 
steps in Appendix \ref{AppSec:Step2ICChnlMtrCorrectPMF}, where we prove
\begin{eqnarray}
 \label{Eqn:Step2ICChnlMtrCorrectPMFConcl}
 P\left(\!\!\!  
 \begin{array}{c}
\boldu\{ \ulineA_{j} \} = \boldu_{j}, \boldV_{j}\{ \ulineB_{j} \} = \boldv_{j}\\
\boldY_{j}=\boldy_{j} \boldX_{j}\{ \ulineA_{j},\ulineB_{j} 
\}=\boldx_{j}\\\boldu_{j}\{ \ulinehatA_{j} \} = \boldhatu_{j} : j \in [2]
 \end{array}
 \!\!\!\right) = \prod_{t=1}^{m} 
p_{\ulinesfU^{l}\ulinesfV^{l}\ulinesfX^{l}\ulinesfY^{l} 
\ulinesfhatU^{l}} 
 \!\!\left(\!\!\!\begin{array}{c}\boldu_{1}(t,1:l),\boldu_{2}(t,1:l),
 \boldv_{1}(t,1:l),\boldv_{2}(t,1:l),
 \boldx_{1}(t,1:l)\\\boldx_{2}(t,1:l),
 \boldy_{1}(t,1:l),\boldy_{2}(t,1:l),
 \boldhatu_{1}(t,1:l),\boldhatu_{2}(t,1:l)\end{array} \!\!\!\right).
\end{eqnarray}
Indeed, (\ref{Eqn:Step2ICChnlMtrCorrectPMFConcl}) is analogous to 
(\ref{Eqn:Step1MACPrelim2-13Conclusion}). As the reader 
might guess, we now prove rows of 
$\boldS_{1},\boldS_{2},\boldhatK_{1},\boldhatK_{2}$ are IID with pmf
$p_{S_{1}^{l} S_{2}^{l} \hatK_{1}^{l} \hatK_{2}^{l} }$. Once again, this can 
be proved by following arguments analogous to those presented in establishing 
(\ref{Eqn:Step1S1S2KHatPMF}). We provide these arguments in Appendix 
\ref{AppSec:Step2ICS1S2HatK1HatK2CorrectPMF}, where we prove
\begin{eqnarray}
\label{Eqn:Step2ICS1S2HatK1HatK2CrctPMF}
P\left(\!\!\!  
 \begin{array}{c}
\boldS_{1}=\bolds_{1}, \boldS_{2}=\bolds_{2},
\boldhatK_{1} = \boldhatk_{1},,
\boldhatK_{2} = \boldhatk_{2}
 \end{array}
 \!\!\!\right)
 = \prod_{t=1}^
{ m } p_{S_{1}^{l}S_{2}^{l}\hatK_{1}^{l}\hatK_{2}^{l}}\left(\!\!\!  
 \begin{array}{c}\bolds_{1}(t,1:l),\bolds_{2} (t , 
1:l),\boldhatk_{1}(t,1:l),\boldhatk_{2}(t,1:l)\end{array}
 \!\!\!
 \right)
\end{eqnarray}
We therefore have $m$ sub-blocks of the source and reconstructions to be IID 
with pmf $p_{S_{1}^{l}S_{2}^{l}\hatK_{1}^{l}\hatK_{2}^{l}}$. We can now appeal 
to standard arguments 
pertaining to Slepian Wolf decoding. In particular, using techniques presented 
in \cite[Chap 10]{201201NIT_ElgKim}, it can be verified that 
there exists $\xi > 0$, such that
\begin{eqnarray}
 \max{\{ P(\mathscr{E}_{2}),P(\mathscr{E}_{3})\}} \leq \exp \{ -m\xi \}~~\mbox{ 
if 
}~~\frac{\log M_{V_{j}}^{l}}{m} > H(S_{j}^{l}|\hatK_{j}^{l}) : j 
\in 
[2].
\end{eqnarray}
We are now concerned with $P(\mathscr{E}_{1j})$ and in particular upper bound 
$\sum_{i=1}^{l}P(\hatB_{ji} \neq B_{ji})$. Since
\begin{eqnarray}
\lefteqn{P ( \!
 \begin{array}{c}
  B_{ji} \neq  \hatB_{ji}
 \end{array}\!
 )  \leq 
 P (
  (\!
 \begin{array}{c}
  V_{ji}^{m}(B_{ji}),
  \boldY_{j}^{\Pi}(1:m,i),\boldu \{ \ulinehatA_{j}\}^{\Pi} (1:m,i)
 \end{array}\!) \ntypical \prod_{t=1}^{m} 
p_{\mathscr{V}_{j}\mathscr{Y}_{j}\hat{\mathscr{U}}_{j}}
 ) }\nonumber\\ 
  \label{Eqn:Step2ICOuterCodeError-1}
  &&+ \displaystyle P \left( \bigcup_{\substack{\hatb_{ji} \in 
[M_{V_{j}}] }}  \! \! \!\left\{
 \!
 \begin{array}{c}
  B_{ji}  \neq \hatb_{ji}
 \end{array}\!, ~( \! \!
 \begin{array}{c}
  V_{ji}^{m}(\hatb_{ji}) ,
  \boldY_{j}^{\Pi}(1:m,i),\boldu \{ \ulinehatA_{j} \}^{\Pi} (1:m,i)
 \end{array}  \! \!) \typical \prod_{t=1}^{m} 
p_{\mathscr{V}_{j}\mathscr{Y}_{j}\hat{\mathscr{U}}_{j}} \right\}
 \right) 
\end{eqnarray}
aim to derive upper bounds on the latter terms. With regard to the first term 
in (\ref{Eqn:Step2ICOuterCodeError-1}), we prove
\begin{eqnarray}\label{Eqn:Step2ICOuterCodeError-2}
\left(\boldV_{j}\{ \ulineB_{j} \}, 
\boldY_{j},\boldu \{ \ulinehatA_{j}\} \right) \!\!\begin{array}{c}\mbox{is 
distributed}\\\mbox{ with pmf }\end{array}\!\!\prod_{t=1}^{m} 
p_{V_{j}^{l}Y_{j}^{l}\hatU_{j}^{l}}\mbox{ and }\Pi_{1},\cdots,\Pi_{m}\mbox{ is 
independent of } \boldV_{j}\{ \ulineB_{j} \},\boldY_{j},\boldu \{ 
\ulinehatA_{j} \}.
\end{eqnarray}
in Appendix \ref{AppSec:Step2ICOuterCode1stErrEvent}. 
(\ref{Eqn:Step2ICOuterCodeError-2}) is 
analogous to (\ref{Eqn:Step1MAC-E1FirstEventSuff2}) and our proof in Appendix 
\ref{AppSec:Step2ICOuterCode1stErrEvent} will closely follow the steps in 
Appendix \ref{AppSec:Step2ICChnlMtrCorrectPMF} that established 
(\ref{Eqn:Step1MAC-E1FirstEventSuff2}). Having established 
(\ref{Eqn:Step2ICOuterCodeError-2}), we conclude that there exists a $\xi >0$ 
such that
\begin{eqnarray}
P ((\!
 \begin{array}{c}
  V_{ji}^{m}(B_{ji}),
  \boldY_{j}^{\Pi}(1:m,i),\boldu \{ \ulinehatA_{j}\}^{\Pi} (1:m,i)
 \end{array}\!) \ntypical \prod_{t=1}^{m} 
p_{\mathscr{V}_{j}\mathscr{Y}_{j}\hat{\mathscr{U}}_{j}}
 ) \leq \exp \{ -m \xi\}
\end{eqnarray}
and hence the first term in (\ref{Eqn:Step2ICOuterCodeError-1}) can be made 
arbitrarily small by choosing $m$ sufficiently large. We are now concerned with
\begin{eqnarray}
 \label{Eqn:ICStep2OutCodeIllCdwrd}
P\!\left(\! \bigcup_{\substack{b_{1i}}} ~
\bigcup_{\substack{\hatb_{1i}\neq b_{1i} 
}}\!\!\! \left\{\!\!\! \begin{array}{l} 
B_{1i}=b_{1i},
\end{array}\!\!\!\!\left(\!\!\! \begin{array}{c} 
V_{1i}^{m}(\hatb_{1i}),\boldY_{1}^{\Pi}(1:m,i),\boldu\{ 
\ulinehatA_{1}\}^{\Pi}(1:m,i) \end{array} 
\!\!\! \right)\!\! \in\! 
T_{\beta}^{m}(p_{\mathscr{V}_{1}\mathscr{Y}\hat{\mathscr{U}}_{1}})  
\!\right\} 
\! \right).
\end{eqnarray}
By the union bound and the law of total probability, the above quantity is at 
most
\begin{eqnarray}
 \label{Eqn:ICStep2OutCodeIllCdwrd1}
 \sum_{\substack{\ulinea_{1},\ulineb_{1}\\\ulinea_{2},\ulineb_{2}}}
  \sum_{\substack{\hatb_{1i} : \\\hatb_{1i}\neq b_{1i} }}
\sum_{\substack{v_{1}^{m}\\v_{2}^{m}}}
 \sum_{\substack{\boldx_{1} \in \boldCalX_{1}\\\boldx_{2} \in \boldCalX_{2}}}
 \sum_{\boldhatu_{1} \in \boldCalU}
 \sum_{\substack{\hatv_{1}^{m} \in \\\mathcal{V}_{1}^{m}}}~
 \sum_{\substack{\boldy_{1
} \in \\ \boldCalY_{1}}}\!\!
P\left( \!\!\!\begin{array}{c}
\ulineA_{j}=\ulinea_{j},\boldV_{j}\{ \ulineb_{j} 
\}^{\Pi}(1:m,i)=v_{j}^{m}\\\ulineB_{j}=\ulineb_{j},V_{1i}^{m}(\hatb_{1i})= 
\hatv_{1}^{m},\boldu\{ 
\ulinehatA_{1}\} = \boldhatu_{1}\\
\boldX_{j}\{ \ulinea_{j},\ulineb_{j} \} = \boldx_{j} 
: j \in [2],\boldY_{1} = \boldy_{1},\\
\end{array}\!\!\!  \right)\mathds{1}_{\left\{  \!\!\! 
\begin{array}{c}   (\hatv_{1}^{m},[\boldy_{1} 
\boldhatu_{1}]^{\Pi}(1:m,i))\\ \in  
T_{\beta}^{m}(p_{\mathscr{V}_{1}\mathscr{Y}\hat{\mathscr{U}}_{1}}) \end{array} 
\!\!\! 
\right\}}.
\end{eqnarray}
Consider a generic term above. Since $\hatb_{1i}\neq b_{1i}$, $\boldV_{j}\{ 
\ulineb_{j} 
\}^{\Pi}(1:m,i)=V_{ji}( 
b_{ji})$ for $j \in [2]$ and $V_{ji}( 
b_{ji}): j \in [2]$ is independent of $V_{1i}^{m}(\hatb_{1i})$, we have 
\begin{eqnarray}
 \label{Eqn:ICStep2OutCodeIllCdwrd1}
 P \left( \!\!\!\begin{array}{c}
\ulineA_{j}=\ulinea_{j},\boldV_{j}\{ \ulineb_{j} 
\}^{\Pi}(1:m,i)=v_{j}^{m}\\\ulineB_{j}=\ulineb_{j}: j \in 
[2],V_{1i}^{m}(\hatb_{1i})= 
\hatv_{1}^{m} 
\end{array}\!\!\! \right) &=& 
P \left( \!\!\!\begin{array}{c}
\ulineA_{j}=\ulinea_{j},\ulineB_{j}=\ulineb_{j},V_{ji}( 
b_{ji})=v_{j}^{m}:j \in [2]\\V_{1i}^{m}(\hatb_{1i})= 
\hatv_{1}^{m}  
\end{array}\!\!\!\right) \nonumber\\
&=& P \left( \!\!\!\begin{array}{c}
\ulineA_{j}=\ulinea_{j},\ulineB_{j}=\ulineb_{j}\\V_{ji}( 
b_{ji})=v_{j}^{m}:j \in [2]
\end{array}\!\!\!\right)P(V_{1i}^{m}(\hatb_{1i})= 
\hatv_{1}^{m}  ) \nonumber\\
&=& 
P \left( \!\!\! \begin{array}{c}
\ulineA_{j}=\ulinea_{j},\boldV_{j}\{ \ulineb_{j} 
\}^{\Pi}(1:m,i)=v_{j}^{m}\\\ulineB_{j}=\ulineb_{j}: j \in 
[2]
\end{array}\!\!\!\right)P(V_{1i}^{m}(\hatb_{1i})= 
\hatv_{1}^{m}  ). \nonumber
\end{eqnarray}
Next, we claim
\begin{eqnarray}
 \label{Eqn:ICStep2OutCodeIllCdwrd2}
 P \left(\!\!\! \begin{array}{c}
\boldX_{j}\{ \ulinea_{j},\ulineb_{j} \} \\= \boldx_{j} 
: j \in [2]
 \end{array}\!\!\!
\left|\!\!\!\begin{array}{c}
\ulineA_{j}=\ulinea_{j},\boldV_{j}\{ \ulineb_{j} 
\}^{\Pi}(1:m,i)=v_{j}^{m}\\\ulineB_{j}=\ulineb_{j}: j \in 
[2],V_{1i}^{m}(\hatb_{1i})= 
\hatv_{1}^{m} 
\end{array}\!\!\!\right. \right) = P \left(\!\!\! \begin{array}{c}
\boldX_{j}\{ \ulinea_{j},\ulineb_{j} \} \\= \boldx_{j} 
: j \in [2]
 \end{array}\!\!\!
\left|\!\!\!\begin{array}{c}
\boldV_{j}\{ \ulineb_{j} 
\}^{\Pi}(1:m,i)=v_{j}^{m}\\\ulineA_{j}=\ulinea_{j},\ulineB_{j}=\ulineb_{j}: j 
\in 
[2]
\end{array}\!\!\!\right. \right).
\end{eqnarray}
The above follows from the fact that conditioned on the event
\begin{eqnarray}\label{Eqn:ICStep2OutCodeIllCdwrd3}\{ 
V_{j\iota}^{m}(b_{j\iota})= v_{j\iota}^{m} : j \in [2], \iota \in [l]\setminus 
\{i\}, V_{ji}^{m}(b_{ji})=v_{j}^{m}, \Pi_{t}=\pi_{t}: t \in 
[m]\},\end{eqnarray}we have $(\boldX\{\ulinea_{j},\ulineb_{j}\}: j \in [2])$ 
independent of 
$V_{1i}^{m}(\hatb_{1i})$ and the random variables in 
(\ref{Eqn:ICStep2OutCodeIllCdwrd3}) are independent of 
$V_{1i}^{m}(\hatb_{1i})$. Indeed, recall that 
$\boldX\{\ulinea_{j},\ulineb_{j}\}$ is conditionally independent of the rest of 
the variables, given the random variables in 
(\ref{Eqn:ICStep2OutCodeIllCdwrd3}) and $\boldu\{\ulinea_{j}\}$, which is a 
deterministic 
function of $\ulinea_{j}$. Finally,
\begin{eqnarray}
 \label{Eqn:ICStep2OutCodeIllCdwrd4}
 P \left(\!\!\!\begin{array}{c}
          \boldu\{ 
\ulinehatA_{1}\} = \boldhatu_{1}\\\boldY_{1} = \boldy_{1}
         \end{array}\!\!\!
\left|
\!\!\!\begin{array}{c}\boldV_{j}\{ \ulineb_{j} 
\}^{\Pi}(1:m,i)=v_{j}^{m}\\\ulineA_{j}=\ulinea_{j},\boldX_{j}\{ 
\ulinea_{j},\ulineb_{j} \}= 
\boldx_{j}\\\ulineB_{j}=\ulineb_{j}: j 
\in [2],V_{1i}^{m}(\hatb_{1i})= 
\hatv_{1}^{m} 
\end{array}\!\!\!\right. \right) = P \left(\!\!\!\begin{array}{c}
          \boldu\{ 
\ulinehatA_{1}\} = \boldhatu_{1}\\\boldY_{1} = \boldy_{1}
         \end{array}\!\!\!
\left|
\!\!\!\begin{array}{c}
\boldV_{j}\{ \ulineb_{j} 
\}^{\Pi}(1:m,i)=v_{j}^{m}\\\ulineA_{j}=\ulinea_{j},\boldX_{j}\{ 
\ulinea_{j},\ulineb_{j} \}= 
\boldx_{j}\\\ulineB_{j}=\ulineb_{j}: j 
\in [2]\end{array}\!\!\!\right. \right)
\end{eqnarray}
holds because the IC ignores the rest given 
$(\boldX\{\ulineA_{j},\ulineB_{j}\}: j \in [2])$ and the $\boldu\{ 
\ulinehatA_{1}\}$ is a deterministic function of $\boldY_{1}$. Substituting 
(\ref{Eqn:ICStep2OutCodeIllCdwrd2}) - (\ref{Eqn:ICStep2OutCodeIllCdwrd4}) in 
(\ref{Eqn:ICStep2OutCodeIllCdwrd1}), we have
\begin{eqnarray}
 \sum_{\substack{\ulinea_{1},\ulineb_{1}\\\ulinea_{2},\ulineb_{2}}}
  \sum_{\substack{\hatb_{1i} : \\\hatb_{1i}\neq b_{1i} }}
\sum_{\substack{v_{1}^{m}\\v_{2}^{m}}}
 \sum_{\substack{\boldx_{1} \in \boldCalX_{1}\\\boldx_{2} \in \boldCalX_{2}}}
 \sum_{\boldhatu_{1} \in \boldCalU}
 \sum_{\substack{\hatv_{1}^{m} \in \\\mathcal{V}_{1}^{m}}}~
 \sum_{\substack{\boldy_{1
} \in \\ \boldCalY_{1}}}\!\!
P\left( \!\!\!\begin{array}{c}
\ulineA_{j}=\ulinea_{j},\boldV_{j}\{ \ulineb_{j} 
\}^{\Pi}(1:m,i)=v_{j}^{m},\ulineB_{j}=\ulineb_{j}\\\boldu\{ 
\ulinehatA_{1}\} = \boldhatu_{1},
\boldX_{j}\{ \ulinea_{j},\ulineb_{j} \} = \boldx_{j} 
: j \in [2],\boldY_{1} = \boldy_{1}
\end{array}\!\!\!  \right)P\left(\!\!\!\begin{array}{c} 
V_{1i}^{m}(\hatb_{1i})\\= 
\hatv_{1}^{m} \end{array}\!\!\!\right)\nonumber\\
  \times \mathds{1}_{\left\{  \!\!\! 
\begin{array}{c}   (\hatv_{1}^{m},[\boldy_{1} 
\boldhatu_{1}]^{\Pi}(1:m,i)) \in  
T_{\beta}^{m}(p_{\mathscr{V}_{1}\mathscr{Y}\hat{\mathscr{U}}_{1}}) \end{array} 
\!\!\! 
\right\}}\nonumber\\
=  \sum_{\substack{\hatb_{1i} : \\\hatb_{1i}\neq b_{1i} }}\sum_{\boldhatu_{1} 
\in \boldCalU}
 \sum_{\substack{\hatv_{1}^{m} \in \\\mathcal{V}_{1}^{m}}}~
 \sum_{\substack{\boldy_{1
} \in \\ \boldCalY_{1}}}P\left( \!\!\!\begin{array}{c}
\boldu\{ 
\ulinehatA_{1}\} = \boldhatu_{1}\\
\boldY_{1} = \boldy_{1}
\end{array}\!\!\!  \right)P\left(\!\!\!\begin{array}{c} 
V_{1i}^{m}(\hatb_{1i})= 
\hatv_{1}^{m} \end{array}\!\!\!\right)\mathds{1}_{\left\{  \!\!\! 
\begin{array}{c}   (\hatv_{1}^{m},[\boldy_{1} 
\boldhatu_{1}]^{\Pi}(1:m,i)) \in  
T_{\beta}^{m}(p_{\mathscr{V}_{1}\mathscr{Y}\hat{\mathscr{U}}_{1}}) \end{array} 
\!\!\! 
\right\}}\nonumber\\
= \sum_{\substack{\hatb_{1i} : \\\hatb_{1i}\neq b_{1i} }}\sum_{\hatu_{1}^{m} \in 
\mathcal{U}^{m}}\sum_{y_{1}^{m} \in \mathcal{Y}_{1}^{m}} \sum_{\hatv_{1}^{m} 
\in \mathcal{V}_{1}^{m}}\!\!\!P\left(\!\!\!\begin{array}{c}  [\boldu\{ 
\ulinehatA_{1}\} \boldY_{1}]^{\Pi}(1:m,i)\\ = (\hatu_{1}^{m},y^{m} ) 
\end{array}\!\!\! \right)P\left(\!\!\!\begin{array}{c} 
V_{1i}^{m}(\hatb_{1i})= 
\hatv_{1}^{m} \end{array}\!\!\!\right)\mathds{1}_{\left\{  \!\!\! 
\begin{array}{c}   (\hatv_{1}^{m},y^{m},\hatu_{1}^{m}) \in  
T_{\beta}^{m}(p_{\mathscr{V}_{1}\mathscr{Y}\hat{\mathscr{U}}_{1}}) \end{array} 
\!\!\! 
\right\}}\nonumber\\
\label{Eqn:ICStep2OutCodeIllCdwrd5}
= \sum_{\substack{\hatb_{1i} : \\\hatb_{1i}\neq b_{1i} }}\sum_{\substack{ 
(\hatv_{1}^{m},y_{1}^{m},\hatu_{1}^{m}) \\\in  
T_{\beta}^{m}(p_{\mathscr{V}_{1}\mathscr{Y}\hat{\mathscr{U}}_{1}})}}
\prod_{t=1}^{m}p_{\mathscr{Y}_{1}\hat{\mathscr{U}_{1}}}(y_{1t},\hatu_{1
t} )p_ {V_{1}}(\hatv_{1t}) = \sum_{\substack{\hatb_{1i} : \\\hatb_{1i}\neq 
b_{1i} }}\sum_{\substack{ (\hatv_{1}^{m},y_{1}^{m},\hatu_{1}^{m}) \\\in  
T_{\beta}^{m}(p_{\mathscr{V}_{1}\mathscr{Y}\hat{\mathscr{U}}_{1}})}}
\prod_{t=1}^{m}p_{\mathscr{Y}_{1}\hat{\mathscr{U}_{1}}}(y_{1t},\hatu_{1
t} )p_ { \mathscr{V}_{1}}(\hatv_{1t})
\end{eqnarray}
as an upper bound on 
(\ref{Eqn:ICStep2OutCodeIllCdwrd}). The first equality in 
(\ref{Eqn:ICStep2OutCodeIllCdwrd5}) follows from 
(\ref{Eqn:Step2ICOuterCodeError-2}) and the Lemma 
\ref{Lem:FullInterleavingLemma} and the second equality therein follows 
from 
$p_{V_{j}}=p_{\mathscr{V}_{j}}$ (Lemma 
\ref{Lem:SimplePropDecodingLem}). Using standard typicality, we conclude that 
there exists $\xi > 0$ such that (\ref{Eqn:ICStep2OutCodeIllCdwrd}) is smaller 
than $\exp \{-m\xi\}$ if $\frac{\log M_{V_{j}}}{m} < 
I(\mathscr{V}_{1};\mathscr{Y}_{j},\mathscr{U}_{j})$.

We summarize our proof thus far. We have proved that if
\begin{eqnarray}
 \label{Eqn:Step2IC:SummaryStep}
 H(S_{j}^{l}|\hat{K}_{j}^{l}) < \frac{\log M_{V_{j}}^{l}}{m} < l 
I(\mathscr{V}_{j};\mathscr{Y}_{j},\hat{\mathscr{U}_{j}})\mbox{ for } j \in [2]
\end{eqnarray}
where $S_{1}^{l},S_{2}^{l},\hatK_{1}^{l},\hatK_{2}^{l}$ and 
$\mathscr{V}_{j},\mathscr{Y}_{j},\hat{\mathscr{U}_{j}}: j \in [2]$ are 
distributed as in (\ref{Eqn:Step2ICSourceCodeDecodingPMF}), 
(\ref{Eqn:Step2ICPMFInterleavedVec}), then the proposed coding scheme can 
enable decoder $j$ recover $\boldS_{j}$ with arbitrarily high reliability for 
sufficiently large $m$. As in the proof of Theorem \ref{Thm:MACStep1}, we are 
left to quantify the upper and lower bounds in 
(\ref{Eqn:Step2IC:SummaryStep}) in terms of the pmf 
$\mathbb{W}_{\ulineS}p_{\ulineU\ulineV\ulineX}\mathbb{W}_{\ulineY|\ulineX}$ 
provided in the theorem statement. We consider the mutual information terms.

\textit{Lower Bounds on 
$I(\mathscr{V}_{j};\mathscr{Y}_{j},\hat{\mathscr{U}_{j}})$}: Suppose 
$(\ulineU^{l},\ulineV^{l},\ulineX^{l},Y^{l}) 
= (U_{1}^{l},U_{2}^{l},V_{1}^{l},V_{2}^{l } ,X_{1}^{l } , X_{ 2 }^{ l}, 
Y^{l}_{1},Y_{2}^{l},\hatU_{1}^{l},\hatU_{2}^{l}) $ 
is distributed with pmf (\ref{Eqn:Step2ICPMFForDecRule}), and $\mfI \in 
\{1,\cdots, l \}$ is a random index independent of the collection 
$\ulineU^{l},\ulineV^{l},\ulineX^{l},Y^{l}$, then 
$U_{1\mfI},U_{2\mfI},V_{1\mfI},V_{2\mfI},X_{1\mfI},X_{2\mfI},Y_{1\mfI},Y_{2\mfI}
, \hatU_{1\mfI}, \hatU_{2\mfI} $ is
distributed with PMF (\ref{Eqn:Step1ICPMFInterleavedVec}). Hence we study 
$I(V_{j\mfI};Y_{\mfI},\hat{U}_{j\mfI}) = I(\mathscr{V}_{j};\mathscr{Y}_{j} 
\hat{\mathscr{U}_{i}})$. Suppose $\frac{1}{2}\geq \phi \geq P(\{ U_{1}^{l} \neq 
U_{2}^{l} \} \cup \{ U_{1}^{l} \neq \hat{U}_{1}^{l} \})$, then
\begin{eqnarray}
\lefteqn{I(V_{j\mfI};Y_{\mfI},\hat{U}_{j\mfI}) = 
H(V_{j\mfI})-H(V_{j\mfI}|Y_{\mfI},\hat{U}_{j\mfI})\geq 
H(V_{j\mfI})-H(V_{j\mfI},\mathds{1}_{\{U_{1}^{l}=U_{2}^{l 
} = \hatU_{1}^{l} \}}|Y_{\mfI},\hat{U}_{j\mfI})}
\label{Eqn:ICSTep2IndChnlBnd-1}
\nonumber\\
&\geq& 
H(V_{j}) -H(V_{j\mfI}|Y_{\mfI},\hat{U}_{j\mfI},\mathds{1}_{\{U_{1}^{l}=U_{2}^{l 
} = \hatU_{1}^{l} \}}) -h_{b}(\phi)
\label{Eqn:ICSTep2IndChnlBnd-3}
\\
&=&H(V_{j})- P(U_{1}^{l} = 
U_{2}^{l}=\hatU_{1}^{l})  
\left [ H(V_{j\mfI} , Y_{\mfI},\hat{U}_{j\mfI}| 
\mathds{1}_{\{U_{1}^{l}=U_{2}^{l 
} = \hatU_{1}^{l} \}}
=1) - H(Y_{\mfI},\hat{U}_{j\mfI}| \mathds{1}_{\{U_{1}^{l}=U_{2}^{l 
} = \hatU_{1}^{l} \}}
=1 ) \right] \nonumber\\&&-P(\{ U_{1}^{l} \neq 
U_{2}^{l} \} \cup \{ U_{1}^{l} \neq \hat{U}_{1}^{l} \})  
H(V_{j\mfI}|Y_{\mfI},\hat{U}_{j\mfI},\mathds{1}_{\{U_{1}^{l}=U_{2}^{l 
} = \hatU_{1}^{l} \}}
=0) -h_{b}(\phi) \label{Eqn:ICSTep2IndChnlBnd-4}
\\
&=&H(V_{j})- P(U_{1}^{l} = 
U_{2}^{l}=U_{1}^{l})  
\left [ H(V_{j\mfI} , Y_{\mfI},{U}_{j\mfI}| 
\mathds{1}_{\{U_{1}^{l}=U_{2}^{l 
} = \hatU_{1}^{l} \}}
=1) - H(Y_{\mfI},{U}_{j\mfI}| \mathds{1}_{\{U_{1}^{l}=U_{2}^{l 
} = \hatU_{1}^{l} \}}
=1 ) \right] \nonumber\\&&-P(\{ U_{1}^{l} \neq 
U_{2}^{l} \} \cup \{ U_{1}^{l} \neq \hat{U}_{1}^{l} \})  
H(V_{j\mfI}|Y_{\mfI},\hat{U}_{j\mfI},\mathds{1}_{\{U_{1}^{l}=U_{2}^{l 
} = \hatU_{1}^{l} \}}
=0) -h_{b}(\phi) \label{Eqn:ICSTep2IndChnlBnd-5}
\nonumber\\
&\geq& H(V_{j}) -\left[ H(V_{j},Y_{j},U_{j}) - H(U_{j},Y_{j}) \right] 
-\phi 
\log|\mathcal{V}_{j}| -h_{b}(\phi)
\label{Eqn:ICSTep2IndChnlBnd-6}
=I(V_{j};Y_{j},U_{j})-\phi 
\log|\mathcal{V}_{j}| -h_{b}(\phi)\\
&=&I(V_{j};Y_{j}|U_{j})-\mathcal{L}(\phi,|\mathcal{V}_{j}|)
\end{eqnarray}
where (\ref{Eqn:ICSTep2IndChnlBnd-3}) follows from 
$p_{V_{j}}=p_{V_{jI}}=p_{\mathscr{V}_{j}}$ (Lemma 
\ref{Lem:SimplePropDecodingLem}) and $\frac{1}{2}\geq \phi \geq P(\{ U_{1}^{l} 
\neq 
U_{2}^{l} \} \cup \{ U_{1}^{l} \neq \hat{U}_{1}^{l} \})$, 
(\ref{Eqn:ICSTep2IndChnlBnd-5}) follows from 
Lemma 
\ref{Lem:PMFOfUnifAndRandCo-Ordinate} in Appendix 
\ref{AppSec:PropOfDecodingPMF} and from $\frac{1}{2} \geq \epsilon \geq 
P(U_{1}^{l}\neq U_{2}^{l})$.

The last part involves deriving upper bound on $H(S_{j}^{l}|\hatK_{j}^{l})$. We 
follow steps identical to that adopted in proof of Theorem \ref{Thm:MACStep1}. 
Recall from 
(\ref{Eqn:Step2ICS1S2HatK1HatK2CrctPMF}) that $p_{S_{1}^{l},S_{2}^{l}, 
\hatK_{1}^{l}, 
\hatK_{2}^{l}}$ is the pmf 
of any row of the quadruple $\boldS_{1},\boldS_{2},\boldhatK_{1},\boldhatK_{1}$ 
of matrices. 
Appealing to the sequence of steps from 
(\ref{Eqn:DueckExExUpperBoundOnBinningRates}) 
through (\ref{Eqn:DueckExExUpperBoundOnSumBinningRate}) we 
recognize that it suffices to characterize an upper bound $\phi$ on 
$P(\boldhatK_{j}(t,1:l) \neq \boldK_{1}(t,1:l))$, that is at most 
$\frac{1}{2}$. Towards that end, recall that our typical set 
source code ensures $d_{k}(e_{k}(k_{1}^{l})) = k_{1}^{l}$ for every $k_{1}^{l} 
\in T_{\delta}^{l}(K_{1})$. This guarantees $\{ \boldhatK_{j} (t,1:l) \neq 
\boldK_{1} (t,1:l) \} \subseteq \{ A_{1t} \neq \hatA_{jt} \}$. In 
order to derive an upper bound on the latter event, we are required to 
characterize the channel $p_{Y_{j}^{l}|U_{j}^{l}}$ experienced by codewords of 
$C_{U}$. 
In particular, since \begin{eqnarray}\label{Eqn:Step1MACFixedBLCodeError}P( 
A_{1t} \neq \hatA_{t}) \leq P(A_{1t} \neq A_{2t}) + P(\hatA_{jt} \neq A_{jt}, 
A_{1t} = A_{2t}) \leq \epsilon + P(\hatA_{t} \neq A_{1t}, A_{1t} = 
A_{2t}),\end{eqnarray}
we are required to characterize the channel $p_{Y_{j}^{l}|U_{j}^{l}}$ 
experienced by those commonly selected codewords. At the end of proof of 
Theorem \ref{Thm:MACStep1}, we proved this for the MAC. Here the additional 
element of conditional decoding plays no role in the arguments. Hence, the 
steps provided therein can be adopted for the IC case without any changes. In 
the interest of brevity, we do not repeat the arguments here and refer the 
reader to the steps provided from (\ref{Eqn:Step1MACFixedB-LChnlCodeError2}) - 
(\ref{Eqn:Step1MACFixedB-LChnlCodeError8}). This completes the proof.
\end{IEEEproof}

\subsection{IC problem : Conditional decoding via Han Kobayashi technique}
\label{SubSec:ICStep2HKDecoding}
In communicating the $\infty-$B-L information stream, we can employ the 
Han-Kobayashi technique of message splitting via superposition coding. Each 
encoder builds outer codes on $\mathcal{W}_{j},\mathcal{V}_{j}$ with the former 
carrying the public part and the latter, the private part. The output of the 
Slepian Wolf binning code is split into two parts, each indexing one of the 
above codes. A conditional Han Kobayashi decoding technique utilizing the 
interleaved vectors of the decoded fixed B-L code is employed. We present the 
following set of sufficient conditions. Techniques developed in Section 
\ref{SubSec:ICStep2GeneralizationJointDecoding}, in conjunction with 
Han-Kobayashi technique are employed to prove achievability. The following 
characterization of the Han-Kobayashi region is from \cite{200702ITA_KobHan}.
\begin{definition}
 \label{Defn:TestChannelsForHK}
 Let $\mathbb{D}(\mathbb{W}_{\ulineY|\ulineX})$ denote the collection of pmfs 
$p_{U}p_{V_{1}}p_{W_{1}}p_{V_{2}}p_{W_{2}}p_{X_{1}|UW_{1}V_{1}}p_{X_{2}|UW_{2}V_
{2}}\mathbb{W}_{Y_{1}Y_{2}|X_{1}X_{2}}$ defined on $\mathcal{U} \times 
\mathcal{V}_{1}\times\mathcal{W}_{1}\times 
\mathcal{V}_{2}\times\mathcal{W}_{2}\times 
\mathcal{X}_{1}\times\mathcal{X}_{2}\times 
\mathcal{Y}_{1}\times\mathcal{Y}_{2}$ such that 
$\mathcal{U},\mathcal{V}_{j},\mathcal{W}_{j}: j \in [2]$ are finite sets. For 
$p_{\ulineU\ulineV\ulineW\ulineX\ulineY} \in 
\mathbb{D}(\mathbb{W}_{\ulineY|\ulineX})$, let 
$\alpha_{CHK}(p_{\ulineU\ulineV\ulineW\ulineX\ulineY} )$ be defined as the set 
of pairs $(R_{1},R_{2})$ that satisfy
\begin{eqnarray}
 \label{Eqn:ICStep3CHKRegion}
 \begin{array}{rclrcl}
  R_{j}&\leq&d_{j}&R_{j}&\leq&a_{j}+e_{\msout{j}}\\
  R_{j}&\leq&a_{j}+f_{\msout{j}}&R_{1}+R_{2}&\leq&a_{j}+g_{\msout{j}}\\
  R_{1}+R_{2}&\leq&e_{1}+e_{2}&2R_{j}+R_{\msout{j}}& \leq & a_{j} + g_{j} + 
e_{\msout{j}} \\
 2R_{j}+R_{\msout{j}}&\leq&2a_{j} + f_{\msout{j}} + 
e_{\msout{j}}&-R_{j}& \leq &0  \\
\end{array}\mbox{ for } j \in [2]\end{eqnarray}where
\begin{eqnarray}\begin{array}{lcl}
 a_{j}&=&I(Y_{j};V_{j}|U,\ulineW)- \mathcal{L} (\phi,| \mathcal{V}_{j} | )\\
 b_{j}& = &I(Y_{j}; W_{j}| U, V_{j}, W_{\msout{j}}) -\mathcal{L} (\phi, 
| \mathcal{W }_{j } |)\\
 c_{j}& = &I(Y_{j}; W_{\msout{j}}| U, V_{j}, W_{{j}}) -\mathcal{L} 
(\phi, 
| \mathcal{W}_{\msout{j} } |)\\
\end{array}
\begin{array}{lcl}
 d_{j}& = &I(Y_{j}; V_{j},W_{j}| U,  W_{\msout{j}}) -\mathcal{L} 
(\phi, 
| \mathcal{V }_{j } ||\mathcal{W}_{j}|),\\
e_{j}& = &I(Y_{j}; 
V_{j},W_{\msout{j}}| U, W_{{j}}) 
-\mathcal{L} (\phi, 
|\mathcal{V }_{j } ||\mathcal{W}_{\msout{j}}|),\\
 f_{j}& = &I(Y_{j}; W_{j},W_{\msout{j}}| U, V_{j}) 
-\mathcal{L} (\phi, 
| \mathcal{W}_{j }||\mathcal{W}_{\msout{j} } |),\\
 g_{j}& = &I(Y_{j}; V_{j},W_{\msout{j}},W_{j}| U) 
-\mathcal{L} (\phi, 
| \mathcal{W}_{j }||\mathcal{W}_{\msout{j} } ||\mathcal{U}_{j}|)
\end{array}
\end{eqnarray}
for $j \in [2]$. We let
\begin{eqnarray}
 \label{Eqn:CHKRegion}
 \alpha_{CHK}(\mathbb{W}_{\ulineY|\ulineX}) \define \cocl \left( 
\bigcup_{\substack{ p_{\ulineU\ulineV\ulineW\ulineX\ulineY}\\ \in 
\mathbb{D}(\mathbb{W}_{\ulineY|\ulineX}) }} 
\alpha(p_{\ulineU\ulineV\ulineW\ulineX\ulineY}) \right) \nonumber
\end{eqnarray}
where $\mathcal{L}(\mu,|\mathcal{A}|) \define 
h_{b}(\mu)+\mu\log|\mathcal{A}|$ 
for any $\mu \in (0,0.5)$, finite set $\mathcal{A}$, $\cocl(\mathcal{A})$ 
denotes the convex closure of $\mathcal{A} \subseteq \reals^{2}$.
\end{definition}
\begin{thm}
 \label{Thm:ICStep3CHKRegion}
 A pair of sources $(\ulineCalS,\mathbb{W}_{\ulineS})$ is transmissible over an 
IC $(\ulineCalX,\ulineCalY,\mathbb{W}_{\ulineY|\ulineX})$ if there exists 
\begin{enumerate}
 \item[(i)]  finite sets 
$\mathcal{K},\mathcal{U},\mathcal{V}_{1},\mathcal{V}_{2}$,$\mathcal{W}_{1}$, 
$\mathcal{W}_{2}$,
 \item[(ii)] maps $f_{j}:\mathcal{S}_{j}\rightarrow 
\mathcal{K}$, with $K_{j}=f_{j}(S_{j})$ for $j \in [2]$,
 \item[(iii)] $\alpha,\beta \geq 0$, $\rho > 0$, $\delta > 0$, 
 \item[(iv)] $l \in 
\naturals, l \geq \max\{ l^{*}(\rho,\mathcal{U},\mathcal{Y}_{j}): j \in [2] 
\}$, where $l^{*}(\cdot,\cdot,\cdot)$ is defined in (\ref{Eqn:DefnOfLStar}),
\item[(v)] pmf 
$p_{U}p_{V_{1}}p_{W_{1}}p_{V_{2}}p_{W_{2}}p_{X_{1}|UV_{1}W_{1}}p_{X_{2}|UV_{2}
W_{2}} 
\mathbb{W}_{\ulineY|\ulineX}$ defined on $\mathcal{U}\times 
\ulineCalV\times\ulineCalW\times \ulineCalX\times \ulineCalY$, where $p_{U}$ is 
a type of 
sequences in $\mathcal{U}^{l}$, such that for some $a \in [2]$, we have
\end{enumerate}
\ifPeerReviewVersion
 \begin{eqnarray}
(1+\delta)H(K_{a}) \leq \alpha+\beta ,~~
 \label{Eqn:Step2ICIndvdualRateBound}
\left( H(S_{j}|K_{a}) + \beta +
\mathcal{L}_{l}(\phi_{j},|\mathcal{S}_{j}|): j \in [2]\right) \in 
\alpha_{CHK}(\mathbb{W}_{\ulineY|\ulineX})\\
\mbox{where }\phi_{j} \define 
g_{j}(\alpha+\rho,l)+\xi^{[l]}(\ulineK)+\tau_{l,\delta}(K_{a}),
g_{j}(R,l) \define 
(l+1)^{2|\mathcal{U}||\mathcal{Y}_{j}|}\exp\{-lE_{r}(R,p_{U},p_{Y_{j}|U})\}, 
\mbox{ for }
\end{eqnarray}
$j \in [2]$, $\mathcal{L}_{l}(\cdot,\cdot)$ is as defined 
in (\ref{Eqn:AdditionalSourceCodingInfo}).\fi
\ifTITVersion
 \begin{eqnarray}
  (1+\delta)H(K_{1}) \!\!\!\!&\leq&\!\!\!\! \alpha+\beta ,\mbox{ and 
for } j \in [2]\nonumber\\ 
\label{Eqn:Step3ICIndvdualRateBound}
\left(H(S_{j}|K_{1}) + \beta +
\mathcal{L}_{l}(\phi_{j},|\mathcal{S}_{j}|): j \in [2] \right) 
\!\!\!\!&\leq&\!\!\!\!
I(V_{j};Y_{j}|U)-\mathcal{L}(\phi_{j},
|\mathcal {V}_{j} |),
\end{eqnarray}
$\phi_{j} \in [0,0.5)$ where
\begin{eqnarray}\phi_{j} &\define &
\label{Eqn:ICStep1Phij}
g_{\rho,l}^{j}+\xi^{[l]}(\ulineK)+\tau_{l,\delta}(K_{1}),\\
\label{Eqn:ICStep1gRhoLj}
g_{\rho,l}^{j} &\define &
\exp\{-l(E_{r}(\alpha+\rho,p_{U},p_{Y_{j}|U})-\rho)\} 
.\end{eqnarray}\fi
\end{thm}

\subsection{MAC Problem}
\label{SubSec:MACStep2GeneralizationJointDecoding}
We present our second and final coding theorem for the MAC problem, wherein we 
incorporate conditional decoding of the outer code.
\begin{thm}
\label{Thm:MACStep2}
A pair of sources $(\ulineCalS,\mathbb{W}_{\ulineS})$ is transmissible over a 
MAC $(\ulineCalX,\outset,\mathbb{W}_{\Out|\ulineX})$ if there exists 
\begin{enumerate}
 \item[(i)] finite sets 
$\mathcal{K},\mathcal{U},\mathcal{V}_{1},\mathcal{V}_{2}$,
 \item[(ii)] maps $f_{j}:\mathcal{S}_{j}\rightarrow \mathcal{K}$, 
with $K_{j}=f_{j}(S_{j})$ for $j \in [2]$,
\item[(iii)] $\alpha,\beta \geq 0$, $\rho > 0$, $\delta > 0$, 
\item[(iv)] $l \in 
\naturals, l \geq l^{*}(\rho,\mathcal{U},\mathcal{Y})$, where 
$l^{*}(\cdot,\cdot,\cdot)$ is defined in (\ref{Eqn:DefnOfLStar}),
\item[(v)] pmf $p_{U}p_{V_{1}}p_{V_{2}}p_{X_{1}|UV_{1}}p_{X_{2}|UV_{2}} 
\mathbb{W}_{Y|\ulineX}$ defined on $\mathcal{U}\times 
\ulineCalV\times\ulineCalX\times \mathcal{Y}$, where $p_{U}$ is a type of 
sequences in $\mathcal{U}^{l}$, such that for some $a \in [2]$, we have
\end{enumerate}
\ifPeerReviewVersion
 \begin{eqnarray}
 \label{Eqn:TypicalSetSize}
 (1+\delta)H(K_{a}) &<& \alpha+\beta ,\nonumber\\ 
 \label{Eqn:StepIIndvdualRateBound}
H(S_{j}|S_{\msout{j}},K_{a}) + 
\mathcal{L}_{l}(\phi,|\mathcal{S}_{j}|) &<& 
I(V_{j};\Out|U,V_{\msout{j}})-\mathcal{L}(\phi,|\mathcal{V}_{j}|)
\mbox{ for }j \in [2]\mbox{ and} \\
\label{Eqn:StepISumRateBound}
\beta + H(\ulineS|K_{a})+ 
\mathcal{L}_{l}(\phi,|\ulineCalS|) &<& I(\ulineV;Y|U) - 
\mathcal{L}(\phi,|\ulineCalV|),\\
\lefteqn{
\!\!\!\!\!\!\!\!\!\!\!\!\!\!\!\!\!\!\!\!\!\!\!\!\!\!\!\!\!\!\!\!\!\!\!\!\!\!\!\!
\!\!\!\!\!\!\!\!\!\!\!\!\!\!\!\!\!\!\!\!\!\!\!\!\!\!\!\!\!\!\!\!\!\!\!\!\!\!\!\!
\!\!\!\!\!\!\!\!\!\!\!\!\!\!\!\!\!\!\!\!\!\!\!\!\!\!\!\!\!\phi \in 
[0,0.5)\mbox{ where }\phi \define 
g(\alpha+\rho,l)+\xi^{[l]}(\ulineK)+\tau_{l,\delta}(K_{a}),~~ g(R,l) 
\define 
(l+1)^{2|\mathcal{U}||\mathcal{Y}|}\exp\{-lE_{r}(R,p_{U},p_{Y|U})\}}
\end{eqnarray}
$\mathcal{L}_{l}(\cdot,\cdot), \mathcal{L}(\cdot,\cdot)$ is as defined 
in (\ref{Eqn:AdditionalSourceCodingInfo}).\fi
\ifTITVersion
 \begin{eqnarray}
 \label{Eqn:TypicalSetSize}
 (1+\delta)H(K_{1}) \!\!\!\!&\leq&\!\!\!\! \alpha+\beta ,\nonumber\\ 
\label{Eqn:StepIIndvdualRateBound}
H(S_{j}|S_{\msout{j}},K_{1}) + 
\mathcal{L}_{l}(\phi,|\mathcal{S}_{j}|) \!\!\!\!&<&\!\!\!\!
I(V_{j};\Out|V_{\msout{j}})-\mathcal{L}(\phi,
|\mathcal{V}_{j} |)\nonumber\\&&\!\!\!\!~~~
\mbox{ for }j \in [2]\mbox{ and} \\
\label{Eqn:StepISumRateBound}
\beta + H(\ulineS|K_{1})+ 
\mathcal{L}_{l}(\phi,|\ulineCalS|) \!\!\!\!&<&\!\!\!\! I(\ulineV;Y) - 
\mathcal{L}(\phi,|\ulineCalV|),
\end{eqnarray}
$\phi \in [0,0.5)$ where
\begin{eqnarray}\phi &\define &
g_{\rho,l}+\xi^{[l]}(\ulineK)+\tau_{l,\delta}(K_{1}),\nonumber\\
g_{\rho,l} &\define &
\exp\{-l(E_{r}(\alpha+\rho,p_{U},p_{Y|U})-\rho)\}\nonumber .\end{eqnarray}\fi
\end{thm}
\begin{remark}
\label{Rem:MACStep2SubsumesSWMAC}
If the sources have a GKW part $K=K_{1}=K_{2}$, then $\xi(\ulineK)=0$. One can 
choose $l$ arbitrarily large such that $\phi$ can be made 
arbitrarily small. The resulting inner bound corresponds to separation based 
scheme involving a common message communicated over the MAC. 
\end{remark}
There are no new elements beyond those presented in proofs of Theorems 
\ref{Thm:MACStep1}, \ref{Thm:ICStep2}. The reader is referred to 
\cite{201601arXiv_Pad} wherein the key error events have been analyzed from 
first principles. The analysis provided therein is similar to that adopted in 
\ref{AppSec:PE2Event1FromFirstPrin} and has a different flavor from the ones 
provided in Section \ref{SubSec:MACStep1GeneralizationSeparateDecoding}, 
\ref{SubSec:ICStep2GeneralizationJointDecoding} for Theorems \ref{Thm:MACStep1}, 
\ref{Thm:ICStep2} respectively.
\section{Robust Distributed Source Coding}
\label{Sec:RobustDSC}
\subsection{Introduction}
\label{Sec:Introduction}
In the multiple description (MD) scenario, a centralized encoder communicates 
multiple descriptions of the observed source to guard against link failures. 
Each subset of descriptions must enable the decoder reconstruct 
the source within specified fidelity. On the other hand, the classical 
distributed source coding (CDSC) problem models a distributed encoder setup, 
wherein 
each encoder observes one component of a joint source and communicates a 
message 
to the decoder. With all messages at its disposal, the decoder is 
required to reconstruct the sources, or functions thereof, within specified 
fidelity. While the MD problem emphasizes link failures, the CDSC problem 
focuses on (distributed) compression efficiency.

\begin{figure}\centering
\includegraphics[width=3.3in]{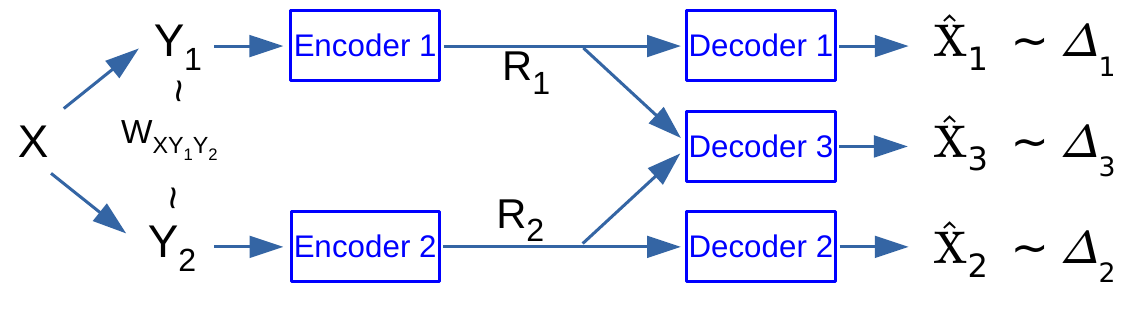}
\caption{Robust Distributed Source Coding Scenario.}
\label{Fig:GeneralRDSCProblem}\vspace{-0.15in}
\end{figure}

Bringing in both features, Chen and Berger \cite{200808TIT_CheBer-Shrt} studied 
the 
robust DSC (RDSC) problem (Fig.~\ref{Fig:GeneralRDSCProblem}). Encoder $j$ 
observes 
component $Y_{j}$ of a triple source $(X,Y_{1},Y_{2})$. The encoders are 
distributed and communicate a message based on their observations. Each 
subset of messages must enable reconstruction of $X$ within a 
specified distortion. In this article, we undertake a Shannon-theoretic study 
of 
the RDSC problem and focus on characterizing inner bounds to the 
rate-distortion (RD) region.

Combining the Zhang-Berger \cite{198707TIT_ZhaBer-Shrt} and 
quantize-and-bin\footnote{Following \cite{201107TIT_WagKelAlt-Shrt}, we refer 
to 
the classical Berger-Tung coding scheme without a common codebook as 
quantize-and-bin coding scheme.} (QB) \cite{Berger-MSC} coding schemes, 
\cite{200808TIT_CheBer-Shrt} has characterized the CB region - the current 
known largest inner bound to the rate-distortion (RD) region for the RDSC 
problem. We derive a new inner bound that subsumes the CB region and strictly 
enlarges the same for identified examples. These findings build on a series of 
works \cite{201307ISIT_ChaSahPra-Shrt, 201406ISIT_ShiPra-2-Shrt, 
201706ISIT_Pad-MAC-Shrt, 
201711arXiv_Pad} that have put forth a new coding scheme for distributed 
information processing.

In 2012, Wagner, Kelly and Altug \cite{201107TIT_WagKelAlt-Shrt} proved, 
via a novel continuity argument, that the QB coding scheme is strictly 
sub-optimal for the CDSC 
problem. Recognizing the QB scheme is inefficient in exploiting the presence of 
highly correlated components, henceforth referred to as near GKW parts, Shirani 
and Pradhan \cite{201406ISIT_ShiPra-2-Shrt} devised a new coding scheme based 
on 
fixed block-length (B-L) quantizers and derived a new inner bound to RD region 
of the CDSC problem. Spurred by their findings \cite{201406ISIT_ShiPra-2-Shrt}, 
we recognized the connection to Dueck's classical work 
\cite{198103TIT_Due-Shrt} 
and devised a \textit{fixed B-L (fBL)} coding scheme 
\cite{201706ISIT_Pad-MAC-Shrt, 201706ISIT_Pad-IC-Shrt, 201711arXiv_Pad} for 
joint source-channel coding over MAC and 
IC that is proven to strictly outperform the previous known best for both 
problems. Analogous to \cite{201406ISIT_ShiPra-2-Shrt}, the fBL coding scheme 
is 
specifically designed to exploit the presence of near GKW parts in distributed 
sources. The fBL scheme devised for joint source-channel coding involves 
certain modifications/improvements over and above those of 
\cite{201406ISIT_ShiPra-2-Shrt}. In this article, we incorporate these to 
design 
a fBL coding scheme for RDSC. The inner bound we derive here 
naturally applies to the CDSC problem, and as 
Rem.~\ref{Rem:ImprovementOverShiraniPradhan} indicates, the above mentioned 
ideas could lead to an improvement of the bound in 
\cite{201406ISIT_ShiPra-2-Shrt}. In the context of the RDSC problem, our 
findings provide a new inner bound to the corresponding RD region that is 
proven 
to strictly enlarge the CB region (Rem.~\ref{Rem:StrictlyBetter}). This work 
specifically answers the questions posed in 
\cite[Rem.~3, Pg 3388]{200808TIT_CheBer-Shrt}.

In the light of \cite{201406ISIT_ShiPra-2-Shrt}, we do not claim our results as 
novel. However, the fBL scheme is a fundamentally new approach. We believe that 
the tools and techniques are being 
crystallized and the bounds improved. We therefore view this 
work, in addition to the above specific contributions, as adding another 
perspective to this new evolving coding scheme.
\subsection{Notation and Problem Statement}

We supplement standard information theory notation with the following. An 
\underline{underline} denotes an appropriate aggregation of related objects. 
For ex., $\ulineCalS$ denotes Cartesian product $\mathcal{S}_{1}\times 
\mathcal{S}_{2}$ of sets and $\ulineS$ denotes the pair $(S_{1},S_{2})$ of 
random variables (RVs). For $m\in \naturals$, 
$[m]\define \{1,\cdots,m\}$. The existence of good fixed B-L covering codes are 
used. We 
will need finite-length quantizer codes that can quantize a source in 
multi-resolution fashion. Consider the Zhang-Berger 
\cite{198707TIT_ZhaBer-Shrt} scheme with a base layer quantizer and 
superposition codes. We will exploit such codes of fixed B-L as below.
\begin{prop}
 \label{Prop:ExistenceOfGoodSourceCodes}
Given finite sets $\mathcal{K},\mathcal{S},\mathcal{V}$, a 
pmf $p_{KSV}$, $l \in \naturals$, there exists (i) an 
$\epsilon_{l}=\epsilon_{l}(p_{KSV}) > 0$, (ii) 
a codebook 
$C_{S}=(s^{l}(m_{s}):m_{s} \in [\mathscr{M}_{s}]) \subseteq \mathcal{S}^{l}$ 
with $\mathscr{M}_{s} \leq \exp\{l(I(K;S)+\epsilon_{l})\}$ codewords and a 
collection $C_{V}(m_{s}) = (v^{l}(m_{s},m_{v}):m_{v} \in [\mathscr{M}_{v}]) 
\subseteq \mathcal{V}^{l} : m_{s} \in [\mathscr{M}_{s}]$ of superposition 
codes, with $\mathscr{M}_{v} \leq 
\exp\{l(I(K;V|S)+\epsilon_{l})\}$, (iii) (codeword index) maps 
$\iota^{l}_{S}:\mathcal{K}^{l}\rightarrow [\mathscr{M}_{s}]$, $\iota_{V}^{l}: 
\mathcal{K}^{l} \rightarrow [\mathscr{M}_{v}]$ and corresponding (codeword) 
maps $\kappa^{l}_{S}:\mathcal{K}^{l} \rightarrow C_{S}$, 
$\kappa^{l}_{V}:\mathcal{K}^{l} \rightarrow C_{V}$, where 
$\kappa^{l}_{S}(k^{l}) 
= s^{l}(\iota_{S}^{l}(k^{l}))$ and $\kappa_{V}^{l}(k^{l}) = 
v^{l}(\iota_{S}^{l}(k^{l}),\iota_{V}^{l}(k^{l}))$ are the codewords indexed by 
the $\iota_{S},\iota_{V}-$maps, such that if 
$\mathscr{P}_{K^{l}S^{l}V^{l}}(k^{l},s^{l},v^{l}) = 
[\prod_{i=1}^{l}p_{K}(k_{i})]\mathds{1}_{\{s^{l} = \kappa^{l}_{S}(k^{l}), v^{l} 
= \kappa^{l}_{V}(k^{l})\}}$ is the pmf induced on $\mathcal{V}^{l} \times 
\mathcal{S}^{l} \times \mathcal{V}^{l}$ by the codes $C_{S},C_{V}$, then
\begin{eqnarray}
\label{Eqn:SourceCodeIsgood}
 \left| \frac{1}{l} \sum_{i=1}^{l} \mathscr{P}_{K_{i}S_{i}V_{i}} (a,b,c) 
-p_{KSV}(a,b, c)\right| \leq \epsilon_{l}
\end{eqnarray}
$\forall (a,b,c) \in \mathcal{K} 
\times \mathcal{S} \times \mathcal{V} $ and $\epsilon_{l}(p_{KSV}) \rightarrow 
0$ as $l \rightarrow \infty$.
\end{prop}
Boldfaced calligraphic letters such as $\boldsymbol{\mathcal{A}}\define 
\mathcal{A}^{m \times l}$ denote the set of all $m \times l$ 
matrices over $\mathcal{A}$. Boldfaced letters 
such as $\bold{a},\bold{A}$ denote 
matrices. For a $\mtimesl$ matrix $\bold{a}$, (i) $\bold{a}(t,i)$ denotes the 
entry in row $t$, column $i$, (ii) $\bold{a}(1:m,i)$ denotes the $i^{th}$ 
column, $\bold{a}(t,1:l)$ denotes $t^{th}$ row. ``with high 
probability'', ``single-letter'', ``long Markov 
chain'', ``block-length'' are abbreviated whp, S-L, LMC, B-L respectively. For 
$K_{1}^{l},K_{2}^{l} 
\in \mathcal{K}^{l}$, we let $\xi^{[l]}(\underline{K}) \define 
P(K_{1}^{l}\neq K_{2}^{l})$, and  $\xi(\underline{K}) \define 
\xi^{[1]}(\underline{K})$. If $(K_{1t},K_{2t}): t \in [l]$, are IID, 
then\footnote{$(1-x)^{l} \geq 1-xl\mbox{ for }x \in 
[0,1]$.} $\xi^{[l]}(\ulineK)=1-(1-\xi(\ulineK))^{l} \leq l\xi(\ulineK)$.

Consider the RDSC scenario in Fig.~\ref{Fig:GeneralRDSCProblem}. Let 
$(X,Y_{1},Y_{2})$ taking values in 
$\mathcal{X}\times \mathcal{Y}_{1}\times 
\mathcal{Y}_{2}$ with pmf $\mathbb{W}_{XY_{1}Y_{2}}$ represent a triple of 
sources. For $j \in [2]$, encoder $j$ observes $Y_{j}$. Let $\mathcal{Z}$ 
be the reconstruction alphabet and $\distfn:\mathcal{X}\times \mathcal{Z} 
\rightarrow [0,\Delta_{\max})$ be a distortion measure. We say $( R_{1}, R_{2}, 
\Delta_{1}, \Delta_{2}, \Delta_{3})$ is an achievable RD vector if 
for every $n \in \naturals$ sufficiently large, there exists (i) encoder maps 
$e_{j}^{(n)}: 
\mathcal{Y}_{j}^{n} \rightarrow [M_{j}^{(n)}] : j \in [2]$, decoders 
$d_{j}^{(n)}\!:\![M_{j}^{(n)}]\! \rightarrow \!\mathcal{Z}^{n} : j \in [2]$, 
$d_{3}^{(n)}:[M_{1}^{(n)}]\times [M_{2}^{(n)}] \rightarrow \mathcal{Z}^{n}$ 
with $Z_{j}^{n}=d_{j}^{(n)}(e_{j}^{(n)}(Y_{j}^{n})): j \in [2]$ and 
$Z_{3}^{n} = d_{3}^{(n)}(e_{1}^{(n)}(Y_{1}^{n}),e_{2}^{(n)}(Y_{2}^{n}))$ such 
that $\displaystyle \lim_{n \rightarrow \infty} \frac{1}{n}\Expectation\{ 
\distfn^{n}(X^{n},Z_{j}^{n}) \} \!\leq \!\Delta_{j} : \!j\! \in [3]$ and 
$\displaystyle 
\lim_{n \rightarrow \infty} \frac{1}{n}\log M_{j}(n) \leq R_{j}$. 
$\mathscr{Q}(\mathbb{W}_{X\ulineY},\distfn)$ denotes the set of achievable RD 
vectors.

\subsection{Enhancing the Chen-Berger Scheme: Step I}
\label{Step1:FBLScheme}
Our first step is to identify \textit{how} the Chen-Berger coding scheme can be 
enhanced. Towards that end, we revisit the latter scheme 
(Sec.~\ref{SubSec:ChenBergerScheme}) and 
state its sub-optimality (Sec.~\ref{SubSec:SubOptimalityOfChenBerger}) through 
an example. Remarks at the end 
of Sec.~\ref{SubSec:SubOptimalityOfChenBerger} provides the ideas for the new 
coding scheme (Sec.~\ref{SubSec:FixedBLWith2Codebooks}).
\begin{comment}{
In this section, we present a new coding scheme that exploits the presence of 
highly correlated components (near GKW parts) via a fixed B-L common code. We 
begin with the characterization of the corresponding sufficient conditions.
}\end{comment}
\subsubsection{The Chen-Berger Coding Scheme}
\label{SubSec:ChenBergerScheme}
In the QB coding scheme, the decoder is unable to obtain any reconstruction 
with any individual message stream. Hence the CB scheme incorporates an 
additional codebook at each encoder to permit a `stand alone' reconstruction. 
We follow the notation in \cite[Thm.~1]{200808TIT_CheBer-Shrt} and let $U_{j}$ 
depict this additional codebook (that is not partitioned into bins). $W_{j}$ 
corresponds to the code in the QB scheme that is partitioned into 
bins. \cite[Thm.~1]{200808TIT_CheBer-Shrt} provides a characterization of 
the CB region.

\begin{remark}
 \label{Rem:ChenBergerWithoutGKWPart} If $U_{j},W_{j}$ 
denote the quantizations identified by encoder $j$, then the CB scheme is 
constrained to a S-L LMC $U_{1}W_{1}-Y_{1}-Y_{2}-U_{2}W_{2}$. Secondly, if 
$\Delta_{1}=\Delta_{2}=\Delta_{\max}$ implying that decoders 1 and 2 are 
redundant, then the CB coding scheme reduces to the QB coding scheme.
\end{remark}

\begin{figure}\centering
\includegraphics[width=3.5in]{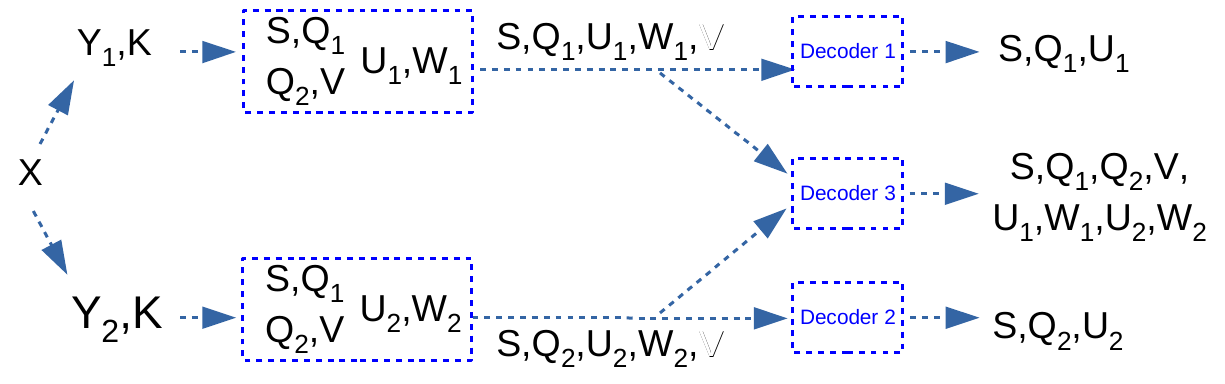}
\caption{Chen-Berger coding scheme with GKW part.}
\label{Fig:RDSCFullConvCoding}\vspace{-0.15in}
\end{figure}

The presence of a GKW part $K=f_{1}(Y_{1})=f_{2}(Y_{2})$ permits a layer of 
\textit{GKW coding}, wherein both encoders share common codebooks to quantize
$K$. In fact, $K$ plays the same role as the source in the centralized MD 
problem \cite{198707TIT_ZhaBer-Shrt}. The CB scheme 
\cite[Thm.~3]{200808TIT_CheBer-Shrt} 
therefore builds $4$ codebooks $S,Q_{1},Q_{2},V$ to code $K$ and these 
are shared by both encoders. Fig.~\ref{Fig:RDSCFullConvCoding} depicts CB 
scheme with GKW coding. At times, we refer to the CB coding scheme with 
common codes as the CBwCC coding scheme.

\begin{remark}
 \label{Rem:ChenBergerWithCommonPart}
GKW coding enable encoders agree on the chosen 
$S-,Q_{1}-,Q_{2}-,V-$codewords. The quantizations $U_{j},W_{j}: j \in 
[2]$ are therefore \underline{not} constrained to a S-L LMC 
$U_{1}W_{1}-Y_{1}-Y_{2}-U_{2}W_{2}$. This enlargement 
in the induced correlation amidst quantizations is strictly more efficient 
(Sec.~\ref{SubSec:SubOptimalityOfChenBerger}). We also note that if 
$\Delta_{1}=\Delta_{2}=\Delta_{\max}$, the CBwCC scheme reduces to the QB 
scheme with one common codebook \cite{201107TIT_WagKelAlt-Shrt}, termed the 
QBwCC scheme.
\end{remark}

}\end{comment}
\subsubsection{Sub-optimality of the Chen-Berger scheme}
\label{SubSec:SubOptimalityOfChenBerger}
If the QB coding scheme is sub-optimal for the CDSC problem, then 
Rem.~\ref{Rem:ChenBergerWithoutGKWPart} implies sub-optimality of the CB scheme 
for the RDSC problem. In the following, we put forth the novel arguments of 
\cite{201107TIT_WagKelAlt-Shrt} that prove the former statement. More 
importantly, this discussion provides us with a roadmap for enhancing the CB 
scheme. We now discuss findings in \cite{201107TIT_WagKelAlt-Shrt}.

\begin{example}
 \label{Ex:WagnerExample}
 Let $\mathcal{B}=\{0,1\}$, $\mathcal{X} =\mathcal{Y}_{2} = \mathcal{B}$ and 
$\mathcal{Y}_{1}=\mathcal{B}\times \mathcal{B}$. Let $A,B,N_{1},N_{2}$ be 
independent Bernoulli RVs with $N_{1} \sim \Bernoulli(\epsilon)$, $N_{2} \sim 
\Bernoulli(\epsilon)$, $A\sim\Bernoulli(\frac{1}{2})$ and $B\sim\Bernoulli(p)$ 
with $\epsilon \in [0,\frac{1}{2}), p \in (0,\frac{1}{2})$. Let $X=A\oplus B$, 
$Y_{1}=(A\oplus N_{1},A\oplus 
B)$ and $Y_{2} = A\oplus N_{2}$. Reconstruction alphabet $\mathcal{Z} = 
\mathcal{B}$, distortion function $\distfn(x,z)=\mathds{1}_{\{ x\neq h \}}$ be 
the usual binary Hamming function and $\Delta_{1}=\Delta_{2}=1$ and $\Delta_{3} 
= \Delta \in (0,p*\epsilon)$.
\end{example}

In essence, encoder $1$ observes $A\oplus N_{1}, A \oplus B$ and encoder $2$ 
observes $A\oplus N_{2}$. Decoder $3$ needs to reconstruct $A\oplus B$ within 
an avg. Hamming distortion $\Delta$. Decoders $1,2$ are absent. 

$\epsilon=0$ implies the presence of GKW part $A$. Let us describe the QBwCC 
scheme in this case. $\delta \in (0,\frac{1}{2})$ is chosen 
and a common quantizer $C_{S}$ of rate $1-h_{b}(\delta)$ that can quantize $A$ 
within an avg. Hamming distortion $\delta$ is employed at both encoders. Let 
$S^{n}(a^{n}) \in C_{S}$ denote the quantization of $a^{n} \in \mathcal{B}^{n}$ 
and $Q^{n}(a^{n}) = S^{n}(a^{n}) \oplus a^{n}$ denote the quantization noise 
that both encoders 
identify. Encoder $2$ communicates $S^{n}(A^{n})$ using a rate 
$1-h_{b}(\delta)$. Encoder $1$ quantizes $U^{n} \define S^{n}(A^{n})\oplus (
A^{n} \oplus B^{n}) = Q^{n}(A^{n}) \oplus B^{n}$ - an IID Ber$(p*\delta)$ 
sequence - to within an avg. Hamming distortion $\Delta$ and communicates the 
resulting quantization $V^{n}(U^{n})$ to the decoder using a rate 
$h_{b}(p*\delta)-h_{b}(\Delta)$. Note that $S^{n}(A^{n}) 
\oplus V^{n}(U^{n}) = S^{n}(A^{n}) \oplus U^{n} \oplus 
[U^{n}\oplus V^{n}(U^{n})] = A^{n}\oplus B^{n} \oplus [U^{n}\oplus 
V^{n}(U^{n})]$ and the term in square braces is IID Ber$(\Delta)$ and hence the 
decoders' reconstruction $S^{n}(A^{n}) \oplus V^{n}(U^{n})$ meets the 
distortion constraint. 

While the above analysis provides achievability, 
\cite{201107TIT_WagKelAlt-Shrt} 
goes onto prove optimality of the above coding scheme.

\noindent\textit{Fact 1:} $(R_{1},R_{2},1,1,\Delta)$ with $\Delta < 
\frac{1}{2}$ 
is achievable for Ex.~\ref{Ex:WagnerExample} with $\epsilon=0$ if and only if 
there exists a $\delta \in (0,\frac{1}{2})$ for which $ R_{1} \geq 
[h_{b}(\delta*p)-h_{b}(\Delta)]^{+}$, $R_{2} \geq 1-h_{b}(\delta)$. Moreover, 
the (true) RD region for Ex.~\ref{Ex:WagnerExample} is continuous in 
$\epsilon$ at $\epsilon=0$.

Going further, \cite{201107TIT_WagKelAlt-Shrt} proves the following facts. 

\noindent\textit{Fact 2:} For any $\delta \in (0,\frac{1}{2})$, the RD vector 
$(h_{b}(\delta*p)-h_{b}(\Delta), 1-h_{b}(\delta),1,1,\Delta)$ is not achievable 
by the QB scheme for Ex.~\ref{Ex:WagnerExample} with $\epsilon=0$. Moreover, 
the 
QB achievable RD region for Ex.~\ref{Ex:WagnerExample} is monotonically 
increasing with decreasing $\epsilon$.

The strict sub-optimality of the QB scheme for the case $\epsilon=0$, the 
continuity of the true RD region at $\epsilon=0$, and the fact that the QB 
achievable region is monotonically shrinking with increasing $\epsilon >0$ 
imply the strict sub-optimality of the QB scheme for 
sufficiently small values of $\epsilon>0$. Since the CB scheme reduces 
to the QB scheme for Ex.~\ref{Ex:WagnerExample}, we have thus verified the 
sub-optimality of the latter scheme.
\begin{comment}{
Why is the QB scheme (strictly) sub-optimal in the presence of a GKW part as 
evidenced in Ex.~\ref{Ex:WagnerExample} with $\epsilon=0$? Consider a general 
CDSC problem with a GKW part $K=f_{1}(Y_{1})=f_{2}(Y_{2})$. In the QB scheme, 
encoder $j$ quantizes $Y_{j}$ to $W_{j}$ and the quantizations are constrained 
to a S-L long Markov chain (LMC) $W_{1}-Y_{1}-Y_{2}-W_{2}$. However, in the 
QBwCC scheme, $K$ is quantized using a common codebook to $S$ which both 
encoders agree on, referred to as GKW coding. $S$ is employed in choosing 
quantization $W_{j}$ of $Y_{j}$ resulting in Markov chain 
$W_{1}-SY_{1}-SY_{2}-W_{2}$. In other words, the correlation induced on the 
quantizations is \textit{not} constrained to an S-L LMC 
$W_{1}-Y_{1}-Y_{2}-W_{2}$, thereby enabling more efficient communication.

% The fixed B-L coding scheme we propose is aimed at coding highly correlated 
% components (near GKW parts) $K_{1}=f_{1}(S_{1})$, $K_{2}=f_{2}(S_{2})$ using 
% common codebooks.
}\end{comment}
\subsubsection{Fixed B-L Coding scheme with $2$ common codes}
\label{SubSec:FixedBLWith2Codebooks}
The sub-optimality of the QB scheme for small values of $\epsilon >0$ in 
Ex.~\ref{Ex:WagnerExample} and the strict enlargement of the RD region obtained 
by the QBwCC coding scheme for $\epsilon=0$ indicates that an 
efficient coding scheme has to be able to exploit the presence of highly 
correlated components (near GKW parts) analogous to the GKW coding technique 
and 
thereby induce enhanced correlation. The codewords chosen by the encoders for 
quantizing $K^{n}$ being identical is central to this enhanced correlation and 
efficiency of the GKW coding technique. This is brought about by the choice of 
common codebooks for quantizing $K^{n}$. The coding scheme we propose builds on 
this premise and crucially modifies the GKW coding layer to leverage the above 
mentioned efficiency even in the absence of a GKW part. To convey the ideas, we 
begin with a simplified coding scheme that lets us explain all the new elements.
\begin{thm}
$(R_{1},R_{2},\Delta_{1},\Delta_{2},\Delta_{3}) \in 
\mathscr{Q}(\mathbb{W}_{X\ulineY},\distfn)$ if there exists (i) $l \in 
\naturals$, (ii) sets 
$\mathcal{K},\mathcal{S},\mathcal{V},\mathcal{U}_{j},\mathcal{W}_{j}: j \in 
[2]$, (iii) maps $f_{j}:\!\!\!\mathcal{Y}_{j}\!\!\!\rightarrow\!\!\! 
\mathcal{K}\!$ , 
reconstruction maps $h_{j}:\mathcal{U}_{j}\!\! \rightarrow\!\! \mathcal{Z}$, 
$g:\mathcal{U}_{1}\!\!\times \!\!\mathcal{W}_{1}\!\!\times\! 
\!\mathcal{U}_{2}\!\!\times\!\! \mathcal{W}_{2} \rightarrow \mathcal{Z}$, (iv) 
pmf $\mathbb{W}_{X\ulineY}\mathds{1}_{\{ K_{1}=f_{1}(Y_{1}) ,K_{2}=f_{2}(Y_{2}) 
\}}p_{SV|K_{1}}  \prod_{j=1}^{2}p_{U_{j}|SY_{j}}  p_{W_{j} | 
U_{j}SVY_{j}}$ defined on 
$\mathcal{A}\define 
\mathcal{X}\times\ulineCalY\times\mathcal{K}\times\mathcal{S} \times\mathcal { 
V 
} \times\ulineCalU\times \ulineCalW$ , such that, for $j \in [2]$
\begin{eqnarray}
\lefteqn{\!R_{j} \geq 
I(S;K_{1})+I(U_{j};Y_{j}|S)+I(W_{j};V Y_{j}|U_{j},S)+E_{l}}\nonumber\\
 &&\!\!\!\!\!\!\!\!\!\!-I(W_{j};W_{\msout{j}},U_{\msout{j}}V|U_{j}S), 
~\Delta_{j} \geq \Expectation\{ \distfn(X,h_{j}(U_{j})) \} + 
\Delta_{\max}\phi_{l}|\mathcal{A}| \nonumber\\
\lefteqn{\! R_{1}+R_{2} \geq 
I(S;K_{1})+I(SV;K_{1})-I(W_{1};W_{2}|U_{1}U_{2}SV) +2E_{l}}\nonumber\\
&&\!\!\!\!\!\!\!\!\!\!\!\!\! + 
\sum_{j=1}^{2}[I(U_{j};Y_{j}|S)+I(W_{j};V,Y_{j}|U_{j}S)  
-I(W_{j};U_{\msout{j}},V|U_{j},S)],~  \nonumber
\end{eqnarray}
$\Delta_{3} \geq 
\Expectation\{ \distfn(X,g(\ulineU,\ulineW)) 
\} + \Delta_{\max}\phi_{l}|\mathcal{A}| $, where $\phi_{l} = 
\xi^{[l]}(\ulineK)+\epsilon_{l}(p_{K_{1}SV})$ with $\epsilon_{l}(p_{K_{1}SV})$ 
as given 
in Prop.~\ref{Prop:ExistenceOfGoodSourceCodes}, $E_{l} = 
\epsilon_{l}(p_{K_{1}SV})+2h_{b}(\phi_{l})+\phi_{l}\log|\mathcal{A}
|+2|\mathcal{A}|\phi_{l}\log\left(\frac{1}{\phi_{l}}\right)$.
\end{thm}
\begin{remark}
\label{Rem:ThmCollapsesWithGKWPart}
If $\xi(\ulineK) = 0$, i.e., $K\!\! \define\! K_{1}\!=\!K_{2}$, then 
$\xi^{[l]}(\ulineK) 
= 0$ for $l \in \naturals$. Since $\epsilon_{l},\phi_{l} {\rightarrow} 
0$ as $l \rightarrow \infty$ choose very large $l$. The 
above bounds reduce to those in \cite[Thm.~3]{200808TIT_CheBer-Shrt} with 
$S=Z_{0}, 
V=Z_{3}$ and $Z_{1}=Z_{2}=\phi$. For this case, the CB scheme 
(Fig.~\ref{Fig:RDSCFullConvCoding}) with two codebooks $S,V$ achieve the 
stated bound.
\end{remark}
\begin{remark}
\label{Rem:StrictlyBetter}
If $\mathcal{S}=\mathcal{U}_{j}=\phi$, we obtain an inner bound for the RD 
region of the CDSC problem. If $\mathcal{K}=\phi$, we obtain inner bound 
achievable using the QB coding scheme, and if $\xi(\ulineK) = 0$, i.e., $K 
\define K_{1}=K_{2}$, by choosing $l$ large, we can recover the inner bound 
obtained from the QBwCC coding scheme. Finally, the above inner bound is 
continuous in $\mathbb{W}_{XY_{1}Y_{2}}$ and hence it strictly outperforms the 
QB coding scheme for Ex.~\ref{Ex:WagnerExample} with $\epsilon>0$ sufficiently 
small. Since the CB scheme reduces to the QB scheme for 
Ex.~\ref{Ex:WagnerExample}, the above inner bound strictly enlarges that 
achievable by the CB coding scheme.
\end{remark}
\begin{remark}
\label{Rem:ImprovementOverShiraniPradhan}
As the reader will note, our coding scheme differs from 
\cite{201406ISIT_ShiPra-2-Shrt}. In \cite{201406ISIT_ShiPra-2-Shrt}, both 
encoders communicate the quantizations of the near GKW parts $K_{1},K_{2}$ and 
they employ Slepian-Wolf binning to communicate the same. In the 
QBwCC coding scheme, communication of the quantized codewords of $K$ is shared 
by both encoders. We take the latter approach, and inspite of the two encoders 
disagreeing in their quantization of the near GKW part, they share their 
transmissions. As the reader will note from the bounds, we do not employ 
conditional coding for $W_{j}$ to permit for this disagreement between the 
terminals. This idea stems from our work in \cite{201711arXiv_Pad}, wherein it 
is impossible for the terminals to agree on the information communicated 
through the fixed B-L codes. Performance characterization of the fBL scheme, 
particularly the fixed B-L GKW layer involves multiple approximations (loss due 
to loose bounds). In this regard, we believe the proposed coding scheme can be 
beneficial over \cite{201406ISIT_ShiPra-2-Shrt}.
\end{remark}
\begin{remark}
 \label{Rem:AFirstCutRateRegion}
For simplicity of description, we have employed simple bounds - $E_{l}, 
2|\mathcal{A}|\phi_{l}\log(\phi_{l}^{-1})$.
\end{remark}
\begin{IEEEproof}
As mentioned earlier, we design a \textit{fBL coding scheme} wherein near GKW 
parts\footnote{fBL scheme and the stated inner bound are applicable for any 
$\xi(\ulineK) < \frac{1}{2}$. However, for the sake of intuition consider 
$\xi(\ulineK)$ is very small.} $K_{1},K_{2}$ are 
quantized with a common code. Rem.~\ref{Rem:ThmCollapsesWithGKWPart} indicates 
that we need only two common codes - $S,V$ in 
Fig.~\ref{Fig:RDSCFullConvCoding}.

An alternate interpretation of CBwCC coding scheme will enable us explain the 
fBL coding scheme and its analysis. Essentially, GKW coding can be viewed as a 
technique to enable all terminals agree on a common (correlated) side 
information which can form the basis for higher level communication. Let us 
consider the CBwCC scheme with $2$ GKW codebooks $S$ and $V$ as in 
Rem.~\ref{Rem:ThmCollapsesWithGKWPart} and understand this interpretation.

The CBwCC scheme builds $C_{S} = (s^{n}(m_{s}):m_{s} 
\in [\mathscr{M}_{s}])\subseteq \mathcal{S}^{n}$ by picking each 
codeword IID $p_{S}$. For each chosen codeword $s^{n}(m_{s})$, a codebook 
$C_{V}(m_{s}) = (v^{n}(m_{s},m_{v}) : m_{v} \in [\mathscr{M}_{v}]) \subseteq 
\mathcal{V}^{n}$ is built by picking each codeword IID $\prod 
p_{V|S}^{n}(\cdot|s^{n}(m_{s}))$. Both encoders employ a common map 
$\kappa^{n}_{SV}:\mathcal{K}^{n} \rightarrow C_{S} \times C_{V}$ such that, if 
$(S^{n}(K^{n}),V^{n}(K^{n})) = \kappa_{SV}^{n}(K^{n})$, then 
$(K^{n},S^{n}(K^{n}),V^{n}(K^{n}))$ is jointly typical wrt $p_{KSV}$. The index 
of $S^{n}(K^{n})$ is communicated by both encoders and the index of 
$V^{n}(K^{n})$ is communicated by only one of the encoders. This ensures that 
(i) all terminals share $S^{n}(K^{n})$, and in addition (ii) the encoders and 
decoder $3$ share $V^{n}(K^{n})$. These vectors are (i) IID and (ii) are 
correlated to the observed source wrt the chosen test channel pmf. The rest of 
the coding scheme is designed for an RDSC problem, wherein (i) encoders observe 
$Y_{1},Y_{2}$ stripped of the common part, $S$ and $K$, (i) Decoders $1,2$ 
observe $S$, and (iii) Decoder $3$ observes $S,V$. The GKW coding layer has 
therefore succeeded in communicating common correlated information that can 
facilitate further communication.

The fBL coding scheme is designed to exploit the presence of near GKW parts and 
communicate analogous common correlated information that (i) all terminals can 
agree upon, and (ii) can facilitate further communication. As we shall see, 
$\xi(\ulineK) > 0$ limits our ability to communicate information that all 
terminals can agree upon. Secondly, we are also unable to precisely 
characterize pmf of this correlated information. In the sequel, 
we refer back to this discussion.\begin{comment}{ We develop new 
approaches to 

As we describe the coding 
scheme
We can only ensure that the 
terminals share highly correlated information.  Yet, as we demonstrate, we
exploit this high correlation in the higher layers of communication. 
Secondly, though we are unable precisely characterize pmf of this `shared 
information', we are able to bound its variation from the chosen 
test channel and thereby characterize an achievable RD region.

However, we we shall 
see $\xi(\ulineK) > 0$, we will note that we are unable to communicate 
information that all terminals can agree upon. 

this shared information is not necessarily identical at all 
terminals, but highly correlated. Secondly, we are also unable to precisely 
characterize its pmf. Yet, we are able to exploit this high correlation in the 
highler layers of communication. }\end{comment}

Let us now describe the fBL coding scheme. The first layer is the fixed B-L GKW 
coding layer which employs identical codes to quantize $K_{1},K_{2}$.

\noindent\textit{Fixed B-L GKW Coding Layer}: We design quantization codes 
for $K_{1}$. Encoder $2$ employs the same codes for quantizing $K_{2}$ too. 
Since $K_{j}=f_{j}(Y_{j}) \in \mathcal{K}: j \in [2]$, one can employ common 
codes $C_{S},C_{V}$ and quantization map $\kappa_{SV}^{n}:\mathcal{K}^{n} 
\rightarrow C_{S}\times 
C_{V}$ at both encoders. However, note that $\displaystyle \lim_{n \rightarrow 
\infty}P(K_{1}^{n} \neq K_{2}^{n})= \lim_{n 
\rightarrow \infty}  1 - ( 1 - \xi(\ulineK))^{n} = 1$, no matter how small 
$\xi(\ulineK) > 0$. Moreover, $K_{2}^{n}$ is uniformly 
distributed on an exponentially large set $T_{\delta}^{n}(K_{2}|K_{1}^{n})$ for 
large $n$. We conclude that, even if both encoders share common codes
$C_{S},C_{V}$ and map $\kappa_{SV}^{n}:\mathcal{K}^{n} \rightarrow C_{S}\times 
C_{V}$, the conventional approach of arbitrarily large B-L codes will result in 
choice of different $C_{S},C_{V}-$codewords. 

We are thus led to coding of near GKW parts $K_{1},K_{2}$ with codes and maps 
of 
fixed B-L $l$. We intend to choose $l-$length quantizers $C_{S},C_{V}$ 
that can cover $K_{1}-$typical sequences whp. For this we leverage 
Prop.~\ref{Prop:ExistenceOfGoodSourceCodes} and choose codes $C_{S},C_{V}$ and 
the $\kappa-,\iota-$maps as stated there.

Since $l$ is fixed and a Shannon-theoretic study requires coding 
over an arbitrarily large number of symbols, we will code over an arbitrarily 
large number $m$ of these $l-$length codewords. The overall coding scheme is of 
B-L $lm$. We employ a matrix notation to describe this.

Encoder $j$ populates $\boldY_{j} \in \boldCalY_{j}$, where $\boldY_{j}(t,i)$ 
is the symbol received during symbol-interval $(t-1)m+i$ for $(t,i) \in [m] 
\times [l]$. $\boldK_{j} \in \boldCalK$ is defined as $\boldK_{j}(t,i) = 
f_{j}(\boldY_{j}(t,i))$ for $(t,i) \in [m] 
\times [l]$. Encoder $j$ quantizes rows of $\boldK_{j}$ separately 
using common codes $C_{S},C_{V}$ into rows of 
$\boldS_{j}\in \boldCalS,\boldV_{j} \in \boldCalV$. For $t \in [m]$, let 
$(\boldS_{j}(t,1:l),\boldV_{j}(t,1:l)) \in C_{S} \times C_{V}$ denote the 
quantizations of $\boldK_{j}(t,1:l)$. In other words, 
$(\boldS_{j}(t,1:l),\boldV_{j}(t,1:l)) = \kappa^{l}_{SV}(\boldK_{j}(t,1:l))$. 
We 
have thus quantized $\boldK_{j}$ into $(\boldS_{j},\boldV_{j})$. For $t \in 
[m]$, let $M_{s}^{j}(t) = \iota_{S}^{l}(\boldK_{j}(t,1:l))$ denote the index of 
the codeword chosen in $C_{S}$. Similarly, for $t \in [m]$, let 
$(M_{s}^{j}(t),M_{v}^{j}(t))$ denote the index of the codeword chosen in 
$C_{V}$. $\boldS_{j},\boldV_{j}$ and indices $\underline{M_{s}^{j}} \define 
(M_{s}^{j}(t):t \in [m])$, $\underline{M_{v}^{j}} \define (M_{v}^{j}(t):t \in 
[m])$ is the output of the fixed B-L coding layer at encoder $j$. Encoder $j$
communicates $\underline{M_{s}^{j}}$. In addition, encoder $1$ communicates a 
fraction $\lambda \in [0,1]$ of the $m$ indices in $\underline{M_{v}^{1}}$ and 
encoder $2$ communicates the rest of the $(1-\lambda)m$ indices in 
$\underline{M_{v}^{2}}$.

The fixed B-L GKW coding layer communicates $\underline{M_{s}^{j}}$ to Decoder 
$j$ for $j \in [2]$. Decoder $3$ receives 
$\underline{M_{s}^{1}},\underline{M_{s}^{2}}$ and a 
selection of indices in $\underline{M_{v}^{1}},\underline{M_{v}^{2}}$. Let 
$\underline{{M}_{v}^{3}}$ denote this selection of indices. For $j \in [2]$, 
Decoder $j$ can reconstruct $\boldS_{j}$. Decoder $3$ can reconstruct 
$\boldS_{1},\boldS_{2}$. Let $\boldV_{3} \in \boldCalV$ denote Decoder $3$'s 
reconstruction based on $\underline{M_{s}^{1}}, \underline{M_{s}^{2}}, 
\underline{{M}_{v}^{3}}$. The reader may recall that encoder $j$ has 
$\boldS_{j},\boldV_{j}$.

The reader is referred back to the discussion prior to describing the fixed B-L 
coding layer. As we stated there, the fixed B-L coding layer has enabled 
sharing of information that the terminals do not agree upon, but as we show, is 
highly correlated. Indeed, owing to the use of common codes,
\begin{eqnarray} 
\lefteqn{
\!\!\!\!\!\!\!\!\!\!\!\!\!\!\!\!\!\!\!\!\!\!\!
P\left(\!\!\left(\!\!\!\begin{array} {c}\boldK_{1} (t,1:l),\boldS_{1}(t,
1:l)\\\boldV_{ 1 } (t,1:l)\end{array}\!\!\!\right) \neq  \left( 
\!\!\!\begin{array}{c} \boldK_{2}(t,1:l), \boldS_{2}(t,1:l)\\\boldV_{2}(t,1:l) 
\end{array}\!\!\!\right)\!\!\right)} \nonumber \\ 
\label{Eqn:DisgareementProbability} \leq 
&&\!\!\!\!\!\!\!\!\!\!\!P\left(\left(\!\!\!\begin{array}{c}\boldK_{1}(t,
1:l)\neq 
\boldK_{2}(t,1:l)\end{array}\!\!\!\right)\right) \leq \xi^{[l]}(\ulineK).
\end{eqnarray}

Our goal now is to quantify how much information has been communicated through 
the fBL GKW coding layer and devise how to communicate the rest of the 
necessary information via efficient S-L coding techniques.\footnote{The 
emphasis on S-L techniques is to enable us characterize a S-L expression for 
the performance.} This can be accomplished if we can identify IID vectors that 
have been communicated to the decoders, and their correlation to the sources. 
Such IID vectors can then be treated as side information in the next layer of 
communication. The difficulty here is that owing to the $l-$letter maps 
$\boldK_{j}(t,1:l) \rightarrow (\boldS_{j}(t,1:l), \boldV_{j}(t,1:l))$, the 
symbols of $\boldS_{j},\boldV_{j}$ are not IID. The elegant technique of 
\textit{interleaving} devised by Shirani and Pradhan 
\cite{201406ISIT_ShiPra-2-Shrt} comes to our rescue. We choose $m$ 
surjective maps (permutations) $\Pi_{t}:[l] \rightarrow [l]$ uniformly at 
random. Since the $m$ rows
\begin{eqnarray}
 \label{Eqn:RowsOfMatrices}
 \left(\!\!\!\begin{array}{c}\boldX(t,1:l),\boldY_{1}(t,1:l),\boldY_{2}(t,
1:l)\\\boldK_{1}(t,1:l),\boldK_{2}(t,1:l),\boldS_{1}(t,
1:l)\\\boldS_{2}(t,1:l),\boldV_{1}(t,1:l),\boldV_{2}(t,
1:l),\boldV_{3}(t,
1:l)\end{array}\!\!\!\right): t \in [m] \nonumber
\end{eqnarray}
are IID (with\footnote{The pmf of $\boldV_{3}(t,1:l)$ conditioned on 
$\boldV_{j}(t,1:l)$ can be made invariant with $t$ by choosing the $\lambda$ 
fraction uniformly at random.} an unknown $l-$letter pmf) 
$p_{X^{l}Y_{1}^{l}Y_{2}^{l}\ulineK^{l} \ulineS^ {l} \ulineV^{ l } }$, 
the\footnote{We know certain marginals of $p_{X^{l}Y_{1}^{l}Y_{2}^{l}\ulineK^{l}
\ulineS^ {l} \ulineV^{ l } }$ such as 
$p_{X^{l}Y_{1}^{l}Y_{2}^{l}}=\prod_{i=1}^{l}\mathbb{W}_{XY_{1}Y_{2}}$. Since 
we do not have a characterization of the $\kappa^{l}_{SV}-$map, we do not have 
a characterization of several of these marginals.} $m-$length vector
\begin{eqnarray}
 \label{Eqn:InterleavedVector}
\left(\!\!\!\begin{array}{c}\boldK_{1}(t,\Pi_{t}(i)), 
\boldK_{2}(t,\Pi_{t}(i)),\boldS_{1 } 
(t,\Pi_{t}(i))\\\boldS_{2}(t,\Pi_{t}(i)),\boldV_{j}(t,\Pi_{t}(i)):j=1,2,3
\end{array} \!\!\!\right): t \in [m] \nonumber
\end{eqnarray}
is IID with pmf 
$p_{\mathscr{X}\mathscr{Y}_{1}\mathscr{Y}_{2}\mathscr{K}_{1}\mathscr{K}_{2}
\mathscr{S}_{1}\mathscr{S}_{2}\mathscr{V}_{1} \mathscr{V}_{2}\mathscr{V}_{3}} = 
$
\begin{eqnarray}
 \label{Eqn:PMFOfInterleavedVector}
p_{\mathscr{X}\underline{\mathscr{Y}\mathscr{K
} \mathscr{S}\mathscr{V}}} = \frac{1}{l}\sum_ {i=1}^{l}p_{ X_{ 
i}Y_{1i}Y_{2i} K_{ 1i}K_{2i}S_{1i} S_{2i}V_{1i}V_{ 2i}V_{ 3i}} 
 \end{eqnarray}
 \begin{figure}\centering
\includegraphics[width=3.5in]{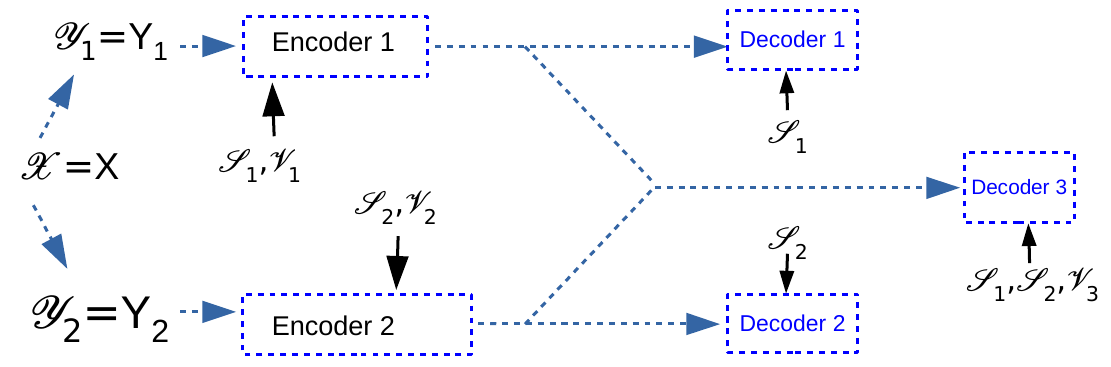}
\caption{Matrices available at the terminals after Fixed B-L GKW coding.}
\label{Fig:IIDSourcesAfterFBLGKWCoding}
\end{figure}
for every $i \in [l]$. We have thus identified IID vectors at each terminal 
that are correlated with the source and the joint pmf of these IID vectors is 
given by (\ref{Eqn:PMFOfInterleavedVector}). We now have the RDSC problem in 
Fig.~\ref{Fig:IIDSourcesAfterFBLGKWCoding} with the sources therein having 
joint pmf (\ref{Eqn:PMFOfInterleavedVector}) and our goal is to characterize an 
achievable RD region wrt test channel 
$p_{U_{1}|SY_{1}}p_{U_{2}|SY_{2}}p_{W_{1}|SVU_{1}Y_{1}}p_{W_{2}|SVU_{2}Y_{2}}$.

Let us provide a broad outline of how we achieve this goal. Define 
$p_{\mathscr{U}_{j} | \mathscr{S}_{j} \mathscr{Y}_{j} } = p_{U_{j}|SY_{j}}$ and 
$p_{\mathscr{W}_{j} | \mathscr{S}_{j} \mathscr{V}_{j} \mathscr{U}_{j} 
\mathscr{Y}_{j} } = p_{ W_ { j } |SVU_ { j } Y_ { j } } $ for $j \in [2]$. 
First, we characterize an achievable RD region $\mathcal{A}$ for the RDSC 
problem in 
Fig.~\ref{Fig:IIDSourcesAfterFBLGKWCoding} wrt the test channel $
\prod_{j=1}^{2}p_{\mathscr{U}_{j} | \mathscr{S}_{j} \mathscr{Y}_{j} 
}p_{\mathscr{W}_{j} | \mathscr{S}_{j} \mathscr{V}_{j} 
\mathscr{U}_{j} \mathscr{Y}_{j} } $. This is provided in Appendix 
\ref{AppSec:RDForSideInfoProblem}. Naturally, this 
characterization for $\mathcal{A}$ is in terms of the pmf 
$\mu=p_{\mathscr{X}\underline{\mathscr{Y}\mathscr{K
} \mathscr{S}\mathscr{V}}} \prod_{j=1}^{2}p_{\mathscr{U}_{j} | 
\mathscr{S}_{j} 
\mathscr{Y}_{j} }p_{\mathscr{W}_{j} | \mathscr{S}_{j} \mathscr{V}_{j} 
\mathscr{U}_{j} \mathscr{Y}_{j} } $. Unfortunately, we have no characterization 
of this pmf. We therefore upper bound the deviation between $\mu$ and the 
chosen test channel 
$\nu=\mathbb{W}_{X\ulineY}p_{SV|K_{1}}p_{U_{1}|SY_{1}}p_{U_{2}|SY_{2}}p_{W_{1}
|SVU_ {1}Y_{1}}p_{W_{2}|SVU_{2}Y_{2}}$. This is provided in Appendix 
\ref{AppSec:BoundDevBetweenInducedAndTestChannel}. The approach for Appendix 
\ref{AppSec:BoundDevBetweenInducedAndTestChannel} is borrowed from 
\cite{201406ISIT_ShiPra-2-Shrt}. The informational functionals characterizing 
$\mathcal{A}$, which are in terms of $\mu$ are then translated into 
informational functions in terms of $\nu$ using the bounds in Appendix 
\ref{AppSec:BoundDevBetweenInducedAndTestChannel}. In other words, we 
characterize an inner bound to $\mathcal{A}$ in terms of $\nu$. We provide this 
last step in \cite[Sec. VI]{201711arXiv_Pad}
\end{IEEEproof}
\appendices
\subsection{Inner bound to RD region for RDSC in 
Fig.~\ref{Fig:IIDSourcesAfterFBLGKWCoding}}
\label{AppSec:RDForSideInfoProblem}\vspace{-0.05in}
This follows by standard information-theoretic arguments. We characterize error 
events and obtain a set of bounds and perform Fourier-Motzkin elimination. In 
the following, $j \in [2]$ and $\msout{j}$ denotes complement index, i.e., 
$\{ j,\msout{j}\}=\{1,2\}$. An inner bound to the RD region for the RDSC 
problem 
in Fig.~\ref{Fig:IIDSourcesAfterFBLGKWCoding} wrt test channel 
$p_{\mathscr{X}\underline{\mathscr{Y}\mathscr{K
} \mathscr{S}\mathscr{V}}} \prod_{j=1}^{2}p_{\mathscr{U}_{j} | 
\mathscr{S}_{j} 
\mathscr{Y}_{j} }p_{\mathscr{W}_{j} | \mathscr{S}_{j} \mathscr{V}_{j} 
\mathscr{U}_{j} \mathscr{Y}_{j} } $ and functions 
$f_{j}:\mathcal{U}_{j}\rightarrow \mathcal{Z}: j \in [2], f_{3}: 
\mathcal{U}_{1}\times \mathcal{U}_{2}\times \mathcal{W}_{1} \times 
\mathcal{W}_{2} \rightarrow \mathcal{Z}$ consists of the set $\mathcal{A}$ of 
all $(R_{1},R_{2},\Delta_{1},\Delta_{2},\Delta_{3})$ that satisfy $\Delta_{j} 
\geq \Expectation\{ \distfn(\mathscr{X},f_{j}(\mathscr{U}_{j}))\}: j \in [2], 
\Delta_{3} \geq \Expectation\{ 
\distfn(\mathscr{X},f_{3}(\mathscr{U}_{1},\mathscr{U}_{2},\mathscr{W}_{1},
\mathscr{W}_{2}))\} $,
\begin{eqnarray}
 \lefteqn{R_{j} \geq  \alpha_{j}
\label{Eqn:IndividualBound}
-I(\mathscr{W}_{j};\mathscr{W}_{\msout{j}},\mathscr{S}_{\msout{j}}\mathscr{U}_{
\msout{j}}\mathscr{V}_{3}|\mathscr{U}_{{j}},\mathscr{S}_{{j}} ): j \in 
[2],}\\
\lefteqn{R_{1}+R_{2} \geq \alpha_{1}+\alpha_{2}-I(\mathscr{W}_{1}; 
\mathscr{W}_{2}|
\mathscr{U}_{1},\mathscr{S}_{1},\mathscr{U}_{2},\mathscr{S}_{2}  
,\mathscr{V}_{3})}\nonumber\\
&&\!\!\!\!\!\!\!\! -I(  \mathscr{W}_{2}; 
\mathscr{U}_{1},\mathscr{S}_{1}
\mathscr{V}_{3} | \mathscr{U}_{2},\mathscr{S}_{2}  )-I(  \mathscr{W}_{1}; 
\mathscr{U}_{2},\mathscr{S}_{2}
\mathscr{V}_{3} | \mathscr{U}_{1},\mathscr{S}_{1}  )
\label{Eqn:SumRateBound}
\end{eqnarray}
where $\alpha_{j} = I(\mathscr{U}_{j};\mathscr{Y}_{j}|\mathscr{S}_{j})+ 
I(\mathscr{W}_{j};\mathscr{Y}_{j}\mathscr{V}_{j}|\mathscr{U}_{j},\mathscr{S}_{j}
)$.
\subsection{Bounding the Deviation between PMFs}
\label{AppSec:BoundDevBetweenInducedAndTestChannel}
We employ the \underline{underline} extensively to group 
related RVs. For ex., $\ulinemathscrU$ abbreviates 
$\mathscr{U}_{1},\mathscr{S}_{2}$, $\uliney$ abbreviates $y_{1},y_{2}$ and so 
on. $\xi^{[l]}$ abbreviates $\xi^{[l]}(K)$ in 
(\ref{Eqn:DisgareementProbability}). We recall 
(\ref{Eqn:PMFOfInterleavedVector}) and
\begin{eqnarray}
p_{\mathscr{U}_{j} | \mathscr{S}_{j} \mathscr{Y}_{j} }\! = p_{U_{j}|SY_{j}}, ~
p_{\mathscr{W}_{j} | \mathscr{S}_{j} \mathscr{V}_{j} \mathscr{U}_{j} 
\mathscr{Y}_{j} }\! = p_{ W_ { j } |SVU_ { j } Y_ { j } } : j \in 
[2].\!\nonumber
\end{eqnarray}
Our goal is to bound deviation between 
$p_{\ulinemathscrY\mathscr{K}_{1}\mathscr{S}_{1}\mathscr{V}_{1}
\ulinemathscrU 
\ulinemathscrW}$ and \vspace{-0.18in}
\begin{eqnarray}
 \displaystyle p_{\ulinemathscrY \mathscr{K}_{1} 
\mathscr{S}_{1}\mathscr{V}_{1}}\! \prod_{j=1}^{2}\!\!p_{ U_ 
{j}|SY_{j}}(u_{j}|s_{1}y_{j})p_{ W_ { j } |SVU_ { j } Y_ { j } } 
(w_{j}|s_{1}v_{1}u_{j}y_{j}) \nonumber.
\end{eqnarray}\vspace{-0.18in}
We begin with
\begin{eqnarray}
 p_{\ulinemathscrY\mathscr{K}_{1}\mathscr{S}_{1}\mathscr{V}_{1}\ulinemathscrU 
\ulinemathscrW}\left( \!\!\!
\begin{array}{c}\uliney,k_{1},s_{1}\\v_{1},\ulineu,\ulinew 
\end{array}\!\!\!\right) \geq 
p_{\ulinemathscrY\mathscr{K}_{1}\ulinemathscrS\ulinemathscrV \ulinemathscrU 
\ulinemathscrW} \left( \!\!\!
\begin{array}{c}\uliney,k_{1},s_{1},s_{1}\\v_{1},v_{1},\ulineu,\ulinew 
\end{array}\!\!\!\right)\!\!\!\!\!\!\!\!\!\!\!\nonumber\\
\label{Eqn:Bound1}
\!\!\!= \!p_{\ulinemathscrY \mathscr{K}_{1} 
\ulinemathscrS\ulinemathscrV}\! \prod_{j=1}^{2}\!\!p_{ U_ 
{j}|SY_{j}}(u_{j}|s_{1}y_{j})p_{ W_ { j } |SVU_ { j } Y_ { j } } 
(w_{j}|s_{1}v_{1}u_{j}y_{j}) \!\!\!\!
\end{eqnarray}
where we have not specified the arguments in the first term for compactness. 
Since
\begin{eqnarray}
\!\!\!\!\!\!\!\!\lefteqn{ p_{\ulinemathscrY \mathscr{K}_{1} \mathscr{S}_{1}  
\mathscr{V}_{1} } (\uliney,k_{1},s_{1},v_{1} ) =  
\sum_{s_{2},v_{2}}\!\!p_{\ulinemathscrY \mathscr{K}_{1} 
\ulinemathscrS\ulinemathscrV} (\uliney,k_{1},s_{1},s_{2},v_{1},v_{2}) }
\nonumber\\
&&\!\!\!\!\!\!\!\!\!\leq p_{\ulinemathscrY \mathscr{K}_{1} 
\ulinemathscrS\ulinemathscrV} (\uliney,k_{1},s_{1},s_{1},v_{1},v_{1}) + 
\xi^{[l]}(\ulineK),\mbox{ where the first}\nonumber
% &&\!\!\!\!\!\!\!\!\!\!\!\!\!\!\!\!\!\!p_{\ulinemathscrY \mathscr{K}_{1} 
% \ulinemathscrS\ulinemathscrV} (\uliney,k_{1},s_{1},s_{1},v_{1},v_{1}) \geq 
% p_{\ulinemathscrY \mathscr{K}_{1} \mathscr{S}_{1}  
% \mathscr{V}_{1} } (\uliney,k_{1},s_{1},v_{1} ) - \xi^{[l]}(\ulineK)
\end{eqnarray}
term in this bound is the first term in (\ref{Eqn:Bound1}), we have
\begin{eqnarray}
 p_{\ulinemathscrY\mathscr{K}_{1}\mathscr{S}_{1}\mathscr{V}_{1}\ulinemathscrU 
\ulinemathscrW}\left( \!\!\!
\begin{array}{c}\uliney,k_{1},s_{1},v_{1},\ulineu,\ulinew 
\end{array}\!\!\!\right)\! \geq \! \left[ p_{\ulinemathscrY \mathscr{K}_{1} 
\mathscr{S}_{1}  \mathscr{V}_{1} } (\uliney,k_{1},s_{1},v_{1} )\right. 
\!\!\!\!\!\!\! \nonumber\\ 
\label{Eqn:AppBound1}
\!\!\!\!\times \prod_{j=1}^{2}p_{ U_ 
{j}|SY_{j}}(u_{j}|s_{1},y_{j})p_{ W_ { j } |SVU_ { j } Y_ { j } } 
(w_{j}|s_{1},v_{1},u_{j},y_{j})]  - \xi^{[l]}. \nonumber
\end{eqnarray}\vspace{-0.1in}

\noindent We now derive an upper bound on the the previous LHS 
term.\vspace{-0.08in}
\begin{eqnarray}
 \lefteqn{p_{\ulinemathscrY\mathscr{K}_{1}\mathscr{S}_{1}\mathscr{V}_{1}
\ulinemathscrU 
\ulinemathscrW} \leq 
p_{\ulinemathscrY\mathscr{K}_{1}\ulinemathscrS\ulinemathscrV
\ulinemathscrU 
\ulinemathscrW}\!\cdot\!\mathds{1}_{\{ \substack{s_{2}=s_{1} \\ 
v_{2}=v_{1}}\}}\!+\!\! \!\!\!\!\!\!\! \!\!\sum_{\substack{(s_{2},v_{2})\neq 
(s_{1},v_{1})}}\!\!\!\!\!\! \!\!\!\! 
p_{\ulinemathscrY\mathscr{K}_{1}\ulinemathscrS\ulinemathscrV \ulinemathscrU 
\ulinemathscrW} }\nonumber\vspace{-0.1in}\\\vspace{-0.18in}
&&\!\!\!\!\!\!\!\!\!\!\!\!\!\!\!
= \!p_{\ulinemathscrY \mathscr{K}_{1} 
\ulinemathscrS\ulinemathscrV}\! \prod_{j=1}^{2}\!\!p_{ U_ 
{j}|SY_{j}}(u_{j}|s_{1}y_{j})p_{ W_ { j } |SVU_ { j } Y_ { j } } 
(w_{j}|s_{1}v_{1}u_{j}y_{j}) +\xi^{[l]}\nonumber\\
 &&\!\!\!\!\!\!\!\!\!\!\!\!\!\!\!
\leq\!p_{\ulinemathscrY \mathscr{K}_{1} 
\mathscr{S}_{1}\mathscr{V}_{1}}\! \prod_{j=1}^{2}\!\!p_{ U_ 
{j}|SY_{j}}(u_{j}|s_{1}y_{j})p_{ W_ { j } |SVU_ { j } Y_ { j } } 
(w_{j}|s_{1}v_{1}u_{j}y_{j}) +\xi^{[l]}.\nonumber
\end{eqnarray}
Our first goal fulfilled, to bound deviation between $p_{\ulinemathscrY 
\mathscr{K}_{1} \mathscr{S}_{1}\mathscr{V}_{1}} \mbox{ and } 
p_{Y_{1}Y_{2}K_{1}SV}$ we refer to (\ref{Eqn:SourceCodeIsgood}).

\section{Concluding Remarks}
\label{Sec:Conclusion}

We have presented one step in a new direction towards deriving S-L admissible 
regions to joint source channel coding problems with distributed and correlated 
information sources. There are several ways in which one can generalize the 
findings presented in this article and thereby enlarge the admissible regions 
presented in Theorems \ref{Thm:MACStep2}, \ref{Thm:ICStep3CHKRegion}. With the 
aim of deriving a S-L characterization for the performance, we have adopted the 
approach of modifying GKW coding and break free from the S-L LMC constraint. 
This has led us to map sub-blocks of $K_{1},K_{2} \in \mathcal{K}^{l}$ via 
common maps. It is worth exploring other approaches. Secondly, we have focused 
on communicating a S-L function $f_{1}:\mathcal{S}_{1}\rightarrow \mathcal{K}$ 
of the sources to the decoder via the fixed B-L coding scheme. One can 
generalize this to communicating a common quantized version of the sources via 
the fixed B-L coding. Thirdly, we can incorporate the 
CES technique of inducing the source correlation onto channel inputs 
\cite{198011TIT_CovGamSal} in communicating the $\infty-$B-L information 
stream. In the second part, we pursue the latter two ways of enlarging the 
admissible region presented in this article. As we will see this will lead to a 
new admissible region that subsume the current known largest for the MAC
\cite{198011TIT_CovGamSal} and IC problems \cite{201112TIT_LiuChe} and strictly 
enlarge the same for identified examples.
\section*{Acknowledgement}
The author is thankful to (i) Prof. Sandeep Pradhan, Farhad Shirani for sharing 
their insights on \cite{201406ISIT_ShiPra-2-Shrt}, (ii) Deepanshu Vasal for 
technical discussions and (iii) Prof. P R Kumar for his support and 
encouragement. The author is particularly very grateful to Prof. Wojciech 
Szpankowski for the inspiration, his support and encouragement.
\appendices
\section{Interleaving results in IID distributions}
\label{AppSec:PMFOfInterleavedVector}

\begin{lemma}
 \label{Lem:SimpleInterleavingLemma}
 Let $\mathcal{A}$ be a finite set and $p_{A^{l}}$ be a pmf on 
$\mathcal{A}^{l}$. Let $A(1,1:l), A(2,1:l),\cdots, A(m,1:l) \in \mathcal{A}^{l}$ 
be independent and identically distributed vectors with pmf $p_{A^{l}}$. Let 
$\Lambda_{1},\cdots,\Lambda_{m}$ be independent and uniformly distributed 
indices taking values in $\{1,\cdots,l\}$. Moreover, 
$\Lambda_{1},\cdots,\Lambda_{m}$ is independent of the collection $A(1,1:l), 
A(2,1:l),\cdots, A(m,1:l)$. Then the components $A(t,\Lambda_{t}): t \in [m]$ 
are independent and identically distributed with pmf 
$\frac{1}{l}\sum_{i=1}^{l}p_{A_{i}}$, 
where $p_{A_{i}}$ is the pmf of $A(t,i)$.
\end{lemma}
\begin{IEEEproof}
 Note that
 \begin{eqnarray}
 \label{Eqn:SummingOverPermutationsSmall}
 \!\!\!\!\!\!\!\!\!\!\!\!\!\!\!\!\lefteqn{ 
\!\!\!\!\!\!\!\!\!\!\!\!\!\!\!\!P(A(t,\Lambda_{t})=a_{t}:t\in[m]) =
\sum_{j_{1}\in[l]}\cdots\sum_{j_{m}\in[l]}P(A(t,j_{t})=a_{
t},\Lambda_{t}=j_{t}:t\in[m])}\nonumber\\
 \label{Eqn:InterleavingLemmaStep1Small}
&&=\frac{1}{l^{m}}\sum_{j_{1}\in[l]}\cdots\sum_{j_{m}\in[l]}P(A(t,j_
{t})=a_{t}:t\in[m])\\
 \label{Eqn:InterleavingLemmaStep2Small}
&&=\frac{1}{l^{m}}\sum_{j_{1}\in[l]}\cdots\sum_{j_{m}\in[l]}\prod_{
t=1}^{m}P(A(t,j_{t})=a_{t})\\
 \label{Eqn:InterleavingLemmaStep3Small}
&&=\prod_{t=1}^{m}\left(\frac{1}{l}\sum_{j_{t}\in[l]}P(A(t,j_{t})=a_{t}
)\right)
 \label{Eqn:InterleavingLemmaStep4Small}
 =\prod_{t=1}^{m}\left(\frac{1}{l}\sum_{i\in[l]}p_{A_{i}}(a_{t})\right),
\end{eqnarray}
where (i) (\ref{Eqn:InterleavingLemmaStep1Small}) follows from independence of 
$(\Lambda_{1},\cdots,\Lambda_{m})$ and $A(1:m,1:l)$, (ii) 
(\ref{Eqn:InterleavingLemmaStep2Small}) follows from the independence of the 
vectors 
$A(1,1:l), A(2,1:l),\cdots, A(m,1:l) \in \mathcal{A}^{l}$, (iii) 
(\ref{Eqn:InterleavingLemmaStep4Small}) follows from $A(1,1:l), 
A(2,1:l),\cdots, 
A(m,1:l) \in \mathcal{A}^{l}$ being identically distributed, and moreover, 
$p_{A_{i}}(a) = P(A(t,i)=a)$.
\end{IEEEproof}
\begin{lemma}
\label{Lem:FullInterleavingLemma}
Let $\mathcal{A}$ be a finite set and $p_{A^{l}}$ be a pmf on $\mathcal{A}^{l}$. 
Let $A(1,1:l), A(2,1:l),\cdots, A(m,1:l) \in \mathcal{A}^{l}$ be independent and 
identically distributed vectors with pmf $p_{A^{l}}$. Let $\Theta_{l}$ be the 
set of all surjective maps on the set $\{1,2,\cdots,l\}$. Let surjective maps 
$\Lambda_{1},\Lambda_{2},\cdots,\Lambda_{m}$ be chosen uniformly and 
independently from $\Theta_{l}$. For $i=1,2,\cdots,l$, let
\begin{equation}
 \label{Eqn:PermutedVectorDefn}
 B(t,i) = A(t,\Lambda_{t}(i)):t \in [m], i \in [l]. \nonumber
\end{equation}
The $l$ vectors $B(1:m,i):i=1,2,\cdots,l$ are identically distributed with pmf 
$\prod_{t=1}^{m}\frac{1}{l}\sum_{i=1}^{l}p_{A_{i}}$, where 
\begin{equation}
 \label{Eqn:DefnOfComponentPmf}
 p_{A_{i}}(a)\!=\! \sum_{\substack{a_{1}\\\in \mathcal{A}}}\!\!\! \cdots\!\!\! 
\sum_{\substack{a_{i-1}\\\in \mathcal{A}}}\sum_{\substack{a_{i+1}\\\in 
\mathcal{A}}}\!\!\!\cdots\! \sum_{\substack{a_{l}\\\in 
\mathcal{A}}}p_{A^{l}}(a_{1},\!\cdots\!,a_{i-1},a,a_{i+1},\!\cdots\!,a_{l}
).\nonumber
\end{equation} 
\end{lemma}
\begin{IEEEproof}
For any $i\in[l]$, note that
\ifTITVersion\begin{eqnarray}
 \label{Eqn:SummingOverPermutations}
 \!\!\!\!\!\!\!\!\!\!\!\!\!\!\!\!\lefteqn{ 
\!\!\!\!\!\!\!\!\!\!\!\!\!\!\!\!P(B(t,i)=a_{t}:t\in[m]) =\!\!\! 
\sum_{j_{1}\in[l]}\!\!\!\cdots\!\!\!\sum_{j_{m}\in[l]}P(\substack{A(t,j_{t})=a_{
t},\\\Lambda_{t}(i)=j_{t}:t\in[m]})}\nonumber\\
 \label{Eqn:InterleavingLemmaStep1}
&&=\frac{1}{l^{m}}\sum_{j_{1}\in[l]}\!\!\!\cdots\!\!\!\sum_{j_{m}\in[l]}P(A(t,j_
{t})=a_{t}:t\in[m])\\
 \label{Eqn:InterleavingLemmaStep2}
&&=\frac{1}{l^{m}}\sum_{j_{1}\in[l]}\!\!\!\cdots\!\!\!\sum_{j_{m}\in[l]}\prod_{
t=1}^{m}P(A(t,j_{t})=a_{t})\\
 \label{Eqn:InterleavingLemmaStep3}
&&=\prod_{t=1}^{m}\left(\frac{1}{l}\sum_{j_{t}\in[l]}P(A(t,j_{t})=a_{t}
)\right)\nonumber\\
 \label{Eqn:InterleavingLemmaStep4}
 &&=\prod_{t=1}^{m}\left(\frac{1}{l}\sum_{i\in[l]}p_{A_{i}}(a_{t})\right),
\end{eqnarray}\fi
\ifPeerReviewVersion\begin{eqnarray}
 \label{Eqn:SummingOverPermutations}
 \!\!\!\!\!\!\!\!\!\!\!\!\!\!\!\!\lefteqn{ 
\!\!\!\!\!\!\!\!\!\!\!\!\!\!\!\!P(B(t,i)=a_{t}:t\in[m]) =
\sum_{j_{1}\in[l]}\cdots\sum_{j_{m}\in[l]}P(A(t,j_{t})=a_{
t},\Lambda_{t}(i)=j_{t}:t\in[m])}\nonumber\\
 \label{Eqn:InterleavingLemmaStep1}
&&=\frac{1}{l^{m}}\sum_{j_{1}\in[l]}\cdots\sum_{j_{m}\in[l]}P(A(t,j_
{t})=a_{t}:t\in[m])\\
 \label{Eqn:InterleavingLemmaStep2}
&&=\frac{1}{l^{m}}\sum_{j_{1}\in[l]}\cdots\sum_{j_{m}\in[l]}\prod_{
t=1}^{m}P(A(t,j_{t})=a_{t})\\
 \label{Eqn:InterleavingLemmaStep3}
&&=\prod_{t=1}^{m}\left(\frac{1}{l}\sum_{j_{t}\in[l]}P(A(t,j_{t})=a_{t}
)\right)
 \label{Eqn:InterleavingLemmaStep4}
 =\prod_{t=1}^{m}\left(\frac{1}{l}\sum_{i\in[l]}p_{A_{i}}(a_{t})\right),
\end{eqnarray}\fi
where (i) (\ref{Eqn:InterleavingLemmaStep1}) follows from independence of the 
surjective maps $(\Lambda_{1},\cdots,\Lambda_{m})$ and $A(1:m,1:l)$, (ii) 
(\ref{Eqn:InterleavingLemmaStep2}) follows from the independence of the vectors 
$A(1,1:l), A(2,1:l),\cdots, A(m,1:l) \in \mathcal{A}^{l}$, (iii) 
(\ref{Eqn:InterleavingLemmaStep4}) follows from $A(1,1:l), A(2,1:l),\cdots, 
A(m,1:l) \in \mathcal{A}^{l}$ being identically distributed, and moreover, 
$p_{A_{i}}(a) = P(A(t,i)=a)$.
\end{IEEEproof}

\begin{lemma}
\label{Lem:ConstantCompostion}
 Let $C_{U}$ be a constant composition code of type $p_{U}$ with message index 
set $[M^{u}]$, encoder map $e_{u}: [M_{u}] \rightarrow \mathcal{U}^{l}$ with 
codewords $u^{l}(a): a \in [M_{u}]$. Let $A_{t} \in [M_{u}]$ be a (random) 
message and $U^{l} \define u^{l}(A_{t})$ denote the corresponding codeword. 
Suppose $I \{1,\cdots, l\}$ is uniformly distributed and independent of 
$U^{l}$, then $p_{U_{I}} = \frac{1}{l}\sum_{i=1}^{l}p_{U_{i}} = p_{U}$.
\end{lemma}
\begin{IEEEproof}
 Finally, let us identify $\prod_{t=1}^{m}\frac{1}{l}\sum_{i=1}^{l}p_{U_{ji}}$, 
the pmf of these sub-vectors. Observe that $C_{U}$ is a constant composition 
code of type $p_{U}$. Irrespective of the pmf of the messages $A_{jt}$ indexing 
this codebook, the indexed codeword $u^{l}(A_{jt})$ has type $p_{U}$. A 
uniformly chosen symbol from $u^{l}(A_{jt})$ will therefore have pmf $p_{U}$. 
We 
make this formal through the following identities. Note that
\ifPeerReviewVersion\begin{eqnarray}
 \label{Eqn:PMFInterleavedSeqIspU}
\lefteqn{\frac{1}{l}\sum_{i=1}^{l}p_{U_{ji}}(c) =
\frac{1}{l}\sum_{i=1}^{l}\sum_{u^{l} \in 
\mathcal{U}^{l}}p_{U^{l}}(u^{l})\mathds{1}_{\{ u_{i}=c \}}=
\frac{1}{l}\sum_{u^{l} \in 
\mathcal{U}^{l}}\sum_{i=1}^{l}p_{U^{l}}(u^{l})\mathds{1}_{\{ u_{i}=c 
\}}}\nonumber\\
&=&\frac{1}{l}\sum_{u^{l} \in 
\mathcal{U}^{l}}\sum_{i=1}^{l}P(u(A_{jt})=u^{l})\mathds{1}_{\{ u_{i}=c 
\}}=\frac{1}{l}\sum_{u^{l} \in 
\mathcal{U}^{l}}\sum_{i=1}^{l}P(u(A_{jt})=u^{l})\mathds{1}_{\{u^{l}\mbox{{
\small has type} }p_{U} \}}\mathds{1}_{\{ u_{i}=c \}}\nonumber\\
&=&\frac{1}{l}\sum_{u^{l} \in 
\mathcal{U}^{l}}P(u(A_{jt})=u^{l})\mathds{1}_{\{u^{l}\mbox{{
\small has type} }p_{U} \}}lp_{U}(c)= p_{U}(c)\sum_{u^{l} \in 
\mathcal{U}^{l}}P(u(A_{jt})=u^{l})\mathds{1}_{\{u^{l}\mbox{{
\small has type} }p_{U} \}} = p_{U}(c),\nonumber
 \end{eqnarray}\fi
\ifTITVersion\begin{eqnarray}
 \label{Eqn:PMFInterleavedSeqIspU}
\lefteqn{\frac{1}{l}\sum_{i=1}^{l}p_{U_{ji}}(c) =
\frac{1}{l}\sum_{i=1}^{l}\sum_{u^{l} \in 
\mathcal{U}^{l}}p_{U^{l}}(u^{l})\mathds{1}_{\{ u_{i}=c \}}}\nonumber\\&=&
\frac{1}{l}\sum_{u^{l} \in 
\mathcal{U}^{l}}\sum_{i=1}^{l}p_{U^{l}}(u^{l})\mathds{1}_{\{ u_{i}=c 
\}}\nonumber\\
&=&\frac{1}{l}\sum_{u^{l} \in 
\mathcal{U}^{l}}\sum_{i=1}^{l}P(u(A_{jt})=u^{l})\mathds{1}_{\{ u_{i}=c 
\}}\nonumber\\&=&\frac{1}{l}\sum_{u^{l} \in 
\mathcal{U}^{l}}\sum_{i=1}^{l}P(u(A_{jt})=u^{l})\mathds{1}_{\{u^{l}\mbox{{
\small has type} }p_{U} \}}\mathds{1}_{\{ u_{i}=c \}}\nonumber\\
&=&\frac{1}{l}\sum_{u^{l} \in 
\mathcal{U}^{l}}P(u(A_{jt})=u^{l})\mathds{1}_{\{u^{l}\mbox{{
\small has type} }p_{U} \}}lp_{U}(c)\nonumber\\&=& 
p_{U}(c)\sum_{u^{l} \in 
\mathcal{U}^{l}}P(u(A_{jt})=u^{l})\mathds{1}_{\{u^{l}\mbox{{
\small has type} }p_{U} \}}= p_{U}(c),\nonumber
 \end{eqnarray}\fi
 and hence conclude sub-vector 
$(\boldU_{j}(t,\Lambda_{1}(t)): t \in [m])$ has pmf
$  \prod_{t=1}^{m}\frac{1}{l}\sum_{i=1}^{l}p_{U_{ji}} = 
\prod_{t=1}^{m}p_{U}$ .
\end{IEEEproof}

\section{Properties of PMFs (\ref{Eqn:Step1MACPMFForDecodingRule}), 
(\ref{Eqn:Step1MACPMFOfInterleavedVector}) employed in Decoding Rule}
\label{AppSec:PropOfDecodingPMF}
Let us recall

\ifTITVersion\begin{eqnarray}
\lefteqn{p_{\ulineU^{l}\ulineV^{l}\ulineX^{l}Y^{l}}\!\!\left(\!\!\!\begin{array}
{c }
\ulineu^ { l
}, \ulinev^ { l } ,\\ \ulinex^ {l},
y^{l}\end{array}\!\!\!\right) \!=\! \left[ \sum_{\substack{(a_{1},a_{2}) \in \\ 
[M_{u}]\times [M_{u}]}} 
\!\!\!\!\!\!\!\!\!P\left(\!\!\!
 \begin{array}{c}
 A_{1}=a_{1}\\A_{2}=a_{2}
 \end{array}\!\!\!\right)\mathds{1}_{\left\{\substack{ 
u^{l}(a_{j})=\\u_{j}^{l}:j \in [2]}\right\}}\right]\!\!\!\!\!\!\!\!} \nonumber\\
&& \times \left[ \prod_{j=1}^{2} \left\{ 
\prod_{i=1}^{l}p_{V_{j}}(v_{ji})p_{X_{j}|UV_{j}}(x_{ji}|u_{ji},v_{ji}) \right\} 
\right]~~~~~~~~~~~~~~~\nonumber\\
 \label{Eqn:AppSecPMFForDecodingRule}
&& \times \left[ \prod_{i=1}^{l} 
\mathbb{W}_{Y|X_{1}X_{2}}(y_{i}|x_{1i},x_{2i})\right]
\end{eqnarray}\fi
\ifPeerReviewVersion\begin{eqnarray}
p_{\ulineU^{l}\ulineV^{l}\ulineX^{l}Y^{l}}(\ulineu^{l},\ulinev^{l},\ulinex^{l},
y^{l}) = \left[ \sum_{\substack{(a_{1},a_{2}) \in \\ [M_{u}]\times [M_{u}]}} 
\!\!\!\!\!\!\!P(
 \begin{array}{c}
 A_{1}=a_{1}\\A_{2}=a_{2}
 \end{array})\mathds{1}_{\left\{\substack{ 
u^{l}(a_{j})=\\u_{j}^{l}:j \in [2]}\right\}}\right]
 \times \left[ \prod_{j=1}^{2} \left\{ 
\prod_{i=1}^{l}p_{V_{j}}(v_{ji})p_{X_{j}|UV_{j}}(x_{ji}|u_{ji},v_{ji}) \right\} 
\right]\nonumber\\
 \label{Eqn:AppSecPMFForDecodingRule}
 \times \left[ \prod_{i=1}^{l} 
\mathbb{W}_{Y|X_{1}X_{2}}(y_{i}|x_{1i},x_{2i})\right]
\end{eqnarray}\fi
be a pmf\footnote{In (\ref{Eqn:AppSecPMFForDecodingRule}), 
$\ulineU^{l}\ulineV^{l}\ulineX^{l}Y^{l}$ abbreviates 
$U_{1}^{l}U_{2}^{l}V_{1}^{l}V_{2}^{l}X_{1}^{l} X_{2}^{l}Y^{l} $ and similarly 
$\ulineu^{l},\ulinev^{l},\ulinex^{l},
y^{l}$ abbreviates $u_{1}^{l},u_{2}^{l},v_{1}^{l},v_{2}^{l},x_{1}^{l}, 
x_{2}^{l},y^{l}$.} on $\ulineCalU^{l}\times 
\ulineCalV^{l}\times\ulineCalX^{l}\times \mathcal{Y}^{l}$, and
\ifTITVersion\begin{eqnarray}
&p_{\mathscr{U}_{1}\mathscr{U}_{2}\mathscr{V}_{1}\mathscr{V}_{2}\mathscr{X}_{
1}\mathscr{X}_{2}\mathscr{Y}}({\ulinea},{\ulineb},{
\ulinec},{d}) \define& 
\nonumber\\&\label{Eqn:AppSecPMFOfInterleavedVector}\displaystyle\frac{1}{l}
\sum_ { i=1 }
^{l}p_{U_{1i}U_{2i}V_{1i}V_{2i}X_{
1i}X_{2i}Y_{i}}(a_{1},a_{2},b_{1},b_{2},c_{1},c_{2},d).& 
\end{eqnarray}\fi
\ifPeerReviewVersion\begin{eqnarray}
\label{Eqn:AppSecPMFOfInterleavedVector}
p_{\mathscr{U}_{1}\mathscr{U}_{2}\mathscr{V}_{1}\mathscr{V}_{2}\mathscr{X}_{
1}\mathscr{X}_{2}\mathscr{Y}}({\ulinea},{\ulineb},{
\ulinec},{d}) \define 
\displaystyle\frac{1}{l}\sum_{i=1}^{l}p_{U_{1i}U_{2i}V_{1i}V_{2i}X_{
1i}X_{2i}Y_{i}}(a_{1},a_{2},b_{1},b_{2},c_{1},c_{2},d).
\end{eqnarray}\fi
\begin{lemma}
 \label{Lem:SimplePropDecodingLem}
Let $U_{1}^{l},U_{2}^{l},V_{1}^{l},V_{2}^{l},X_{1}^{l},X_{2}^{l},Y^{l}$ take 
values in $\ulineCalU^{l}\times 
\ulineCalV^{l}\times\ulineCalX^{l}\times \mathcal{Y}^{l}$ with pmf 
(\ref{Eqn:AppSecPMFForDecodingRule}) and 
consider the pmf 
(\ref{Eqn:AppSecPMFOfInterleavedVector}) defined on $\ulineCalU\times 
\ulineCalV\times\ulineCalX\times \mathcal{Y}$. Suppose 
$I$ is a random index taking values in $\{1,\cdots,l \}$ that is uniformly 
distributed and independent of 
$U_{1}^{l},U_{2}^{l},V_{1}^{l},V_{2}^{l},X_{1}^{l},X_{2}^{l},Y^{l}$. The 
following are true.
\begin{enumerate}
 \item $p_{V_{1}^{l}V_{2}^{l}} = \prod_{i=1}^{l}p_{V_{1}}p_{V_{2}}$,
 \item $U_{1I},U_{2I},V_{1I},V_{2I},X_{1I},X_{2I},Y_{I}$ has pmf 
(\ref{Eqn:AppSecPMFOfInterleavedVector})
 \item $p_{\mathscr{V}_{j}}=p_{V_{j}}$ for $j \in [2]$,
 \item The marginals
 \begin{eqnarray}
  p_{U_{1}^{l}U_{2}^{l}}(u_{1}^{l},u_{2}^{l}) &= &
\sum_{\substack{(a_{1},a_{2}) ~\in~
[M_{u}]\times [M_{u}]}} 
\!\!P\left(\!\!\!
 \begin{array}{c}
 A_{1}=a_{1},A_{2}=a_{2}
 \end{array}\!\!\!\right)\mathds{1}_{\left\{\substack{ 
u^{l}(a_{j})= u_{j}^{l}:j \in [2]}\right\}}, 
\label{Eqn:AppSecPropDecPMFU1lU2lMar}\\
%%%%%%%%%%%%%%%%%%newline
\lefteqn{
\!\!\!\!\!\!\!\!\!\!\!\!\!\!\!\!\!\!\!\!\!\!\!\!\!\!\!\!\!\!\!\!\!\!\!\!p_
{ \ulineV^ {l }
\ulineX^{l} Y^{l
} |\ulineU^ {l }} (\ulinev^{ l} , \ulinex^{ l} ,
y^{l}|\ulineu^{l}) = \left[ \prod_{i=1}^{l} \left\{ 
\prod_{j=1}^{2}p_{V_{j}}(v_{ji})p_{X_{j}|UV_{j}}(x_{ji}|u_{ji},v_{ji}) \right\} 
\mathbb{W}_{Y|\ulineX}(y_{i}|x_{1i},x_{2i})\right]\!\!,\mbox{ and in 
particular} 
}~~~~~~~~~~~~~~~~~~~~~~~~~\label{Eqn:AppSecPropDecPMFCondU1lU2}\end{eqnarray}
\begin{eqnarray}
\!\!\!\!\!\!\!\!\!\!\!\!\!\!\!\!\!\!\!\!\!\!\!\!\!\!\!\!\!\!\!\!\!\!\!\!\!\!\!\!
\!\!\!\!\!\!p_ { \ulineV^{ l} \ulineX^{l }Y^{ l} |\ulineU^{ l}} (\ulinev^{l}, 
\ulinex^ {l },
y^{l}|u^{l},u^{l}) &= &
\left[ \prod_{i=1}^{l} \left\{ 
\prod_{j=1}^{2}p_{V_{j}}(v_{ji})p_{X_{j}|UV_{j}}(x_{ji}|u_{i},v_{ji}) \right\} 
\mathbb{W}_{Y|\ulineX}(y_{i}|x_{1i},x_{2i})\right]
\label{Eqn:AppSecPropDecPMFCondEqualU1lU2}\\
\label{Eqn:AppSecPropDecPMFCondEqualU1lU22}
&=& \prod_{i=1}^{l}p_{V_{1}V_{2}X_{1}X_{2}Y|U}(v_{1i},v_{2i},x_{1i},x_{2i},y_{i}
| u_{i}), \mbox{ and hence}\\
\label{Eqn:AppSecPropDecPMFCommonCuCode}
p_{\ulineY^{l}|U_{1}^{l}U_{2}^{l}}(y^{l}|u^{l},u^{l})&=&\prod_{i=1}^{l}p_{Y|U}
(y_{i}|u_{i})
 \end{eqnarray}
\end{enumerate}
\end{lemma}
\begin{IEEEproof}
1) Follows by just computing the marginal $p_{V_{1}^{l}V_{2}^{l}}$ wrt 
(\ref{Eqn:AppSecPMFForDecodingRule}). 2) Straightforward to verify. 3) Follows 
from previous two assertions. 4) Follows by just evaluating the LHSs wrt to 
(\ref{Eqn:AppSecPMFForDecodingRule}).
\end{IEEEproof}

\begin{lemma}
 \label{Lem:PMFOfUnifAndRandCo-Ordinate}
Given $l \in \naturals$, finite alphabet sets 
$\mathcal{A},\mathcal{B}_{1},\mathcal{B}_{2},\mathcal{C}$ and a pmf 
$p_{AB_{1}B_{2}C} = p_{A}p_{B_{1}|A}p_{B_{2}|A}p_{C|B_{1}B_{2}}$ on 
$\mathcal{A}\times\mathcal{B}_{1}\times\mathcal{B}_{2}\times\mathcal{C}$ such 
that $p_{A}$ is a type of sequences in $\mathcal{A}^{l}$. Suppose 
$(A_{1}^{l},A_{2}^{l},B_{1}^{l},B_{2}^{l},C^{l})$ take values in 
$\mathcal{A}^{l} \times \mathcal{B}_{1}^{l}\times \mathcal{B}_{2}^{l}\times 
\mathcal{C}^{l}$ with pmf $p_{A_{1}^{l}A_{2}^{l}B_{1}^{l}B_{2}^{l}C^{l}}$ given 
by
\begin{eqnarray}
p_{A_{1}^{l}A_{2}^{l}B_{1}^{l}B_{2}^{l}C^{l}}( a_{1}^{l}, a_{2}^{l}, b_{1}^{l}, 
b_{2}^{l},c^{l} ) = \left[ \sum_{m_{1},m_{2}}P\left( \!\!\! \begin{array}{c} 
M_{1}=m_{1}\\M_{2}=m_{2} \end{array} \!\!\!\right) \mathds{1}_{\left\{  
\begin{array}{c} u^{l}(m_{1}) = a_{1}^{l}\\ u^{l}(m_{2}) = a_{2}^{l} 
\end{array} \right\}}\right]\prod_{i=1}^{l} \left\{ \begin{array}{c}
p_{B_{1}|A}(b_{1i}|a_{1i})p_{B_{2}|A}(b_{2i}|a_{2i})\\P_{C|B_{1}B_{2}}(c_{i}|b_{
1i } ,b_{2i}) \end{array}\right\},\nonumber
\nonumber
\end{eqnarray}
where (i) $u^{l}:[M_{u}] \rightarrow \mathcal{A}^{l}$ is a map such that 
$u^{l}(m) \in \mathcal{A}^{l}$ is of type $p_{A}$ for every $m$, (ii) 
$(M_{1},M_{2}) \in [M_{u}] \times [M_{u}]$ are a pair of (message) random 
variables with pmf $P(M_{1}=\cdot,M_{2}=\cdot)$. Suppose 
$I$ is a random index taking values in $\{1,\cdots,l \}$ that is uniformly 
distributed and independent of $A_{1}^{l}A_{2}^{l}B_{1}^{l}B_{2}^{l}C^{l}$, then
\begin{eqnarray}
 \label{Eqn:Whatever1}
P\left( A_{1I}=x,A_{2I}=x,B_{1I}=y_{1},B_{2I}=y_{2},C_{I}=z | 
\mathds{1}_{\{ A_{1}^{l}=A_{2}^{l} \}}=1\right) = 
p_{AB_{1}B_{2}C}(x,y_{1},y_{2},z).
\end{eqnarray}
\end{lemma}
\begin{IEEEproof}
 Let $J = \mathds{1}_{\{ A_{1}^{l}=A_{2}^{l} \}}$. It can be verified by 
summing over $b_{1}^{l},b_{2}^{l},c^{l}$ that 
\[p_{A_{1}^{l}A_{2}^{l}}(a_{1}^{l},a_{2}^{l}) = \left[ 
\sum_{m_{1},m_{2}}p\left( \!\!\! \begin{array}{c} 
M_{1}=m_{1}\\M_{2}=m_{2} \end{array} \!\!\!\right) \mathds{1}_{\left\{  
\begin{array}{c} u^{l}(m_{1}) = a_{1}^{l}\\ u^{l}(m_{2}) = a_{2}^{l} 
\end{array} \right\}}\right],\]
and hence $
p_{B_{1}^{l} B_{2}^{l} C^{l} | A_{1}^{l} 
A_{2}^{l}}(b_{1}^{l},b_{2}^{l},c^{l}|a^{l},a^{l}) = 
\prod_{i=1}^{l}p_{B_{1}B_{2}C|A}(b_{1i},b_{2i},c_{i}|a_{i})$. Since 
$B_{1}^{l}B_{2}^{l}C^{l}-A_{1}^{l}A_{2}^{l}-J$ forms a Markov chain, we have
\begin{eqnarray}
 \label{Eqn:Whatever2}
p_{A_{1}^{l} A_{2}^{l}B_{1}^{l} B_{2}^{l} C^{l}|J} ( a^{l},a^{l},b_{1}^{l}, 
b_{2}^{l}, c^{l}| 1) = p_{A_{1}^{l}A_{2}^{l}|J}(a^{l},a^{l}|1)\prod_{i=1}^{l} 
 p_{B_{1}B_{2}C|A}(b_{1i},b_{2i},c_{i}|a_{i}).
\nonumber
\end{eqnarray}
The PMF of a randomly chosen co-ordinate is given by
\begin{eqnarray}
 \label{Eqn:PMFOfRandomlyChosenCoOr}
 \lefteqn{p_{A_{1I}A_{2I}B_{1I}B_{2I}C_{I}|J}(x,x,y_{1},y_{2},z|1) = 
\frac{1}{l}\sum_{i=1}^{l}
p_{A_{1i}A_{2i}B_{1i}B_{2i}C_{i}|J}(x,x,y_{1},y_{2},
z|1)} \nonumber\\
&=&\frac{1}{l}
\sum_{i=1}^{l}
\sum_{\substack{ s^{l-1} \in \\ \mathcal{A}^{l-1} } } ~
\sum_{\substack{ t^{l-1} \in \\ \mathcal{B}_{1}^{l-1} }} ~
\sum_ {\substack{ u^{l-1} \in \\ \mathcal{B}_{2}^{l-1} }} ~
\sum_ {\substack{ v^{l-1} \in \\ \mathcal{C}^{l-1} }} \!\!
p_{A_{1}^{l}A_{2}^{l}B_{1}^{l}B_{2}^{l}C^{l}|J}\left( 
\!\!\!\left.\begin{array}{c} s^{i-1}x s^{l-i}, s^{i-1}x s^{l-i}, 
 t^{i-1} y_{1} t^{l-i},
u^{i-1} y_{2} u^{l-i}, v^{i-1} z v^{l-i} \end{array}\!\!\!\right|1\right) 
\nonumber\\
&=& \frac{1}{l}
\sum_{i=1}^{l}
\sum_{\substack{ s^{l-1} \in \\ \mathcal{A}^{l-1} } } ~
\sum_{\substack{ t^{l-1} \in \\ \mathcal{B}_{1}^{l-1} }} ~
\sum_ {\substack{ u^{l-1} \in \\ \mathcal{B}_{2}^{l-1} }} ~
\sum_ {\substack{ v^{l-1} \in \\ \mathcal{C}^{l-1} }} \!\!
p_{A_{1}^{l}A_{2}^{l}|J}(s^{i-1}x s^{l-i},s^{i-1}x 
s^{l-i}|1)p_{B_{1}B_{2}C|A}(y_{1},y_{2},z|x)\prod_{\substack{j=1\\j \neq 
i}}^{l} p_{B_{1}B_{2}C|A}(t_{j},u_{j},v_{j}|s_{j}) \nonumber\\
&=& \frac{1}{l}
\sum_{i=1}^{l}\sum_{\substack{ s^{l-1} \in \\ \mathcal{A}^{l-1} } 
}p_{A_{1}^{l}A_{2}^{l}|J}(s^{i-1}x s^{l-i},s^{i-1}x 
s^{l-i}|1)p_{B_{1}B_{2}C|A}(y_{1},y_{2},z|x)\nonumber\\
&=& p_{B_{1}B_{2}C|A}(y_{1},y_{2},z|x)\frac{1}{l}
\sum_{i=1}^{l}p_{A_{1i}A_{2i}|J}(x ,x 
|1) = p_{B_{1}B_{2}C|A}(y_{1},y_{2},z|x)p_{A}(x) = 
p_{AB_{1}B_{2}C}(x,y_{1},y_{2},z),
\label{Eqn:PMFRandCo-OrdConstComp}
\end{eqnarray}
where, $\sum_{i=1}^{l}p_{A_{1i}A_{2i}|J}(x ,x 
|1) = lp_{A}(x)$ is argued as follows. Since $u^{l}(m) \in \mathcal{A}^{l}$ is 
of type $p_{A}$ for every $m \in [M_{u}]$,
\begin{eqnarray}
 \label{Eqn:AppSecPropOfDecPMFConditional}
\lefteqn{\!\!\!\!\!\!\!\!\!\!\!\!\!\!\!\!\!\!\!\!\!\!\!\!\!\!\!\!\!\!\sum_{
\substack { a_{ 1} ^{ l} , a_{ 2} ^{ l} \in \mathcal{A}^{l}: \\a_{1}^{l}, 
a_{2}^{l}\mbox{\small{ 
are of type 
}}p_{A}}}\!\!\!\!\!\!\!\!\!p_{A_{1}^{l}A_{2}^{l}}(a_{1}^{l},a_{2}^{l}) = 1,
~\mbox{ and hence }~ \sum_{\substack{a^{l} \in \mathcal{A}^{l}: 
\\a^{l}\mbox{\small{ 
is of type 
}}p_{A}}}\!\!\!\!\!\!\!\!\!p_{A_{1}^{l}A_{2}^{l}|J}(a^{l},a^{l}|1) = 1,\mbox{ 
which implies}~~~~~~~}\nonumber\\
 \label{Eqn:PMFRandCo-OrdConstCompFinalStep}
 \sum_{i=1}^{l}p_{A_{1i}A_{2i}|J}(x ,x 
|1) &=& 
\sum_{i=1}^{l}\sum_{\substack{a^{l} \in \mathcal{A}^{l}: \\a^{l}\mbox{\small{ 
is of type 
}}p_{A}}}\!\!\!\!\!\!\!\!\!p_{A_{1}^{l}A_{2}^{l}|J}(a^{l},a^{l}|1)\mathds{ 1 } 
_ { \left\{ a_{i}=x \right\}} = \!\!\!\!\!\sum_{\substack{a^{l} \in 
\mathcal{A}^{l}: \\a^{l}\mbox{\small{ 
is of type 
}}p_{A}}}\sum_{i=1}^{l}p_{A_{1}^{l}A_{2}^{l}|J}(a^{l},a^{l}|1)\mathds{ 1 } _ { 
\left\{ a_{i}=x \right\}} \nonumber\\
&=& \sum_{\substack{a^{l} \in 
\mathcal{A}^{l}: \\a^{l}\mbox{\small{ 
is of type 
}}p_{A}}}p_{A_{1}^{l}A_{2}^{l}|J}(a^{l},a^{l}|1)lp_{A}(x) = lp_{A}(x). \nonumber
\end{eqnarray}

\end{IEEEproof}

\section{Proof of Equation (\ref{Eqn:Step2ICChnlMtrCorrectPMFConcl})}
\label{AppSec:Step2ICChnlMtrCorrectPMF}

Note that
\begin{eqnarray}
 \label{Eqn:Step2ICChnlMtrHaveCorrectPMF}
 P\left(\!\!\!  
 \begin{array}{c}
\boldu\{ \ulineA_{j} \} = \boldu_{j}, \boldV_{j}\{ \ulineB_{j} \} = \boldv_{j}\\
\boldY_{j}=\boldy_{j}, \boldX_{j}\{ \ulineA_{j},\ulineB_{j} 
\}=\boldx_{j}\\\boldu\{ \ulinehatA_{j} \} = \boldhatu_{j} : j \in [2]
 \end{array}
 \!\!\!\right) =
\sum_{\substack{\ulinea_{1},\ulinea_{2}\\\ulineb_{1},\ulineb_{2}}}\!
P \left(  \!\!\!
\begin{array}{c}
\ulineA_{j}=\ulinea_{j}\\\ulineB_{j}=\ulineb_{j} \\: j \in [2]
\end{array}
 \!\!\! \right)
 P\left(\!\!\!  \left.
 \begin{array}{c}
\boldu\{ \ulinea_{j} \} = \boldu_{j}, \boldV_{j}\{ \ulineb_{j} \} = \boldv_{j} 
\\
\boldY_{j}=\boldy_{j}, \boldX_{j}\{ \ulinea_{j},\ulineb_{j} 
\}=\boldx_{j}\\
\boldu\{ \ulinehatA_{j} \} = \boldhatu_{j} : j \in [2]
 \end{array}
 \!\!\!\right| \!\!\!
\begin{array}{c}
\ulineA_{j}=\ulinea_{j}\\\ulineB_{j}=\ulineb_{j} \\:j \in [2]
\end{array}
 \!\!\!
 \right).
\end{eqnarray}
We break down the second factor in a generic term above just as we did for the 
analogous term in (\ref{Eqn:Step1MACPrelim2-1}). Essentially 
(\ref{Eqn:Step1MACPrelim1-Prelim1}), (\ref{Eqn:Step1MACPrelim1-Prelim1Mid}),  
(\ref{Eqn:Step1MACPrelim1-Prelim2}), and in addition
\begin{eqnarray}
 \label{Eqn:Step2ICChnlMtrHaveCorrectPMF2}
P\left(\!\!\!  \left.
 \begin{array}{c}
\boldu\{ \ulinehatA_{j} \} = \boldhatu_{j} 
\\
\boldY_{j}=\boldy_{j}, : j \in [2]
 \end{array}
 \!\!\!\right| \!\!\!
\begin{array}{c}
\boldu\{ \ulinea_{j} \} = \boldu_{j},\ulineA_{j}=\ulinea_{j}\\ \boldV_{j}\{ 
\ulineb_{j} \} = \boldv_{j},\ulineB_{j}=\ulineb_{j} \\\boldX_{j}\{ 
\ulinea_{j},\ulineb_{j} 
\}=\boldx_{j}:j \in [2]
\end{array}
 \!\!\!
 \right) = \prod_{t=1}^{m}\mathds{1}_{ \left\{
\!\!\!
\begin{array}{c}
 \boldhatu_{j}(t,1:l) = \\d_{u,j}^{l}(\boldy_{j}(t,1:l)) 
\end{array}
\!\!\!
\right\} } \prod_{i=1}^{l}\mathbb{W}_{\ulineY|\ulineX}\left(\!\!\! 
\begin{array}{c} \boldy_{1}(t,i)\\\boldy_{2} (t,i) \end{array} 
\!\!\!\left|\!\!\! \begin{array}{c}
\boldx_{1}(t,i)\\\boldx_{2}(t,i) \end{array}\!\!\!\right.\right)
 \end{eqnarray}
leads us to breaking down the second factor in a generic term of 
(\ref{Eqn:Step1MACPrelim2-1}) as
\begin{eqnarray}
 \label{Eqn:Step2ICChnlMtrHaveCorrectPMF-GenTerm1}
 \lefteqn{P\left(\!\!\!  \left.
 \begin{array}{c}
\boldu\{ \ulinea_{j} \} = \boldu_{j}, \boldV_{j}\{ \ulineb_{j} \} = \boldv_{j} 
\\
\boldY_{j}=\boldy_{j}, \boldX_{j}\{ \ulinea_{j},\ulineb_{j} 
\}=\boldx_{j}\\
\boldu\{ \ulinehatA_{j} \} = \boldhatu_{j} : j \in [2]
 \end{array}
 \!\!\!\right| \!\!\!
\begin{array}{c}
\ulineA_{j}=\ulinea_{j}\\\ulineB_{j}=\ulineb_{j} \\:j \in [2]
\end{array}
 \!\!\!
 \right) =
 \prod_{t=1}^{m}\left\{  \mathds{1}_{\left\{\!\!\! 
\begin{array}{c} \boldu_{j}(t,1:l) =\\ u^{l}(a_{jt}): j \in [2] 
\end{array}\!\!\!\right\}}\left\{ \prod_{i=1}^{l}\left\{ \prod_{j=1}^{2}
p_{V_{j}}(\boldv_{j}(t,\Pi_{t}(i)))\right. \right.  \right.}\nonumber\\
\label{Eqn:Step2ICChnlMtrHaveCorrectPMF-GenTerm2}
&&\!\!\!\!\!\!\!\!\!\!\!\!\!\!\left. 
\left. \left.p_{X_{j}|UV_{j}}  \left(   \boldx_{j} (t, 
i)\left|\!\!\!  \begin{array}{c}  \boldu_{j}(t,i),\boldv_{j}(t,i) 
\end{array}\!\!\! \right. \right) \right\}
\mathbb{W}_{\ulineY|\ulineX}\left(\!\!\! 
\begin{array}{c} \boldy_{1}(t,i)\\\boldy_{2} (t,i) \end{array} 
\!\!\!\left|\!\!\! \begin{array}{c}
\boldx_{1}(t,i)\\ \boldx_{2}(t,i) \end{array}\!\!\!\right.\right)\right\} 
\mathds{1}_{ \left\{
\!\!\!
\begin{array}{c}
 \boldhatu_{j}(t,1:l) = d_{u,j}^{l}(\boldy_{j}(t,1:l)) 
\end{array}
\!\!\!
\right\} }\right\}.
\end{eqnarray}
Substituting (\ref{Eqn:Step2ICChnlMtrHaveCorrectPMF-GenTerm2}) in 
(\ref{Eqn:Step2ICChnlMtrHaveCorrectPMF}) and following a sequence of steps that 
took us from (\ref{Eqn:Step1MACPrelim2-3}) to (\ref{Eqn:Step1MACPrelim2-12}), 
we have
\begin{eqnarray}
 \label{Eqn:Step2ICPrelim2-1}
 P\left(\!\!\!  
 \begin{array}{c}
\boldu\{ \ulineA_{j} \} = \boldu_{j}, \boldV_{j}\{ \ulineB_{j} \} = \boldv_{j}\\
\boldY_{j}=\boldy_{j} \boldX_{j}\{ \ulineA_{j},\ulineB_{j} 
\}=\boldx_{j}\\\boldu_{j}\{ \ulinehatA_{j} \} = \boldhatu_{j} : j \in [2]
 \end{array}
 \!\!\!\right) =
\sum_{\substack{\ulinea_{1},\ulinea_{2}\\\ulineb_{1},\ulineb_{2}}}\!
P \left(  \!\!\!
\begin{array}{c}
\ulineA_{j}=\ulinea_{j}\\\ulineB_{j}=\ulineb_{j} \\: j \in [2]
\end{array}
 \!\!\! \right)
 P\left(\!\!\!  \left.
 \begin{array}{c}
\boldu\{ \ulinea_{j} \} = \boldu_{j}, \boldV_{j}\{ \ulineb_{j} \} = \boldv_{j} 
\\
\boldY_{j}=\boldy_{j}, \boldX_{j}\{ \ulinea_{j},\ulineb_{j} 
\}=\boldx_{j}\\
\boldu_{j}\{ \ulinehatA_{j} \} = \boldhatu_{j} : j \in [2]
 \end{array}
 \!\!\!\right| \!\!\!
\begin{array}{c}
\ulineA_{j}=\ulinea_{j}\\\ulineB_{j}=\ulineb_{j} \\:j \in [2]
\end{array}
 \!\!\!
 \right) \nonumber\\
 \label{Eqn:Step2ICPrelim2-2}
 = \sum_{\substack{\ulinea_{1},\ulinea_{2}\\\ulineb_{1},\ulineb_{2}}}\!
P \left(  \!\!\!
\begin{array}{c}
\ulineA_{j}=\ulinea_{j}\\\ulineB_{j}=\ulineb_{j} \\: j \in [2]
\end{array}
 \!\!\! \right) \prod_{t=1}^{m}\left\{  \mathds{1}_{\left\{\!\!\! 
\begin{array}{c} \boldu_{j}(t,1:l) =\\ u^{l}(a_{jt}): j \in [2] 
\end{array}\!\!\!\right\}}\left\{ \prod_{i=1}^{l}\left\{ \prod_{j=1}^{2}
p_{V_{j}}(\boldv_{j}(t,i))p_{X_{j}|UV_{j}}(\boldx_{j}(t, 
i)|\boldu_{j}(t,i)\boldv_{j}(t,i)) \right\} \right.  \right.\nonumber\\
\label{Eqn:Step2ICPrelim2-3}
\left. 
\left. 
\mathbb{W}_{\ulineY | \ulineX} ( \boldy_{1}(t,i), \boldy_{2}(t,i) 
| \boldx_{1}(t, i), \boldx_{2}(t,i))\right\} \mathds{1}_{ \left\{
\!\!\!
\begin{array}{c}
 \boldhatu_{j}(t,1:l) = d_{u,j}^{l}(\boldy_{j}(t,1:l)) 
\end{array}
\!\!\!
\right\} }\right\}\end{eqnarray}\begin{eqnarray}
\label{Eqn:Step2ICPrelim2-4}
= \sum_{\substack{\ulinea_{1},\ulinea_{2}}}\!
P \left(  \!\!\!
\begin{array}{c}
\ulineA_{j}=\ulinea_{j}\\: j \in [2]
\end{array}
 \!\!\! \right) \prod_{t=1}^{m}\left\{  \mathds{1}_{\left\{\!\!\! 
\begin{array}{c} \boldu_{j}(t,1:l) =\\ u^{l}(a_{jt}): j \in [2] 
\end{array}\!\!\!\right\}}\left\{ \prod_{i=1}^{l}\left\{ \prod_{j=1}^{2}
p_{V_{j}}(\boldv_{j}(t,i))p_{X_{j}|UV_{j}}(\boldx_{j}(t, 
i)|\boldu_{j}(t,i)\boldv_{j}(t,i)) \right\} \right.  \right.\nonumber\\
\label{Eqn:Step2ICPrelim2-5}
\left. 
\left. 
\mathbb{W}_{\ulineY | \ulineX} ( \boldy_{1}(t,i), \boldy_{2}(t,i) 
| \boldx_{1}(t, i), \boldx_{2}(t,i))\right\} \mathds{1}_{ \left\{
\!\!\!
\begin{array}{c}
 \boldhatu_{j}(t,1:l) = d_{u,j}^{l}(\boldy_{j}(t,1:l)) 
\end{array}
\!\!\!
\right\} }\right\}\nonumber\\
\label{Eqn:Step2ICPrelim2-6}
=
\sum_{\substack{\ulinea_{1},\ulinea_{2}}}\!
\prod_{t=1}^{m}\left\{P\left(\!\!\!
\begin{array}{c}A_{j}=a_{jt}\\: j \in [2]
\end{array}
 \!\!\! \right)  
\mathds{1}_{\left\{\!\!\! 
\begin{array}{c} \boldu_{j}(t,1:l) =\\ u^{l}(a_{jt}): j \in [2] 
\end{array}\!\!\!\right\}}\left\{ \prod_{i=1}^{l}\left\{ \prod_{j=1}^{2}
p_{V_{j}}(\boldv_{j}(t,i))p_{X_{j}|UV_{j}}(\boldx_{j}(t, 
i)|\boldu_{j}(t,i)\boldv_{j}(t,i)) \right\} \right.  \right.\nonumber\\
\label{Eqn:Step2ICPrelim2-7}
\left. 
\left. 
\mathbb{W}_{\ulineY | \ulineX} ( \boldy_{1}(t,i), \boldy_{2}(t,i) 
| \boldx_{1}(t, i), \boldx_{2}(t,i))\right\} \mathds{1}_{ \left\{
\!\!\!
\begin{array}{c}
 \boldhatu_{j}(t,1:l) = d_{u,j}^{l}(\boldy_{j}(t,1:l)) 
\end{array}
\!\!\!
\right\} }\right\}\\
\label{Eqn:Step2ICPrelim2-8}
=
\prod_{t=1}^{m}\left\{
\left[ \sum_{\substack{a_{1} , a_{2}}}
P\left(\!\!\!
\begin{array}{c}A_{j}=a_{j}\\: j \in [2]
\end{array}
 \!\!\! \right)  
\mathds{1}_{\left\{\!\!\! 
\begin{array}{c} \boldu_{j}(t,1:l) =\\ u^{l}(a_{j}): j \in [2] 
\end{array}\!\!\!\right\}}\right]\left\{ \prod_{i=1}^{l}\left\{ \prod_{j=1}^{2}
p_{V_{j}}(\boldv_{j}(t,i))p_{X_{j}|UV_{j}}\left(\boldx_{j}(t, 
i)\left|\!\!\!\begin{array}{c}\boldu_{j}(t,i)\\\boldv_{j}(t,i) 
\end{array}\!\!\!\right)\right\}\right. \right.  \right.\nonumber\\
\label{Eqn:Step2ICPrelim2-9}
\left. 
\left. 
\mathbb{W}_{\ulineY | \ulineX} ( \boldy_{1}(t,i), \boldy_{2}(t,i) 
| \boldx_{1}(t, i), \boldx_{2}(t,i))\right\} \mathds{1}_{ \left\{
\!\!\!
\begin{array}{c}
 \boldhatu_{j}(t,1:l) = d_{u,j}^{l}(\boldy_{j}(t,1:l)) 
\end{array}
\!\!\!
\right\} }\right\}\nonumber\\
\label{Eqn:Step2ICPrelim2-10}
= \prod_{t=1}^{m} p_{\ulinesfU^{l}\ulinesfV^{l}\ulinesfX^{l}\ulinesfY^{l} 
\ulinesfhatU^{l}} 
 \!\!\left(\!\!\!\begin{array}{c}\boldu_{1}(t,1:l),\boldu_{2}(t,1:l),
 \boldv_{1}(t,1:l),\boldv_{2}(t,1:l),
 \boldx_{1}(t,1:l),\\\boldx_{2}(t,1:l),
 \boldy_{1}(t,1:l),\boldy_{2}(t,1:l),
 \boldhatu_{1}(t,1:l),\boldhatu_{2}(t,1:l)\end{array} \!\!\!\right) \nonumber
\end{eqnarray}
where (\ref{Eqn:Step2ICPrelim2-3}) follows from the fact that 
$\prod_{t=1}^{m}\prod_{i=1}^{l}p_{V_{j}}(\boldv_{j}(t,\Pi_{t}(i))) = 
\prod_{t=1}^{m}\prod_{i=1}^{l}p_{V_{j}}(\boldv_{j}(t,i))$ and 
(\ref{Eqn:Step2ICPrelim2-7}) follows from the invariance of the distribution of 
$A_{jt} = d_{K}(\boldk_{j}(t,1:l))$ with $t \in [m]$.

\section{Proof of Equation (\ref{Eqn:Step2ICS1S2HatK1HatK2CrctPMF})}
\label{AppSec:Step2ICS1S2HatK1HatK2CorrectPMF}

Our proof will closely mimic steps that took us from 
(\ref{Eqn:Step1MACEpsilon1}) 
to (\ref{Eqn:Step1S1S2KHatPMF}). We first note the following. Firstly,
\begin{eqnarray}
 \label{Eqn:Step2ICS1S2HatK1HatK2CorrectPMF3}
 P\left(\!\!\!
 \begin{array}{c}
 \ulineB_{j} = \ulineb_{j}:j \in [2]
 \end{array}\!\!\!\left|
 \!\!
 \begin{array}{c}
 \boldS_{1}=\bolds_{1},\boldS_{2}=\bolds_{2}
 \end{array}\!\!\!\right.
 \right) = P\left(\!\!\!
 \begin{array}{c}
 \beta_{j}(\bolds_{j}) = \ulineb_{j}:j \in [2]
 \end{array}\!\!\!\left|
 \!\!
 \begin{array}{c}
 \boldS_{1}=\bolds_{1},\boldS_{2}=\bolds_{2}
 \end{array}\!\!\! \right.
 \right) = \frac{1}{M_{V_{1}}^{l}M_{V_{2}}^{l}}
\end{eqnarray}
owing to the uniform distribution of $\beta_{j}(\ulines_{j}) : j \in [2]$ and 
its independence from the source realization. Secondly, 
\begin{eqnarray}
\label{Eqn:Step2ICS1S2HatK1HatK2CorrectPMF4}
 P\left( \!\!\!
 \begin{array}{c}
 \boldV_{j}\{\ulineb_{j}\} = \boldv_{j} :j \in [2]
 \end{array}\!\!\!\left|
 \!\!
 \begin{array}{c}
 \boldS_{j}=\bolds_{j},\ulineB_{j} = \ulineb_{j}: j \in [2]
 \end{array}\!\!\!\right.
 \right) = \prod_{t=1}^{m}\prod_{i=1}^{l}p_{V_{j}}(\boldv_{j}(t,\Pi_{t}(i)))
\end{eqnarray}
since $\boldV_{j}\{ \ulineb_{j} \}$ is independent of the 
$\boldS_{j},\beta_{j}(\boldS_{j}): j \in [2]$. Thirdly, suppose 
$e_{u}(a_{jt})_{i} 
$ denotes the $i$-th symbol in $e_{u}(a_{jt}) = u^{l} (a_{jt}) \in 
\mathcal{U}^{l}$, then 
\begin{eqnarray}
\label{Eqn:Step2ICS1S2HatK1HatK2CorrectPMF5}
 P\left(  \!\!\!
 \begin{array}{c}
 \boldX_{j}\{\ulinea_{j},\ulineb_{j}\} = \boldx_{j} \\:j \in [2]
 \end{array}\!\!\!\left|
 \!\!
 \begin{array}{c}
 \boldV_{j}\{ \ulineb_{j}\} = \boldv_{j},\boldS_{j}=\bolds_{j},\ulineB_{j} = 
\ulineb_{j}\\: j \in [2]
 \end{array}\!\!\!\right.
 \right) = \prod_{t=1}^{m}\prod_{i=1}^{l}p_{X_{j}|UV_{j}}( 
\boldx_{j} (t,i) | e_{u}(a_{jt})_{i},\boldv_{j}(t,i) ).
\end{eqnarray}
Fourthly, suppose $a_{jt} = 
e_{k}(\boldk_{j}(t,1:l)) : t \in [m]$ and $\boldk_{j}(t,i) = 
f_{j}(\bolds_{j}(t,i))$, then 
\begin{eqnarray}
\label{Eqn:Step2ICS1S2HatK1HatK2CorrectPMF6}
 P\left(  \!\!\!
 \begin{array}{c}
 \boldY_{1} = \boldy_{1}\\ \boldY_{2} = \boldy_{2}
 \end{array}\!\!\!\left|
 \!\!
 \begin{array}{c}
 \boldV_{j}\{ \ulineb_{j}\} = \boldv_{j},\boldS_{j}=\bolds_{j},\ulineB_{j} = 
\ulineb_{j}\\ \boldX_{j}\{\ulinea_{j},\ulineb_{j}\} = \boldx_{j}: j \in [2]
 \end{array}\!\!\!\right.
 \right) = 
\prod_{t=1}^{m}\prod_{i=1}^{l}\mathbb{W}_{\ulineY|\ulineX}\left(\!\!\! 
\begin{array}{c} \boldy_{1}(t,i)\\\boldy_{2}(t,i) \end{array} 
\!\!\!\left|\!\!\! \begin{array}{c}
\boldx_{1}(t,i)\\ \boldx_{2}(t,i) \end{array}\!\!\!\right.\right).
\end{eqnarray}
Suppose for $j \in [2], t \in [m], i \in [l]$, 
we have $\boldk_{j}(t,i) = 
f_{j}(\bolds_{j}(t,i))$ and $a_{jt} = e_{k}(\boldk_{j}(t,1:l))$, then 
substituting for factors from (\ref{Eqn:Step2ICS1S2HatK1HatK2CorrectPMF3}) - 
(\ref{Eqn:Step2ICS1S2HatK1HatK2CorrectPMF6}), we have
\begin{eqnarray}
 \label{Eqn:Step2ICS1S2HatK1HatK2CorrectPMF1}
 \lefteqn{P\!\left(\!\!\! 
 \begin{array}{c}
\boldS_{j}=\bolds_{j},\boldX_{j}\{ \ulinea_{j},\ulineb_{j} \} = 
\boldx_{j},\boldV_{j}\{ 
\ulineb_{j} \} = \boldv_{j}\\
\ulineB_{j}=\ulineb_{j},\boldY_{j} = \boldy_{j}, 
\boldhatK_{j} = \boldhatk_{j}: j \in [2]
\end{array}
 \!\!\!\right) = P\left(\!\!\!\begin{array}{c} 
\boldS_{1}=\bolds_{1}\\\boldS_{2}=\bolds_{2}
\end{array}\!\!\!\right) 
\frac{1}{M_{V_{1}}^{l}M_{V_{2}}^{l}}\prod_{t=1}^{m}\left\{  \left\{ 
\prod_{i=1}^{l}  \mathbb{W}_{\ulineY|\ulineX}\left(\!\!\! 
\begin{array}{c} \boldy_{1}(t,i)\\\boldy_{2}(t,i) \end{array} 
\!\!\!\left|\!\!\! \begin{array}{c}
\boldx_{1}(t,i)\\ \boldx_{2}(t,i) \end{array}\!\!\!\right.\right)\right.  
\right.}\nonumber\\
&& ~~~~~~~~~~~~~~~~~~~~~~~~\left. \left.
\label{Eqn:Step2ICS1S2HatK1HatK2CorrectPMF2}
\left\{\prod_{j=1}^{2}p_{V_{j}}(\boldv_{j}(t,\Pi_{t}(i))) 
  p_{X_{j}|UV_{j}}\left( 
\boldx_{j} (t,i) \left|\! \!\!
\begin{array}{c}e_{u}(a_{jt})_{i}\\\boldv_{j}(t,i)\end{array} 
\!\!\!\right.\right)\right\} 
\right\}\mathds{1}_{\left\{  \!\!\! 
\begin{array}{c}
d_{k}( d_{u}( \boldy_{j}(t,1:l) ) )\\= \boldhatk_{j}(t,1:l)  : j \in [2]
\end{array}
\!\!\!\right\} }\!\!\right\}\\
\label{Eqn:Step2ICS1S2HatK1HatK2App7}
&&\!\!\!\!\!\!\!\!\!\!\!\!=P\left(\!\!\!\begin{array}{c} 
\boldS_{1}=\bolds_{1}\\\boldS_{2}=\bolds_{2}\end{array}\!\!\!\right)
\frac{1}{M_{V_{1}}^{l}M_{V_{2}}^{l}}\prod_{t=1}^{m}\left\{  \left\{ 
\prod_{i=1}^{l}  
\mathbb{W}_{\ulineY|\ulineX}\left(\!\!\! 
\begin{array}{c} \boldy_{1}(t,i)\\\boldy_{2}(t,i) \end{array} 
\!\!\!\left|\!\!\! \begin{array}{c}
\boldx_{1}(t,i)\\ \boldx_{2}(t,i) \end{array}\!\!\!\right.\right)\right.  
\!\!\!\left\{\prod_{j=1}^{2}p_{V_{j}}(\boldv_{j}(t,i)) 
  p_{X_{j}|UV_{j}}\left( 
\boldx_{j} (t,i) \left| \!\!\! 
\begin{array}{c} e_{u}(a_{jt})_{i}\\\boldv_{j}(t,i) \end{array}\!\!\!\right. 
\right)\right\}\right\} \nonumber\\ 
\label{Eqn:Step2ICS1S2HatK1HatK2App8}
&&~~~~~~~~~~~~~~~~~~~~~~~~~~~~~~~~~~~~~~~~~~~~~~~~~~~~~~~~~~~~~~~~~~
\!\!\!\!\left.
\mathds{1}_{\left\{  \!\!\! 
\begin{array}{c}
\boldhatk_{j}(t,1:l) = d_{k}( d_{u}( \boldy_{j}(t,1:l) ) )  : j \in [2]
\end{array}
\!\!\!\right\} }\right\}
\end{eqnarray}
wherein $e_{u}(a_{jt})_{i} $ denotes the $i$-th symbol in $e_{u}(a_{jt}) = u^{l} 
(a_{jt}) \in \mathcal{U}^{l}$, (\ref{Eqn:Step2ICS1S2HatK1HatK2App8}) 
is obtained by re-ordering the 
product 
$\prod_{i=1}^{l}\prod_{t=1}^{m}p_{V_{j}}(\boldv_{j}(t,\Pi_{t}(i))) = 
\prod_{i=1}^{l}\prod_{t=1}^{m}p_{V_{j}}(\boldv_{j}(t,i))$. We now note that the 
marginal 
$p_{\ulineU^{l}}$ wrt pmf in (\ref{Eqn:Step2ICPMFForDecRule}) is given by
\begin{eqnarray}
\label{Eqn:Step2ICS1S2HatK1HatK2App10}
p_{\ulineU^{l}}(u_{1}^{l},u_{2}^{l}) = 
\sum_{\substack{(a_{1},a_{2}) \in \\ 
[M_{u}]\times [M_{u}]}} 
\!\!P(
 \begin{array}{c}
 A_{1}=a_{1},A_{2}=a_{2}
  \end{array})\mathds{1}_{\left\{\substack{ 
 u^{l}(a_{j})=u_{j}^{l}:j \in [2]}\right\}}\mbox{ and hence }
 \nonumber\\
 \label{Eqn:Step2ICS1S2HatK1HatK2App17}
 p_{\ulineV^{l}\ulineX^{l} 
\ulineY^{l} | \ulineU^{l}} ( \ulinev^{l}, \ulinex^{l}, \uliney^{l}|\ulineu^{l}) 
= 
\left[ \prod_{j=1}^{2} \left\{ 
\prod_{i=1}^{l}p_{V_{j}}(v_{ji})p_{X_{j}|UV_{j}}(x_{ji}|u_{ji},v_{ji})\right\} 
\right]\left[ \prod_{i=1}^{l} 
\mathbb{W}_{Y_{1}Y_{2}|X_{1}X_{2}}(y_{1i},y_{2i}| x_{1i},x_{2i}) \right]
 \end{eqnarray}
Using (\ref{Eqn:Step2ICS1S2HatK1HatK2App17}), expression 
(\ref{Eqn:Step2ICS1S2HatK1HatK2App8}) is equal to
\begin{eqnarray}
\label{Eqn:Step2ICS1S2HatK1HatK2App9}
P\left(\!\!\!\begin{array}{c} 
\boldS_{1}=\bolds_{1}\\\boldS_{2}=\bolds_{2}\end{array}\!\!\!\right)\!\frac{1}{
M_{ V_{1}}^{l}M_{V_{2}}^{l}}\!\prod_{t=1}^{m}\! p_{\ulineV^{l}\ulineX^{l} 
\ulineY^{l}|\ulineU^{l}}\!\left( \!\!\!\! \left.
\begin{array}{c}
\boldv_{1}(t,1:l),\boldv_{2}(t,1:l),\boldx_{1}(t,1:l)\\
\boldx_{2}(t,1:l),\boldy_{1}(t,1:l),\boldy_{2}(t,1:l)
\end{array}\!\!\!\right| \!\!\!
\begin{array}{c}
 e_{u}(a_{1t})\\e_{u}(a_{2t})
\end{array}
\!\!\!
\right)
\mathds{1}_{\left\{  \!\!\! 
\begin{array}{c}
d_{k}( d_{u}( \boldy_{j}(t,1:l) ) )\\= \boldhatk_{j}(t,1:l)  : j \in [2]
\end{array}
\!\!\!\right\} }.
\end{eqnarray}
Following from (\ref{Eqn:Step2ICS1S2HatK1HatK2CorrectPMF2}) to 
(\ref{Eqn:Step2ICS1S2HatK1HatK2App9}), 
we conclude that if
\begin{eqnarray}
\label{Eqn:Step2ICS1S2HatK1HatK2App11}
\lefteqn{ \begin{array}{c}
\boldk_{j}(t,i) = 
f_{j}(\bolds_{j}(t,i))\mbox{ and }a_{jt} = 
e_{k}(\boldk_{j}(t,1:l))\\ 
\mbox{for }j \in [2], t \in [m], i \in [l],\mbox{ 
we have}\end{array}\mbox{ 
then }
P\left(\!\!\! 
 \begin{array}{c}
\boldS_{j}=\bolds_{j},\boldX_{j}\{ \ulinea_{j},\ulineb_{j} \} = 
\boldx_{j},\ulineB_{j}=\ulineb_{j}\\
\boldV_{j}\{ 
\ulineb_{j} \} = \boldv_{j},\boldY_{j} = \boldy_{j}, 
\boldhatK_{j} = \boldhatk_{j}: j \in [2]
\end{array}
 \!\!\!\right) =}\nonumber\\&&\!\!\!\!\!\! \!\!P\left(\!\!\!\begin{array}{c} 
\boldS_{1}=\bolds_{1}\\\boldS_{2}=\bolds_{2}\end{array}\!\!\!\right)\!\frac{1}{
M_{ V_{1}}^{l}M_{V_{2}}^{l}}\!\prod_{t=1}^{m}\! p_{\ulineV^{l}\ulineX^{l} 
\ulineY^{l}|\ulineU^{l}}\!\left( \!\!\!\! \left.
\begin{array}{c}
\boldv_{1}(t,1:l),\boldv_{2}(t,1:l),\boldx_{1}(t,1:l)\\
\boldx_{2}(t,1:l),\boldy_{1}(t,1:l),\boldy_{2}(t,1:l)
\end{array}\!\!\!\right| \!\!\!
\begin{array}{c}
 e_{u}(a_{1t})\\e_{u}(a_{2t})
\end{array}
\!\!\!
\right)
\mathds{1}_{\left\{  \!\!\! 
\begin{array}{c} d_{k}( d_{u}( \boldy_{j}(t,1:l) )\\
=\boldhatk_{j}(t,1:l): j \in [2])
\end{array}
\!\!\!\right\} }\!.~~
\end{eqnarray}
Equipped with (\ref{Eqn:Step2ICS1S2HatK1HatK2App11}), we now characterize pmf 
of $\boldS_{1},\boldS_{2},\boldhatK_{1},\boldhatK_{2}$. Note that if 
$\boldk_{j}(t,i) = 
f_{j}(\bolds_{j}(t,i))$ and $a_{jt} = e_{k}(\boldk_{j}(t,1:l))$ for $j \in [2], 
t \in [m], i \in [l]$, we have 
\begin{eqnarray}
 \label{Eqn:Step2ICS1S2HatK1HatK2App12}
 \lefteqn{\!\!P\left(\!\!\!  
 \begin{array}{c}
\boldS_{j}=\bolds_{j},\boldhatK_{j} = \boldhatk_{j}\\: j \in [2]
 \end{array}
 \!\!\!\right) = 
\sum_{\ulineb_{1},\ulineb_{2}}
\sum_{\substack{\boldv_{1} \in 
\boldCalV_{1} \\ \boldv_{2} \in \boldCalV_{2}}}
\sum_{\substack{\boldx_{1} \in 
\boldCalX_{1} \\ \boldx_{2} \in \boldCalX_{2}}}
\sum_{\substack{\boldy_{1} \in 
\boldCalY_{1}\\ \boldy_{2} \in 
\boldCalY_{2} }}
P \left(
\!\!\!  
\begin{array}{c}
\boldS_{j}=\bolds_{j},\boldX_{j}\{ \ulinea_{j},\ulineb_{j} \} = 
\boldx_{j},\boldV_{j}\{ 
\ulineb_{j} \} = \boldv_{j}\\
\ulineB_{j}=\ulineb_{j},\boldY_{j} = \boldy_{j}, 
\boldhatK_{j} = \boldhatk_{j}: j \in [2]
\end{array}
 \!\!\!
\right)}\nonumber\\
&\!\!\!\!\!\!\!\!=&\!\! \sum_{\ulineb_{1},\ulineb_{2}}
\sum_{\substack{\boldv_{1} \in 
\boldCalV_{1} \\ \boldv_{2} \in \boldCalV_{2}}}
\sum_{\substack{\boldx_{1} \in 
\boldCalX_{1} \\ \boldx_{2} \in \boldCalX_{2}}}
\sum_{\substack{\boldy_{1} \in 
\boldCalY_{1}\\ \boldy_{2} \in 
\boldCalY_{2} }}
\!\!\!
\frac{P\left(\!\!\!\begin{array}{c} 
\boldS_{1}=\bolds_{1}\\\boldS_{2}=\bolds_{2}\end{array}\!\!\!\right)}{
M_{ V_{1}}^{l}M_{V_{2}}^{l}}\!\prod_{t=1}^{m}\! p_{\ulineV^{l}\ulineX^{l} 
\ulineY^{l}|\ulineU^{l}}\!\left( \!\!\!\! \left.
\begin{array}{c}
\boldv_{1}(t,1:l),\boldv_{2}(t,1:l)\\\boldx_{1}(t,1:l)
\boldx_{2}(t,1:l)\\\boldy_{1}(t,1:l), \boldy_{2}(t,1:l)
\end{array}\!\!\!\right| \!\!\!
\begin{array}{c}
 e_{u}(a_{1t})\\e_{u}(a_{2t})
\end{array}
\!\!\!
\right)
\mathds{1}_{\left\{  \!\!\! 
\begin{array}{c} d_{k}( d_{u}( \boldy_{j}(t,1:l) )\\
=\boldhatk_{j}(t,1:l): j \in [2])
\end{array}
\!\!\!\right\} }\nonumber\\
\label{Eqn:Step2ICS1S2HatK1HatK2App13}
&\!\!\!\!\!\!\!\!=&\!\! \sum_{\substack{\boldy_{1} \in 
\boldCalY_{1}\\ \boldy_{2} \in 
\boldCalY_{2} }}\!\!\!\!P\left(\!\!\!\begin{array}{c} 
\boldS_{1}=\bolds_{1}\\\boldS_{2}=\bolds_{2}\end{array}\!\!\!\right)\prod_{t=1}^
{ m } p_{\ulineY^{l}|\ulineU^{l}}\!\left( \!\!\!\!
\begin{array}{c}
\boldy_{1}(t,1:l)\\\boldy_{2}(t,1:l)
\end{array}\!\!\left| \!\!
\begin{array}{c}
 e_{u}(e_{k}(\boldk_{1}(t,1:l)))\\e_{u}(e_{k}(\boldk_{2}(t,1:l)))
\end{array}\right.
\!\!\!
\right)
\mathds{1}_{\left\{  \!\!\! 
\begin{array}{c}
\boldhatk_{j}(t,1:l) =d_{k}( d_{u}( \boldy_{j}(t,1:l) ) ): j \in [2] 
\end{array}
\!\!\!\right\} }.\end{eqnarray}
Since the above sum is over all of 
$\boldCalY_{1}\times \boldCalY_{2}$, we rename dummy variables 
$\boldy_{1}(t,1:l),\boldy_{2}(t,1:l)$ and we use 
(\ref{Eqn:Step2ICSourceCodeDecodingPMFAux}), 
(\ref{Eqn:Step2ICSourceCodeDecodingPMF}) to conclude that 
(\ref{Eqn:Step2ICS1S2HatK1HatK2App13}) is equal to
\begin{eqnarray}
\lefteqn{\sum_{\substack{\boldy_{1} \in 
\boldCalY_{1}\\ \boldy_{2} \in 
\boldCalY_{2} }}\prod_{t=1}^
{ m } \left\{ \prod_{i=1}^{l} \mathbb{W}_{S_{1}S_{2}}\left(\!\!\!  
 \begin{array}{c}\bolds_{1}(t,i)\\\bolds_{2} (t ,i)\end{array}
 \!\!\!
 \right)\right\} 
p_{\ulineY^{l}
|\ulineU^{l}}\!\left( \!\!\!\! 
\begin{array}{c}
\boldy_{1}(t,1:l)\\ \boldy_{2}(t,1:l)
\end{array}\!\!\left| \!\!
\begin{array}{c}
 e_{u}(e_{k}(\boldk_{1}(t,1:l)))\\e_{u}(e_{k}(\boldk_{2}(t,1:l)))
\end{array}\right.
\!\!\!
\right)
\mathds{1}_{\left\{  \!\!\! 
\begin{array}{c} d_{k}( d_{u}( \boldy_{j}(t,1:l) )\\
=\boldhatk_{j}(t,1:l): j \in [2])
\end{array}
\!\!\!\right\} }}
\nonumber\\
&=& \prod_{t=1}^
{ m } \left[\left\{ \prod_{i=1}^{l} \mathbb{W}_{S_{1}S_{2}}\left(\!\!\!  
 \begin{array}{c}\bolds_{1}(t,i)\\\bolds_{2} (t ,i)\end{array}
 \!\!\!
 \right)\right\}  \left\{ \sum_{y^{l} \in \mathcal{Y}^{l}}
p_{\ulineY^{l}
|\ulineU^{l}}\!\left( \!\!\!\! 
\begin{array}{c}
\boldy_{1}(t,1:l)\\ \boldy_{2}(t,1:l)
\end{array}\!\!\left| \!\!
\begin{array}{c}
 e_{u}(e_{k}(\boldk_{1}(t,1:l)))\\e_{u}(e_{k}(\boldk_{2}(t,1:l)))
\end{array}\right.
\!\!\!
\right)
\mathds{1}_{\left\{  \!\!\! 
\begin{array}{c} d_{k}( d_{u}( \boldy_{j}(t,1:l) )\\
=\boldhatk_{j}(t,1:l): j \in [2])
\end{array}
\!\!\!\right\} } \right\}\right]
\nonumber\nonumber\\
&=& \prod_{t=1}^
{ m } \left[\left\{ \prod_{i=1}^{l} \mathbb{W}_{S_{1}S_{2}}\!\!\left(\!\!\!  
 \begin{array}{c}\bolds_{1}(t,i)\\\bolds_{2} (t ,i)\end{array}
 \!\!\!
 \right)\right\}  \left\{  p_{\hatK_{1}^{l}\hatK_{2}^{l}|K_{1}^{l}K_{2}^{l}} 
\left(\!\!\!\begin{array}{c}\boldhatk_{1}(t,1:l)\\\boldhatk_{2}(t,1:l) 
\end{array}\!\!\!\left| \!\!\begin{array}{c} 
f_{1}(\bolds_{1}(t,1))\cdots f_{1}(\bolds_{1}(t,l))\\ 
f_{2}(\bolds_{2}(t,1))\cdots f_{2}(\bolds_{2}(t,l))  \end{array}\!\! \!\right. 
\right) \right\}\right]
\nonumber\nonumber\\
&=& \prod_{t=1}^
{ m } p_{S_{1}^{l}S_{2}^{l}\hatK_{1}^{l}\hatK_{2}^{l}}\left(\!\!\!  
 \begin{array}{c}\bolds_{1}(t,1:l),\bolds_{2} (t , 
1:l),\boldhatk_{1}(t,1:l),\boldhatk_{2}(t,1:l)\end{array}
 \!\!\!
 \right). \nonumber
\end{eqnarray}
We therefore have
\begin{eqnarray}
 \label{Eqn:AppStep1S1S2KHatPMF}
  P\left(\!\!\!  
 \begin{array}{c}
\boldS_{1}=\bolds_{1}, \boldS_{2}=\bolds_{2},
\boldhatK_{1} = \boldhatk_{1},
\boldhatK_{2} = \boldhatk_{2}
 \end{array}
 \!\!\!\right)
 = \prod_{t=1}^
{ m } p_{S_{1}^{l}S_{2}^{l}\hatK_{1}^{l}\hatK_{2}^{l}}\left(\!\!\!  
 \begin{array}{c}\bolds_{1}(t,1:l),\bolds_{2} (t , 
1:l),\boldhatk_{1}(t,1:l),\boldhatk_{2}(t,1:l)\end{array}
 \!\!\!
 \right)
\end{eqnarray}
\section{Proof of (\ref{Eqn:Step2ICOuterCodeError-2})}
\label{AppSec:Step2ICOuterCode1stErrEvent}
Note that
\begin{eqnarray}
\label{Eqn:Step2ICOutCodeErr1stEventApp1}
 P\left(\!\!\!  
 \begin{array}{c}
\boldu\{ \ulineA_{j} \} = \boldu_{j}, \boldV_{j}\{ \ulineB_{j} \} = \boldv_{j}\\
\boldX_{j}\{ \ulineA_{j},\ulineB_{j} 
\}=\boldx_{j},\boldY_{j}=\boldy_{j}\\\boldu \{ \hatA_{j}\} = \boldhatu_{j}: j 
\in [2] \\\Pi_{t} =\pi_{t}: t \in [m]
 \end{array}
 \!\!\!\right) =
\sum_{\substack{\ulinea_{1},\ulinea_{2}\\\ulineb_{1},\ulineb_{2}}}\!
P \left(  \!\!\!
\begin{array}{c}
\ulineA_{j}=\ulinea_{j}\\\ulineB_{j}=\ulineb_{j} \\: j \in [2]
\end{array}
 \!\!\! \right)
 P\left(\!\!\!  \left.
 \begin{array}{c}
\boldu\{ \ulinea_{j} \} = \boldu_{j}, \boldV_{j}\{ \ulineb_{j} \} = \boldv_{j} 
\\
\boldX_{j}\{ \ulinea_{j},\ulineb_{j} 
\}=\boldx_{j} ,\boldY_{j}=\boldy_{j}\\\boldu \{ \hatA_{j}\} = \boldhatu_{j}: j 
\in [2] \\\Pi_{t} =\pi_{t}: t \in [m]
 \end{array}
 \!\!\!\right| \!\!\!
\begin{array}{c}
\ulineA_{j}=\ulinea_{j}\\\ulineB_{j}=\ulineb_{j} \\:j \in [2]
\end{array}
 \!\!\!
 \right),
 \end{eqnarray}
We consider the second factor of a generic term in the above sum and break it 
up using (\ref{Eqn:Step1MAC-E1First-0}), (\ref{Eqn:Step1MAC-E1FirstMid}), 
(\ref{Eqn:Step1MAC-E1First-2}) and in addition
\begin{eqnarray}
 \label{Eqn:Step2ICOutCodeErr1stEventApp2}
P\left(\!\!\!  \left.
 \begin{array}{c}
\boldu\{ \ulinehatA_{j} \} = \boldhatu_{j} 
\\
\boldY_{j}=\boldy_{j}\\ : j \in [2]
 \end{array}
 \!\!\!\right| \!\!\!
\begin{array}{c}
\boldu\{ \ulinea_{j} \} = \boldu_{j},\boldX_{j}\{ 
\ulinea_{j},\ulineb_{j} 
\}=\boldx_{j}\\ 
\boldV_{j}\{ 
\ulineb_{j} \} = \boldv_{j},\ulineA_{j}=\ulinea_{j},\ulineB_{j}=\ulineb_{j} 
\\:j \in [2], \Pi_{t}=\pi_{t} : t \in [m]
\end{array}
 \!\!\!
 \right) = \prod_{t=1}^{m}\mathds{1}_{ \left\{
\!\!\!
\begin{array}{c}
 \boldhatu_{j}(t,1:l) = \\d_{u,j}^{l}(\boldy_{j}(t,1:l)) 
\end{array}
\!\!\!
\right\} } \prod_{i=1}^{l}\mathbb{W}_{\ulineY|\ulineX}\left(\!\!\! 
\begin{array}{c} \boldy_{1}(t,i)\\\boldy_{2} (t,i) \end{array} 
\!\!\!\left|\!\!\! \begin{array}{c}
\boldx_{1}(t,i)\\\boldx_{2}(t,i) \end{array}\!\!\!\right.\right)
\end{eqnarray}
The RHS of (\ref{Eqn:Step1MAC-E1FirstMid}), (\ref{Eqn:Step1MAC-E1First-2}) and 
\ref{Eqn:Step2ICOutCodeErr1stEventApp2} are invariant with $\pi_{t}: t \in [m]$ 
and hence
\begin{eqnarray}
 \label{Eqn:Step2ICOutCodeErr1stEventApp3}
 P\left(\!\!\!  \left.
 \begin{array}{c}
\boldu\{ \ulinea_{j} \} = \boldu_{j}, \boldV_{j}\{ \ulineb_{j} \} = \boldv_{j} 
, \boldY_{j}=\boldy_{j}\\\boldX_{j}\{ \ulinea_{j},\ulineb_{j} 
\}=\boldx_{j}, \boldu \{ \hatA_{j}\} = \boldhatu_{j}: 
j 
\in [2] 
 \end{array}
 \!\!\!\right| \!\!\!
\begin{array}{c}
\ulineA_{j}=\ulinea_{j},\ulineB_{j}=\ulineb_{j} :j \in [2]\\\Pi_{t} 
=\pi_{t}: t \in [m]
\end{array}
 \!\!\!
 \right)
\end{eqnarray}
is invariant with $\pi_{t}: t \in [m]$. Moreover, from 
(\ref{Eqn:Step1MAC-E1First-0}), we have
\begin{eqnarray}
 \label{Eqn:Step2ICOutCodeErr1stEventApp4}
 P(\Pi_{t} = \pi_{t}: t \in [m] | 
\ulineA_{j}=\ulinea_{j},\ulineB_{j}=\ulineb_{j} :j \in [2] ) = \frac{1}{l!}^{m} 
 =  P(\Pi_{t} = \pi_{t}: t \in [m]). \nonumber
\end{eqnarray}
It is now easy to verify
\begin{eqnarray}
 \label{Eqn:Step2ICOutCodeErr1stEventApp5}
P\left(\!\!\!  
 \begin{array}{c}
\boldu\{ \ulineA_{j} \} = \boldu_{j}, \boldV_{j}\{ \ulineB_{j} \} = \boldv_{j}\\
\boldX_{j}\{ \ulineA_{j},\ulineB_{j} 
\}=\boldx_{j},\boldY_{j}=\boldy_{j}\\\boldu \{ \hatA_{j}\} = \boldhatu_{j}: j 
\in [2], \Pi_{t} =\pi_{t}: t \in [m]
 \end{array}
 \!\!\!\right) = P\left(\!\!\!  
 \begin{array}{c}
\boldu\{ \ulineA_{j} \} = \boldu_{j}, \boldV_{j}\{ \ulineB_{j} \} = \boldv_{j}\\
\boldX_{j}\{ \ulineA_{j},\ulineB_{j} 
\}=\boldx_{j},\boldY_{j}=\boldy_{j}\\\boldu \{ \hatA_{j}\} = \boldhatu_{j}: j 
\in [2]  \end{array}
 \!\!\!\right) P\left(\!\!\! \begin{array}{c}\Pi_{t} = \pi_{t}\\: t \in [m] 
\end{array} \!\!\!\right).
\end{eqnarray}
The fact that
\begin{eqnarray}
\label{Eqn:Step2ICOutCodeErr1stEventApp6}
 \boldU_{j} \define 
\boldu 
\{\ulineA_{j} \}, \boldV_{j} \define 
\boldV_{j}\{ \ulineB_{j} \}, \boldX_{j} \define \boldX_{j}\{ 
\ulineA_{j},\ulineB_{j}\},\boldY_{j},\boldhatU_{j} \define \boldu \{ 
\ulinehatA_{j}\} : j \in [2]
\end{eqnarray}
is IID with pmf 
$p_{\ulineU^{l}\ulineV^{l}\ulineX^{l}\ulineY^{l}\ulinehatU^{l}}$ follows from 
(\ref{Eqn:Step2ICChnlMtrCorrectPMFConcl}) which is proven in Appendix 
\ref{AppSec:Step2ICChnlMtrCorrectPMF}. This completes proof of 
(\ref{Eqn:Step2ICOuterCodeError-2}).
\section{Analysis of $\mathscr{E}_{1}$ in Theorem \ref{Thm:MACStep1} : Joint 
typicality of the Legitimate Codewords}
\label{AppSec:PE2Event1FromFirstPrin}
Recall that our goal is to prove, for sufficiently large $m$, existence of a 
specific code $\boldv_{j}\{\cdot 
\}, \boldx_{j}\{ 
\cdot,\cdot\} : j \in [2] \pi_{t}: t \in [m]$ such that 
$P((v_{1i}^{m}(B_{1i}),v_{2i}^{m}(B_{2i}),\boldY^{\pi}(1:m,i))\notin 
T_{\alpha}^{m}(p_{\underline{\mathscr{V}}\mathscr{Y}}))$ for that code can be 
made arbitrarily small. Instead of just analyzing the probability of this 
event with respect to a random code (as we have done in Section 
\ref{SubSec:MACStep1GeneralizationSeparateDecoding}), in the following 
analysis, we first characterize an upper bound on the probability of the event 
in question for the \textit{specific code}. (\ref{Eqn:PE2Full-ParticularCode2}) 
characterizes such an upper bound. This is just the above error event for a 
specific code. We then average this upper bound on the ensemble of codes with 
respect to the pmf of the code (\ref{Eqn:CodePMF}). We prove that this average 
shrinks to $0$ exponentially in $m$. We concede that, in principle, there is no 
difference in the approach employed in Section 
\ref{SubSec:MACStep1GeneralizationSeparateDecoding} and here. However, since we 
have not undertaken, in any prior information theory work, analysis of any 
coding technique similar to that one proposed here, and moreover, our coding 
technique involves certain codes that remain fixed through the randomization, 
we deem it necessary to present the following analysis from first principles.

\ifTITVersion\begin{eqnarray}
\lefteqn{P\left( \!\!\left(\!\!\! 
\begin{array}{c}v_{1i}^{m}(B_{1i}),v_{2i}^{m}(B_{2i})\\\boldY^{\pi}(1:m,i)
\end{array}\!\!\!\right) \notin 
T_{\alpha}^{m}(p_{\underline{\mathscr{V}}\mathscr{Y}}) 
\right)}
\label{Eqn:PE2Defn}
\\
 %==============nextline
&=&\!\!\!\sum_{\substack{\ulinea_{1},\ulinea_{2}\\\ulineb_{1},\ulineb_{2}}}
P\left(  
\!\!\!\begin{array}{c}\ulineA_{j}=\ulinea_{j}\\\ulineB_{j}=\ulineb_{j}\\:j \in 
[2]\end{array},\!\left(\!\!\!\begin{array}{c}v_{1i}^{m}(b_{1i}),v_{2i}^{m}(b_{
2i 
})\\\boldY^{\pi}(1:m,i)\end{array}\!\!\!\right) \notin 
T_{\alpha}^{m}(p_{\underline{\mathscr{V}}\mathscr{Y}}) \right)\nonumber\\
%==============nextline
&=&\!\!\! \sum_{\substack{\ulinea_{1},\ulinea_{2}\\\ulineb_{1},\ulineb_{2}}}
P\left(  
\!\!\!\begin{array}{c}\ulineA_{j}=\ulinea_{j}\\\ulineB_{j}=\ulineb_{j}\\:j \in 
[2]\end{array},\!
\left[\!\!\!
\begin{array}{c}\boldv_{1}\{\ulineb_{1}\} 
\\\boldv_{ 2}\{\ulineb_{2}\} 
\\\boldY
\end{array} 
\!\!\!\right]^{\pi}\!\!\!\!(1:m,i) ~\notin 
T_{\alpha}^{m}(p_{\underline{\mathscr{V}}\mathscr{Y}}) \right)\nonumber\\
 %==================nextline
&\leq&\!\!\! 
\sum_{\substack{\ulinea_{1},\ulinea_{2}\\\ulineb_{1},\ulineb_{2}}}P\left(  
\!\!\!\begin{array}{c}\ulineA_{j}=\ulinea_{j}\\\ulineB_{j}=\ulineb_{j}\\:j \in 
[2]\end{array}\!\begin{array}{c}\left[\!\!\!\begin{array}{c}\boldu\{\ulinea_{j}
\},\boldv_{j}\{\ulineb_{j}\}\\\boldx_{j}\{ \ulinea_{j},\ulineb_{j}\}:j \in 
[2]\\\boldY(1:m,i)\end{array}\!\!\!\right]^{\pi}\!\!\!\!(1:m,i)\\\notin 
T_{\alpha}^{m}(p_{\underline{\mathscr{U}\mathscr{V}\mathscr{X}}\mathscr{Y}})\end
{array}  
\right)\nonumber
\end{eqnarray}\fi
\ifPeerReviewVersion\begin{eqnarray}
\lefteqn{P\left( \!\!\left(\!\!\! 
\begin{array}{c}v_{1i}^{m}(B_{1i}),v_{2i}^{m}(B_{2i})\\\boldY^{\pi}(1:m,i)
\end{array}\!\!\!\right) \notin 
T_{\alpha}^{m}(p_{\underline{\mathscr{V}}\mathscr{Y}}) 
\right)
\label{Eqn:PE2Defn}
 %==============nextline
=\!\!\!\sum_{\substack{\ulinea_{1},\ulinea_{2}\\\ulineb_{1},\ulineb_{2}}}
P\left(  
\!\!\!\begin{array}{c}\ulineA_{j}=\ulinea_{j}\\\ulineB_{j}=\ulineb_{j}\\:j \in 
[2]\end{array},\!\left(\!\!\!\begin{array}{c}v_{1i}^{m}(b_{1i}),v_{2i}^{m}(b_{
2i 
})\\\boldY^{\pi}(1:m,i)\end{array}\!\!\!\right) \notin 
T_{\alpha}^{m}(p_{\underline{\mathscr{V}}\mathscr{Y}}) \right)}\\
%==============nextline
&=&\!\!\! \sum_{\substack{\ulinea_{1},\ulinea_{2}\\\ulineb_{1},\ulineb_{2}}}
P\left(  
\!\!\!\begin{array}{c}\ulineA_{j}=\ulinea_{j}\\\ulineB_{j}=\ulineb_{j}\\:j \in 
[2]\end{array}\!\!,\!
\left[\!\!\!
\begin{array}{c}\boldv_{1}\{\ulineb_{1}\} 
\\\boldv_{ 2}\{\ulineb_{2}\} 
\\\boldY
\end{array} 
\!\!\!\right]^{\pi}\!\!\!\!(1:m,i) ~\notin 
T_{\alpha}^{m}(p_{\underline{\mathscr{V}}\mathscr{Y}}) \right)\nonumber
 %==================nextline
\leq\!\!\! 
\sum_{\substack{\ulinea_{1},\ulinea_{2}\\\ulineb_{1},\ulineb_{2}}}P\left(  
\!\!\!\begin{array}{c}\ulineA_{j}=\ulinea_{j}\\\ulineB_{j}=\ulineb_{j}\\:j \in 
[2]\end{array}\!\begin{array}{c}\left[\!\!\!\begin{array}{c}\boldu\{\ulinea_{j}
\},\boldv_{j}\{\ulineb_{j}\}\\\boldx_{j}\{ \ulinea_{j},\ulineb_{j}\}:j \in 
[2]\\\boldY(1:m,i)\end{array}\!\!\!\right]^{\pi}\!\!\!\!(1:m,i)\\\notin 
T_{\alpha}^{m}(p_{\underline{\mathscr{U}\mathscr{V}\mathscr{X}}\mathscr{Y}})\end
{array}  
\!\!\!\right)\nonumber
\end{eqnarray}\fi
where the last inequality follows from the fact that projections of jointly 
typical sets are jointly typical. In the next step, we sum over all 
possibilities for the matrix of received symbols and employ the indicator 
function to count only those matrices that satisfy the event of our interest. 
In particular, an upper bound on the previous expression is
\ifPeerReviewVersion\begin{eqnarray}
  %==================nextline
\sum_{\substack{\ulinea_{1},\ulinea_{2}\\\ulineb_{1},\ulineb_{2}}}
\sum_{\substack{\boldy\in\\\mathcal{Y}^{m \times l}}}\!\! P\left(  
\!\!\!
\begin{array}{c}
\ulineA_{j}=\ulinea_{j}\\\ulineB_{j}=\ulineb_{j}\\:j \in 
[2]\end{array}
\!\!\right)\mathds{1}_{\left\{\!\!\! 
\begin{array}{c}\left[\!\!\!\begin{array}{c}\boldu\{\ulinea_{j}\},\boldv_{j}\{
\ulineb_{j}\}\\\boldx_{j}\{ \ulinea_{j},\ulineb_{j}\}\\:j \in 
[2],\boldy\end{array}\!\!\!\right]^{\pi}\!\!\!\!(1:m,i)\\\notin 
T_{\alpha}^{m}(p_{\underline{\mathscr{U}\mathscr{V}\mathscr{X}}\mathscr{Y}})\end
{array}
\!\!\!\right\}}
 %==========nextline
 \label{Eqn:PE2Full-ParticularCode1}
\prod_{t=1}^{m}\prod_{i=1}^{l}\mathbb{W}_{Y|X_{1}X_{2}}\left( 
\boldy(t,i)\left| 
\!\!\begin{array}{c}\boldx_{1}\{\ulinea_{1},\ulineb_{1}\}(t,i)\\\boldx_{2}\{
\ulinea_{2},\ulineb_{2}\}(t,i)\end{array}\!\!\right.\right)~~~
 %===========nextline
\end{eqnarray}\fi
\ifTITVersion\begin{eqnarray}
  %==================nextline
\lefteqn{\sum_{\substack{\ulinea_{1},\ulinea_{2}\\\ulineb_{1},\ulineb_{2}}}
\sum_{\substack{\boldy\in\\\mathcal{Y}^{m \times l}}}\!\! P\left(  
\!\!\!
\begin{array}{c}
\ulineA_{j}=\ulinea_{j}\\\ulineB_{j}=\ulineb_{j}\\:j \in 
[2]\end{array}
\!\!\right)\mathds{1}_{\left\{\!\!\! 
\begin{array}{c}\left[\!\!\!\begin{array}{c}\boldu\{\ulinea_{j}\},\boldv_{j}\{
\ulineb_{j}\}\\\boldx_{j}\{ \ulinea_{j},\ulineb_{j}\}\\:j \in 
[2],\boldy\end{array}\!\!\!\right]^{\pi}\!\!\!\!(1:m,i)\\\notin 
T_{\alpha}^{m}(p_{\underline{\mathscr{U}\mathscr{V}\mathscr{X}}\mathscr{Y}})\end
{array}
\!\!\!\right\}}}\nonumber\\
 %==========nextline
 \label{Eqn:PE2Full-ParticularCode1}
&&\times~\prod_{t=1}^{m}\prod_{i=1}^{l}\mathbb{W}_{Y|X_{1}X_{2}}\left( 
\boldy(t,i)\left| 
\!\!\begin{array}{c}\boldx_{1}\{\ulinea_{1},\ulineb_{1}\}(t,i)\\\boldx_{2}\{
\ulinea_{2},\ulineb_{2}\}(t,i)\end{array}\!\!\right.\right)~~~
 %===========nextline
\end{eqnarray}\fi
(\ref{Eqn:PE2Full-ParticularCode1}) is an upper bound on (\ref{Eqn:PE2Defn}) 
for 
the particular code of interest. Before we average over the ensemble of codes, 
we express this as a sum over the possible $\boldu_{1},\boldu_{2}$ matrices. To 
include only terms corresponding to matrices $\boldu_{j}\{ \ulinea_{j}\}: j \in 
[2]$ in our count, we employ the indicator function. 
(\ref{Eqn:PE2Full-ParticularCode1}) is equal to
\ifPeerReviewVersion\begin{eqnarray}
  %==================nextline
\!\!\!\!\!\sum_{\substack{\ulinea_{1},\ulinea_{2}\\\ulineb_{1},
\ulineb_ {2 }} }
\sum_{\substack{\boldy\in\\\mathcal{Y}^{m \times l}}}\sum_{\substack{\boldu_{1} 
\in \mathcal{U}^{m \times l}\\\boldu_{2} \in \mathcal{U}^{m \times 
l}}}\!\!\!\!\!\! 
P\!\left(  
\!\!\!
\begin{array}{c}
\ulineA_{j}=\ulinea_{j}\\\ulineB_{j}=\ulineb_{j}\\:j \in 
[2]\end{array}
\!\!\right)\!\mathds{1}_{\left\{\!\!\!\!\!
\begin{array}{c}\left[\!\!\!\begin{array}{c}\boldu_{j},\boldv_{j}\{
\ulineb_{j}\}\\\boldx_{j}\{ \ulinea_{j},\ulineb_{j}\}\\:j \in 
[2],\boldy\end{array}\!\!\!\right]^{\pi}\!\!\!\!(1:m,i)\\\notin 
T_{\alpha}^{m}(p_{\underline{\mathscr{U}\mathscr{V}\mathscr{X}}\mathscr{Y}})\end
{array}
\!\!\!\right\}}
 %==========nextline
 \label{Eqn:PE2Full-ParticularCode2}
\!\!\mathds{1}_{\left\{ \!\!\!\!
\begin{array}{c}\boldu\{\ulinea_{j}\}=\\\boldu_{j}:j \in [2] 
\end{array}\!\!\!\!\right\}}\!  
\prod_{t=1}^{m}\prod_{i=1}^{l}\!\mathbb{W}_{Y|\ulineX}\!\left( 
\boldy(t,i)\left| 
\!\!\begin{array}{c}\boldx_{1}\{\ulinea_{1},\ulineb_{1}\}(t,i)\\\boldx_{2}\{
\ulinea_{2},\ulineb_{2}\}(t,i)\end{array}\!\!\right.\!\!\right)\!\!.
 %===========nextline
\end{eqnarray}\fi
\ifTITVersion\begin{eqnarray}
  %==================nextline
\lefteqn{\!\!\!\!\!\sum_{\substack{\ulinea_{1},\ulinea_{2}\\\ulineb_{1},
\ulineb_ {2 }} }
\sum_{\substack{\boldy\in\\\mathcal{Y}^{m \times l}}}\sum_{\substack{\boldu_{1} 
\in \mathcal{U}^{m \times l}\\\boldu_{2} \in \mathcal{U}^{m \times 
l}}}\!\!\!\! 
P\!\left(  
\!\!\!
\begin{array}{c}
\ulineA_{j}=\ulinea_{j}\\\ulineB_{j}=\ulineb_{j}\\:j \in 
[2]\end{array}
\!\!\right)\!\mathds{1}_{\left\{\!\!\!\!\!
\begin{array}{c}\left[\!\!\!\begin{array}{c}\boldu_{j},\boldv_{j}\{
\ulineb_{j}\}\\\boldx_{j}\{ \ulinea_{j},\ulineb_{j}\}\\:j \in 
[2],\boldy\end{array}\!\!\!\right]^{\pi}\!\!\!\!(1:m,i)\\\notin 
T_{\alpha}^{m}(p_{\underline{\mathscr{U}\mathscr{V}\mathscr{X}}\mathscr{Y}})\end
{array}
\!\!\!\right\}}}\nonumber\\
 %==========nextline
 \label{Eqn:PE2Full-ParticularCode2}
\!\!&\!\!\!\!\!\!\!\!\!&\!\!\!\!\!\!\!\!\!\!\!\!\!\mathds{1}_{\left\{ \!\!\!\!
\begin{array}{c}\boldu\{\ulinea_{j}\}=\\\boldu_{j}:j \in [2] 
\end{array}\!\!\!\!\right\}}\!  
\prod_{t=1}^{m}\prod_{i=1}^{l}\!\mathbb{W}_{Y|\ulineX}\!\left( 
\boldy(t,i)\left| 
\!\!\begin{array}{c}\boldx_{1}\{\ulinea_{1},\ulineb_{1}\}(t,i)\\\boldx_{2}\{
\ulinea_{2},\ulineb_{2}\}(t,i)\end{array}\!\!\right.\!\!\right)\!\!.
 %===========nextline
\end{eqnarray}\fi
We now average the above quantity over the ensemble of codes. In particular, we 
multiply the above with the probability of the code and sum. For ease of 
reference, we recall the pmf on the ensemble 
of codes. Denoting the components of the random code via upper case letters, we 
recall that (i) the $m$ surjective maps $\Pi_{t}:t \in [m]$ are mutually 
independent and uniformly distributed over the entire collection of surjective 
maps over $[l]$, (ii) each codeword in the collection $(V_{ji}^{m}(b_{j})\in 
\mathcal{V}_{j}^{m}:b_{j}\in [M_{V_{j}}],i \in [l], j\in [2])$ is mutually 
independent of the others and $V_{ji}^{m}(b_{ji}) \sim 
\prod_{t=1}^{m}p_{V_{j}}(\cdot)$, and (iii) $(\boldX_{j}(u,v_{j})\in 
\mathcal{X}_{j}^{m \times 
l}:u \in \mathcal{U}^{m \times l}, v_{j} \in \mathcal{V}_{j}^{m \times l})$ is 
mutually independent and $\boldX_{j}(u,v_{j}) \sim 
\prod_{t=1}^{m}\prod_{i=1}^{l}p_{X_{j}|UV_{j}}(\cdot|u(t,i),v_{j}(t,i))$. 
Recall that $\boldV_{j}\{ \ulineb_{i}\}^{\Pi}(1:m,i) = V_{ji}^{m}(b_{ji})$, and 
hence, conditioned on $\{ V_{ji}(b_{ji})_{t}=v,u(a_{1t})_{\Pi_{t}(i)}=u\}$, 
$\boldX_{j}(t,\Pi_{t}(i)) \sim p_{X_{j}|UV_{j}}(\cdot|u,v)$
We therefore have
\ifTITVersion\begin{eqnarray}
 \label{Eqn:ProbOfRandomCodeCorrespondingToRelevantComponents}
P\left(\!\!\!\begin{array}{c}V_{ji}^{m}(b_{ji})=v_{ji}^{m}(b_{ji}),\Pi_{t}
(i)=\pi_{t}(i),\\\boldX_{j}\left( 
\left[
\begin{array}{c}u^{l}(a_{j1})\\\vdots\\u^{l}(a_{jm})
\end{array}
\right],[v_{j1}^{m}(b_{j1})\cdots 
v_{jl}^{m}(b_{jl})]^{\Pi^{-1}}\right)\\=\boldx_{j}\{\ulinea_{j},\ulineb_{j}\}:j 
\in [2],t \in [m], i\in [l]
\end{array}
\!\!\!\right)\nonumber\\
=\prod_{t=1}^{m}\prod_{i=1}^{l}\left\{ p_{X_{1}|UV_{1}}\left( 
\boldx_{1}\{\ulinea_{1},\ulineb_{1}\}(t,\pi_{t}(i)) 
\left| 
\begin{array}{c} v_{1i}(b_{1i})_{t}\\ 
u(a_{1t})_{\pi_{t}(i)}
\end{array}  
\right.\!\!\!\right)\right.\nonumber\\  p_{X_{2}|UV_{2}}\left( 
\boldx_{2}\{\ulinea_{2},\ulineb_{2}\}(t,\pi_{t}(i)) 
\left| 
\begin{array}{c} v_{2i}(b_{2i})_{t}\\ 
u(a_{2t})_{\pi_{t}(i)}
\end{array}  
\right.\!\!\!\right)\nonumber\\\left. 
p_{V_{1}}(v_{1i}(b_{1i})_{t}) 
p_{V_{2}}(v_{2i}(b_{2i})_{t}) \right\}\left(\frac{1}{l!}\right)^{m}\nonumber\\
=\prod_{t=1}^{m}\prod_{i=1}^{l}\left\{ p_{X_{1}|UV_{1}}\left( 
\boldx_{1}\{\ulinea_{1},\ulineb_{1}\}(\gamma_{t},\pi_{\gamma_{t}}(i)) 
\left| 
\begin{array}{c} v_{1i}(b_{1i})_{\gamma_{t}}\\ 
u(a_{1\gamma_{t}})_{\pi_{\gamma_{t}}(i)}
\end{array}  
\right.\!\!\!\right)\right.\nonumber\\  p_{X_{2}|UV_{2}}\left( 
\boldx_{2}\{\ulinea_{2},\ulineb_{2}\}(\gamma_{t},\pi_{\gamma_{t}}(i)) 
\left| 
\begin{array}{c} v_{2i}(b_{2i})_{\gamma_{t}}\\ 
u(a_{2\gamma_{t}})_{\pi_{\gamma_{t}}(i)}
\end{array}  
\right.\!\!\!\right)\nonumber\\
\label{Eqn:CodePMF}
\left. 
p_{V_{1}}(v_{1i}(b_{1i})_{\gamma_{t}}) 
p_{V_{2}}(v_{2i}(b_{2i})_{\gamma_{t}}) 
\right\}\left(\frac{1}{l!}\right)^{m}.
\end{eqnarray}\fi
\ifPeerReviewVersion\begin{eqnarray}
 \label{Eqn:ProbOfRandomCodeCorrespondingToRelevantComponents}
\lefteqn{P\left(\!\!\!\begin{array}{c}V_{ji}^{m}(b_{ji})=v_{ji}^{m}(b_{ji}),\Pi_
{ t }
(i)=\pi_{t}(i),\\\boldX_{j}\left( 
\left[
\begin{array}{c}u^{l}(a_{j1})\\\vdots\\u^{l}(a_{jm})
\end{array}
\right],[v_{j1}^{m}(b_{j1})\cdots 
v_{jl}^{m}(b_{jl})]^{\Pi^{-1}}\right)\\=\boldx_{j}\{\ulinea_{j},\ulineb_{j}\}:j 
\in [2],t \in [m], i\in [l]
\end{array}
\!\!\!\right)}\nonumber\\
&=&\prod_{t=1}^{m}\prod_{i=1}^{l}\left\{ p_{X_{1}|UV_{1}}\left( 
\boldx_{1}\{\ulinea_{1},\ulineb_{1}\}(t,\pi_{t}(i)) 
\left| 
\begin{array}{c} v_{1i}(b_{1i})_{t}\\ 
u(a_{1t})_{\pi_{t}(i)}
\end{array}  
\right.\!\!\!\right)\right.  p_{X_{2}|UV_{2}}\left( 
\boldx_{2}\{\ulinea_{2},\ulineb_{2}\}(t,\pi_{t}(i)) 
\left| 
\begin{array}{c} v_{2i}(b_{2i})_{t}\\ 
u(a_{2t})_{\pi_{t}(i)}
\end{array}  
\right.\!\!\!\right)\nonumber\\&&~~~~~~~~~~~~~~~~~~~~~~~~~~~~~~~~~~\left. 
p_{V_{1}}(v_{1i}(b_{1i})_{t}) 
p_{V_{2}}(v_{2i}(b_{2i})_{t}) \right\}\left(\frac{1}{l!}\right)^{m}\nonumber\\
&=&\prod_{t=1}^{m}\prod_{i=1}^{l}\left\{ p_{X_{1}|UV_{1}}\left( 
\boldx_{1}\{\ulinea_{1},\ulineb_{1}\}(\gamma_{t},\pi_{\gamma_{t}}(i)) 
\left| 
\begin{array}{c} v_{1i}(b_{1i})_{\gamma_{t}}\\ 
u(a_{1\gamma_{t}})_{\pi_{\gamma_{t}}(i)}
\end{array}  
\right.\!\!\!\right)\right.   p_{X_{2}|UV_{2}}\left( 
\boldx_{2}\{\ulinea_{2},\ulineb_{2}\}(\gamma_{t},\pi_{\gamma_{t}}(i)) 
\left| 
\begin{array}{c} v_{2i}(b_{2i})_{\gamma_{t}}\\ 
u(a_{2\gamma_{t}})_{\pi_{\gamma_{t}}(i)}
\end{array}  
\right.\!\!\!\right)\nonumber\\&&\label{Eqn:CodePMF}
~~~~~~~~~~~~~~~~~~~~~~~~~~~~~~~~~~\left. 
p_{V_{1}}(v_{1i}(b_{1i})_{\gamma_{t}}) 
p_{V_{2}}(v_{2i}(b_{2i})_{\gamma_{t}}) 
\right\}\left(\frac{1}{l!}\right)^{m}.
\end{eqnarray}\fi

Multiplying (\ref{Eqn:PE2Full-ParticularCode2}) with the pmf of the code 
stated in (\ref{Eqn:CodePMF}), and after the 
crucial interchange of summations, we have
\ifPeerReviewVersion\begin{eqnarray}
 \sum_{\substack{\ulinea_{1},\ulinea_{2}\\\ulineb_{1},\ulineb_{2}}}~\sum_{
\substack{\boldy\in\mathcal{Y}^{m \times l}}}~\sum_{\substack{\boldu_{1} \in 
\mathcal{U}^{m \times l}\\\boldu_{2} \in \mathcal{U}^{m \times 
l}}}~\sum_{\pi_{1}}\cdots\sum_{\pi_{m}}\sum_{\substack{\boldv_{1}\{\ulineb_{1}\}
\\\in \mathcal{V}_{1}^{m \times 
l}}}~\sum_{\substack{\boldv_{2}\{\ulineb_{2}\}\\\in \mathcal{V}_{2}^{m \times 
l}}}\sum_{\substack{\boldx_{1}\{\ulinea_{1},\ulineb_{1}\}\\\in 
\mathcal{X}_{1}^{m 
\times l}}}
 %=============nextline
  \sum_{\substack{\boldx_{2}\{\ulinea_{2},\ulineb_{2}\}\\\in 
\mathcal{X}_{2}^{m \times l}}} \!\!\!\!\!\!P\!\left(  
\!\!\!\begin{array}{c}\ulineA_{j}=\ulinea_{j}\\\ulineB_{j}=\ulineb_{j}\\:j \in 
[2]\end{array}\!\!\right)\!\mathds{1}_{\left\{\!\!\! 
\begin{array}{c}\left[\!\!\!\begin{array}{c}\boldu_{j},\boldv_{j}\{\ulineb_{j}\}
\\\boldx_{j}\{ \ulinea_{j},\ulineb_{j}\}\\:j \in 
[2],\boldy
\end{array}\!\!\!\right]^{\pi}\!\!\!\!(1:m,i)\\\notin 
T_{\alpha}^{m}(p_{\underline{\mathscr{U}\mathscr{V}\mathscr{X}}\mathscr{Y}})
\end{array}
\!\!\!\right\}}\frac{1}{\left(l!\right)^{m}}~~\nonumber\\
 %=============nextline
 \prod_{i=1}^{l}\prod_{t=1}^{m}\left[ 
p_{V_{1}}\left( \boldv_{1}\{\ulineb_{1} \}(t,\pi_{t}(i))\right)p_{V_{2}}\left( 
\boldv_{2}\{\ulineb_{2} \}(t,\pi_{t}(i))\right) \right]
 %=============nextline
 \prod_{i=1}^{l}\prod_{t=1}^{m}\left[ p_{X_{1}|V_{1}U}\left( \boldx_{1}\{ 
\ulinea_{1},\ulineb_{1}\}(t,i) |\boldv_{1}\{\ulineb_{1} 
\}(t,i),\boldu\{\ulinea_{1}\}(t,i)\right)\right.\nonumber\\
 %=====================nextline
 \left.p_{X_{2}|V_{2}U}\left( \boldx_{2}\{ \ulinea_{2},\ulineb_{2}\}(t,i) 
|\boldv_{2}\{\ulineb_{2} \}(t,i),\boldu\{\ulinea_{2}\}(t,i)\right) 
\right]\times
 %==========nextline
 \prod_{t=1}^{m}\prod_{i=1}^{l}\mathbb{W}_{Y|X_{1}X_{2}}\left( 
\boldy(t,i)\left| 
\!\!\begin{array}{c}\boldx_{1}\{\ulinea_{1},\ulineb_{1}\}(t,i)\\\boldx_{2}\{
\ulinea_{2},\ulineb_{2}\}(t,i)\end{array}\!\!\right.\right)\mathds{1}_{\left\{ 
\substack{\boldu_{j}=\boldu\{\ulinea_{j}\}\\:j \in [2]}\right\}}
\label{Eqn:PE2FullMultByCodePMF}
\nonumber
\end{eqnarray}\fi
\ifTITVersion\begin{eqnarray}
 \sum_{\substack{\ulinea_{1},\ulinea_{2}\\\ulineb_{1},\ulineb_{2}}}~\sum_{
\substack{\boldy\in\mathcal{Y}^{m \times l}}}~\sum_{\substack{\boldu_{1} \in 
\mathcal{U}^{m \times l}\\\boldu_{2} \in \mathcal{U}^{m \times 
l}}}~\sum_{\pi_{1}}\cdots\sum_{\pi_{m}}\sum_{\substack{\boldv_{1}\{\ulineb_{1}\}
\\\in \mathcal{V}_{1}^{m \times 
l}}}~\sum_{\substack{\boldv_{2}\{\ulineb_{2}\}\\\in \mathcal{V}_{2}^{m \times 
l}}}\sum_{\substack{\boldx_{1}\{\ulinea_{1},\ulineb_{1}\}\\\in 
\mathcal{X}_{1}^{m 
\times l}}}\nonumber\\
 %=============nextline
  \sum_{\substack{\boldx_{2}\{\ulinea_{2},\ulineb_{2}\}\\\in 
\mathcal{X}_{2}^{m \times l}}} \!\!\!\!\!\!P\!\left(  
\!\!\!\begin{array}{c}\ulineA_{j}=\ulinea_{j}\\\ulineB_{j}=\ulineb_{j}\\:j \in 
[2]\end{array}\!\!\right)\!\mathds{1}_{\left\{\!\!\! 
\begin{array}{c}\left[\!\!\!\begin{array}{c}\boldu_{j},\boldv_{j}\{\ulineb_{j}\}
\\\boldx_{j}\{ \ulinea_{j},\ulineb_{j}\}\\:j \in 
[2],\boldy
\end{array}\!\!\!\right]^{\pi}\!\!\!\!(1:m,i)\\\notin 
T_{\alpha}^{m}(p_{\underline{\mathscr{U}\mathscr{V}\mathscr{X}}\mathscr{Y}})
\end{array}
\!\!\!\right\}}~~~~~~~~~~\nonumber\\
 %=============nextline
 \frac{1}{\left(l!\right)^{m}} \prod_{i=1}^{l}\prod_{t=1}^{m}\left[ 
p_{V_{1}}\left( \boldv_{1}\{\ulineb_{1} \}(t,\pi_{t}(i))\right)p_{V_{2}}\left( 
\boldv_{2}\{\ulineb_{2} \}(t,\pi_{t}(i))\right) \right]\nonumber\\
 %=============nextline
 \prod_{i=1}^{l}\prod_{t=1}^{m}\left[ p_{X_{1}|V_{1}U}\left( \boldx_{1}\{ 
\ulinea_{1},\ulineb_{1}\}(t,i) |\boldv_{1}\{\ulineb_{1} 
\}(t,i),\boldu\{\ulinea_{1}\}(t,i)\right)\right.\nonumber\\
 %=====================nextline
 \left.p_{X_{2}|V_{2}U}\left( \boldx_{2}\{ \ulinea_{2},\ulineb_{2}\}(t,i) 
|\boldv_{2}\{\ulineb_{2} \}(t,i),\boldu\{\ulinea_{2}\}(t,i)\right) 
\right]\times\nonumber\\
 %==========nextline
 \prod_{t=1}^{m}\prod_{i=1}^{l}\mathbb{W}_{Y|X_{1}X_{2}}\left( 
\boldy(t,i)\left| 
\!\!\begin{array}{c}\boldx_{1}\{\ulinea_{1},\ulineb_{1}\}(t,i)\\\boldx_{2}\{
\ulinea_{2},\ulineb_{2}\}(t,i)\end{array}\!\!\right.\right)\mathds{1}_{\left\{ 
\substack{\boldu_{j}=\boldu\{\ulinea_{j}\}\\:j \in [2]}\right\}}
\label{Eqn:PE2FullMultByCodePMF}
\nonumber
\end{eqnarray}\fi
The expressions being involved, we paraphrase each step. Since a term in the 
sum is positive only when $\boldu_{j}=\boldu \{ \ulinea_{j}\}: j \in [2]$, we 
replace $\boldu \{ \ulinea_{j}\}$ with $\boldu_{j}$ in the terms evaluating the 
probability of the code. Moreover, since
\begin{eqnarray}
 \prod_{t=1}^{m}\prod_{i=1}^{l}p_{V_{j}}\left( \boldv_{j}\{\ulineb_{j} 
\}(t,\pi_{t}(i))\right) = \prod_{t=1}^{m}\prod_{i=1}^{l}p_{V_{j}}\left( 
\boldv_{j}\{\ulineb_{1} \}(t,i)\right) \nonumber
\end{eqnarray}
for $j \in [2]$, we have the earlier expression to be
\ifTITVersion\begin{eqnarray}
 \sum_{\substack{\ulinea_{1},\ulinea_{2}\\\ulineb_{1},\ulineb_{2}}}~\sum_{
\substack{\boldy\in\mathcal{Y}^{m \times l}}}~\sum_{\substack{\boldu_{1} \in 
\mathcal{U}^{m \times l}\\\boldu_{2} \in \mathcal{U}^{m \times 
l}}}~\sum_{\pi_{1}}\cdots\sum_{\pi_{m}}\sum_{\substack{\boldv_{1}\{\ulineb_{1}\}
\\\in \mathcal{V}_{1}^{m \times 
l}}}~\sum_{\substack{\boldv_{2}\{\ulineb_{2}\}\\\in \mathcal{V}_{2}^{m \times 
l}}}\sum_{\substack{\boldx_{1}\{\ulinea_{1},\ulineb_{1}\}\\\in 
\mathcal{X}_{1}^{m 
\times l}}}\nonumber\\
 %=============nextline
  \sum_{\substack{\boldx_{2}\{\ulinea_{2},\ulineb_{2}\}\\\in 
\mathcal{X}_{2}^{m \times l}}} \!\!\!\!\!\!P\!\left(  
\!\!\!\begin{array}{c}\ulineA_{j}=\ulinea_{j}\\\ulineB_{j}=\ulineb_{j}\\:j \in 
[2]\end{array}\!\!\right)\!\mathds{1}_{\left\{\!\!\! 
\begin{array}{c}\left[\!\!\!\begin{array}{c}\boldu_{j},\boldv_{j}\{\ulineb_{j}\}
\\\boldx_{j}\{ \ulinea_{j},\ulineb_{j}\}\\:j \in 
[2],\boldy
\end{array}\!\!\!\right]^{\pi}\!\!\!\!(1:m,i)\\\notin 
T_{\alpha}^{m}(p_{\underline{\mathscr{U}\mathscr{V}\mathscr{X}}\mathscr{Y}})
\end{array}
\!\!\!\right\}}~~~~~~~~~~\nonumber\\
 %=============nextline
\frac{1}{\left(l!\right)^{m}} \prod_{i=1}^{l}\prod_{t=1}^{m}\left[ 
p_{V_{1}}\left( \boldv_{1}\{\ulineb_{1} \}(t,i)\right)p_{V_{2}}\left( 
\boldv_{2}\{\ulineb_{2} \}(t,i)\right) \right]\nonumber\\
 %=============nextline
 \prod_{i=1}^{l}\prod_{t=1}^{m}\left[ p_{X_{1}|V_{1}U}\left( \boldx_{1}\{ 
\ulinea_{1},\ulineb_{1}\}(t,i) |\boldv_{1}\{\ulineb_{1} 
\}(t,i),\boldu_{1}(t,i)\right)\right.\nonumber\\
 %=====================nextline
 \left.p_{X_{2}|V_{2}U}\left( \boldx_{2}\{ \ulinea_{2},\ulineb_{2}\}(t,i) 
|\boldv_{2}\{\ulineb_{2} \}(t,i),\boldu_{2}(t,i)\right) 
\right]\times\nonumber\\
 %==========nextline
 \prod_{t=1}^{m}\prod_{i=1}^{l}\mathbb{W}_{Y|X_{1}X_{2}}\left( 
\boldy(t,i)\left| 
\!\!\begin{array}{c}\boldx_{1}\{\ulinea_{1},\ulineb_{1}\}(t,i)\\\boldx_{2}\{
\ulinea_{2},\ulineb_{2}\}(t,i)\end{array}\!\!\right.\right)\mathds{1}_{\left\{ 
\substack{\boldu_{j}=\boldu\{\ulinea_{j}\}\\:j \in [2]}\right\}}
\label{Eqn:PE2FullAfterCodePMFReplaceUj}
\nonumber
\end{eqnarray}\fi
\ifPeerReviewVersion\begin{eqnarray}
 \sum_{\substack{\ulinea_{1},\ulinea_{2}\\\ulineb_{1},\ulineb_{2}}}~\sum_{
\substack{\boldy\in\mathcal{Y}^{m \times l}}}~\sum_{\substack{\boldu_{1} \in 
\mathcal{U}^{m \times l}\\\boldu_{2} \in \mathcal{U}^{m \times 
l}}}~\sum_{\pi_{1}}\cdots\sum_{\pi_{m}}\sum_{\substack{\boldv_{1}\{\ulineb_{1}\}
\\\in \mathcal{V}_{1}^{m \times 
l}}}~\sum_{\substack{\boldv_{2}\{\ulineb_{2}\}\\\in \mathcal{V}_{2}^{m \times 
l}}}\sum_{\substack{\boldx_{1}\{\ulinea_{1},\ulineb_{1}\}\\\in 
\mathcal{X}_{1}^{m 
\times l}}}
 %=============nextline
  \sum_{\substack{\boldx_{2}\{\ulinea_{2},\ulineb_{2}\}\\\in 
\mathcal{X}_{2}^{m \times l}}} \!\!\!\!\!\!P\!\left(  
\!\!\!\begin{array}{c}\ulineA_{j}=\ulinea_{j}\\\ulineB_{j}=\ulineb_{j}\\:j \in 
[2]\end{array}\!\!\right)\!\mathds{1}_{\left\{\!\!\! 
\begin{array}{c}\left[\!\!\!\begin{array}{c}\boldu_{j},\boldv_{j}\{\ulineb_{j}\}
\\\boldx_{j}\{ \ulinea_{j},\ulineb_{j}\}\\:j \in 
[2],\boldy
\end{array}\!\!\!\right]^{\pi}\!\!\!\!(1:m,i)\\\notin 
T_{\alpha}^{m}(p_{\underline{\mathscr{U}\mathscr{V}\mathscr{X}}\mathscr{Y}})
\end{array}
\!\!\!\right\}}~~~~~~~~~~\nonumber\\
 %=============nextline
\frac{1}{\left(l!\right)^{m}} \prod_{i=1}^{l}\prod_{t=1}^{m}\left[ 
p_{V_{1}}\left( \boldv_{1}\{\ulineb_{1} \}(t,i)\right)p_{V_{2}}\left( 
\boldv_{2}\{\ulineb_{2} \}(t,i)\right) \right]
 %=============nextline
 \prod_{i=1}^{l}\prod_{t=1}^{m}\left[ p_{X_{1}|V_{1}U}\left( \boldx_{1}\{ 
\ulinea_{1},\ulineb_{1}\}(t,i) |\boldv_{1}\{\ulineb_{1} 
\}(t,i),\boldu_{1}(t,i)\right)\right.\nonumber\\
 %=====================nextline
 \left.p_{X_{2}|V_{2}U}\left( \boldx_{2}\{ \ulinea_{2},\ulineb_{2}\}(t,i) 
|\boldv_{2}\{\ulineb_{2} \}(t,i),\boldu_{2}(t,i)\right) 
\right]\times
 %==========nextline
 \prod_{t=1}^{m}\prod_{i=1}^{l}\mathbb{W}_{Y|X_{1}X_{2}}\left( 
\boldy(t,i)\left| 
\!\!\begin{array}{c}\boldx_{1}\{\ulinea_{1},\ulineb_{1}\}(t,i)\\\boldx_{2}\{
\ulinea_{2},\ulineb_{2}\}(t,i)\end{array}\!\!\right.\right)\mathds{1}_{\left\{ 
\substack{\boldu_{j}=\boldu\{\ulinea_{j}\}\\:j \in [2]}\right\}}
\label{Eqn:PE2FullAfterCodePMFReplaceUj}
\nonumber
\end{eqnarray}\fi
We perform the simple manipulation of pulling 
$P(\ulineA_{j}=\ulinea_{j},\ulineB_{j}=\ulineb_{j}: j \in [2])$ out of the 
internal summations. We therefore have
\ifTITVersion\begin{eqnarray}
 \sum_{\substack{\ulinea_{1},\ulinea_{2}\\\ulineb_{1},\ulineb_{2}}}
 \!\!P\!\left(  
\!\!\!\begin{array}{c}\ulineA_{j}=\ulinea_{j}\\\ulineB_{j}=\ulineb_{j}\\:j \in 
[2]\end{array}\!\!\right)\!
\sum_{
\substack{\boldy\in\mathcal{Y}^{m \times l}}}~\sum_{\substack{\boldu_{1} \in 
\mathcal{U}^{m \times l}\\\boldu_{2} \in \mathcal{U}^{m \times 
l}}}~\sum_{\pi_{1}}\cdots\sum_{\pi_{m}}\sum_{\substack{\boldv_{1}\{\ulineb_{1}\}
\\\in \mathcal{V}_{1}^{m \times 
l}}}~\nonumber\\
 %=============nextline
\sum_{\substack{\boldv_{2}\{\ulineb_{2}\}\\\in \mathcal{V}_{2}^{m \times 
l}}}\sum_{\substack{\boldx_{1}\{\ulinea_{1},\ulineb_{1}\}\\\in 
\mathcal{X}_{1}^{m 
\times l}}}  \sum_{\substack{\boldx_{2}\{\ulinea_{2},\ulineb_{2}\}\\\in 
\mathcal{X}_{2}^{m \times l}}} \!\!\!\mathds{1}_{\left\{\!\!\! 
\begin{array}{c}\left[\!\!\!\begin{array}{c}\boldu_{j},\boldv_{j}\{\ulineb_{j}\}
\\\boldx_{j}\{ \ulinea_{j},\ulineb_{j}\}\\:j \in 
[2],\boldy
\end{array}\!\!\!\right]^{\pi}\!\!\!\!(1:m,i)\\\notin 
T_{\alpha}^{m}(p_{\underline{\mathscr{U}\mathscr{V}\mathscr{X}}\mathscr{Y}})
\end{array}
\!\!\!\right\}}~~~~\nonumber\\
 %=============nextline
\frac{1}{\left(l!\right)^{m}} \prod_{i=1}^{l}\prod_{t=1}^{m}\left[ 
p_{V_{1}}\left( \boldv_{1}\{\ulineb_{1} \}(t,i)\right)p_{V_{2}}\left( 
\boldv_{2}\{\ulineb_{2} \}(t,i)\right) \right]\nonumber\\
 %=============nextline
 \prod_{i=1}^{l}\prod_{t=1}^{m}\left[ p_{X_{1}|V_{1}U}\left( \boldx_{1}\{ 
\ulinea_{1},\ulineb_{1}\}(t,i) |\boldv_{1}\{\ulineb_{1} 
\}(t,i),\boldu_{1}(t,i)\right)\right.\nonumber\\
 %=====================nextline
 \left.p_{X_{2}|V_{2}U}\left( \boldx_{2}\{ \ulinea_{2},\ulineb_{2}\}(t,i) 
|\boldv_{2}\{\ulineb_{2} \}(t,i),\boldu_{2}(t,i)\right) 
\right]\times\nonumber\\
 %==========nextline
 \prod_{t=1}^{m}\prod_{i=1}^{l}\mathbb{W}_{Y|X_{1}X_{2}}\left( 
\boldy(t,i)\left| 
\!\!\begin{array}{c}\boldx_{1}\{\ulinea_{1},\ulineb_{1}\}(t,i)\\\boldx_{2}\{
\ulinea_{2},\ulineb_{2}\}(t,i)\end{array}\!\!\right.\right)\mathds{1}_{\left\{ 
\substack{\boldu_{j}=\boldu\{\ulinea_{j}\}\\:j \in [2]}\right\}}
\label{Eqn:PE2FullAfterCodePMFInterchangeAjSum}
\nonumber
\end{eqnarray}\fi
\ifPeerReviewVersion\begin{eqnarray}
 \sum_{\substack{\ulinea_{1},\ulinea_{2}\\\ulineb_{1},\ulineb_{2}}}
 \!\!P\!\left(  
\!\!\!\begin{array}{c}\ulineA_{j}=\ulinea_{j}\\\ulineB_{j}=\ulineb_{j}\\:j \in 
[2]\end{array}\!\!\right)\!
\sum_{
\substack{\boldy\in\mathcal{Y}^{m \times l}}}~\sum_{\substack{\boldu_{1} \in 
\mathcal{U}^{m \times l}\\\boldu_{2} \in \mathcal{U}^{m \times 
l}}}~\sum_{\pi_{1}}\cdots\sum_{\pi_{m}}\sum_{\substack{\boldv_{1}\{\ulineb_{1}\}
\\\in \mathcal{V}_{1}^{m \times 
l}}}~
 %=============nextline
\sum_{\substack{\boldv_{2}\{\ulineb_{2}\}\\\in \mathcal{V}_{2}^{m \times 
l}}}\sum_{\substack{\boldx_{1}\{\ulinea_{1},\ulineb_{1}\}\\\in 
\mathcal{X}_{1}^{m 
\times l}}}  \sum_{\substack{\boldx_{2}\{\ulinea_{2},\ulineb_{2}\}\\\in 
\mathcal{X}_{2}^{m \times l}}} \!\!\!\mathds{1}_{\left\{\!\!\! 
\begin{array}{c}\left[\!\!\!\begin{array}{c}\boldu_{j},\boldv_{j}\{\ulineb_{j}\}
\\\boldx_{j}\{ \ulinea_{j},\ulineb_{j}\}\\:j \in 
[2],\boldy
\end{array}\!\!\!\right]^{\pi}\!\!\!\!(1:m,i)\\\notin 
T_{\alpha}^{m}(p_{\underline{\mathscr{U}\mathscr{V}\mathscr{X}}\mathscr{Y}})
\end{array}
\!\!\!\right\}}~~~~\nonumber\\
 %=============nextline
\frac{1}{\left(l!\right)^{m}} \prod_{i=1}^{l}\prod_{t=1}^{m}\left[ 
p_{V_{1}}\left( \boldv_{1}\{\ulineb_{1} \}(t,i)\right)p_{V_{2}}\left( 
\boldv_{2}\{\ulineb_{2} \}(t,i)\right) \right]
 %=============nextline
 \prod_{i=1}^{l}\prod_{t=1}^{m}\left[ p_{X_{1}|V_{1}U}\left( \boldx_{1}\{ 
\ulinea_{1},\ulineb_{1}\}(t,i) |\boldv_{1}\{\ulineb_{1} 
\}(t,i),\boldu_{1}(t,i)\right)\right.\nonumber\\
 %=====================nextline
 \left.p_{X_{2}|V_{2}U}\left( \boldx_{2}\{ \ulinea_{2},\ulineb_{2}\}(t,i) 
|\boldv_{2}\{\ulineb_{2} \}(t,i),\boldu_{2}(t,i)\right) 
\right]\times
 %==========nextline
 \prod_{t=1}^{m}\prod_{i=1}^{l}\mathbb{W}_{Y|X_{1}X_{2}}\left( 
\boldy(t,i)\left| 
\!\!\begin{array}{c}\boldx_{1}\{\ulinea_{1},\ulineb_{1}\}(t,i)\\\boldx_{2}\{
\ulinea_{2},\ulineb_{2}\}(t,i)\end{array}\!\!\right.\right)\mathds{1}_{\left\{ 
\substack{\boldu_{j}=\boldu\{\ulinea_{j}\}\\:j \in [2]}\right\}}
\label{Eqn:PE2FullAfterCodePMFInterchangeAjSum}
\nonumber
\end{eqnarray}\fi
In our first expression, we denoted our dummy variables as 
$\boldx_{1}\{\ulinea_{1},\ulineb_{1}\}$ etc to illustrate the component of the 
code they corresponded to, and we have carried the same names for the dummy 
variables. We note that the ranges of dummy variables 
$\ulinex_{j}\{\ulinea_{j},\ulineb_{j}\}, \boldv_{j}\{ \ulineb_{j}\}$ do not 
depend on $\ulinea_{j},\ulineb_{j}$, we therefore rename these dummy 
variables without the parenthesis. As the reader will note, this reduces 
clutter and enables us recognize the invariance of the inner sum with respect 
to $\ulineb_{1},\ulineb_{2}$. Specifically, the above expression is equal to
\ifTITVersion\begin{eqnarray}
 \sum_{\substack{\ulinea_{1},\ulinea_{2}\\\ulineb_{1},\ulineb_{2}}}
 \!\!P\!\left(  
\!\!\!\begin{array}{c}\ulineA_{j}=\ulinea_{j}\\\ulineB_{j}=\ulineb_{j}\\:j \in 
[2]\end{array}\!\!\right)\!
\sum_{
\substack{\boldy\in\mathcal{Y}^{m \times l}}}~\sum_{\substack{\boldu_{1} \in 
\mathcal{U}^{m \times l}\\\boldu_{2} \in \mathcal{U}^{m \times 
l}}}~\sum_{\pi_{1}}\cdots\sum_{\pi_{m}}\sum_{\substack{\boldv_{1}
\\\in \mathcal{V}_{1}^{m \times 
l}}}~\nonumber\\
 %=============nextline
\sum_{\substack{\boldv_{2}\\\in \mathcal{V}_{2}^{m \times 
l}}}\sum_{\substack{\boldx_{1}\\\in 
\mathcal{X}_{1}^{m 
\times l}}}  \sum_{\substack{\boldx_{2}\\\in 
\mathcal{X}_{2}^{m \times l}}} \!\!\!\mathds{1}_{\left\{\!\!\! 
\begin{array}{c}\left[\!\!\!\begin{array}{c}\boldu_{j},\boldv_{j}
\\\boldx_{j}:j \in 
[2],\\\boldy
\end{array}\!\!\!\right]^{\pi}\!\!\!\!(1:m,i)\\\notin 
T_{\alpha}^{m}(p_{\underline{\mathscr{U}\mathscr{V}\mathscr{X}}\mathscr{Y}})
\end{array}
\!\!\!\right\}}~~~~\nonumber\\
 %=============nextline
\frac{1}{\left(l!\right)^{m}} \prod_{i=1}^{l}\prod_{t=1}^{m}\left[ 
p_{V_{1}}\left( \boldv_{1}(t,i)\right)p_{V_{2}}\left( 
\boldv_{2}(t,i)\right) \right]\nonumber\\
 %=============nextline
 \prod_{i=1}^{l}\prod_{t=1}^{m}\left[ p_{X_{1}|V_{1}U}\left( \boldx_{1}(t,i) 
|\boldv_{1}(t,i),\boldu_{1}(t,i)\right)\right.\nonumber\\
 %=====================nextline
 \left.p_{X_{2}|V_{2}U}\left( \boldx_{2}(t,i) 
|\boldv_{2}(t,i),\boldu_{2}(t,i)\right) 
\right]\times\nonumber\\
 %==========nextline
 \prod_{t=1}^{m}\prod_{i=1}^{l}\mathbb{W}_{Y|X_{1}X_{2}}\left( 
\boldy(t,i)\left| 
\!\!\begin{array}{c}\boldx_{1}(t,i)\\\boldx_{2}(t,i)\end{array}
\!\!\right.\right)\mathds{1}_{\left\{ 
\substack{\boldu_{j}=\boldu\{\ulinea_{j}\}\\:j \in [2]}\right\}}
\label{Eqn:PE2FullAfterCodePMFRenAllDumVars}
\nonumber
\end{eqnarray}\fi
\ifPeerReviewVersion\begin{eqnarray}
 \sum_{\substack{\ulinea_{1},\ulinea_{2}\\\ulineb_{1},\ulineb_{2}}}
 \!\!P\!\left(  
\!\!\!\begin{array}{c}\ulineA_{j}=\ulinea_{j}\\\ulineB_{j}=\ulineb_{j}\\:j \in 
[2]\end{array}\!\!\right)\!
\sum_{
\substack{\boldy\in\mathcal{Y}^{m \times l}}}~\sum_{\substack{\boldu_{1} \in 
\mathcal{U}^{m \times l}\\\boldu_{2} \in \mathcal{U}^{m \times 
l}}}~\sum_{\pi_{1}}\cdots\sum_{\pi_{m}}\sum_{\substack{\boldv_{1}
\\\in \mathcal{V}_{1}^{m \times 
l}}}
 %=============nextline
\sum_{\substack{\boldv_{2}\\\in \mathcal{V}_{2}^{m \times 
l}}}\sum_{\substack{\boldx_{1}\\\in 
\mathcal{X}_{1}^{m 
\times l}}}  \sum_{\substack{\boldx_{2}\\\in 
\mathcal{X}_{2}^{m \times l}}} \!\!\!\mathds{1}_{\left\{\!\!\! 
\begin{array}{c}\left[\!\!\!\begin{array}{c}\boldu_{j},\boldv_{j}
\\\boldx_{j}:j \in 
[2],\\\boldy
\end{array}\!\!\!\right]^{\pi}\!\!\!\!(1:m,i)\\\notin 
T_{\alpha}^{m}(p_{\underline{\mathscr{U}\mathscr{V}\mathscr{X}}\mathscr{Y}})
\end{array}
\!\!\!\right\}}~~~~\nonumber\\
 %=============nextline
\frac{1}{\left(l!\right)^{m}} \prod_{i=1}^{l}\prod_{t=1}^{m}\left[ 
p_{V_{1}}\left( \boldv_{1}(t,i)\right)p_{V_{2}}\left( 
\boldv_{2}(t,i)\right) \right]
 %=============nextline
 \prod_{i=1}^{l}\prod_{t=1}^{m}\left[ p_{X_{1}|V_{1}U}\left( \boldx_{1}(t,i) 
|\boldv_{1}(t,i),\boldu_{1}(t,i)\right)\right.\nonumber\\
 %=====================nextline
 \left.p_{X_{2}|V_{2}U}\left( \boldx_{2}(t,i) 
|\boldv_{2}(t,i),\boldu_{2}(t,i)\right) 
\right]\times
 %==========nextline
 \prod_{t=1}^{m}\prod_{i=1}^{l}\mathbb{W}_{Y|X_{1}X_{2}}\left( 
\boldy(t,i)\left| 
\!\!\begin{array}{c}\boldx_{1}(t,i)\\\boldx_{2}(t,i)\end{array}
\!\!\right.\right)\mathds{1}_{\left\{ 
\substack{\boldu_{j}=\boldu\{\ulinea_{j}\}\\:j \in [2]}\right\}}
\label{Eqn:PE2FullAfterCodePMFRenAllDumVars}
\nonumber
\end{eqnarray}\fi
Summing over $\ulineb_{1},\ulineb_{2}$, we have
\ifTITVersion\begin{eqnarray}
 \sum_{\substack{\ulinea_{1},\ulinea_{2}}}
 \!\!P\!\left(  
\!\!\!\begin{array}{c}\ulineA_{j}=\ulinea_{j}\\:j \in 
[2]\end{array}\!\!\right)\!
\sum_{
\substack{\boldy\in\mathcal{Y}^{m \times l}}}~\sum_{\substack{\boldu_{1} \in 
\mathcal{U}^{m \times l}\\\boldu_{2} \in \mathcal{U}^{m \times 
l}}}~\sum_{\pi_{1}}\cdots\sum_{\pi_{m}}\sum_{\substack{\boldv_{1}
\\\in \mathcal{V}_{1}^{m \times 
l}}}~\nonumber\\
 %=============nextline
\sum_{\substack{\boldv_{2}\\\in \mathcal{V}_{2}^{m \times 
l}}}\sum_{\substack{\boldx_{1}\\\in 
\mathcal{X}_{1}^{m 
\times l}}}  \sum_{\substack{\boldx_{2}\\\in 
\mathcal{X}_{2}^{m \times l}}} \!\!\!\mathds{1}_{\left\{\!\!\! 
\begin{array}{c}\left[\!\!\!\begin{array}{c}\boldu_{j},\boldv_{j}
\\\boldx_{j}:j \in 
[2],\\\boldy
\end{array}\!\!\!\right]^{\pi}\!\!\!\!(1:m,i)\\\notin 
T_{\alpha}^{m}(p_{\underline{\mathscr{U}\mathscr{V}\mathscr{X}}\mathscr{Y}})
\end{array}
\!\!\!\right\}}~~~~\nonumber\\
 %=============nextline
\frac{1}{\left(l!\right)^{m}} \prod_{i=1}^{l}\prod_{t=1}^{m}\left[ 
p_{V_{1}}\left( \boldv_{1}(t,i)\right)p_{V_{2}}\left( 
\boldv_{2}(t,i)\right) \right]\nonumber\\
 %=============nextline
 \prod_{i=1}^{l}\prod_{t=1}^{m}\left[ p_{X_{1}|V_{1}U}\left( \boldx_{1}(t,i) 
|\boldv_{1}(t,i),\boldu_{1}(t,i)\right)\right.\nonumber\\
 %=====================nextline
 \left.p_{X_{2}|V_{2}U}\left( \boldx_{2}(t,i) 
|\boldv_{2}(t,i),\boldu_{2}(t,i)\right) 
\right]\times\nonumber\\
 %==========nextline
 \prod_{t=1}^{m}\prod_{i=1}^{l}\mathbb{W}_{Y|X_{1}X_{2}}\left( 
\boldy(t,i)\left| 
\!\!\begin{array}{c}\boldx_{1}(t,i)\\\boldx_{2}(t,i)\end{array}
\!\!\right.\right)\mathds{1}_{\left\{ 
\substack{\boldu_{j}=\boldu\{\ulinea_{j}\}\\:j \in [2]}\right\}}
\label{Eqn:PE2FullAfterCodePMFSumB1B2}
\nonumber
\end{eqnarray}\fi
\ifPeerReviewVersion\begin{eqnarray}
 \sum_{\substack{\ulinea_{1},\ulinea_{2}}}
 \!\!P\!\left(  
\!\!\!\begin{array}{c}\ulineA_{j}=\ulinea_{j}\\:j \in 
[2]\end{array}\!\!\right)\!
\sum_{
\substack{\boldy\in\mathcal{Y}^{m \times l}}}~\sum_{\substack{\boldu_{1} \in 
\mathcal{U}^{m \times l}\\\boldu_{2} \in \mathcal{U}^{m \times 
l}}}~\sum_{\pi_{1}}\cdots\sum_{\pi_{m}}\sum_{\substack{\boldv_{1}
\\\in \mathcal{V}_{1}^{m \times 
l}}}
 %=============nextline
\sum_{\substack{\boldv_{2}\\\in \mathcal{V}_{2}^{m \times 
l}}}\sum_{\substack{\boldx_{1}\\\in 
\mathcal{X}_{1}^{m 
\times l}}}  \sum_{\substack{\boldx_{2}\\\in 
\mathcal{X}_{2}^{m \times l}}} \!\!\!\mathds{1}_{\left\{\!\!\! 
\begin{array}{c}\left[\!\!\!\begin{array}{c}\boldu_{j},\boldv_{j}
\\\boldx_{j}:j \in 
[2],\\\boldy
\end{array}\!\!\!\right]^{\pi}\!\!\!\!(1:m,i)\\\notin 
T_{\alpha}^{m}(p_{\underline{\mathscr{U}\mathscr{V}\mathscr{X}}\mathscr{Y}})
\end{array}
\!\!\!\right\}}~~~~\nonumber\\
 %=============nextline
\frac{1}{\left(l!\right)^{m}} \prod_{i=1}^{l}\prod_{t=1}^{m}\left[ 
p_{V_{1}}\left( \boldv_{1}(t,i)\right)p_{V_{2}}\left( 
\boldv_{2}(t,i)\right) \right]
 %=============nextline
 \prod_{i=1}^{l}\prod_{t=1}^{m}\left[ p_{X_{1}|V_{1}U}\left( \boldx_{1}(t,i) 
|\boldv_{1}(t,i),\boldu_{1}(t,i)\right)\right.\nonumber\\
 %=====================nextline
 \left.p_{X_{2}|V_{2}U}\left( \boldx_{2}(t,i) 
|\boldv_{2}(t,i),\boldu_{2}(t,i)\right) 
\right]\times
 %==========nextline
 \prod_{t=1}^{m}\prod_{i=1}^{l}\mathbb{W}_{Y|X_{1}X_{2}}\left( 
\boldy(t,i)\left| 
\!\!\begin{array}{c}\boldx_{1}(t,i)\\\boldx_{2}(t,i)\end{array}
\!\!\right.\right)\mathds{1}_{\left\{ 
\substack{\boldu_{j}=\boldu\{\ulinea_{j}\}\\:j \in [2]}\right\}}
\label{Eqn:PE2FullAfterCodePMFSumB1B2}
\nonumber
\end{eqnarray}\fi
Noting that
\ifTITVersion
\begin{eqnarray}P(\ulineA_{j}=\ulinea_{j}: j \in [2]) = 
\prod_{t=1}^{m}P(A_{jt}=a_{jt}: j \in [2])\nonumber\mbox{ and}\\
\mathds{1}_{\left\{\!\!\! \begin{array}{c}
\boldu_{j}=\boldu\{ \ulinea_{j} \}\\:j \in [2] \end{array}\!\!\right\}} = 
\prod_{t=1}^{m}\mathds{1}_{\left\{ \begin{array}{c}\boldu_{j}(t,1:l) = 
u^{l}(a_{jt})\\: j \in [2]\end{array} \right\}}\nonumber
\end{eqnarray}\fi
\ifPeerReviewVersion
\begin{eqnarray}P(\ulineA_{j}=\ulinea_{j}: j \in [2]) = 
\prod_{t=1}^{m}P(A_{jt}=a_{jt}: j \in [2])\nonumber\mbox{ and }
\mathds{1}_{\left\{\!\!\! \begin{array}{c}
\boldu_{j}=\boldu\{ \ulinea_{j} \}\\:j \in [2] \end{array}\!\!\right\}} = 
\prod_{t=1}^{m}\mathds{1}_{\left\{ \begin{array}{c}\boldu_{j}(t,1:l) = 
u^{l}(a_{jt})\\: j \in [2]\end{array} \right\}}\nonumber
\end{eqnarray}\fi
the reader may verify that the above expression is given by
\ifTITVersion\begin{eqnarray}
 \sum_{\substack{\ulinea_{1},\ulinea_{2}}}
\sum_{
\substack{\boldy\in\mathcal{Y}^{m \times l}}}~\sum_{\substack{\boldu_{1} \in 
\mathcal{U}^{m \times l}\\\boldu_{2} \in \mathcal{U}^{m \times 
l}}}~\sum_{\pi_{1}}\cdots\sum_{\pi_{m}}\sum_{\substack{\boldv_{1}
\\\in \mathcal{V}_{1}^{m \times 
l}}}~\sum_{\substack{\boldv_{2}\\\in \mathcal{V}_{2}^{m \times 
l}}}~~~~~~~~~~\nonumber\\
 %=============nextline
\sum_{\substack{\boldx_{1}\\\in 
\mathcal{X}_{1}^{m 
\times l}}}  \sum_{\substack{\boldx_{2}\\\in 
\mathcal{X}_{2}^{m \times l}}} \!\!\!\mathds{1}_{\left\{\!\!\! 
\begin{array}{c}\left[\!\!\!\begin{array}{c}\boldu_{j},\boldv_{j}
\\\boldx_{j}:j \in 
[2],\\\boldy
\end{array}\!\!\!\right]^{\pi}\!\!\!\!(1:m,i)\\\notin 
T_{\alpha}^{m}(p_{\underline{\mathscr{U}\mathscr{V}\mathscr{X}}\mathscr{Y}})
\end{array}
\!\!\!\right\}}\frac{1}{\left(l!\right)^{m}}~~~~~~~~~~\nonumber\\
 %=============nextline
\left\{\prod_{t=1}^{m}\left[ P\left(  \!\!\!\begin{array}{c}A_{jt}=a_{jt}\\:j 
\in [2]\end{array}\!\!\right)\mathds{1}_{\left\{\substack{ 
\boldu_{j}(t,1:l)\\=u^{l}(a_{jt})\\:j \in [2]}\right\}}\prod_{i=1}^{l}\left\{ 
\prod_{j=1}^{2}p_{V_{j}}\left( \boldv_{j}(t,i)\right)\right. 
\right.\right.\nonumber\\
 %=============nextline
 \left. p_{X_{j}|V_{j}U}\left( \boldx_{j}(t,i) 
\left|\!\!\!\begin{array}{c}\boldv_{j}(t,i)\\\boldu_{j}(t,i)\end{array}\!\!\!
\right.\right)\right\}
 \left.\left.\mathbb{W}_{Y|\ulineX}\!\left( \boldy(t,i)\left| 
\!\!\begin{array}{c}\boldx_{1}(t,i)\\\boldx_{2}(t,i)\end{array}\!\!\right. 
\right)\right]\right\}
\label{Eqn:PE2FullAfterCodePMFWriteProdNotation}
\nonumber
\end{eqnarray}\fi
\ifPeerReviewVersion\begin{eqnarray}
 \sum_{\substack{\ulinea_{1},\ulinea_{2}}}
\sum_{
\substack{\boldy\in\mathcal{Y}^{m \times l}}}~\sum_{\substack{\boldu_{1} \in 
\mathcal{U}^{m \times l}\\\boldu_{2} \in \mathcal{U}^{m \times 
l}}}~\sum_{\pi_{1}}\cdots\sum_{\pi_{m}}\sum_{\substack{\boldv_{1}
\\\in \mathcal{V}_{1}^{m \times 
l}}}~\sum_{\substack{\boldv_{2}\\\in \mathcal{V}_{2}^{m \times 
l}}}
 %=============nextline
\sum_{\substack{\boldx_{1}\\\in 
\mathcal{X}_{1}^{m 
\times l}}}  \sum_{\substack{\boldx_{2}\\\in 
\mathcal{X}_{2}^{m \times l}}} \!\!\!\mathds{1}_{\left\{\!\!\! 
\begin{array}{c}\left[\!\!\!\begin{array}{c}\boldu_{j},\boldv_{j}
\\\boldx_{j}:j \in 
[2],\\\boldy
\end{array}\!\!\!\right]^{\pi}\!\!\!\!(1:m,i)\\\notin 
T_{\alpha}^{m}(p_{\underline{\mathscr{U}\mathscr{V}\mathscr{X}}\mathscr{Y}})
\end{array}
\!\!\!\right\}}\frac{1}{\left(l!\right)^{m}}~~~~~~~~~~\nonumber\\
 %=============nextline
\left\{\prod_{t=1}^{m}\left[ P\left(  \!\!\!\begin{array}{c}A_{jt}=a_{jt}\\:j 
\in [2]\end{array}\!\!\right)\mathds{1}_{\left\{\substack{ 
\boldu_{j}(t,1:l)\\=u^{l}(a_{jt})\\:j \in [2]}\right\}}\prod_{i=1}^{l}\left\{ 
\prod_{j=1}^{2}p_{V_{j}}\left( \boldv_{j}(t,i)\right)\right. 
\right.\right.
 %=============nextline
 \left. p_{X_{j}|V_{j}U}\left( \boldx_{j}(t,i) 
\left|\!\!\!\begin{array}{c}\boldv_{j}(t,i)\\\boldu_{j}(t,i)\end{array}\!\!\!
\right.\right)\right\}
 \left.\left.\mathbb{W}_{Y|\ulineX}\!\left( \boldy(t,i)\left| 
\!\!\begin{array}{c}\boldx_{1}(t,i)\\\boldx_{2}(t,i)\end{array}\!\!\right. 
\right)\right]\right\}
\label{Eqn:PE2FullAfterCodePMFWriteProdNotation}
\nonumber
\end{eqnarray}\fi
Interchanging the order of summations, and in particular moving the summation 
over $\ulinea_{1},\ulinea_{2}$ from being an outer sum to an inner sum, we have
\ifTITVersion\begin{eqnarray}
&&\!\!\!\!\!\!\!\!\!\!\sum_{
\substack{\boldy\in\mathcal{Y}^{m \times l}}}~\sum_{\substack{\boldu_{1} \in 
\mathcal{U}^{m \times l}\\\boldu_{2} \in \mathcal{U}^{m \times 
l}}}~\sum_{\pi_{1}}\cdots\sum_{\pi_{m}}\sum_{\substack{\boldv_{1}
\\\in \mathcal{V}_{1}^{m \times 
l}}}~\sum_{\substack{\boldv_{2}\\\in \mathcal{V}_{2}^{m \times 
l}}}\sum_{\substack{\boldx_{1}\\\in 
\mathcal{X}_{1}^{m 
\times l}}}\sum_{\substack{\boldx_{2}\\\in 
\mathcal{X}_{2}^{m \times l}}}\nonumber\\
 %=============nextline
 &&\!\!\!\!\!\!\!\!\!\!\!\!\mathds{1}_{\left\{\!\!\! 
\begin{array}{c}\left[\!\!\!\begin{array}{c}\boldu_{j},\boldv_{j}
,\boldx_{j}\\:j \in 
[2],\\\boldy
\end{array}\!\!\!\right]^{\pi}\!\!\!\!(1:m,i)\\\notin 
T_{\alpha}^{m}(p_{\underline{\mathscr{U}\mathscr{V}\mathscr{X}}\mathscr{Y}})
\end{array}
\!\!\!\!\right\}}\frac{1}{\left(l!\right)^{m}}\!\!\sum_{\substack{\ulinea_{1},
\ulinea_ {2}}}\!\!\left\{ \prod_{t=1}^{m} \left[ P\left(  
\!\!\!\begin{array}{c}A_{jt}=a_{jt}\\:j 
\in [2]\end{array}\!\!\!\right)\right. \right.
\nonumber\\
 %=============nextline
&&\!\!\!\!\!\!\!\!\!\!\mathds{1}_{\left\{\substack{ 
\boldu_{j}(t,1:l)\\=u^{l}(a_{jt})\\:j \in [2]}\right\}}\!\prod_{i=1}^{l} 
\!  \mathbb{W}_{Y|\ulineX}\!\left(\! \boldy(t,i)\left| 
\!\!\begin{array}{c}\boldx_{1}(t,i)\\\boldx_{2}(t,i)\end{array}\!\!\!\!\right. 
\right)\!\! \left\{ \! \prod_{j=1}^{2} p_{V_{j}} ( \boldv_{j}(t,i)) 
\right. 
\nonumber\\
 %=============nextline
  \lefteqn{\left.\left.\left.p_{X_{j}|V_{j}U}\!\left( \boldx_{j}(t,i) 
\left|\!\!\!\begin{array}{c}\boldv_{j}(t,i)\\\boldu_{j}(t,i)\end{array}\!\!\!\!
\right.\right)\right\}\right]\right\}}
\label{Eqn:PE2FullAfterCodePMFIntChOrdSum}
\end{eqnarray}\fi
\ifPeerReviewVersion
\begin{eqnarray}
&&\!\!\!\!\!\!\!\!\!\!\sum_{
\substack{\boldy\in\mathcal{Y}^{m \times l}}}~\sum_{\substack{\boldu_{1} \in 
\mathcal{U}^{m \times l}\\\boldu_{2} \in \mathcal{U}^{m \times 
l}}}~\sum_{\pi_{1}}\cdots\sum_{\pi_{m}}\sum_{\substack{\boldv_{1}
\\\in \mathcal{V}_{1}^{m \times 
l}}}~\sum_{\substack{\boldv_{2}\\\in \mathcal{V}_{2}^{m \times 
l}}}\sum_{\substack{\boldx_{1}\\\in 
\mathcal{X}_{1}^{m 
\times l}}}\sum_{\substack{\boldx_{2}\\\in 
\mathcal{X}_{2}^{m \times l}}}
 %=============nextline
 \mathds{1}_{\left\{\!\!\! 
\begin{array}{c}\left[\!\!\!\begin{array}{c}\boldu_{j},\boldv_{j}
,\boldx_{j}\\:j \in 
[2],\\\boldy
\end{array}\!\!\!\right]^{\pi}\!\!\!\!(1:m,i)\\\notin 
T_{\alpha}^{m}(p_{\underline{\mathscr{U}\mathscr{V}\mathscr{X}}\mathscr{Y}})
\end{array}
\!\!\!\!\right\}}\frac{1}{\left(l!\right)^{m}}\!\!\sum_{\substack{\ulinea_{1},
\ulinea_ {2}}}\!\!\left\{ \prod_{t=1}^{m} \left[ P\left(  
\!\!\!\begin{array}{c}A_{jt}=a_{jt}\\:j 
\in [2]\end{array}\!\!\!\right)\right. \right.
\nonumber\\
 %=============nextline
&&\mathds{1}_{\left\{\substack{ 
\boldu_{j}(t,1:l)\\=u^{l}(a_{jt})\\:j \in [2]}\right\}}\!\prod_{i=1}^{l} 
\!  \mathbb{W}_{Y|\ulineX}\!\left(\! \boldy(t,i)\left| 
\!\!\begin{array}{c}\boldx_{1}(t,i)\\\boldx_{2}(t,i)\end{array}\!\!\!\!\right. 
\right)\!\! \left\{ \! \prod_{j=1}^{2} p_{V_{j}} ( \boldv_{j}(t,i)) 
\right.
 %=============nextline
  \left.\left.\left.p_{X_{j}|V_{j}U}\!\left( \boldx_{j}(t,i) 
\left|\!\!\!\begin{array}{c}\boldv_{j}(t,i)\\\boldu_{j}(t,i)\end{array}\!\!\!\!
\right.\right)\right\}\right]\right\}
\label{Eqn:PE2FullAfterCodePMFIntChOrdSum}
\end{eqnarray}\fi
Note that
\ifTITVersion\begin{eqnarray}
\lefteqn{\sum_{\ulinea_{1},\ulinea_{2}}  \prod_{t=1}^{m} \left[ P\left(  
\!\!\!\begin{array}{c}A_{jt}=a_{jt}\\:j 
\in [2]\end{array}\!\!\!\right)\mathds{1}_{\left\{\!\!\! \begin{array}{c}
\boldu_{j}(t,1:l)=u^{l}(a_{jt})\\:j \in [2] \end{array}\!\!\!  \right\}} 
\right]}\nonumber\\
\!\!\!\!&\!\!\!\!\!\!\!\!=&\!\!\!\!\prod_{t=1}^{m}\left\{ 
\sum_{a_{1t},a_{2t}}\!\!\! P\left(  
\!\!\!\begin{array}{c}A_{1t}=a_{1t}\\A_{2t}=a_{2t}\end{array}
\!\!\!\right)\mathds{1}_{\left\{\!\!\! \begin{array}{c}
\boldu_{j}(t,1:l)=u^{l}(a_{jt})\\:j \in [2] \end{array}\!\!\!  
\right\}}\right\}\nonumber\\
\!\!\!\!&\!\!\!\!\!\!\!\!=&\!\!\!\!\prod_{t=1}^{m}\left\{ 
\sum_{a_{1},a_{2}}\!\!\! P\left(  
\!\!\!\begin{array}{c}A_{1}=a_{1}\\A_{2}=a_{2}\end{array}
\!\!\!\right)\mathds{1}_{\left\{\!\!\! \begin{array}{c}
\boldu_{j}(t,1:l)=u^{l}(a_{j})\\:j \in [2] \end{array}\!\!\!  
\right\}}\right\}
\label{Eqn:PE2FullDummyVariableRenaming}
\end{eqnarray}\fi
\ifPeerReviewVersion\begin{eqnarray}
\sum_{\ulinea_{1},\ulinea_{2}}  \prod_{t=1}^{m} \left[ P\left(  
\!\!\!\begin{array}{c}A_{jt}=a_{jt}\\:j 
\in [2]\end{array}\!\!\!\right)\mathds{1}_{\left\{\!\!\! \begin{array}{c}
\boldu_{j}(t,1:l)=u^{l}(a_{jt})\\:j \in [2] \end{array}\!\!\!  \right\}} 
\right]
&=&\prod_{t=1}^{m}\left\{ 
\sum_{a_{1t},a_{2t}}\!\!\! P\left(  
\!\!\!\begin{array}{c}A_{1t}=a_{1t}\\A_{2t}=a_{2t}\end{array}
\!\!\!\right)\mathds{1}_{\left\{\!\!\! \begin{array}{c}
\boldu_{j}(t,1:l)=u^{l}(a_{jt})\\:j \in [2] \end{array}\!\!\!  
\right\}}\right\}\nonumber\\
&=&\prod_{t=1}^{m}\left\{ 
\sum_{a_{1},a_{2}}\!\!\! P\left(  
\!\!\!\begin{array}{c}A_{1}=a_{1}\\A_{2}=a_{2}\end{array}
\!\!\!\right)\mathds{1}_{\left\{\!\!\! \begin{array}{c}
\boldu_{j}(t,1:l)=u^{l}(a_{j})\\:j \in [2] \end{array}\!\!\!  
\right\}}\right\}
\label{Eqn:PE2FullDummyVariableRenaming}
\end{eqnarray}\fi
where (\ref{Eqn:PE2FullDummyVariableRenaming}) follows from (i) the pairs 
$(A_{1t},A_{2t}): t \in [m]$ being independent and identically distributed and 
(ii) renaming the dummy variable $a_{jt}$ and $a_{j}$ for $j\in [2], t \in 
[m]$. With this, (\ref{Eqn:PE2FullAfterCodePMFIntChOrdSum}) evaluates to
\ifTITVersion\begin{eqnarray}
 \sum_{
\substack{\boldy\in\mathcal{Y}^{m \times l}}}~\sum_{\substack{\boldu_{1} \in 
\mathcal{U}^{m \times l}\\\boldu_{2} \in \mathcal{U}^{m \times 
l}}}\sum_{\pi_{1}}\!\!\cdots\!\sum_{\pi_{m}}\sum_{\substack{\boldv_{1}
\in\\ \mathcal{V}_{1}^{m \times 
l}}}\sum_{\substack{\boldv_{2}\in\\ \mathcal{V}_{2}^{m \times 
l}}}\sum_{\substack{\boldx_{1}\in\\ 
\mathcal{X}_{1}^{m 
\times l}}}\sum_{\substack{\boldx_{2}\in\\ 
\mathcal{X}_{2}^{m \times l}}}\!\!\frac{1}{\left(l!\right)^{m}}\nonumber\\
 %=============nextline
 \!\!\!\mathds{1}_{\left\{\!\!\! 
\begin{array}{c}\left[\!\!\!\begin{array}{c}\boldu_{j},\boldv_{j}
,\boldx_{j}\\:j \in 
[2],\boldy
\end{array}\!\!\!\right]^{\pi}\!\!\!\!(1:m,i)\\\notin 
T_{\alpha}^{m}(p_{\underline{\mathscr{U}\mathscr{V}\mathscr{X}}\mathscr{Y}})
\end{array}
\!\!\!\!\right\}} \prod_{t=1}^{m}   \left[ \sum_{\substack{a_{1},a_{2}}} 
   \! \left\{ P\!\left(  
\!\!\!\begin{array}{c}A_{j}=a_{j}\\:j \in 
[2]\end{array}\!\!\right)\right.\right.\nonumber\\
\left. \mathds{1}_{\left\{ \!\!\!\begin{array}{c} 
\boldu_{j}(t,1:l)=u^{l}(a_{j})\\:j \in [2] \end{array}\!\! \right\}} 
\right\}\prod_{i=1}^{l}\left[\left\{ \prod_{j=1}^{2}p_{V_{j}}\left( 
\boldv_{j}(t,i)\right)\right.\right.\nonumber\\
\left.\left. p_{X_{j}|V_{j}U}\left( \boldx_{j}(t,i) 
\left|\!\!\!\begin{array}{c} \boldv_{j}(t,i)\\\boldu_{j}(t,i)\end{array} 
\!\!\!\! \right.\right)\!\!\right\}\left.\!\!\mathbb{W}_{Y|\ulineX}\left( 
\boldy(t,i)\left| 
\!\!\begin{array}{c}\boldx_{1}(t,i)\\\boldx_{2}(t,i)\end{array}\!\!\!\!\right. 
\right)\!\right]\right]\nonumber
\label{Eqn:PE2FullSubstituteDecPMF}
\end{eqnarray}\fi
\ifPeerReviewVersion\begin{eqnarray}
 \sum_{
\substack{\boldy\in\mathcal{Y}^{m \times l}}}~\sum_{\substack{\boldu_{1} \in 
\mathcal{U}^{m \times l}\\\boldu_{2} \in \mathcal{U}^{m \times 
l}}}\sum_{\pi_{1}}\!\!\cdots\!\sum_{\pi_{m}}\sum_{\substack{\boldv_{1}
\in\\ \mathcal{V}_{1}^{m \times 
l}}}\sum_{\substack{\boldv_{2}\in\\ \mathcal{V}_{2}^{m \times 
l}}}\sum_{\substack{\boldx_{1}\in\\ 
\mathcal{X}_{1}^{m 
\times l}}}\sum_{\substack{\boldx_{2}\in\\ 
\mathcal{X}_{2}^{m \times l}}}\!\!\frac{1}{\left(l!\right)^{m}}
 %=============nextline
 \!\!\!\mathds{1}_{\left\{\!\!\! 
\begin{array}{c}\left[\!\!\!\begin{array}{c}\boldu_{j},\boldv_{j}
,\boldx_{j}\\:j \in 
[2],\boldy
\end{array}\!\!\!\right]^{\pi}\!\!\!\!(1:m,i)\\\notin 
T_{\alpha}^{m}(p_{\underline{\mathscr{U}\mathscr{V}\mathscr{X}}\mathscr{Y}})
\end{array}
\!\!\!\!\right\}} \prod_{t=1}^{m}   \left[ \sum_{\substack{a_{1},a_{2}}} 
   \! \left\{ P\!\left(  
\!\!\!\begin{array}{c}A_{j}=a_{j}\\:j \in 
[2]\end{array}\!\!\right)\right.\right.\nonumber\\
\left. \mathds{1}_{\left\{ \!\!\!\begin{array}{c} 
\boldu_{j}(t,1:l)=u^{l}(a_{j})\\:j \in [2] \end{array}\!\! \right\}} 
\right\}\prod_{i=1}^{l}\left[\left\{ \prod_{j=1}^{2}p_{V_{j}}\left( 
\boldv_{j}(t,i)\right)\right.\right.
\left.\left. p_{X_{j}|V_{j}U}\left( \boldx_{j}(t,i) 
\left|\!\!\!\begin{array}{c} \boldv_{j}(t,i)\\\boldu_{j}(t,i)\end{array} 
\!\!\!\! \right.\right)\!\!\right\}\left.\!\!\mathbb{W}_{Y|\ulineX}\left( 
\boldy(t,i)\left| 
\!\!\begin{array}{c}\boldx_{1}(t,i)\\\boldx_{2}(t,i)\end{array}\!\!\!\!\right. 
\right)\!\right]\right]
\label{Eqn:PE2FullSubstituteDecPMF}
\end{eqnarray}\fi
We recognize the term inside the outer square parenthesis in 
(\ref{Eqn:PE2FullSubstituteDecPMF}) is indeed the pmf defined in 
(\ref{Eqn:Step1MACPMFForDecodingRule}) evaluated on the $t$-th row of the 
corresponding 
matrices. Using (\ref{Eqn:Step1MACPMFForDecodingRule}), 
(\ref{Eqn:PE2FullSubstituteDecPMF}) is equal to
\ifTITVersion\begin{eqnarray}
\sum_{\substack{\boldy\in\mathcal{Y}^{m \times 
l}}}~\sum_{\substack{\boldu_{1} \in \mathcal{U}^{m \times l}\\\boldu_{2} \in 
\mathcal{U}^{m \times 
l}}}~\sum_{\pi_{1}}\cdots\sum_{\pi_{m}}\sum_{\substack{\boldv_{1}\\\in 
\mathcal{V}_{1}^{m \times l}}}~\sum_{\substack{\boldv_{2}\\\in 
\mathcal{V}_{2}^{m \times l}}}\nonumber\\
 %=============nextline
 \sum_{\substack{\boldx_{1}\\\in \mathcal{X}_{1}^{m \times l}}} 
\sum_{\substack{\boldx_{2}\\\in \mathcal{X}_{2}^{m \times 
l}}}\!\!\mathds{1}_{\left\{\!\!\! 
\begin{array}{c}\left[\!\!\!\begin{array}{c}\boldu_{j},\boldv_{j},\boldx_{j}
\\:j 
\in [2]\\\boldy\end{array}\!\!\!\right]^{\pi}\!\!\!\!(1:m,i)\\\notin 
T_{\alpha}^{m}(p_{\underline{\mathscr{U}\mathscr{V}\mathscr{X}}\mathscr{Y}})\end
{array}
\!\!\!\right\}}\left(\frac{1}{l!}\right)^{m}\nonumber\\
 %=============nextline
  \prod_{t=1}^{m}\left[ 
p_{\ulineU^{l}\ulineV^{l}\ulineX^{l}Y^{l}}\left(\begin{array}{c}\boldu_{1}(t,
1:l),\boldu_{2}(t,1:l),\\\boldv_{1}(t,1:l),\boldv_{2}(t,1:l),\\\boldx_{1}(t,1:l)
,\boldx_{2}(t,1:l),\\\boldy(t,1:l)\end{array}\right)\right].
\label{Eqn:PE2FullBeforeInterleavingResult}
\end{eqnarray}\fi
\ifPeerReviewVersion\begin{eqnarray}
\sum_{\substack{\boldy\in\mathcal{Y}^{m \times 
l}}}~\sum_{\substack{\boldu_{1} \in \mathcal{U}^{m \times l}\\\boldu_{2} \in 
\mathcal{U}^{m \times 
l}}}~\sum_{\pi_{1}}\cdots\sum_{\pi_{m}}\sum_{\substack{\boldv_{1}\\\in 
\mathcal{V}_{1}^{m \times l}}}~\sum_{\substack{\boldv_{2}\\\in 
\mathcal{V}_{2}^{m \times l}}}
 %=============nextline
 \sum_{\substack{\boldx_{1}\\\in \mathcal{X}_{1}^{m \times l}}} 
\sum_{\substack{\boldx_{2}\\\in \mathcal{X}_{2}^{m \times 
l}}}\!\!\mathds{1}_{\left\{\!\!\! 
\begin{array}{c}\left[\!\!\!\begin{array}{c}\boldu_{j},\boldv_{j},\boldx_{j}
\\:j 
\in [2]\\\boldy\end{array}\!\!\!\right]^{\pi}\!\!\!\!(1:m,i)\\\notin 
T_{\alpha}^{m}(p_{\underline{\mathscr{U}\mathscr{V}\mathscr{X}}\mathscr{Y}})\end
{array}
\!\!\!\right\}}\left(\frac{1}{l!}\right)^{m}\nonumber\\
 %=============nextline
  \prod_{t=1}^{m}\left[ 
p_{\ulineU^{l}\ulineV^{l}\ulineX^{l}Y^{l}}\left(\begin{array}{c}\boldu_{1}(t,
1:l),\boldu_{2}(t,1:l),\boldv_{1}(t,1:l),\boldv_{2}(t,1:l),\boldx_{1}(t,1:l)
,\boldx_{2}(t,1:l),\boldy(t,1:l)\end{array}\right)\right].
\label{Eqn:PE2FullBeforeInterleavingResult}
\end{eqnarray}\fi
The last step follows from recognizing that 
(\ref{Eqn:PE2FullBeforeInterleavingResult}) is indeed what we have been 
seeking. Appealing to Appendix \ref{AppSec:InterleavingConstruct}, we have 
(\ref{Eqn:PE2FullBeforeInterleavingResult}) equal to
\ifTITVersion\begin{eqnarray}
\sum_{\substack{u_{1}^{m} \in \mathcal{U}^{m}\\u_{2}^{m} \in \mathcal{U}^{m}}}
\sum_{\substack{v_{1}^{m} \in \mathcal{V}_{1}^{m}\\v_{2}^{m} \in 
\mathcal{V}_{2}^{m}}}
\sum_{\substack{x_{1}^{m} \in \mathcal{X}_{1}^{m}\\x_{2}^{m} \in 
\mathcal{X}_{2}^{m}}}
\sum_{y^{m} \in \mathcal{Y}^{m}} \!\!\!
\mathds{1}_{\left\{ 
\!\!\!\!\begin{array}{c}(\ulineu^{m},\ulinev^{m},\ulinex^{m},y^{m}) \\ \notin 
T_{\beta}^{m}(p_{\underline{\hat{U}}\hat{\ulineV}\hat{\ulineX}\hat{Y}}) 
\end{array}\!\!\!\!\right\} }\times\nonumber\\
\times 
\prod_{t=1}^{m}p_{\underline{\hatU\hatV\hatX Y}}( 
u_{1t},u_{2t},v_{1t},v_{2t},x_{1 t},x_{2t},y_{t} ) \nonumber
\nonumber
\end{eqnarray}\fi
\ifPeerReviewVersion\begin{eqnarray}
\sum_{\substack{u_{1}^{m} \in \mathcal{U}^{m}\\u_{2}^{m} \in \mathcal{U}^{m}}}
\sum_{\substack{v_{1}^{m} \in \mathcal{V}_{1}^{m}\\v_{2}^{m} \in 
\mathcal{V}_{2}^{m}}}
\sum_{\substack{x_{1}^{m} \in \mathcal{X}_{1}^{m}\\x_{2}^{m} \in 
\mathcal{X}_{2}^{m}}}
\sum_{y^{m} \in \mathcal{Y}^{m}} \!\!\!
\mathds{1}_{\left\{ 
\!\!\!\!\begin{array}{c}(\ulineu^{m},\ulinev^{m},\ulinex^{m},y^{m}) \\ \notin 
T_{\beta}^{m}(p_{\underline{\hat{U}}\hat{\ulineV}\hat{\ulineX}\hat{Y}}) 
\end{array}\!\!\!\!\right\} }
\prod_{t=1}^{m}p_{\underline{\hatU\hatV\hatX Y}}( 
u_{1t},u_{2t},v_{1t},v_{2t},x_{1 t},x_{2t},y_{t} ) \nonumber
\nonumber
\end{eqnarray}\fi
which falls to $0$ exponentially in $m$.
\section{The Interleaving Construct}
\label{AppSec:InterleavingConstruct}
Let $p_{A^{l}B^{l}}$ be a pmf on $\mathcal{A}^{l}\times \mathcal{B}^{l}$. We 
will prove
\ifTITVersion\begin{eqnarray}
\sum_{\bolda \in \mathcal{A}^{m \times l}}\sum_{\boldb \in \mathcal{B}^{ m 
\times 
l}}\prod_{t=1}^{m}p_{A^{l}B^{l}}\left(\!\!\!\begin{array}{c}\bolda(t,
1:l)\\\boldb(t , 1:l)\end{array}\!\!\!\right)\frac { 1 } { l^ { m } }
\sum_{\pi_{1}(i)=1}^{l}\!\!\!\cdots\!\!\! \sum_{\pi_{m}(i)=1}^{l}
\nonumber\\
%===================newline
\mathds{1}_{\left\{ \left( \begin{array}{c}\bolda(1,\pi_{1}(i))\cdots 
\bolda(m,\pi_{m}(i))\\\boldb(1,\pi_{1}(i))\cdots 
\boldb(m,\pi_{m}(i))\end{array} 
\right) \notin T_{\delta}^{m}(p_{\hat{A}\hat{B}})\right\}}\nonumber\\
%===================newline
=\sum_{a^{m} \in \mathcal{A}^{m}}\sum_{b^{m} \in 
\mathcal{B}^{m}}\mathds{1}_{\left\{(a^{m},b^{m}) \notin 
T_{\delta}^{m}(p_{\hat{A}\hat{B}})  
\right\}}\prod_{t=1}^{m}p_{\hat{A}\hat{B}}(a_{t},b_{t}).
\end{eqnarray}\fi
\ifPeerReviewVersion\begin{eqnarray}
\sum_{\substack{\bolda \in \\\mathcal{A}^{m \times l}}} 
\sum_{\substack{\boldb \in \\\mathcal{B}^{ m \times l}}}
\prod_{t=1}^{m}p_{A^{l}B^{l}}\left(\!\!\!\begin{array}{c}\bolda(t,
1:l)\\\boldb(t , 1:l)\end{array}\!\!\!\right)\frac { 1 } { l^ { m } }
\sum_{\pi_{1}(i)=1}^{l}\!\!\!\cdots\!\!\! \sum_{\pi_{m}(i)=1}^{l}
%===================newline
\!\!\mathds{1}_{\left\{  \!\!\!\!\begin{array}{c}\left[\bolda 
\boldb \right]^{\pi}(1:m,i)\\
\notin T_{\delta}^{m}(p_{\hat{A}\hat{B}})\end{array}\!\!\!\!\right\}}
%===================newline
=\sum_{\substack{a^{m} \in \\\mathcal{A}^{m}}}~\sum_{\substack{b^{m} 
\in \\\mathcal{B}^{ m}}}\mathds{1}_{\left\{\!\!\!\begin{array}{c}(a^{m},b^{m}) 
\notin \\
T_{\delta}^{m}(p_{\hat{A}\hat{B}})  
\end{array}\!\!\!\right\}}\!\!\prod_{t=1}^{m}p_{\hat{A}\hat{B}}(a_{t},b_{t}).
\end{eqnarray}\fi
where
\begin{eqnarray}
 \label{Eqn:PMFOfRandomCo-ordinate}
 p_{\hatA\hatB}(u,v) = \frac{1}{l}\sum_{i=1}^{l}p_{A_{i}B_{i}}(u,v)
 \nonumber
\end{eqnarray}
$p_{A_{i}B_{i}}$ is the pmf of the $i-$th component of $A^{l},B^{l}$.
Observe that
\ifTITVersion\begin{eqnarray}
\sum_{\bolda \in \mathcal{A}^{m \times l}}\sum_{\boldb \in \mathcal{B}^{ m 
\times 
l}}\prod_{t=1}^{m}p_{A^{l}B^{l}}\left(\!\!\!\begin{array}{c}\bolda(t,
1:l)\\\boldb(t , 1:l)\end{array}\!\!\!\right)\frac{1}{l^{m}}
\sum_{\pi_{1}(i)=1}^{l}\!\!\!\cdots\!\!\! \sum_{\pi_{m}(i)=1}^{l}
\nonumber\\
%===================newline
\mathds{1}_{\left\{ \left( \begin{array}{c}a(1,\pi_{1}(i))\cdots 
a(m,\pi_{m}(i))\\b(1,\pi_{1}(i))\cdots b(m,\pi_{m}(i))\end{array} \right) 
\notin 
T_{\delta}^{m}(p_{\hat{A}\hat{B}})\right\}}\nonumber\\
%===================NEWEQUATION
%===================newline
=\sum_{\bolda \in \mathcal{A}^{m \times l}}\sum_{\boldb \in \mathcal{B}^{ m 
\times 
l}}\prod_{t=1}^{m}p_{A^{l}B^{l}}\left(\!\!\!\begin{array}{c}\bolda(t,
1:l)\\\boldb(t , 1:l)\end{array}\!\!\!\right)\frac{1}{l^{m}}
\sum_{\pi_{1}(i)=1}^{l}\!\!\!\cdots\!\!\! \sum_{\pi_{m}(i)=1}^{l}
\nonumber\\
%===================newline
\sum_{x^{m} \in \mathcal{A}^{m}}\sum_{y^{m} \in 
\mathcal{B}^{m}}\mathds{1}_{\left\{ \left( 
\begin{array}{c}a(1,\pi_{1}(i))\cdots 
a(m,\pi_{m}(i))\\b(1,\pi_{1}(i))\cdots b(m,\pi_{m}(i))\end{array} \right) 
\notin 
T_{\delta}^{m}(p_{\hat{A}\hat{B}})\right\}}\nonumber\\
%===================newline
\mathds{1}_{\left\{ \left( \begin{array}{c}a(1,\pi_{1}(i))\cdots 
a(m,\pi_{m}(i))\\b(1,\pi_{1}(i))\cdots b(m,\pi_{m}(i))\end{array} \right) = 
\left( \begin{array}{c}x^{m}\\y^{m}\end{array} \right) \right\}}\nonumber\\
%===================NEWEQUATION
%===================newline
=\sum_{\bolda \in \mathcal{A}^{m \times l}}\sum_{\boldb \in \mathcal{B}^{ m 
\times 
l}}\prod_{t=1}^{m}p_{A^{l}B^{l}}\left(\!\!\!\begin{array}{c}\bolda(t,
1:l)\\\boldb(t , 1:l)\end{array}\!\!\!\right)\frac{1}{l^{m}}
\sum_{\pi_{1}(i)=1}^{l}\!\!\!\cdots\!\!\! \sum_{\pi_{m}(i)=1}^{l}
\nonumber\\
%===================newline
\sum_{x^{m} \in \mathcal{A}^{m}}\sum_{y^{m} \in 
\mathcal{B}^{m}}\mathds{1}_{\left\{ \left( x^{m},y^{m} \right) \notin 
T_{\delta}^{m}(p_{\hat{A}\hat{B}})\right\}}\nonumber\\
%===================newline
\mathds{1}_{\left\{ \left( \begin{array}{c}a(1,\pi_{1}(i))\cdots 
a(m,\pi_{m}(i))\\b(1,\pi_{1}(i))\cdots b(m,\pi_{m}(i))\end{array} \right) = 
\left( \begin{array}{c}x^{m}\\y^{m}\end{array} \right) \right\}}\nonumber\\
%===================NEWEQUATION
%===================newline
=\sum_{x^{m} \in \mathcal{A}^{m}}\sum_{y^{m} \in 
\mathcal{B}^{m}}\sum_{\pi_{1}(i)=1}^{l}\!\!\!\cdots\!\!\! 
\sum_{\pi_{m}(i)=1}^{l}\sum_{\bolda \in \mathcal{A}^{m \times l}}\sum_{\boldb 
\in \mathcal{B}^{ m \times l}}\frac{1}{l^{m}}\nonumber\\
%===================newline
\mathds{1}_{\left\{ \left( x^{m},y^{m} \right) \notin 
T_{\delta}^{m}(p_{\hat{A}\hat{B}})\right\}}\prod_{t=1}^{m}p_{A^{l}B^{l}}
(\bolda(t,1:l),\boldb(t,1:l))
\nonumber\\
%===================newline
\mathds{1}_{\left\{ \left( \begin{array}{c}a(1,\pi_{1}(i))\cdots 
a(m,\pi_{m}(i))\\b(1,\pi_{1}(i))\cdots b(m,\pi_{m}(i))\end{array} \right) = 
\left( \begin{array}{c}x^{m}\\y^{m}\end{array} \right) \right\}}\nonumber
\end{eqnarray}\fi
\ifPeerReviewVersion\begin{eqnarray}
\lefteqn{\sum_{\substack{\bolda \in \\\mathcal{A}^{m \times l}}} 
\sum_{\substack{\boldb \in \\\mathcal{B}^{ m \times 
l}}}\prod_{t=1}^{m}p_{A^{l}B^{l}}\left(\!\!\!\begin{array}{c}\bolda(t,
1:l)\\\boldb(t , 1:l)\end{array}\!\!\!\right)\frac{1}{l^{m}}
\sum_{\pi_{1}(i)=1}^{l}\!\!\!\cdots\!\!\! \sum_{\pi_{m}(i)=1}^{l}
%===================newline
\mathds{1}_{\left\{  \!\!\!\!\begin{array}{c}\left[\bolda 
\boldb \right]^{\pi}(1:m,i)\\
\notin 
T_{\delta}^{m}(p_{\hat{A}\hat{B}})\end{array}\!\!\!\!\right\}}}\nonumber\\
%===================NEWEQUATION
%===================newline
&=&\sum_{\substack{\bolda \in \\\mathcal{A}^{m \times l}}} 
\sum_{\substack{\boldb \in \\\mathcal{B}^{ m \times 
l}}}\prod_{t=1}^{m}p_{A^{l}B^{l}}\left(\!\!\!\begin{array}{c}\bolda(t,
1:l)\\\boldb(t , 1:l)\end{array}\!\!\!\right)\frac{1}{l^{m}}
\sum_{\pi_{1}(i)=1}^{l}\!\!\!\cdots\!\!\! \sum_{\pi_{m}(i)=1}^{l}
%===================newline
\sum_{x^{m} \in \mathcal{A}^{m}}\sum_{y^{m} \in 
\mathcal{B}^{m}}\mathds{1}_{\left\{  \!\!\!\!\begin{array}{c}\left[\bolda 
\boldb \right]^{\pi}(1:m,i)\\
=(x^{m},y^{m})\end{array}\!\!\!\!\right\}}
%===================newline
\mathds{1}_{\left\{  \!\!\!\!\begin{array}{c}\left[\bolda 
\boldb \right]^{\pi}(1:m,i)\\
\notin T_{\delta}^{m}(p_{\hat{A}\hat{B}})\end{array}\!\!\!\!\right\}}\nonumber\\
%===================NEWEQUATION
%===================newline
&=&\sum_{\substack{\bolda \in \\\mathcal{A}^{m \times l}}} 
\sum_{\substack{\boldb \in \\\mathcal{B}^{ m \times 
l}}}\prod_{t=1}^{m}p_{A^{l}B^{l}}\left(\!\!\!\begin{array}{c}\bolda(t,
1:l)\\\boldb(t , 1:l)\end{array}\!\!\!\right)\frac{1}{l^{m}}
\sum_{\pi_{1}(i)=1}^{l}\!\!\!\cdots\!\!\! \sum_{\pi_{m}(i)=1}^{l}
%===================newline
\sum_{x^{m} \in \mathcal{A}^{m}}\sum_{y^{m} \in 
\mathcal{B}^{m}}\mathds{1}_{\left\{  \!\!\!\!\begin{array}{c}(x^{m},y^{m})\\
\notin T_{\delta}^{m}(p_{\hat{A}\hat{B}})\end{array}\!\!\!\!\right\}}
%===================newline
\mathds{1}_{\left\{  \!\!\!\!\begin{array}{c}\left[\bolda 
\boldb \right]^{\pi}(1:m,i)\\
=(x^{m},y^{m})\end{array}\!\!\!\!\right\}}\nonumber\\
%===================NEWEQUATION
%===================newline
&=&\sum_{x^{m} \in \mathcal{A}^{m}}\sum_{y^{m} \in 
\mathcal{B}^{m}}\sum_{\pi_{1}(i)=1}^{l}\!\!\!\cdots\!\!\! 
\sum_{\pi_{m}(i)=1}^{l}\sum_{\substack{\bolda \in \\\mathcal{A}^{m \times l}}} 
\sum_{\substack{\boldb \in \\\mathcal{B}^{ m \times l}}}\frac{1}{l^{m}}
%===================newline
\mathds{1}_{\left\{  \!\!\!\!\begin{array}{c}(x^{m},y^{m})\\
\notin 
T_{\delta}^{m}(p_{\hat{A}\hat{B}})\end{array}\!\!\!\!\right\}}\prod_{t=1}^{m}p_{
A^{l}B^{l}} \left(\!\!\!\begin{array}{c}\bolda(t,
1:l)\\\boldb(t , 1:l)\end{array}\!\!\!\right)
%===================newline
\mathds{1}_{\left\{  \!\!\!\!\begin{array}{c}\left[\bolda 
\boldb \right]^{\pi}(1:m,i)\\
=(x^{m},y^{m})\end{array}\!\!\!\!\right\}}\nonumber
\end{eqnarray}\fi
\ifPeerReviewVersion\begin{eqnarray}
%===================NEWEQUATION
%===================newline
\lefteqn{=\sum_{x^{m} \in \mathcal{A}^{m}}\sum_{y^{m} \in 
\mathcal{B}^{m}}\sum_{\pi_{1}(i)=1}^{l}\!\!\!\cdots\!\!\! 
\sum_{\pi_{m}(i)=1}^{l}\mathds{1}_{\left\{  
\!\!\!\!\begin{array}{c}(x^{m},y^{m})\\
\notin 
T_{\delta}^{m}(p_{\hat{A}\hat{B}})\end{array}\!\!\!\!\right\}}
%===================newline
\sum_{\substack{\bolda \in \\\mathcal{A}^{m \times l}}} 
\sum_{\substack{\boldb \in \\\mathcal{B}^{ m \times 
l}}}\frac{1}{l^{m}}\prod_{t=1}^{m}
p_{A^{l}B^{l}} \left(\!\!\!\begin{array}{c}\bolda(t,
1:l)\\\boldb(t , 1:l)\end{array}\!\!\!\right)
%===================newline
\mathds{1}_{\left\{  \!\!\!\!\begin{array}{c}\left[\bolda 
\boldb \right]^{\pi}(1:m,i)\\
=(x^{m},y^{m})\end{array}\!\!\!\!\right\}}}\nonumber\\
%===================NEWEQUATION
%===================newline
&=&\sum_{x^{m} \in \mathcal{A}^{m}}\sum_{y^{m} \in 
\mathcal{B}^{m}}\sum_{\pi_{1}(i)=1}^{l}\!\!\!\cdots\!\!\! 
\sum_{\pi_{m}(i)=1}^{l}\mathds{1}_{\left\{  
\!\!\!\!\begin{array}{c}(x^{m},y^{m})\\
\notin 
T_{\delta}^{m}(p_{\hat{A}\hat{B}})\end{array}\!\!\!\!\right\}}
%===================newline
\frac{1}{l^{m}}\prod_{t=1}^{m}p_{A_{\pi_{t}(i)}B_{\pi_{t}(i)}}(x_{t},y_{t}
)\nonumber\\
%===================NEWEQUATION
%===================newline
&=&\sum_{x^{m} \in \mathcal{A}^{m}}\sum_{y^{m} \in 
\mathcal{B}^{m}}\prod_{t=1}^{m}\left\{ 
\frac{1}{l}\sum_{\alpha=1}^{l}p_{A_{\alpha}B_{\alpha}}(x_{t},y_{t}) 
\right\}
%===================newline
\mathds{1}_{\left\{  
\!\!\!\!\begin{array}{c}(x^{m},y^{m})\\
\notin 
T_{\delta}^{m}(p_{\hat{A}\hat{B}})\end{array}\!\!\!\!\right\}}
%===================NEWEQUATION
%===================newline
=\sum_{x^{m} \in \mathcal{A}^{m}}\sum_{y^{m} \in 
\mathcal{B}^{m}}\prod_{t=1}^{m}p_{\hat{A}\hat{B}}(x_{t},y_{t})
\mathds{1}_{\left\{  
\!\!\!\!\begin{array}{c}(x^{m},y^{m})\\
\notin 
T_{\delta}^{m}(p_{\hat{A}\hat{B}})\end{array}\!\!\!\!\right\}}\nonumber
\end{eqnarray}\fi

\ifTITVersion\begin{eqnarray}
%===================NEWEQUATION
%===================newline
=\sum_{x^{m} \in \mathcal{A}^{m}}\sum_{y^{m} \in 
\mathcal{B}^{m}}\sum_{\pi_{1}(i)=1}^{l}\!\!\!\cdots\!\!\! 
\sum_{\pi_{m}(i)=1}^{l}\mathds{1}_{\left\{  
\!\!\!\!\begin{array}{c}(x^{m},y^{m})\\
\notin 
T_{\delta}^{m}(p_{\hat{A}\hat{B}})\end{array}\!\!\!\!\right\}}\nonumber\\
%===================newline
\sum_{\substack{\bolda \in \\\mathcal{A}^{m \times l}}} 
\sum_{\substack{\boldb \in \\\mathcal{B}^{ m \times 
l}}}\frac{1}{l^{m}}\prod_{t=1}^{m}
p_{A^{l}B^{l}} \left(\!\!\!\begin{array}{c}\bolda(t,
1:l)\\\boldb(t , 1:l)\end{array}\!\!\!\right)
%===================newline
\mathds{1}_{\left\{  \!\!\!\!\begin{array}{c}\left[\bolda 
\boldb \right]^{\pi}(1:m,i)\\
=(x^{m},y^{m})\end{array}\!\!\!\!\right\}}\nonumber\\
%===================NEWEQUATION
%===================newline
=\sum_{\substack{x^{m} \in 
\mathcal{A}^{m} \\ y^{m} \in
\mathcal{B}^{m}}}\sum_{\pi_{1}(i)=1}^{l}\!\!\!\cdots\!\!\! 
\sum_{\pi_{m}(i)=1}^{l}\!\!\!
\mathds{1}_{\!\!\left\{  
\!\!\!\!\begin{array}{c}(x^{m},y^{m})\\
\notin 
T_{\delta}^{m}(p_{\hat{A}\hat{B}})\end{array}\!\!\!\!\right\}}\!
%===================newline
\frac{1}{l^{m}}\!\prod_{t=1}^{m}p_{A_{\alpha_{t}}B_{\alpha_{t}}}(x_{t},y_{t}
)\nonumber\\
%===================NEWEQUATION
%===================newline
=\sum_{\substack{x^{m} \in\\ \mathcal{A}^{m}}}\sum_{\substack{y^{m} \in\\ 
\mathcal{B}^{m}}}\prod_{t=1}^{m}\left\{ 
\frac{1}{l}\sum_{\alpha=1}^{l}p_{A_{\alpha}B_{\alpha}}(x_{t},y_{t}) 
\!\!\!
\right\}
%===================newline
\mathds{1}_{\left\{  
\!\!\!\!\begin{array}{c}(x^{m},y^{m})\\
\notin 
T_{\delta}^{m}(p_{\hat{A}\hat{B}})\end{array}\!\!\!\!\right\}}\nonumber\\
%===================NEWEQUATION
%===================newline
=\sum_{x^{m} \in \mathcal{A}^{m}}\sum_{y^{m} \in 
\mathcal{B}^{m}}\prod_{t=1}^{m}p_{\hat{A}\hat{B}}(x_{t},y_{t})
\mathds{1}_{\left\{  
\!\!\!\!\begin{array}{c}(x^{m},y^{m})\\
\notin 
T_{\delta}^{m}(p_{\hat{A}\hat{B}})\end{array}\!\!\!\!\right\}}\nonumber
\end{eqnarray}\fi
which is what we sought out to prove.
\bibliographystyle{IEEEtran}
{
\bibliography{FixedBLCoding}
\end{document}